\documentclass[11pt,a4paper]{article}
\usepackage{jheppub}
\usepackage{lineno}
\usepackage{float}
\usepackage{macros}
\usepackage{stmaryrd}
\usepackage{theorems}
\usepackage{amsthm}
\usepackage{tcolorbox,subfigure}
\usepackage{lscape}
\usepackage[T1]{fontenc}
\usepackage{capt-of}

\usepackage{tikz}
\usepackage[dvipsnames]{xcolor}
\usepackage{color}
\usepackage{colortbl,graphicx,rotate,fancyhdr}
\usepackage{tikz-3dplot}
\usepackage{multirow}

\usetikzlibrary{arrows}
\usetikzlibrary{3d}
\usetikzlibrary{mindmap,trees,shadows}
\usetikzlibrary{positioning,patterns,patterns.meta}
\usetikzlibrary{intersections} 
\usetikzlibrary{decorations.pathmorphing}
\usetikzlibrary{matrix,decorations.pathreplacing}
\usetikzlibrary{calc,angles,positioning,intersections,quotes,decorations.markings}
\usepackage{tkz-euclide}

\def\coordinatesHexagon{
\protect\coordinate (O) at (0,0);
\protect\coordinate (A) at (1.000, 0.000);
\protect\coordinate (B) at (0.500, 0.866);
\protect\coordinate (C) at (-0.500, 0.866);
\protect\coordinate (D) at (-1.000, 0.000);
\protect\coordinate (E) at (-0.500, -0.866);
\protect\coordinate (F) at (0.500, -0.866);}
\def\drawHexagon{\foreach \i/\j in {A/B,  B/C,  C/D,  D/E,  E/F, F/A}{\protect\draw[dashed,thick] (\i) -- (\j);} }

\def\coordinatesSquare{
\protect\coordinate (O) at (0,0);
\protect\coordinate (A) at (1,1);
\protect\coordinate (B) at (-1,1);
\protect\coordinate (C) at (-1,-1);
\protect\coordinate (D) at (1,-1);}
\def\drawSquare{\foreach \i/\j in {A/B,  B/C,  C/D,  D/A}{\protect\draw[dashed,thick] (\i) -- (\j);} }

\def\coordinatesRectangle{
\protect\coordinate (O) at (0,0);
\protect\coordinate (A) at (1,0.707);
\protect\coordinate (B) at (-1,0.707);
\protect\coordinate (C) at (-1,-0.707);
\protect\coordinate (D) at (1,-0.707);}
\def\drawRectangle{\foreach \i/\j in {A/B,  B/C,  C/D,  D/A}{\protect\draw[dashed,thick] (\i) -- (\j);} }

\def\coordinatesTriangle{
\protect\coordinate (O) at (0,0);
\protect\coordinate (A) at (0,1);
\protect\coordinate (B) at (0.866025,-0.5);
\protect\coordinate (C) at (-0.866025,-0.5);}

\def\drawTriangle{
\foreach \i/\j in {A/B,  B/C,  C/A}{\protect\draw[dashed,thick] (\i) -- (\j);};}

\def\coordinatesOctagon{
\protect\coordinate (O) at (0,0);
\protect\coordinate (A) at (1,-0.5);
\protect\coordinate (B) at (1,0.5);
\protect\coordinate (C) at (0.5,1);
\protect\coordinate (D) at (-0.5,1);
\protect\coordinate (E) at (-1,0.5);
\protect\coordinate (F) at (-1,-0.5);
\protect\coordinate (G) at (-0.5,-1);
\protect\coordinate (H) at (0.5,-1);}
\def\drawOctagon{\foreach \i/\j in {A/B,  B/C,  C/D,  D/E, E/F, F/G, G/H, H/A}{\protect\draw[dashed,thick] (\i) -- (\j);} }

\def\coordinatesIsoTriangle{
\protect\coordinate (A) at (-0.707107,-0.801784);
\protect\coordinate (B) at (0.707107,-0.801784);
\protect\coordinate (C) at (0,1.06904);
\protect\coordinate (O) at (0,0);}
\def\drawIsoTriangle{\foreach \i/\j in {A/B,  B/C,  C/A}{\protect\draw[dashed,opacity=0.7] (\i) -- (\j);} }

\def\coordinatesTriangularPrismB{

\protect\coordinate (O) at (0,0,0);

\protect\coordinate (A) at (0.788675, 0.211325, 0.707107);
\protect\coordinate (B) at (-0.57735, 0.57735, 0.707107);
\protect\coordinate (C) at (-0.211325, -0.788675, 0.707107);
\protect\coordinate (A') at (0.788675, 0.211325, -0.707107);
\protect\coordinate (B') at (-0.57735, 0.57735, -0.707107);
\protect\coordinate (C') at (-0.211325, -0.788675, -0.707107);
\protect\coordinate (X) at (0.788675, 0.211325, 0.707107);
\protect\coordinate (Y) at (0.105662, 0.394338, 0.707107);
\protect\coordinate (Z) at (0.788675, 0.211325, 0.);
\protect\coordinate (W) at (0. , 0., 0.707107);
\protect\coordinate (Q) at (0.105662, 0.394338, 0.);

\protect\coordinate (Xa) at (2., 2., 1.41421);
\protect\coordinate (Ya) at (0.633975, 2.36603, 1.41421);
\protect\coordinate (Xb) at (2.36603, 0.633975, 1.41421);
\protect\coordinate (Za) at (2., 2., 0.);
\protect\coordinate (Wa) at (0., 0., 1.41421);
\protect\coordinate (Qa) at (0.633975, 2.36603, 0.);
\protect\coordinate (Zb) at (2.36603, 0.633975, 0.);

\protect\coordinate (XaN) at (0.894427, 0.894427, 0.632456);

}

\def\coordinatesCubeB{
\coordinate (O) at (0,0,0);
\coordinate (A) at (-.5,-.5,-.5);
\coordinate (B) at (.5,-.5,-.5);
\coordinate (C) at (.5,.5,-.5);
\coordinate (D) at (-.5,.5,-.5);
\coordinate (A') at (-.5,-.5,.5);
\coordinate (B') at (.5,-.5,.5);
\coordinate (C') at (.5,.5,.5);
\coordinate (D') at (-.5,.5,.5);
}
\def\drawCubeB{
\draw[dashed] (D) -- (A) -- (B);
\draw[] (B)-- (C) -- (D);
\draw[] (B) -- (B');
\draw[] (C) -- (C');
\draw[] (D) -- (D');
\draw[dashed] (A) -- (A');
\draw[] (A') -- (B');
\draw[] (A') -- (D');
\draw[] (B') -- (C');
\draw[] (C') -- (D');}

\tdplotsetmaincoords{16}{177}
\tdplotsetrotatedcoords{-2}{0}{-5}

\definecolor{ao(english)}{rgb}{0.0, 0.5, 0.0}
\definecolor{applegreen}{rgb}{0.55, 0.71, 0.0}
\definecolor{colGamma}{rgb}{0.89, 0.27, 0.97}

\def\colone{blue}
\def\coltwo{applegreen}
\def\coltwoALT{teal}
\def\colthree{cyan}
\def\colinf{red}
\def\colGamma{colGamma}

\SetSymbolFont{stmry}{bold}{U}{stmry}{m}{n}

\renewcommand{\b}[1]{\bar{#1}}
\newcommand{\cF}{\mathcal{F}}
\newcommand{\w}[1]{[\omega_#1]}
\newcommand{\ww}[2]{[\omega_#1]\cap[\omega_#2]}
\newcommand{\kVk}[1]{\left(\kappa\C{V}\kappa\right)_{#1}}
\newcommand{\ekV}{\left(e\kappa\C{V}\right)}
\newcommand{\eVe}{\left(e\C{V}e\right)}
\newcommand{\VkV}[1]{\left(\C{V}\kappa\C{V}\right)_{#1}}

\title{Moduli-Space Laplacians, Asymptotic Geometry, and the Emergent String Conjecture}

\author[a]{Christian Aoufia,}
\author[b]{Muldrow Etheredge,}
\author[a]{Bernardo Fraiman,}
\author[c]{Sanjay Raman,}
\author[d]{and Alexander~Stewart}

\emailAdd{christian.aoufia@estudiante.uam.es}
\emailAdd{muldrow.etheredge@mpp.mpg.de} 
\emailAdd{bernardo.fraiman@csic.es}
\emailAdd{sanjayraman@fas.harvard.edu}
\emailAdd{ajstewa1@uci.edu}

\affiliation[a]{Instituto de Física Teórica UAM/CSIC,
Cantoblanco, 28049 Madrid, Spain}

\affiliation[b]{Max-Planck-Institut f\"ur Physik (Werner-Heisenberg-Institut),\\ Bolzmannstr. 8, 85748, Garching, Germany }

\affiliation[c]{Jefferson Physical Laboratory, Harvard University,\\ 17 Oxford St, Cambridge, MA 02138, United States of America}

\affiliation[d]{Department of Physics and Astronomy, University of California, Irvine\\
CA 92697-4575, USA}

\abstract{
At asymptotic limits of moduli spaces of quantum gravity vacua, we argue that the masses of the lightest particle towers are eigenfunctions of the moduli-space Laplacian. The associated eigenvalues are quantized by the Emergent String Conjecture and determine the exponential decay rates of axionic directions in these limits. We also connect the quantization of Laplacian eigenvalues to the spectrum of instantons for the theory. We support our claims with diverse examples that preserve between 4 and 32 supercharges.}
\hypersetup{pageanchor=false}
\makeatletter
\gdef\@fpheader{\\}
\makeatother

\hypersetup{
    pdftitle={Moduli-Space Laplacians, Asymptotic Geometry, and the Emergent String Conjecture},
    pdfauthor={Christian Aoufia, Muldrow Etheredge, Bernardo Fraiman, Sanjay Raman, Alexander Stewart},
    pdfsubject={}}

\begin{document}
\preprint{{
\begin{flushright}\tt IFT-UAM/CSIC-26-91\end{flushright}
}}

\maketitle
\flushbottom

\section{Introduction and summary} \label{sec:intro}
Much of string phenomenology examines how the geometry of \emph{moduli spaces of vacua}---spaces parametrized by the vevs of scalar fields in Effective Field Theories (EFTs)---are related to the spectra of the EFTs. For example, the Swampland \cite{Vafa:2005ui}\footnote{For Swampland reviews, see \cite{Brennan:2017rbf,Palti:2019pca,vanBeest:2021lhn,Grana:2021zvf,Harlow:2022ich,Agmon:2022thq,VanRiet:2023pnx}.} Distance Conjecture (DC) \cite{Ooguri:2006in} proposes that, at sufficiently large distances\footnote{Here, the distance is measured using the metric $g_{ij}$ in the low-energy effective action, $\mathcal L=\frac 12R-\frac 12g_{ij}(\partial \phi^i)\cdot \partial \phi^j+\dots$, and throughout this paper we work in Planck units.} $\Delta$ along any asymptotic limit in the moduli space, there is an infinite tower of particles in the EFT whose masses exponentially decay with the distance $\Delta$. The Emergent String Conjecture (ESC) \cite{Lee:2019wij}\footnote{Much effort has been devoted to understanding the ESC in quantum gravity, both from the top-down in diverse string theory settings \cite{Lee:2018urn, Lee:2019wij, Lee:2019xtm, Marchesano:2019ifh, Baume:2019sry, Lee:2019jan, Klaewer:2020lfg, Xu:2020nlh, Lee:2021qkx,Lee:2021usk, Alvarez-Garcia:2021pxo, Rudelius:2023odg, Alvarez-Garcia:2023gdd, Blumenhagen:2023yws, Alvarez-Garcia:2023qqj, Chen:2024cvc, Basile:2022zee, Aoufia:2024awo, Hassfeld:2025uoy, Monnee:2025ynn, Monnee:2025msf, Aoufia:2026bau} and bottom-up relying on properties of supergravity probes, amplitudes or species \cite{Basile:2023blg, Cribiori:2023ffn, Basile:2024dqq,Herraez:2024kux, Bedroya:2024ubj, Kaufmann:2024gqo}.} strengthens the DC and argues that every infinite-distance limit corresponds to either a decompactification or the emergence of a tensionless perturbative string.

In this paper, we will explore consequences of the ESC for the geometry of the moduli space in a specific context by making the following assumption:\footnote{As will be explained further in Section \ref{sec:2}, one reason we assume that the decompactifications take this form is to avoid an effect called ``sliding'' identified in \cite{Etheredge:2023odp}. It would be interesting to relax this assumption.}
\begin{assumption*}[ESC without running decompactifications]
    All infinite distance limits we consider correspond either to tensionless weakly-coupled string limits or to decompactification limits over manifolds satisfying Einstein's equations in vacuum. 
\end{assumption*}
\noindent In this context, the ESC heavily constrains EFTs. For example, as explored in \cite{Etheredge:2024tok, Etheredge:2025ahf}, this version of the ESC can be used to classify the moduli-dependence of the tensions of branes and particles in asymptotically flat regions of moduli spaces. In this paper, we identify a new type of constraint from the moduli space---namely that the ESC controls the geometry of axionic fibers over these asymptotically flat regions, and connects this with the spectrum of the theory.

To make this more precise, consider a geodesic $\gamma$ in moduli space that asymptotes to an infinite-distance limit. Along a generic geodesic, the ESC tells us that there will be $N \geq 1$ light towers corresponding to either KK-modes or string oscillator modes with masses $m_1 , m_2, \cdots, m_N$ which decay exponentially at a rate faster than the species scale $\Lambda$. (We offer a precise characterization of these towers in Section \ref{ssec:geometry}.) The set of KK-mode and string oscillator towers obtained this way are called \emph{principal towers}, and the logarithms of these tower masses $\log m_i$ parametrize a geodesic cone of the moduli space which we call the \textit{principal plane} \cite{Etheredge:2024tok, Etheredge:2025ahf}. In this paper, we argue that the principal tower masses are eigenfunctions of the moduli-space Laplacian, and that their eigenvalues are quantized and are related to the axionic fibration structure over the principal plane.

Our main proposal is as follows. In asymptotic limits of moduli spaces, the masses of these principal towers are governed by the following condition:
\begin{prop*}[Hyperbolic Towers Quantization]
Let $m$ be the characteristic mass measured in Planck units of a principal tower that is either a KK-mode decompactifying from $d$ to $D$-dimensions, or a string oscillator mode. Then the moduli-space Laplacian acting on the logarithm of this mass is quantized:
\begin{equation}  c \equiv \nabla^2 \log m = \begin{cases} \frac{N}{D-d}, & \text{KK-mode decompactifying from $d$ to $D$ dimensions,} \\ \frac{N}{2},& \text{string oscillator,} \end{cases} \label{eq:hyp-towers-quantization-intro}
\end{equation}
for some nonnegative integer $N$. Throughout the text, we will call this quantized value the hyperbolic towers coefficient, or $c$-value for short.
\end{prop*}
\noindent 

We emphasize that these Laplacian $c$-values are simple rational numbers with small denominator. When studying the behavior of the masses of principal towers, it is convenient to define the scaling vector $\vec{\alpha}$ associated to a tower with mass $m$ as follows, 
\begin{equation}
    \vec{\alpha} = -\nabla \log m\ . \label{eq:alpha-vector-intro}
\end{equation}
The ESC implies that the lengths of $\vec\alpha$-vectors satisfy
\begin{equation} \abs{\vec\alpha} = \begin{cases} \sqrt{\frac{D-2}{(D-d)(d-2)}} ,& \text{KK-mode } d \to D, \\ \frac 1{\sqrt{d-2}}, & \text{string oscillator.}
\end{cases}
\end{equation}
This also implies that the masses of principal towers are themselves eigenfunctions of the moduli-space Laplacian, satisfying the eigenvalue equations\footnote{It was observed in \cite{Etheredge:2023zjk} that the lightest towers, as well as some brane tensions, are eigenfunctions of the moduli-space Laplacian, see also \cite{Ruehle:2024ufw}. Similar observations have been made for higher-derivative operators in \cite{Aoufia:2025ppe} and \cite{Aoufia:2024awo,Aoufia:2026bau} and for the instanton expansion of the prepotential in 4d $\C{N}=2$ theories in \cite{Alvarez-Garcia:2026uaq, Alvarez-Garcia:2026vwq}}
\begin{align}
    \nabla^2 m=\begin{cases}\left(\frac{N}{D-d}+\frac{D-2}{(D-d)(d-2)}\right)m\ ,&\text{KK-mode } d \to D\, ,\\
    \left(\frac N{2}+\frac 1{d-2}\right)m\ ,&\text{string oscillator .}
    \end{cases}\label{eq:laplaceeq}
\end{align}

(Eq. \eqref{eq:laplaceeq}, as eq. \eqref{eq:hyp-towers-quantization-intro}, relies on our assumption of the ESC without running decompactification to avoid issues related to sliding \cite{Etheredge:2023odp}.) Because the Laplacian involves derivatives of the metric, the eigenvalue equation \eqref{eq:laplaceeq} captures information about the axionic fibration structure of moduli spaces in asymptotic limits. As was noted in \cite{Etheredge:2023zjk}, the mass scales for principal towers can be viewed as exponentiations of distance functions $\phi$ by the ESC, i.e.
\begin{equation}
     m= m(\phi_0)e^{-\alpha \phi} ,
\end{equation}
for some positive constant $\alpha$ and reference value $m(\phi_0)$. Along a geodesic with distance parametrized by $\phi$, the transverse geometry of the moduli space can then be represented in the following form: 
\begin{align}
    ds^2=d\phi^2+g_{ij}(\phi,x^i)dx^idx^j\, .
\end{align}
As we will show, the hyperbolic towers quantization encodes the fact that the determinant of the metric $g_{ij}$ in the transverse coordinates $x^i$, and thus the transverse volume of the moduli space, decays with $\phi$ at a rate controlled by the hyperbolic towers coefficient. Namely,
\begin{align}
    \sqrt{\det g_{ij}(\phi,x^i)}\sim \exp\left(-\frac{c}{\alpha }\phi\right)\sim m^{\frac{c}{\alpha^2} }  \, . 
\end{align}
This directly shows that the transverse volume decays exponentially with $\phi$, at a quantized rate, and also scales with the principal tower mass, again at a quantized rate. 

As we explain in detail in Section \ref{sec:2}, the formula in eq. \ref{eq:hyp-towers-quantization-intro} is closely related to the origin of axion decay constants as functions of the principal plane. In decompactification limits, we can often relate axion decay constants to reductions of higher-dimensional gauge fields, and these have quantized radion dependence. In perturbative string limits, we argue that axion decay constants are exponential functions of dilatons, also with quantized exponential rates. It is these quantized exponential rates that we argue in Section \ref{sec:2} are responsible for the hyperbolic towers condition.

Under the assumption of the ESC without running decompactification and two further technical assumptions, we argue that asymptotically the moduli separate into three types: the principal plane moduli $\phi^i$, the axions $\theta_I$, and some other moduli $x^m$, and that the fibration structure over the principal plane is described with the following block-diagonal metric,
\begin{equation}\label{eq:fullmetric_intro} 
\d s^2 = \d s^2_{\C{P}} +  \sum_I f_I^2(\phi^i,x^m) \d \theta_I^2 +g_{mn}(x)\d x^m \d x^n\,, \qquad f_I=g_I^{-1}(x^i) e^{- \vec{\beta}^I \cdot \vec{\phi}},
\end{equation}
Here, $ds_{\mathcal P}^2$ is the metric for the principal plane directions, the $f_I$'s are the axion-decay constants, which are functions of the $x^m$ moduli times exponentials of the principal plane moduli $\phi^i$ (with exponential rates governed by a vector $\vec \beta^I$ on the principal plane), and $g_{mn}(x^m)$ controls the other moduli, and is independent of the axionic and principal plane moduli. As we will explain, the vectors $\vec \beta_I$ are also lattice-valued, satisfying taxonomy rules for instantons worked out in \cite{Etheredge:2025ahf}. Strictly speaking, $\d \theta_I$ are not literally exact one-forms and should be more properly regarded as left-invariant one-forms on a nilmanifold; we say therefore that the no-sliding assumption above implies a \emph{nilmanifold fibration} structure for the axions over the principal plane directions.

The hyperbolic towers quantization condition is closely related to the Laplacians of the logarithms of the principal towers and the decay constants for the axions $\theta_I$. More specifically, for a set of axions with decay constants governed by vectors $\vec \beta_I$, we find that the Laplacian eigenvalues can also be constrained by the following expression,
\begin{align} \label{eq:gamma-vector-laplacian-intro}
    \nabla^2\log m=  \vec \Gamma \cdot \vec\alpha \, , \qquad \vec \Gamma\equiv \sum_I \vec{\beta}_I  \, .
\end{align}
Here, $\vec{\alpha}$ is defined as in eq. \eqref{eq:alpha-vector-intro}, and the $\Gamma$-vector points in the direction in the principal plane where the total volume of the axion fiber decays fastest. (We will explicitly describe the sum over the $\vec{\beta}^I$ in Section \ref{ssec:instantons,laplacians,towers}.) As we will explain in Section \ref{sec:2}, the $\beta_I$-vectors satisfy (under our assumptions) the brane taxonomy rules of \cite{Etheredge:2025ahf} for instantons\footnote{For related work on quantum gravity bounds on instantons and axions, see \cite{Montero:2015ofa,Etheredge:2024amg,Etheredge:2025ahf, Maldacena:2026jqd,DiUbaldo:2026rly,Artime:2026kfq,Etheredge:2026rio,Reece:2026hmp}.}, and thus the $\Gamma$-vector inherits these rules. We will also explain that the $\Gamma$-vector is related to specific components of the trace of the Christoffel connection, and thus explain that taxonomy rules for instantons constrain various components of Christoffel connections of moduli spaces. 
Furthermore, the formula \eqref{eq:gamma-vector-laplacian-intro} implies that, knowledge of $\nabla^2 \log m$ for all of the principal towers can allow one to reconstruct $\vec{\Gamma}$.

The organization of this paper is as follows. In Section \ref{sec:2}, after outlining in detail our assumptions and claims, we provide a detailed derivation of the hyperbolic towers quantization condition from the ESC, as well as an explicit geometric structure for the asymptotic regime of a quantum gravity moduli space. We also study the relationship between the hyperbolic towers property, the axion fibration structure, and the spectrum of electrically-coupled instantons of the theory, referencing the taxonomic analysis of \cite{Etheredge:2025ahf}. 

Sections \ref{s.geq16}, \ref{sec:8charges}, and \ref{sec:4charges} illustrate the hyperbolic towers property in a diverse array of examples. In Section \ref{s.geq16}, we explore simple and pedagogical examples in maximal supergravity. We also explain, using the formalism of \cite{Baines:2025upi,Baines:2026aug}, how our results apply more generally in asymptotic limits of symmetric moduli spaces, thus also covering maximal supergravity in lower dimensions, as well as parts of half-maximal supergravity limits. 

In Section \ref{sec:8charges}, we turn to theories with 8 supercharges with a focus on 5d $\C{N}=1$ supergravity. We compute the relevant Laplacians and prove finiteness of the $c$-values using the bottom-up analysis of \cite{Kaufmann:2024gqo}, and check their quantization in explicit examples. Here there is a subtlety, as the ``axions'' in each infinite-distance limit are not directly visible from a single supergravity description. Our results show that the expected story with Laplacians and $\beta$-vectors persists despite the apparent ``myopia'' of the supergravity. We illustrate the breakdown of these statements in limits where the scalar curvature diverges and rigid field theory descriptions emerge \cite{Marchesano:2023thx,Castellano:2024gwi,Marchesano:2024tod,Blanco:2025qom,Castellano:2026bnx,Aoufia:2026mqb}. We also briefly comment on other theories with 8 supercharges, namely the tensor branch of 6d $\C{N}=(1,0)$ theories and the hypermultiplet sector in 6d, 5d, and 4d theories. The former will be particularly interesting, as some of our assumptions will be violated, but the propositions outlined above will persist. 

In Section \ref{sec:4charges}, we consider 4d EFTs with $\C{N}=1$ supersymmetry, and find more evidence for our claims. We focus on EFT string limits in the moduli spaces of $E_8 \times E_8$ theory compactified on a Calabi-Yau threefold (CY3), and we explicitly compute the Laplacians, axion fibration structure, and instanton spectrum in all of these examples. Interestingly, though there is no axion fibration structure over a global flat slice in 4d $\C{N}=1$ theories, we nevertheless find that the asymptotic structure \textit{locally} over a generic frame in a \textit{growth sector} (where the prepotential is dominated by a single monomial) is exactly as we propose. 

We conclude in Section \ref{sec:conclusions}. 

\section{Hyperbolic towers and asymptotic geometry}\label{sec:2}

In this section, we present and derive the main results of this paper. To make our terminology precise, we begin by defining principal towers and duality frames/principal planes. With a further assumption ---Assumption \ref{ass:principal}--- we see that the asymptotic geometry of moduli space is a compact nilmanifold fibration over an asymptotically flat cone, where the total axion fiber volume shrinks exponentially at a rate controlled by a vector $\vec \Gamma$. By linking these decay rates to tower masses, we derive the hyperbolic towers quantization condition. We explicitly find that \emph{$\beta$-vectors} that measure the moduli-dependence of instanton actions sum to produce the $ \Gamma$-vector. This allows us to relate the hyperbolic towers condition to the spectrum of instantons with small exponentiated action $e^{-S_{\m{inst}}}$ and frame these results in terms of the instanton taxonomy rules derived in \cite{Etheredge:2025ahf}.

In decompactification limits, both the axion fibration and the hyperbolic towers condition follow directly from dimensional reduction. Though this story is classical, we review the salient features here for completeness. In emergent string limits, the axion fibration structure is argued instead from the general principles of string perturbation theory for a tensionless weakly-coupled string. Concretely, we derive the nilmanifold  by assuming that the ---possibly dual--- weakly coupled frame organizes as a perturbative expansion in the emergent string coupling, see Assumption \ref{ass:perturbative-esc}. We may thus deduce constraints on the axion kinetic coefficients/decay constants from the expected form of the string-frame low-energy effective action.

In what follows, we will make the following three assumptions: 

\begin{assumption}[ESC without Running Decompactification]\label{ass:ESC}
    All infinite distance limits we consider correspond either to a weakly-coupled tensionless string limit or a decompactification over a $k$-manifold $M^k$ which solves Einstein's equations in vacuum. 
\end{assumption}

\begin{assumption}[Perturbative String Limits] \label{ass:perturbative-esc}
Every non-decompactification infinite-distance limit in a moduli space of vacua corresponds to the emergence of a tensionless, perturbative, critical fundamental string. The EFT in such a limit can be organized into a string-coupling expansion for the emergent perturbative string.
\end{assumption}

\begin{assumption}[Asymptotically Flat Slice]\label{ass:principal}
Let $\C{M}$ be a moduli space and $p$ a point in its interior. For a duality frame $\C{C} \subset T_p\C{M}$ (defined in Section \ref{ssec:geometry}), the image of $\C{C}$ under the exponential map is an asymptotically flat geodesic cone. That is, each geodesic emanating from $p$ in $\C{P}$ asymptotes to a straight line. 
\end{assumption}

Assumption \ref{ass:ESC} is to avoid issues associated with \textit{sliding} of $\alpha$-vectors \cite{Etheredge:2023odp}. It would be interesting to see if this assumption can be relaxed. Indeed, we will see that tensor moduli limits in 6d $\C{N}=(1,0)$ theories obtained by compactifying heterotic on K3, studied in Section \ref{ssec:other-8-supercharges}, appear to satisfy the hyperbolic towers property despite failing Assumption \ref{ass:ESC}. Assumption \ref{ass:perturbative-esc} is a stronger version of the ESC in emergent string limits, which requires the tensionless string to be perturbative.  Here, we stress the perturbativity of the string limit, as our derivation of the hyperbolic towers property relies extensively on string perturbation theory. The same 6d $\C{N}=(1, 0)$ examples in Section \ref{ssec:other-8-supercharges} will also violate this Assumption, as NS5-branes will lead to the breakdown of string perturbation theory in the 6d heterotic vacuum. Finally, Assumption \ref{ass:principal} allows a convenient presentation of the full global asymptotic geometry as a fibration of periodic axions over a flat principal plane. Though we do not have a proof of the generality of Assumption \ref{ass:principal} (and we expect it to be broken in certain 4d limits, as we discuss in Section \ref{sec:4charges}), we nevertheless find it to hold in a diverse array of examples, including generic limits in a class of theories preserving only 4 supercharges in Section \ref{sec:4charges}. 

With these assumptions in tow, we present the following results: 

\begin{prop}[Axion Fibration]\label{prop:nilmanifold}
Consider a moduli space $\C{M}$ along with a point $p \in \C{M}$ from which the space of infinite-distance limits in a duality frame $\C{C}$ forms an asymptotically flat cone $\C{P}$, which we call the principal plane (Assumption \ref{ass:principal}). The general asymptotic structure of the moduli-space metric is as follows: 
\begin{equation}\label{eq:fullmetric} \d s^2 = \d s^2_{\C{P}} +  \sum_I \frac{1}{g_I(x^m)^2} e^{-2 \vec{\beta}^I \cdot \vec{\phi}} d \theta_I^2 + g_{mn}(x^m) \d x^m \d x^n \, ,\end{equation}
where the $x^m$ are purely higher-dimensional moduli whose metric $g_{mn}$ factorizes from the principal plane metric $\d s_\C{P}^2$ and the metric for the axions $\theta_I$. Here, $\vec\phi$ are the moduli on $\C{P}$ with asymptotically flat metric $\d s_{\C{P}}^2$ and $\vec{\beta}^I \in T_p \C{M}$ characterize the scaling of the axion decay constants for $\theta_I$ with the principal plane moduli. 
\end{prop}

\begin{prop}[Instanton Scaling Vectors] \label{prop:instantons-beta-vectors}
Let $\C{C} \subset T_p \C{M}$ be a duality frame which obeys Assumption \ref{ass:principal} and exponentiates to an asymptotically flat geodesic cone $\C{P} \subset \C{M}$. Then, as vectors in $T_p \C{M}$, the scaling vector for the axion decay constants/kinetic coefficients in Prop. \ref{prop:nilmanifold} are related to the gradients of the (real parts of) electrically-coupled instanton actions via  
    \begin{equation} \vec\beta^I = \vec\nabla \log S^I. \label{eq:beta-nable-log-S} \end{equation}
\end{prop}

\begin{prop}[Hyperbolic Towers]\label{prop:hyperbolic towers} Consider a principal tower with mass $m$ such that $-\vec\nabla \log m = \vec{\alpha}$, and suppose that $\vec{\a}$ belongs to a duality frame $\C{C} \subset T_p \C{M}$ exponentiating to an asymptotically flat cone $\C{P} \subset \C{M}$. We then have
\begin{equation} \label{eq:c-value-gamma-alpha} c \equiv \nabla^2 \log m = \vec{\Gamma} \cdot \vec{\alpha}, \q \vec{\Gamma} = \sum_{\vec{\beta}^I \in \C{C}^\vee} \vec{\beta}^I . \end{equation}
where the expression for $\vec\Gamma$ sums over all $\vec\beta^I$ contained within the dual cone $\C{C}^\vee$ to $\C{C}$. We will call this value the hyperbolic towers coefficient, or $c$-value for short.
\end{prop}

\begin{prop}[Hyperbolic Towers Quantization]\label{prop:hyperbolic towers quantization}
Let $m$ be the characteristic mass measured in Planck units of a principal tower that is either a KK-mode decompactifying from $d$ to $D$-dimensions, or a string oscillator mode. Then the moduli-space Laplacian acting on the logarithm of this mass is quantized:
\begin{equation}  c \equiv \nabla^2 \log m = \begin{cases} \frac{N}{D-d}, & \text{KK-mode decompactifying from $d$ to $D$ dimensions,} \\ \frac{N}{2},& \text{string oscillator,} \end{cases}
\end{equation}
for some nonnegative integer $N$. We will call this quantized value the hyperbolic towers coefficient, or $c$-value for short.
\end{prop}

Propositions \ref{prop:nilmanifold} and \ref{prop:instantons-beta-vectors} are standard and easily derived; we present their derivations in order to make the overall logic clear. Proposition \ref{prop:instantons-beta-vectors} in particular will help make the connection between our results and instanton taxonomy \cite{Etheredge:2025ahf} clear. Proposition \ref{prop:hyperbolic towers} is a central result, relating the Laplacians $c = \nabla^2 \log m$ to tower masses ($\alpha$-vectors) and geometry (sums of $\beta$-vectors). Together with the expected scaling of the $\beta$-vectors presented in Sections \ref{ssec:decomp} and \ref{ssec:emstring}, we derive Proposition \ref{prop:hyperbolic towers quantization}, which is the main computational result of this paper. 

The organization of the remainder of this section is as follows. In Section \ref{ssec:geometry}, we introduce principal towers and the principal plane, which will allow for an easy characterization of a duality frame $\C{C}$ in a moduli space $\C{M}$ as a cone within the tangent space $T_p\C{M}$ about a point $p \in \C{M}$. Then, in Section \ref{ssec:decomp} and Section \ref{ssec:emstring}, we argue for Proposition \ref{prop:nilmanifold} in decompactification and emergent string limits separately. In our derivation, we keep careful track of the scaling vectors $\vec{\beta}^I$ in each asymptotic frame. In Section \ref{ssec:nilmanifold-fibrations}, we put the pieces in Sections \ref{ssec:decomp} and \ref{ssec:emstring} together along with Assumption \ref{ass:principal} to derive Proposition \ref{prop:nilmanifold}. In Section \ref{ssec:axionsinstantons}, we discuss instantons and argue for Proposition \ref{prop:instantons-beta-vectors}, which will motivate our discussion of instanton taxonomy. Finally, we prove Proposition \ref{prop:hyperbolic towers} in Section \ref{ssec:instantons,laplacians,towers} and Proposition \ref{prop:hyperbolic towers quantization} in Section \ref{ssec:hyp-towers-quantization}. Familiar readers may safely skip over the derivations in Sections \ref{ssec:decomp} and \ref{ssec:emstring}.

\subsection{Principal towers}\label{ssec:geometry}

In the following, we give an ansatz for the geometry of an asymptotic duality frame in a quantum gravity moduli space, which we motivate (and prove under some assumptions in Appendix \ref{app:A}) with a strong version of the ESC (i.e.\ Assumptions \ref{ass:ESC} and \ref{ass:perturbative-esc}). In particular, we assume that all decompactification limits are to a higher-dimensional vacuum background without any running profiles for fields in the higher-dimensional theory. (This is Assumption \ref{ass:ESC} in the beginning of the section.)

\begin{assumption*}[ESC without Running Decompactification]
    All infinite distance limits we consider correspond either to a weakly-coupled string limit or a decompactification over a $k$-manifold $M^k$ which solves Einstein's equations in vacuum. 
\end{assumption*}
\noindent This bypasses any issues associated with sliding as described in e.g.\ \cite{Etheredge:2023odp}. It would be interesting to study the asymptotic geometry of moduli spaces in setups including running decompactifications (for example, the dual decompactification of heterotic on $S^1$ to Type I' \cite{Etheredge:2023odp}) but this discussion is beyond the scope of this paper and will be deferred to future work. 

To systematically describe these limits, we now introduce the notions of principal towers, the principal plane, and duality frames. Consider a geodesic $\gamma$ that asymptotes from a point $p$ to an infinite-distance limit in the moduli space of a $d$-dimensional theory. By the DC, the species scale $\Lambda$ ---corresponding to the quantum-gravity cutoff of the EFT--- decays exponentially along $\gamma$. Consequently, at a generic point, a hierarchy of towers $m_1 \ll m_2 \ll \cdots \ll m_N \lesssim \Lambda$ can emerge, becoming asymptotically light. (At a non-generic point, tower scales $m_i, m_j$ may overlap.)

According to the ESC, the lightest of these towers, $m_1$, signals either an emergent string or a decompactification to a $D_1$-dimensional theory.  In the latter scenario, there are two possibilities: either $\gamma$ terminates\footnote{Here, we say that a geodesic terminates in the higher-dimensional theory if it does not approach infinite distance in higher-dimensional Planck units.} in the higher-dimensional theory (e.g.\ if the higher dimensional theory has no moduli, as in M-theory), or it continues to reach infinite distance within the higher-dimensional moduli space. If the geodesic approaches infinite distance in $D_1$-dimensional Planck units then, by the ESC, we will have a second tower $m_2 > m_1$ associated with either an emergent string limit or a further decompactification to $D_2$ dimensions, with $D_2 > D_1$. This iterative process continues until the geodesic either reaches an emergent string limit or eventually terminates at a finite distance within a higher-dimensional theory.

The hierarchy of towers obtained this way, with masses along the geodesic satisfying $m_1<m_2<\dots<m_N \leq \Lambda$ are called the \emph{principal towers} \cite{Etheredge:2024tok}. 
Define the $\alpha$-vector of the $i$-th principal tower as follows,
\begin{equation}
    \vec \alpha_i = - \nabla\log m_i \, ,
\end{equation} 
where $m_i$ are the masses (in Planck units) of the principal towers, and $\vec\nabla$ is the moduli-space gradient for the original $d$-dimensional theory. Given a collection of principal towers $m_i$, we define the \emph{duality frame}, $\mathcal C$ as 
\begin{equation}
    \C{C} = \left\{ v \in T_{p}\C{M} \; \bigg| \; v = \sum_{i} x_i \vec{\alpha}_i, \; x_i \geq 0 \right\} \, .
\end{equation}
In the language of \cite{Etheredge:2024tok}, this cone identifies the so-called \emph{principal plane}. We will use the two terms interchangeably, but care must be taken to emphasize that $\C{C}$ lies within the \textit{tangent} space $T_p \C{M}$ and not the moduli space $\C{M}$ itself. (Later, when we invoke Assumption \ref{ass:principal}, we will be able to extend $\C{C}$ to an asymptotically flat geodesic cone $\C{P} \subset \C{M}$ emanating from $p$, which we will also call the principal plane.) 

Notice that the definition depends in principle on the chosen interior point $p$. We will specialize the discussion to settings where this detail is immaterial in Section \ref{ssec:nilmanifold-fibrations} ---and indeed, Assumption \ref{ass:ESC} allows us to do this, cf. \cite{Etheredge:2023odp}. Finally, we define the \emph{frame simplex} $\Delta$ as the polytope spanned by the convex hull of the vertices of the $\vec{\alpha}_i$ for a duality frame \cite{Etheredge:2024tok}. If all vertices are principal KK towers associated to decompactification, we say that the frame is \emph{geometric}. If one of the principal towers is an oscillator tower for an emergent tensionless string, we say that the frame is \emph{stringy}. In what follows, we will not rely explicitly on the construction of $\Delta$, but we will return to the taxonomic setup in several later examples where it does apply.

For an arbitrary vector $v \in \C{C}$ in some duality frame $\C{C}$, suppose we follow the corresponding geodesic $\gamma_v(t)$ emanating from a point $p = \gamma_v(t_0)$ for $t_0 \gg 1$ arbitrarily far along.  The broad aim of our paper (and this section) is to understand the \emph{transverse} geometry to a family of such geodesics, and to relate it to the Laplacian of principal towers. As we show in Sections \ref{ssec:decomp} and \ref{ssec:emstring}, Assumptions \ref{ass:ESC} and \ref{ass:perturbative-esc} place stringent constraints on the form that the metric can take in this regime. Specifically, we first consider decompactification and emergent string limits separately in Section \ref{ssec:decomp} and \ref{ssec:emstring}, respectively. This way, we derive the general structure of the transverse geometry along a single geodesic. We then derive the axion nilmanifold fibration structure over the principal plane by employing Assumption \ref{ass:principal} in Section \ref{ssec:nilmanifold-fibrations}. 

\subsection{Decompactification limits}\label{ssec:decomp}

Let us first suppose that $\gamma_v(t)$ corresponds to a decompactification of the $d$-dimensional EFT to $D$ dimensions over a manifold $M^k$, where $k = D-d$. Due to Assumption \ref{ass:ESC}, the fluctuation corresponding to the canonically normalized radion $\rho$ is massless. The direction of the limit, which we denote by $\hat\phi$, always has a non-trivial component along the radion direction $\hat\rho$, with $\rho$ defined by 
\begin{equation}\label{eq:radion}
\rho = \sqrt{\frac{D-2}{k(d-2)}}\,\log \mathcal V_k\, 
\end{equation}
in terms of the volume $\C{V}_k$ of the $k$-dimensional manifold. In fact, in any decompactification frame, the Sharpened Distance Conjecture (SDC) \cite{Etheredge:2022opl} states that the alignment of a given direction $\hat{\phi}$ with the canonically normalized radion direction $ \rho$ satisfies
\begin{equation}
    \hat{\phi} \cdot {\rho} \geq \frac{1}{\sqrt{d-2}}. \label{eq:sdc}
\end{equation}
Additionally, there generically are other ``shape moduli'' which decouple and become orthogonal to the radion. Finally, there are flat directions given by the reduction of the higher-dimensional gauge-fields, which are compact due to charge quantization\footnote{It is expected that all continuous gauge symmetry is compact due to the completeness of representations and the finiteness of black hole entropy. If there were to be a non-compact gauge group, then completeness would require a continuous spectrum of representations, which violates the finiteness of black hole entropy at any finite energy.}. Concretely, consider a $p$-form gauge field $C_p$ in the higher-dimensional theory, its kinetic term is given as 
\begin{equation}
S_D\supset -\int \frac{1}{2  g(x^i)^2} F_{p+1}\wedge *_D F_{p+1},
\end{equation}
where $F_{p+1} = \d C_p + \cdots$ is a gauge-invariant field strength constructed from $C_p$ and $g(x^l)$ is the coupling for $C_p$, possibly depending on the higher-dimensional moduli $x^l$. Dimensionally reducing the $p$-form $C_p$ on a $p$-cycle, and switching to Einstein frame, the resulting kinetic term for the axion $\theta_p$ thus obtained is
\begin{equation}
    \C{S}_d \supset - \int \sqrt{-g_d} \frac{1}{2 g(x^l)^2 R^{2p}} (\p \theta_p)^2 \, ,
\end{equation}
where we introduced the radion modulus $R \sim \C{V}_k^{1/k}$, and the axion decay constant goes as $ f_{\theta_p}^2 \sim R^{k-2p}\mathcal V_k^{-1} \sim R^{-2p}$. Now, we note that $R \sim e^{\g \phi}$, where $\g$ is constrained by eq. \eqref{eq:sdc}:
\begin{equation}\label{eq:fcp}
f_{\theta_p}^2 \sim e^{-2p\g \phi}, \q \g \geq \frac{1}{\sqrt{d-2}} \, ,
\end{equation}
so the axion metric decreases exponentially in the decompactification limit.  The interested reader can consult Appendix \ref{app:A} for a detailed derivation of these results. In the following, we directly present the form of the asymptotic metric instead.

The metric on the moduli space at $q$ transverse to a decompactification over $M_k$ can be expressed in the following form (cf. Appendix \ref{app:A}): 
\begin{equation}\label{eq:asympmetric}
    ds^2 \simeq \d\phi^2 + \sum_{0 < p \leq k} \sum_{\omega_i \in H^p(M^k, \B{R})} \frac{1}{g_i(x^l)^2}  e^{-2p \g_{p, i} \rho}  \d \theta_{p,i}^2 + g_{mn}(x^m) \d x^m \d x^n \, ,
\end{equation} 
where $g_{mn}(x^l) \d x^m \d x^n$ is the metric associated to the shape deformations and the higher-dimensional moduli, which we have collectively denoted by $x^i$. As before, $\g_{p, i} \geq 1/\sqrt{d-2}$ for all directions within the decompactification frame $\C{C}$. If the direction of the limit is identified with $\hat{\rho}$ itself, the constants $\gamma_{p, i}$ for the axions $\theta_{p, i}$ are required by the ESC to be
\begin{equation}\label{eq:gammarate}
    \g = \sfc{d-2}{k(D-2)} = \sfc{d-2}{(D-d)(D-2)}\, .
\end{equation} 
We stress that the couplings $g_I(\Phi)$ will in general depend on the shape moduli of $M^k$ as well as the higher-dimensional moduli. However, the dependence of the coefficient of $\d \theta_p^2$ on the radion $\rho$ can be absorbed entirely into the prefactor $e^{-2p\gamma \phi}$. 

We emphasize also that moduli space is compact in the $\theta_{p, i}$ directions due to charge quantization, but need not be a torus. Indeed, this can arise if the gauge-invariant field strengths $F_{p+1}$ for the higher-dimensional $C_p$ have nontrivial modifications due to the presence of higher-dimensional triple Chern-Simons-type interactions. The general structure of the manifold parametrized by the axions $\theta_{p, i}$ is thus not a torus, but a \textit{nilmanifold}, and the one-forms $\d\theta_{p, i}$ are not literally exact one-forms but rather left-invariant one-forms on a nontrivial nilmanifold. An example of this is given in Section \ref{ssec:hyper}, in the case of the hypermultiplet sector of $4$d $\C{N}=2$ theories. In this paper, we retain the notation $\d\theta_{p, i}$ for the axion directions in the metric, since what we compute will anyway not be sensitive to the detailed nilmanifold structure. We plan to study in detail the consequences of the Chern-Simons couplings and the nilmanifold structure on the lower-dimensional moduli space geometry in a subsequent paper \cite{ricci:toappear}. 

\subsection{Emergent string limits}\label{ssec:emstring}

Thus far, the preceding discussion has focused almost exclusively on the asymptotic structure of the moduli space geometry in decompactification limits, and indeed, the expression for the asymptotic decomposition of the metric we have obtained is true only for each decompactification limit. 

However, this still does not explain the dependence of the axions on dilatonic moduli. In principle, we might expect the structure of the moduli space geometry in a string limit in such examples to be significantly different from a decompactification limit. In fact, this is not the case, and the asymptotic fibration structure of axions over a dilaton with exponentially shrinking fibers continues to be preserved. As a simple example, consider Type IIB string theory in 10d as in Section \ref{s.geq16}, where there is a single infinite-distance limit corresponding to taking the axiodilaton $\tau = C_0 + ie^{-\phi} \to +i\infty$. The size of the axion circle, parametrized by $C_0$, decays as $\sim e^{-\phi}$ as $\phi \to -\infty$, exactly in accordance with the general structure we predicted. 

We now wish to derive the asymptotic fibration structure in string limits directly from the structure of the low-energy effective action in string perturbation theory. In doing so, we fully use the \emph{perturbativity} of the emergent string from Assumption \ref{ass:ESC}. Explicitly, we take the following form of the ESC. (This is Assumption \ref{ass:perturbative-esc} in the beginning of the section.)

\begin{assumption*}[Perturbative String Limits]
Every non-decompactification infinite-distance limit in a moduli space of vacua corresponds to the emergence of a tensionless, perturbative, critical fundamental string. The EFT in such a limit can be organized into a string-coupling expansion for the emergent perturbative string.
\end{assumption*}

While the hyperbolic towers proposition holds in decompactification frames without Assumption \ref{ass:perturbative-esc}, it will be necessary to invoke Assumption \ref{ass:perturbative-esc} to justify analogous statement in string limits, and to do this, we will make use of the general low-energy effective action derived from string perturbation theory. We emphasize that Assumption \ref{ass:perturbative-esc} is not true in every string-theory example --- although in every known emergent string limit there is a tensionless, weakly-coupled critical string that is \textit{locally} perturbative, the corresponding vacuum need not admit an EFT that organizes itself via a perturbative string description. We will discuss an example of this in Section \ref{ssec:other-8-supercharges}, where we consider infinite-distance limits in the 6d $\C{N}=(1, 0)$ tensor moduli space with $n_T > 1$ tensor multiplets corresponding to a tensionless heterotic string on K3; these limits do not admit a description in terms of string perturbation theory due to the presence of $N_5 = n_T-1$ NS5-branes over K3. As it turns out, the hyperbolic towers property of Proposition \ref{prop:hyperbolic towers quantization} is still satisfied in these examples, which suggests that it may hold even if assumptions are relaxed.

Let us now describe the most general EFT action derived from string perturbation theory. The most general string-frame effective action for a critical perturbative string contains the following universal closed-string sector: 
\begin{align}\begin{split}
S_{\mathrm{str}}
&=
\frac{1}{2\kappa_D^2}
\int d^D x\,
\sqrt{-g_{\mathrm{s}}}\,
e^{-2\Phi}
\Bigg[
R(g_{\mathrm{s}})
+4\,g_{\mathrm{s}}^{\mu\nu}\partial_\mu\Phi\,\partial_\nu\Phi
-\frac{1}{12}H_{\mu\nu\rho}H^{\mu\nu\rho}
\\
&\hspace{4.5cm}
-\frac{\alpha'}{4}\operatorname{Tr}\!\left(F_{\mu\nu}F^{\mu\nu}\right)
+\mathcal{O}\!\left(\alpha' R^2,\alpha'^2\right)
\Bigg] + S_{\text{non-gravitational}}\, ,
\end{split}\end{align}
where all terms are expressed in terms of the string-frame metric $g_s^{\u\v}$. Here, $S_{\m{non-gravitational}}$ can possibly contain kinetic terms for $p$-form gauge fields, including axions. In general, the string-frame effective action for a $p$-form gauge field is of the following form: 
\begin{equation}
S_{p\text{-form}} = \int \d^D x \, \sqrt{-g_s} \, e^{-a\Phi} \abs{F_{p+1}}^2, \q a \in \B{Z}\, . 
\end{equation}
That $a$ is an integer follows on general grounds from string perturbation theory. Indeed, $g_s = e^{\Phi}$, and the prefactor multiplying any kinetic term in the effective action is associated to the worldsheet topology $\Sigma$ preparing the states which couple to the associated background fields via $a = \chi(\Sigma)$\footnote{Note that the RR kinetic terms for Type II strings pick up an additional factor of $g_s^2$ from the two RR vertex operator insertions; this does not affect the overall integrality of $a$. In general, vertex operator insertions can shift $a$ by a nonnegative integer.}.

Let us now move to the $d$-dimensional Einstein frame. Starting from the \(d\)-dimensional string-frame kinetic term for a $p$-form gauge field $C_p$ with field strength $F_{p+1}$
\begin{equation}
S_s \supset -\frac12 \int d^d x\,\sqrt{-g_s}\,e^{-a\Phi}\lvert F_{p+1}\rvert_s^2\, ,
\end{equation}
perform the Weyl rescaling $g_{\mu\nu}^{(s)}=e^{\frac{4\Phi}{d-2}}g_{\mu\nu}^{(E)}$. The volume element transforms as $\sqrt{-g_s}=e^{\frac{2d}{d-2}\Phi}\sqrt{-g_E}$, while the \((p+1)\) inverse metrics entering the norm of \(F_{p+1}\) give
\begin{equation}
\lvert F_{p+1}\rvert_s^2
=
e^{-\frac{4(p+1)}{d-2}\Phi}\lvert F_{p+1}\rvert_E^2\, .
\end{equation}
Therefore,
\begin{equation}
S_E
\supset
-\frac12
\int d^d x\,\sqrt{-g_E}\,
e^{\left[\frac{2(d-2p-2)}{d-2}-a\right]\Phi}
\lvert F_{p+1}\rvert_E^2.
\end{equation}
In terms of the canonically normalized dilaton $\varphi=\sqrt{4/(d-2)}\,\Phi$ in $d$-dimensional Planck units, this becomes
\begin{equation}\label{eq:normalizeddil}
S_E
\supset
-\frac12
\int d^d x\,\sqrt{-g_E}\,
\exp\!\left[
\frac{d-2 -2p -\frac{a}{2}(d-2)}
{\sqrt{d-2}}\,\varphi
\right]
\lvert F_{p+1}\rvert_E^2.
\end{equation}
For axions $\theta = C_0$ with $p = 0$, the axion decay constant/kinetic coefficient thus scales as
\begin{equation}\label{eq:axiondecaystring1}
f_{\theta} \simeq \exp \left( \frac{2 - a}{4} \sqrt{d-2} \, \varphi \right), \q 2-a \in \B{Z}_{\geq 0}. 
\end{equation}
The last condition $2-a \in \B{Z}_{\geq 0}$ follows from the fact $a = \chi(\Sigma) - N_{\m{cl}} = 2 - 2g - n - N_{\m{cl}} \leq 2$. Here $g, n$ are the number of handles and boundaries of $\Sigma$, respectively, and $N_{\m{cl}}$ is the number of additional RR vertex operators in the Type II case. Alternatively, a bottom-up argument for $2-a \geq 0$ follows from the compactifiability of the moduli space viz. \cite{Delgado:2024skw} in the $\varphi \to -\infty$ limit. 

This takes care of all axions visible directly from string perturbation theory whose kinetic coefficients depend on $\varphi$. A further remaining possibility concerns axions obtained by dualization of $(p-2)$-form gauge fields. For a kinetic coefficient of the form $e^{-b \phi} \abs{F_{p-1}}^2$, the dual axion $\theta$ has kinetic term $e^{+b\phi} \abs{F_1}^2$. We thus obtain the following kinetic coefficient:
\begin{equation}\label{eq:axiondecaystring2}
f_{\theta} \simeq \exp \left( \frac{2+a}{4} \sqrt{d-2} \, \varphi \right), \q 2 + a \in \B{Z}_{\geq 0}\, . 
\end{equation}
The nonnegativity of $2 + a$ this time follows from the fact that $a = 2, 0$ (i.e.\ NSNS- and RR- sectors) for leading-order $p$-form gauge kinetic terms; one may deduce also $2 + a \geq 0$ from compactifiability.

We finally remark that for \textit{vacuum} compactifications without branes (i.e.\ those satisfying Assumption \ref{ass:ESC}), the open-string axion kinetic sectors are absent. This is due to the fact that open string axions necessarily arise from reductions of Chan-Paton gauge fields on 1-cycles, and are T-dual to D-brane positions along the dual circle, which violates the assumption of vacuum decompactification in all duality frames. Moreover, we have $N_{\m{cl}} \in 2 \B{Z}_{\geq 0}$ for a general kinetic term since kinetic terms involve the insertion of two canonically-normalized closed-string worldsheet operators, which may each be accompanied by an integer power of $g_s$. Thus, we additionally have 
\begin{equation}
2 - a \in 2\B{Z}_{\geq 0}, \q 2 + a \in 2\B{Z}_{\geq 0} \,.
\end{equation}
provided that Assumption \ref{ass:ESC} holds, whence also
\begin{equation}
    f_\theta \simeq \exp \left( \frac{k}{2} \sqrt{d-2} \, \varphi \right), \q k \in \B{Z}_{\geq 0}. \label{eq:string-axion-decay-constant}
\end{equation}
In particular, we have $k = 1$ for RR axions and $k = 2$ for dual NS axions. This fact will become relevant when we introduce the hyperbolic towers quantization condition (Proposition \ref{prop:hyperbolic towers quantization}) in Section \ref{ssec:instantons,laplacians,towers}. In examples where Assumption \ref{ass:ESC} fails, which are beyond the scope of the present paper, we expect the quantization condition of Proposition \ref{prop:hyperbolic towers quantization} to be possibly halved due to open string sectors. We will see also an example in Section \ref{ssec:other-8-supercharges} where there are periodic moduli in the 6d $\C{N}=(1, 0)$ tensor branch that arise in M-theory on $\m{K3} \times S^1/\B{Z}_2$ as NS5-brane positions along the Horava-Witten interval in the dual tensionless heterotic string limit after a duality quotient; this setup also violates Assumption \ref{ass:ESC}, but in this case, we find the hyperbolic towers property to be satisfied.

Notice that this asymptotic structure once again agrees precisely with eq. \eqref{eq:asympmetric}. We summarize therefore that the fibration structure of axions over a dilaton modulus are quite universal in perturbative string limits and assume much the same form as axions in decompactification limits, this time with different coefficients in the exponential. 

\subsection{Flat slices and axion fibration}\label{ssec:nilmanifold-fibrations}
We have thus established that by Assumption \ref{ass:ESC} in decompactification limits, and Assumption \ref{ass:perturbative-esc} in emergent string limits, the geometry of moduli space is given by a nilmanifold fibered over the (one-dimensional) geodesic corresponding to the limit. Let us now specialize to an important class of setups for which the structure of duality frames is substantially more rigid. For a point $p \in \C{M}$ in the interior of a moduli space $\C{M}$, we let $\C{C} \subset T_p \C{M}$ be a duality frame of infinite-distance limits of $p$ as defined in Section \ref{ssec:geometry}. We now make a further assumption that will allow us to significantly simplify the analysis. (This is Assumption \ref{ass:principal} at the beginning of the section.) 

\begin{assumption*}[Asymptotically Flat Slice]
The image $\C{P} = \exp_p(\C{C})$ of $\C{C}$ under the exponential map is an asymptotically flat geodesic cone. That is, each geodesic emanating from $p$ in $\C{P}$ asymptotes to a straight line. 
\end{assumption*}

In other words, the above assumption expresses simply the condition that a duality frame/principal plane $\C{C} \subset T_p \C{M}$ give rise to an asymptotically flat cone $\C{P} \subset \C{M}$ of the moduli space $\C{M}$, cf. \cite{Etheredge:2024tok}, which we will also (by a slight abuse of notation) refer to as the \textit{principal plane}. Assumption \ref{ass:principal} holds in all of the locally symmetric moduli spaces of Section \ref{s.geq16}. Even outside the lamppost of symmetric spaces, it will frequently be true that there exists a subset of the non-compact moduli parametrized by $\log m_i$ for principal towers $m_i$ spanning an asymptotically flat cone $\C{P} \subset \C{M}$ in certain scaling regimes; this lies within the regime of validity of Assumption \ref{ass:principal}\footnote{Examples for which rigid sectors decouple asymptotically are automatically of this type. Indeed, due the analysis in \cite{Aoufia:2026mqb}, the $\alpha$-vectors associated to these sectors are orthogonal to the principal plane, as it comprises only extremal towers. Moreover, the decoupled directions source a curvature divergence, which naively spoils the fibration structure we have been discussing up to now. It would be interesting to better understand these cases, which we leave for future work.}. We will see an example of this in Section \ref{sec:4charges} in 4d supergravity. Note that if there is a \textit{single} principal tower ---i.e.\ an isolated decompactification or string limit--- Assumption \ref{ass:principal} is trivially satisfied. Thus, the analysis in this section will turn out to be general for the examples we consider.

However, we are unable to provide a general argument for Assumption \ref{ass:principal} in all of the cases of interest ---i.e.\ assuming the ESC without sliding, cf. Assumption \ref{ass:ESC}--- and indeed, we suspect that Assumption \ref{ass:principal} may not be true in general\footnote{Briefly, consider a 4d $\C{N}=2$ vector multiplet moduli space, which is projective special Kähler specified by a prepotential $\C{F} \sim C_{IJK} X^I X^J X^K$. Along any infinite-distance limit where the volume does not scale as a single monomial of $\C{F}$ (i.e.\ $\C{F} \sim X_1 X_2^2 + X_2 X_2^3$) the Kähler metric in the directions associated to the nonzero terms of $\C{F}$ in the scaling limit is not generally asymptotically flat. Such scaling limits are, however, expected to be of measure zero; the generic saxionic limit is expected to asymptote to a growth sector with a single dominant monomial in the asymptotic prepotential. See Section \ref{sec:4charges} for similar considerations.}. It would be interesting to characterize Laplacians on the moduli space in cases where Assumption \ref{ass:principal} does not hold, but this is beyond the scope of the present paper.

Suppose $\C{C}$ is generated by $\alpha$-vectors $\vec{\alpha}_1, \cdots, \vec{\alpha}_k$ associated to principal towers with mass $m_1, \cdots, m_k$ via
\begin{equation} \vec{\phi} = \sum_i x^i \vec{\alpha}_i \in \C{C}\, , \end{equation}
then the $x^i$ are related to the canonically-normalized coordinates $\phi^i$ associated to each $\alpha$-vector via
\begin{equation} \phi^i = x^i \abs{\vec{\alpha}_i}\, . \end{equation}
We may thus write the canonically-normalized asymptotic metric on the flat cone $\C{P}$ as
\begin{equation} \d s^2_{\C{P}} \simeq g_{ij} \d \phi^i \d \phi^j\, , \q g_{ij} = \frac{\vec{\alpha}_i \cdot \vec{\alpha}_j}{\abs{\vec{\alpha}_i} \abs{\vec{\alpha}_j}}\, . \end{equation}
In a purely geometric frame, a flat cone $\C{P}$ consists only of geometric deformations. A radial deformation corresponds to a geodesic ray in $\C{P}$, while the transverse (flat) directions correspond to shape deformations of the associated decompactifying manifold. The remaining directions within a duality frame $\C{C}$ thus correspond to axionic directions fibering over $\C{P}$ which depend in principle on all of the principal-plane moduli $\phi^i$. Along any ray in $\C{P}$, however, we have seen that the dependence on the geodesic coordinate is precisely exponential in each principal plane modulus in a manner exactly consistent with eq. \eqref{eq:asympmetric}. For a stringy frame, where one of the vertices of $\C{C}$ may correspond to a string tower, we derived in Section \ref{ssec:emstring} that same axion fibration structure is present along any emergent string limit. Since possible dependences on additional moduli (i.e.\ moduli not on the principal plane $\C{P}$) will in any case factorize, we thus arrive at the following general ansatz for the moduli-space metric over an asymptotically flat principal plane. (This is Proposition \ref{prop:nilmanifold} presented at the beginning of the section.)

\begin{prop*}[Axion Fibration]
For a moduli space $\C{M}$ along with a point $p \in \C{M}$ from which the space of infinite-distance limits forms an asymptotically flat cone $\C{P}$ (Assumption \ref{ass:principal}), the general structure of the moduli-space metric is as follows: 
\begin{equation}\label{eq:fullmetric_2} \d s^2 = \d s^2_{\C{P}} +  \sum_I \frac{1}{g(x^m)^2} e^{-2 \vec{\beta}^I \cdot \vec{\phi}} d \theta_I^2 + g_{mn}(x^m) \d x^m \d x^n \, ,\end{equation}
where the $x^m$ are purely higher-dimensional moduli whose metric $g_{mn}$ factorizes from the principal plane metric $\d s_\C{P}^2$ and the metric for the axions $\theta_I$. 
\end{prop*}

Here, the $\theta_I$ are axion fields and the $\vec\beta^I \in T_p \C{M}$ are vectors parametrizing the exponential dependence of the axionic metric coefficients on the principal plane moduli $\phi^i$. From this, we can compute that the determinant of the moduli-space metric $G$, which asymptotically scales as
\begin{equation}\label{eq:detGGamma} \sqrt{\det G} \sim e^{-\vec{\Gamma} \cdot \vec{\phi}}, \q \vec{\Gamma} = \sum_I \vec{\beta}^I\, . \end{equation}
By the compactifiability criterion proposed in \cite{Delgado:2024skw}\footnote{See also \cite{Grimm:2025lip} for a characterization of the criterion in terms of tameness.}, if no curvature divergence is present, we expect $\vec{\Gamma} \cdot \vec{\phi} \geq 0$, and thus $\vec{\Gamma} \in \C{C}^\vee$. Here, $\C{C}^\vee$ is the \textit{dual cone} to $\C{C}$, defined by\footnote{In Section \ref{sec:4charges}, we will see that in the context of EFT strings, this cone will coincide with the cone of EFT instantons \cite{Lanza:2021udy}.}
\begin{equation} 
\C{C}^\vee = \{ \vec{v} \in T_p \C{M} \mid \vec{v} \cdot \vec{w} \geq 0 \q \forall \vec{w} \in \C{C} \}\, .  
\end{equation}
In fact, the arguments in \cite{Delgado:2024skw} can be applied analogously to show a stronger statement: No direction in $\C{C}$ can have an exponentially growing axion fiber, as this would lead to an infinite number of ground states for the 1d supersymmetric sigma model probing the 1d moduli space. Thus, we additionally have
\begin{equation} \vec{\beta}^I \in \C{C}^\vee \, .\end{equation}
This expectation will hold throughout all examples in the text.

\subsection{Axions and instantons}\label{ssec:axionsinstantons}

The vectors $\vec{\beta}^I$ associated to the shrinking axionic fibers are the key ingredients that control the structure of the asymptotic moduli-space geometry and share a superficial resemblance to the $\vec{\alpha}_i$ vectors associated to the principal towers. Thus, it is natural to wonder whether the $\vec{\beta}^I$ admit a similar interpretation as $\vec\nabla \log T$ for some moduli-dependent quantity $T$. Schematically, we can see that the answer is affirmative, as instantons electrically couple to axions, and thus it is natural to surmise that
\begin{equation} \vec{\beta}^I \simeq \vec\nabla \log S^I \, , \end{equation}
for $S^I$ the instanton action of the $I$-th instanton\footnote{For recent work studying constraints on instantons and axions through these vectors, see \cite{Artime:2026kfq, Etheredge:2026rio, Reece:2026hmp}.}. In fact, we will see that this guess is essentially correct, though the exact nature of the instanton in question deserves some elaboration. We now show this in detail for both decompactification and string limits.

Let us first consider the expression in eq. \eqref{eq:fullmetric_2} for the asymptotic moduli-space metric in decompactification limits, where the axionic directions are expressed directly as the dimensional reduction of higher-dimensional gauge fields. The ESC, having allowed us to deduce already this asymptotic structure, predicts additionally non-trivial features of the spectrum of instantons in the lower-dimensional theory.  Consider a $p$-form gauge field $C_p$ in the higher-dimensional theory. Completeness of the spectrum predicts the existence of a $(p-1)$-brane electrically charged under $C_p$, whose tension we will denote by $T_p$. Accordingly, one has Euclidean $(p-1)$-brane instantons wrapping $p$-cycles $[\Sigma^I] \in H_p(X, \B{R})$ in the spectrum of the lower-dimensional theory. These instantons have action
\begin{equation} \C{S}^I = T_p \m{vol}(\Sigma^I) + i \int_{\Sigma^I} C_p := S^I_{\rm KK} + i \theta^I \, .   \end{equation}
The dependence of $\m{vol}(\Sigma^I)$ on the radion $R$ is then 
\begin{equation} S^I_{\rm KK} \sim \m{vol}(\Sigma^I) \simeq R^p,  \end{equation}
so for the canonically-normalized radion $\rho$ in eq. \eqref{eq:radion}, we find that
\begin{equation}\label{eq:ddphilogS} \frac{d}{d\rho} \log S^I_{\rm KK} = p \sfc{d-2}{k(D-2)} \, . \end{equation}
Thus, the (real part of the) instanton action depends on the radion modulus in exactly the same way as does the axion kinetic coefficient. As we will see in Section \ref{ssec:instantons,laplacians,towers}, the relation above can be recast in terms of the instanton taxonomy rules of \cite{Etheredge:2025ahf}.

Under Assumption \ref{ass:principal}, one can show that in a geometric frame that the gradient of $\log S^I$ with respect to the principal plane moduli is exactly the $\beta^I$ vector. Indeed, we note that
\begin{equation}\label{eq:ddphilogS2} 
\frac{d}{d\phi^i} \log S^I = \beta^I_i 
\end{equation}
for each principal direction $\phi^i$ with associated ${\alpha}$-vector $\vec{\alpha}_i$. In a geometric frame, the $\vec{\alpha}_i$ generate the duality frame $\C{C}$, so these equations in components are in fact enough to deduce an equality of vectors on the (tangent space to the) principal plane. 

In a stringy frame, one of the frame generators is an $\alpha$-vector associated to a string tower. Provided we assume a perturbative string description (Assumption \ref{ass:perturbative-esc}), instantons are expected to appear due to saddle-point contributions in string amplitudes, as estimated in \cite{Shenker:1990uf}, and have actions of the type $S_{\rm str}^I \sim g_s^{-k}$ for $k = 1, 2$ (Here $k = 1$ corresponds to D-instantons, while $k = 2$ corresponds to NS instantons)\footnote{Notice that this agrees with the typical tension of a D-instanton, e.g.\ in Type IIB. The general estimate is instead argued for in terms of the leading non-perturbative saddle expected in matrix models. This is much larger than the $\C{O}(e^{-1/g_s^2})$-effects expected from trans-series of closed string amplitudes \cite{Shenker:1990uf}.}. Thus, in terms of the canonically normalized dilaton $\varphi$ above eq. \eqref{eq:normalizeddil}, one gets 
\begin{equation}\label{eq:betastring}
    \frac{d}{d(-\varphi)} \log S^I_{\rm str} = \frac{k}{2} \sqrt{{d-2}} \, .
\end{equation}
This agrees exactly with the $\beta$-vector component computed from the dilaton dependence of RR- and NS axion decay constants in eq. \eqref{eq:string-axion-decay-constant}. Note that we take the gradient with respect to $-\varphi$ since $\varphi \to -\infty$ corresponds to the tensionless string limit. 
In either case, we therefore have the following central result, Proposition \ref{prop:instantons-beta-vectors} as presented in the beginning of the section.

\begin{prop*}
Let $\C{C} \subset T_p \C{M}$ be a duality frame which exponentiates to an asymptotically flat cone $\C{P} \into \C{M}$. Then, as vectors in $\C{C} = T_p \C{P}$, the scaling vector for the axion decay constants/kinetic coefficients are related to the gradients of the (real parts of) the corresponding instanton actions via  
    \begin{equation} \beta^I = \vec\nabla \log S^I. \label{eq:beta-nable-log-S_2} \end{equation}
\end{prop*}
\noindent Unfortunately, it is not \emph{a priori} true that the instanton action is related in a simple way to the axion kinetic coefficient ---like eq. \eqref{eq:ddphilogS2}--- everywhere in the moduli space. Indeed, for a decompactification over $M^k$, we are comparing the scaling of two quantities: 
\begin{equation} f_{c_{I}}^2 = \frac{1}{\C{V}_k} \int_{M^k} \omega_I \wedge *_k \omega_I = \frac{1}{\C{V}_k} |\omega^I|^2, \qquad S^I \simeq \m{vol}(\Sigma^I) . \end{equation}
These quantities need not in general be related by any simple identity.\footnote{As a simple example, consider the smooth Calabi--Yau hypersurface $X \subset \mathbb P^1 \times \mathbb P^3$ of degree $(2,4)$, for which $h^{1,1}(X)=2$. Let $\omega_1,\omega_2$ denote the restrictions of the hyperplane classes from $\mathbb P^1$ and $\mathbb P^3$, respectively, and write the Kähler form as $J=t_1\omega_1+t_2\omega_2$. The nonzero triple intersections are $\kappa_{122}=4$ and $\kappa_{222}=2$. If $\Sigma_2$ is a curve class normalized by $\int_{\Sigma_2}\omega_2=1$ and $\int_{\Sigma_2}\omega_1=0$, then its calibrated volume is simply $\operatorname{Vol}(\Sigma_2)=\int_{\Sigma_2}J=t_2$. By contrast, the Hodge norm of the associated harmonic $(1,1)$-form $\omega_2$ is not determined solely by this curve volume. Using standard Calabi--Yau identities, it follows that
\begin{equation}
\int_X \omega_2\wedge *\omega_2
= g_{22}=
\frac{3(24t_1^2+8t_1t_2+t_2^2)}{t_2^2(6t_1+t_2)^2} \,
\end{equation}
where $g$ is the moduli-space metric. Thus, although the associated $2$-cycle volume is $\m{vol}(\Sigma_2) = t_2$, the Hodge norm depends also on the second K\"ahler modulus $t_1$.}

Despite this, we will see that even in examples without a global principal plane (cf. Assumption \ref{ass:principal}), there are certain scaling regimes for which an asymptotically flat slice $\cal P$ can be constructed as in Assumption \ref{ass:principal} for some of the non-compact moduli, and in these examples, the relation \eqref{eq:ddphilogS2} will hold for those non-compact moduli, see Section \ref{sec:4charges}.

\subsection{Laplacians and hyperbolic towers}\label{ssec:instantons,laplacians,towers}

In this section, we put all of the pieces together and state our main proposition concerning the Laplacians of principal towers over a principal plane $\C{P}$. We recall first the general asymptotic fibration structure over $\C{P}$ that follows from the ESC: 
\begin{equation}\label{eq:metricnilmanifold}  \d s^2 = \d s^2_{\C{P}} +  \sum_I \frac{1}{g_I(x^m)^2} e^{-2 \vec{\beta}^I \cdot \vec{\phi}} d \theta_I^2 + g_{mn}(x^m) \d x^m \d x^n \, ,  \end{equation}
with $\phi^i$ the principal-plane moduli spanning $\mathcal{P}$. The index $I$ sums over all axions whose ${\beta}$-vectors lie within the dual cone $\C{C}^\vee$. Given two principal towers with mass scales $m_i$ and $m_j$, their $\alpha$-vectors are shown, given Assumptions \ref{ass:ESC}, \ref{ass:perturbative-esc}, and \ref{ass:principal}, to satisfy the following taxonomy rule \cite{Etheredge:2024tok}
\begin{align}\label{eq:taxonomyprincipal}
    \vec \alpha_{i}\cdot \vec \alpha_j=\frac 1{d-2}+\frac 1{D_i-d}\delta_{ij}\, ,
\end{align}
where, $D_i$ is the dimension that the $i$-th principal tower decompactifies to if it is a KK-mode, or $D_i=\infty$ if the $i$-th principal tower is a string oscillator. This constrains the metric over the principal plane ${\cal P}$.

We now wish to understand the relationship between the geometry thus described and the Laplacians of principal towers. From the metric in eq. \eqref{eq:metricnilmanifold}, we compute the Laplacian as follows: 
\begin{equation}\label{eq:laplnilmanifold}
\nabla^2
=
g^{ij}\partial_{\phi^i}\partial_{\phi^j}
-
g^{ij}
\Gamma_i
\partial_{\phi^j}
+
\sum_I e^{2\vec\beta^I\cdot \vec\phi}\,
\partial_{\theta_I}^2 , \q \vec{\Gamma} = \sum_{\vec{\beta}^I \in \C{C}^\vee} \vec{\beta}^I \, .
\end{equation}
Let us now consider a tower with mass $m$ such that $-\vec\nabla \log m = \vec{\alpha}$. Since $\vec{\alpha}$ is a constant over $\C{P}$ due to Assumption \ref{ass:ESC} (see \cite{Etheredge:2023odp} for details) we thus obtain the following result. (This is Proposition \ref{prop:hyperbolic towers} as presented in the beginning of the section.)

\begin{prop*}[Hyperbolic Towers] Consider a principal tower with mass $m$ such that $-\vec\nabla \log m = \vec{\alpha}$, and suppose that $\vec{\a}$ belongs to a duality frame $\C{C} \subset T_p \C{M}$ exponentiating to an asymptotically flat cone $\C{P} \subset \C{M}$. We then have
\begin{equation} \label{eq:c-value-gamma-alpha_2} c \equiv \nabla^2 \log m = \vec{\Gamma} \cdot \vec{\alpha}, \q \vec{\Gamma} = \sum_{\vec{\beta}^I \in \C{C}^\vee} \vec{\beta}^I . \end{equation}
where the expression for $\vec\Gamma$ sums over all $\vec\beta^I$ contained within the dual cone $\C{C}^\vee$ to $\C{C}$. We will call this value the hyperbolic towers coefficient, or $c$-value for short.
\end{prop*}

\noindent As we already stressed, each ${\beta}$-vector need not necessarily lie in the duality frame $\C{C}$ itself, but within the dual cone $\C{C}^\vee$. However, in all examples that we consider, for a principal tower $m_i$ one can always find a frame $\C{C}' \ni \vec{\alpha}_i$ (not necessarily the original frame) containing each $\beta$-vector contributing to $\vec{\Gamma}$ in eq. \eqref{eq:c-value-gamma-alpha_2}. It would be interesting to understand more generally if this has to be true, and if so, why. Let us now investigate the geometric content of the above statement.

The $\Gamma$-vector introduced above is directly related to the geometry of the moduli space, as we noted in eq. \eqref{eq:detGGamma}. Using eq. \eqref{eq:laplnilmanifold} and Proposition \ref{prop:hyperbolic towers}, we find that on the principal plane, in components
\begin{equation}\label{eq:gammapartiallogg} 
{\Gamma}_i = -\p_i \log \sqrt{\det G}\, .  
\end{equation}
Though this presentation is no longer tensorial, we can also write $\Gamma_i$ as minus the trace of the Christoffel symbol $\Gamma^i_{jk}$ in a basis of principal towers $\hat \alpha^m$ as
\begin{equation} 
\Gamma_i = -\Gamma^m_{im}\, , 
\end{equation}
which additionally motivates our nomenclature for the $\Gamma$-vectors. Consequently, these vectors can be understood as pointing on the principal plane in the direction of steepest descent for the volume form $\sqrt{\det G}$ as a function of the principal plane moduli.

Let us go back to the case where the principal plane is one-dimensional, i.e.\ consider the geodesic ray generated by one principal tower. Their $\alpha$-vectors have constant length, and as shown in \cite{Etheredge:2023zjk}, any function $F$ such that $|\vec\nabla \log F|^2$ is constant is automatically an exponentiation of a distance function. Thus, the $i$-th principal tower has a mass that is an exponentiation of a distance function $t$,
\begin{align}
    m=m_{0}\exp(-\alpha t),
\end{align}
where $m_{0}$ is a moduli-independent constant, and the metric on moduli-space takes the form
\begin{align}
    \d s^2=\d t^2+g_{ab}(t,\phi^a)\d\phi^a \d\phi^b, \label{eq:distmetric}
\end{align}
where $\phi^a$ are moduli orthogonal to $t$. Indeed, this is precisely the case where we consider a single ray in the principal plane. Depending on whether the tower is a KK-mode or a string oscillator, $t$ is either a radion or a dilaton. Equation
\eqref{eq:distmetric} indeed implies that we have 
\begin{align}
	\nabla^2 \log m=\nabla^2 (-\alpha t)
	=- \frac {\alpha}{2}\partial_t \log \det g_{ab} \ .
\end{align}
Solving this equation, we find that the determinant of the metric on moduli space factors into the following form
\begin{align}\label{eq:detgasympt}
    \sqrt{\det g_{ab}(t,\phi^a)}=g_0(\phi^a) \exp\left(-\frac{c}{\alpha}t\right)\sim m^{\frac {c}{\alpha^2}}
\end{align}
All in all, the determinant of the metric must decay exponentially as a function of $t$, and the exponential rate is $(2/\alpha^2)c$. Moreover, the metric also scales as the mass to a power, so as $m$ decreases, the volume-form decreases\footnote{unless $\nabla^2 \log m=0$, as in Type IIA string theory. See Section \ref{s.geq16} for details.}.

\subsection{Hyperbolic towers quantization} \label{ssec:hyp-towers-quantization} We now turn to showing the main result of this paper, namely that the $c$-values introduced above are quantized. We start from decompactification limits and then turn to emergent string limits. 

In the case that $\vec{\alpha}$ is associated to a KK tower decompactifying over $M^k$ from $d$ to $D=d+k$ dimensions, the ${\alpha}$-vector is
\begin{equation} \abs{ \vec{\alpha} } = \sfc{ D-2 }{(D-d)(d-2)}\, ,  \end{equation} 
so using eq. \eqref{eq:ddphilogS} yielding the component of $\beta^I$ along the canonically-normalized radion direction, we get to
\begin{equation} \vec{\beta}^I \cdot \vec{\alpha} = p  \sfc{d-2}{(D-d)(D-2)} \cdot \sfc{ D-2 }{(D-d)(d-2)} 
= \frac{p}{D-d} \, , \q p > 0 \, . \end{equation}
Accordingly, we find that the $c$-value is necessarily a nonnegative integer multiple of $(D-d)^{-1}$, namely
\begin{align}\label{eq:quantization_c}
c_{\rm KK} = \nabla^2 \log m = \vec{\Gamma} \cdot \vec{\alpha} = \frac{N}{D-d}, \q N \in \B{Z}_{\geq 0} \, . 
\end{align}

To find the $c$-values in emergent string limits, we rely on the expectations laid out in Section \ref{ssec:emstring}. Concretely, the emergent string can be described---in an appropriate duality frame---as a fundamental weakly coupled string. The fiber depends on the $d$-dimensional dilaton $\phi$ as in eqs. \eqref{eq:axiondecaystring1}, \eqref{eq:axiondecaystring2}. By the ESC, we expect that the string scale agrees with the species scale $\Lambda$ and is given by 
\begin{equation}
     m_{\rm osc}\sim \Lambda\sim \exp\left(-\frac {\phi}{\sqrt{d-2}}\right)\, .
\end{equation}
and finally using \eqref{eq:laplnilmanifold}, their Laplacian reads
\begin{align}
    c_{\rm osc}=\nabla^2 \log m_{\rm osc}=\frac{N}{2} \, , \quad N\in \mathbb{Z}_{\geq 0}\, .
\end{align}
Notice this value agrees with the Proposition \ref{prop:hyperbolic towers} with the $\Gamma$-vector given by eq. \eqref{eq:betastring}. We again stress that this argument essentially relies on string perturbation theory with respect to the emergent (dual) string  to determine the couplings of the effective action.

We have thus shown in both decompactification as well as emergent string limits that the Laplacian eigenvalues are quantized. All in all, we have the following result, which is Proposition \ref{prop:hyperbolic towers quantization} as presented in the beginning of the section:

\begin{prop*}[Hyperbolic Towers Quantization]
Let $m$ be the characteristic mass measured in Planck units of a principal tower that is either a KK-mode decompactifying from $d$ to $D$-dimensions, or a string oscillator mode. Then the moduli-space Laplacian acting on the logarithm of this mass is quantized:
\begin{equation}  c \equiv \nabla^2 \log m = \begin{cases} \frac{N}{D-d}, & \text{KK-mode decompactifying from $d$ to $D$ dimensions,} \\ \frac{N}{2},& \text{string oscillator,} \end{cases}
\end{equation}
for some nonnegative integer $N$. We will call this quantized value the hyperbolic towers coefficient, or $c$-value for short.
\end{prop*}

\noindent We remark that the quantized values of $c$ are simple rational numbers with small denominators, despite the fact that they appear to combine a number of irrational quantities.

In fact, we can connect the above quantization of $c$-values with the taxonomy rules of \cite{Etheredge:2025ahf}. The taxonomy rules relating principal towers, which we described above in eq. \eqref{eq:taxonomyprincipal}, were later generalized in \cite{Etheredge:2025ahf} to subleading non-principal towers, and also to brane tensions. For instance, consider a non-oscillator particle tower with mass $m$, or more generally a $p$-dimensional brane with tension $T_p$, and its $(\vec\alpha_p=-\vec\nabla \log T_p)$-vector. Assuming Assumptions \ref{ass:ESC}, \ref{ass:perturbative-esc}, and \ref{ass:principal}, if one has a KK-mode corresponding to decompactification to $D$-dimensions with $\alpha$-vector $\vec \alpha_\text{KK,$D$}$, then the $\alpha$-vector of the brane is shown to satisfy the following \emph{{radion lattice}} rule:
\begin{align}
    \vec \alpha_p\cdot \vec \alpha_{\text{KK,$D$}}=\frac{p(D-2)-P(d-2)}{(D-d)(d-2)}\, ,
\end{align}
where $P$ is an integer that can sometimes be viewed as the spacetime dimension of the brane in the higher-dimensional theory (for KK-modes and monopoles, $P$ can take more exotic values with this interpretation). Meanwhile, for a principal tower that is a string oscillator, non-oscillator $(p-1)$-branes are shown (under the same assumptions) to satisfy the \emph{{dilaton lattice}} rule
\begin{align}
    \vec \alpha_p\cdot \vec \alpha_\text{osc}=\frac{p}{d-2}-\frac{1}{2}(P-p)\, .
\end{align}
Here $P$ is related to the number of $g_s$ factors in the tension of the brane measured in string units.

These two lattice rules control the $\alpha$-vectors of various branes. In particular, given a frame simplex, these rules must be satisfied for each vertex, and this results in the $\alpha$-vectors of branes having principal-plane components that are lattice-valued. Specifically, these rules extend to instantons, where $p=0$ and the $\alpha$-vector of an instanton is defined as 
\begin{align}
    \vec \alpha_0=-\vec\nabla \log S\, .
\end{align}
We will be interested in how axion decay constants $f$ scale. These also often scale inversely with instanton actions. Notice that these are related to the $\beta$-vectors introduced in Section \ref{ssec:axionsinstantons}: In asymptotic limits where Assumption \ref{ass:principal} is satisfied, we argued in eq. \eqref{eq:beta-nable-log-S_2} that 
\begin{align}
    \vec \beta=-\vec\nabla \log f= -\vec \alpha_0\, .
\end{align}
As a consequence, these $\beta$-vectors satisfy taxonomy rules as well, in particular
\begin{align}
    \vec \beta\cdot \vec \alpha_{\text{KK,$D$}}=-\frac{P}{D-d}\, ,\qquad 
    \vec \beta\cdot \hat \alpha_\text{osc}=-\frac{P}{2}\, .\label{e.betadotalpha}
\end{align}
Thus, these dot products of instanton $\alpha$-vectors with principal towers are related to the Laplacians of the logarithms of the masses of these towers by Proposition \ref{prop:hyperbolic towers quantization}.

Now, from Proposition \ref{prop:hyperbolic towers}, we have that for each principal tower
\begin{align}
\vec \Gamma\cdot \vec \alpha^{(i)}=\nabla^2\log m_{i}\, .\label{e.gammadotalpha}
\end{align}
By using the hyperbolic towers quantization condition, which quantizes $\nabla^2 \log m$, we actually find that the $\Gamma$-vector in \eqref{e.gammadotalpha} is satisfying the $p=0$ instanton taxonomy rule, or, equivalently, the $\beta$-taxonomy rule of eq. \eqref{e.betadotalpha}. Indeed, this is as it should be, since from Proposition \ref{prop:hyperbolic towers}, $\Gamma$ is a sum of $\beta$-vectors and thus lies on the $\beta$-vector/instantonic $\alpha$-vector lattice.

To end the discussion, we note that rewriting $\nabla^2 \log m$ in terms of $|\nabla\log m|^2$ and $\nabla^2 m$
allows us to deduce that the masses of principal towers are eigenfunctions of the Laplacians in asymptotic limits of the moduli space. The eigenvalues are positive rational numbers and are given by
\begin{align}\label{eq:laplaceeq2}
    \nabla^2 m= |\vec \alpha|^2 + c=\begin{cases}\left(\frac{N}{D-d}+\frac{D-2}{(D-d)(d-2)}\right)m\ ,&\text{KK-mode } d \to D\, ,\\
    \left(\frac N{2}+\frac 1{d-2}\right)m\ ,&\text{string oscillator ,}
    \end{cases}
\end{align}
where we used that $|\vec \alpha|$ approaches a constant asymptotically, namely
\begin{equation} \abs{\vec\alpha} = \begin{cases} \sqrt{\frac{D-2}{(D-d)(d-2)}}, & \text{KK-mode } d \to D, \\ \frac 1{\sqrt{d-2}}, & \text{string oscillator.} \end{cases}\, 
\end{equation}
It would be interesting to connect this with eigenvalue equation with moduli space quantum mechanics, in the sense of \cite{Anchordoqui:2025izb,Anchordoqui:2026nit}.

This concludes our general overview of the hyperbolic towers property. We devote the remainder of this work to an extensive study of Proposition \ref{prop:hyperbolic towers quantization} in examples. In each example, we compute the Laplacians of (principal) towers and relate the associated $c$-values to the $\Gamma$-vectors in each duality frame. We explicitly identify the contributing instantons and verify that their sum yields $\vec{\Gamma}$ in each case as motivated in Section \ref{ssec:axionsinstantons}. In Section \ref{s.geq16}, we illustrate the hyperbolic towers property in moduli spaces of theories preserving $Q \ge 16$ supercharges. We begin by introducing 10d and 9d maximal supergravity as motivating examples, after which we employ the group-theoretic formalism of symmetric spaces to extract the $\vec{\beta}$- and $\Gamma$-vectors from root lattices in the general case.

In Section \ref{sec:8charges}, we turn to theories with 8 supercharges, with a focus on 5d $\C{N}=1$ supergravity. Here, we find rather surprisingly that the axion circles in decompactification limits to 6d are not always visible within a single supergravity regime. This is explained by the fact that BPS masses in the 6d theory are dependent on the axions and can induce shifts in the structure constants $\C{F}_{IJK}$ as they hit zero. We briefly mention other 8 supercharges theories, namely the tensor multiplet moduli space of 6d $\C{N}=(1,0)$ theories and the hypermultiplet moduli spaces of 4d, 5d, and 6d theories with 8 supercharges. Finally, in Section \ref{sec:4charges}, we turn to 4 dimensions and study 4d $\C{N}=1$ theories using the EFT string framework of \cite{Lanza:2021udy}. We will see explicit examples for which the moduli space does not admit a principal plane that is also a global flat slice of moduli space. Nevertheless, using the technology of EFT strings, in various growth sectors with lower-dimensional principal planes, the hyperbolic towers property does indeed hold. In passing, even though it is not the focus of this paper, we show the same property is obeyed by the species scale. Our assumptions and discussion in both 4d and 5d elide the possibility of infinite-distance limits with divergent curvature recently considered in \cite{Marchesano:2023thx,Castellano:2024gwi,Marchesano:2024tod,Blanco:2025qom,Castellano:2026bnx,Aoufia:2026mqb}. We comment briefly on how this behavior reflects on the Laplacians of rigid sectors, deferring a full analysis to future work.

\section{\texorpdfstring{$Q\ge 16$}{Q>=16} supercharges and symmetric moduli spaces}\label{s.geq16}
We begin by studying examples with 16 and 32 supercharges. Our analysis is pedagogical, and we analyze 10d IIB and 9d maximal supergravity in particular depth. For lower-dimensional examples, and for the case of 16 supercharges, we use the techniques of symmetric moduli spaces as recently employed in \cite{Baines:2025upi,Baines:2026aug}. As a byproduct, the analysis also extends to cases with less or no supersymmetry but whose moduli space is a symmetric space.

\subsection{10d Type IIB}
Type IIB string theory in 10d is a nontrivial example that illustrates many of the main points of our analysis, because it has both a dilaton and axion, with a nontrivial geometry, and the Laplacians of the logarithms of many masses and tensions are nonzero. Let us express the axiodilaton with complexified notation
\begin{align}
    \tau=\tau_1+i\tau_2\, ,\qquad \tau_1\simeq \tau_1+1\, ,
\end{align}
with $\tau_1$ the axion and $\tau_2$ the exponentiated dilaton, and the low-energy effective Lagrangian density and moduli-space metrics are
\begin{align}
    \mathcal L=\frac{1}{2}R-\frac{d\tau_1^2+d\tau_2^2}{2\tau_2^2},\qquad  ds^2=\frac {d\tau_1^2+d\tau_2^2}{\tau_2^2}.
\end{align}

In Planck-units, the 1/2 BPS $(p,q)$ fundamental strings (and also fivebranes) have tensions proportional to
\begin{align}
    T_{p,q}\propto \frac{|p+\tau q|}{\sqrt{\tau_2}}\,.
\end{align}
As was shown in \cite{Etheredge:2023usk,Etheredge:2023zjk}, one can compute the lengths of the $\alpha$-vectors of these branes given their tension and the metric, and these branes have constant-length $\alpha$-vectors everywhere in the moduli space, with lengths given by
\begin{align}
    \alpha_\text{$(p,q)$}^2=\frac 12\,.
\end{align}
This implies, by \cite{Etheredge:2023zjk}, that each brane's tension is an exponentiation of a distance function. Furthermore, as was observed in \cite{Etheredge:2023zjk}, one can compute the Laplacian of the logarithm of each of these branes tension, and the values of these Laplacians are \footnote{There is a small typo in version 1 of \cite{Etheredge:2023zjk}, where one of the Laplacians is off by a factor of $\frac 12$.}
\begin{align}
    \nabla^2 \log T_{p,q}=1\, .
\end{align}
This matches the expectation, for strings, that Laplacians of logarithms of string tensions are integer in perturbative string limits. Furthermore, since the mass of a string oscillator scales with the square root of the tension of the string, $\log m_\text{osc}\sim \frac 12 \log T_\text{str}$, this implies that the oscillator modes of $(p,q)$ strings satisfy
\begin{align}
    \nabla^2 \log m_\text{osc}=\frac 12\,,\qquad \alpha_\text{osc}^2=\frac 18\, .
\end{align}
In the weak coupling limit of IIB string theory, the hyperbolic towers condition is satisfied by the fundamental string oscillator.

Let us now connect this with $\Gamma$-vectors, instantons, and axion decay constants. One can canonically normalize the dilaton at each point in moduli space, and the metric can be expressed in the form
\begin{align}
    ds^2=d\phi^2+e^{-2\Gamma}d\theta^2
\end{align}
where $\phi$ is the canonically normalized dilaton, $\theta$ is the axion, and $\Gamma=\sqrt 2$ controls the shrinking of the axion decay constant. The $\Gamma$-vector associated with this metric, is given by
\begin{align}
    \vec\Gamma=\sqrt 2 \hat \phi\, .
\end{align}
And, since the $\alpha$-vector of the fundamental string oscillator is given by $\vec \alpha_\text{F1 osc}=\frac 1{\sqrt 8}\hat \phi$, we have that the relationship between the $\alpha$-vector of this oscillator and the $\Gamma$-vector are related to the Laplacian of the logarithm of the oscillator's mass,
\begin{align}
    \nabla^2 \log m_\text{F1 osc}=\vec \Gamma \cdot \vec \alpha_\text{F1 osc}=\frac 12\, .
\end{align}

Finally, note that this $\Gamma$-vector is also related to the instantons of the theory. In this example, there is only a single instanton, the D$(-1)$-brane, and the $\Gamma$-vector is simply equal to the $\beta$-vector for this instanton. The $\beta$-vector for the D$(-1)$-brane, and thus D$(-1)$-brane instanton $\alpha$-vector $(-\vec\nabla \log S_\text{D$(-1)$})$, are related by
\begin{align}
    \vec\beta = -\vec\nabla \log f=\vec\nabla \log S_\text{D$(-1)$}\, .
\end{align}
Since in this case $\vec\Gamma = \vec\beta$, this establishes Propositions \ref{prop:hyperbolic towers} and \ref{prop:hyperbolic towers quantization} in this example.

Given the claims of our paper, there is a large amount of rigidity and interdependence between Laplacians, $\Gamma$-vectors, and instanton $\beta$-vectors. In fact, one could ``forget'' that IIB string theory has some objects, and recover them given the claims of our paper. We now explore some of these thought experiments with this example.
\begin{itemize}
    \item One could forget the instantons and their $\beta$-vectors. But, knowing that the fundamental string oscillators have $c=1/2$ and $\vec \alpha=\frac 1{\sqrt 8}\hat \phi$, one could find that there must be an instanton with $-\vec\nabla \log f=\sqrt 2 \hat \phi$.
    \item One could forget the $c$-value of the fundamental string oscillators, but remember that $\vec \alpha=\frac 1{\sqrt 8}\hat \phi$ and that $-\vec\nabla \log f=\sqrt 2 \hat \phi$, and thus rederive that $c=1/2$.
    \item One could forget the $\alpha$-vectors of the fundamental string oscillators, but know that $c=1/2$ and $-\vec\nabla \log f$, and then rederive that $\vec \alpha_\text{osc}=\frac 1{\sqrt 8}\hat \phi$.
\end{itemize}

These three thought-experiments demonstrate how tightly-constrained the objects of $\vec \alpha$, $c$, and $\vec\beta = -\vec\nabla \log f$ are.

\subsection{Other 10d theories}
The other 10d 32 and 16 supercharge string theories are considerably simpler, because here the moduli spaces are just 1d. For example, asymptotically, all of the known branes have tensions that are just exponentiations of the canonically normalized moduli,
\begin{align}
    T=T_0 \exp\left(-\alpha \phi \right),
\end{align}
where $\phi$ is the canonically normalized dilaton.

\textit{}
Since the moduli spaces of these theories are just the real lines, and the metric is just the Euclidean metric, the Laplacian of the logarithm of all of these branes is zero,
\begin{align}
    \nabla^2\log T=0\, .
\end{align}
Thus, since these branes also have constant-length $\alpha$-vectors, all of the branes here are trivially hyperbolic branes. In this case, the exponential decay rate of the volume of moduli space is zero, and thus their volumes are infinite.

\subsection{9d maximal supergravity}
Consider reducing IIB string theory on a circle. The metric on moduli space for the radion-dilaton-axion ($\rho$, $\tau_2$, $\tau_1$) is given by the Lagrangian density
\begin{align}
    \mathcal L=\frac 12 R-\frac{1}{2}(\partial \rho)^2-\frac{(\partial \tau_1)^2+(\partial \tau_2)^2}{2\tau_2^2}.
\end{align}
The 1/4 BPS particles have masses given by \cite{Obers:1998fb, Etheredge:2022opl}
\begin{align}
    m_{p,q,w}=c_{p,q} e^{-\sqrt{\frac{9 }{14}}\rho} \sqrt{\frac{(p+q \tau_1)^2+q^2 \tau_2^2}{\tau_2}}+c_we^{\sqrt{\frac{8}{7}}\rho } w \, ,
\end{align}
where $c_{p,q}$ and $c_w$ are unimportant constants. Here, $p,q$ refer to the $(p,q)$-winding numbers, and $w$ is the KK-mode number (or, from the M-theory perspective, $(p,q)$ are the KK-momenta on the two cycles of the torus, and $w$ is the winding number of the M2-brane). The 1/2 BPS condition is that either $(p,q)$ is zero and $w$ is nonzero, or vice versa. This has the form
\begin{align}
    m_{p,q,w}=m_{p,q}+m_w\, ,
\end{align}
where $m_{p,q}$ is the mass for 1/2 BPS particles from wrapped $(p,q)$-strings, and $m_w$ is the mass formula for KK-modes from IIB on a circle (or fully wrapped M2-branes from the M-theory perspective).

For a general 1/4 BPS particle (where $(p,q)$ and $w$ are simultaneously nonzero), the $\alpha$-vector does not have constant length $\alpha$-vector, and also the Laplacian of the masses are nonzero. However, upon restriction to purely $1/2$-BPS states (where $(p,q)$ is nonzero but $w$ is zero, or vice versa), things are different. One can compute that the $\alpha$-vectors of the 1/2 BPS particles all have length
\begin{align}
    \alpha_\text{1/2 BPS}^2=\frac 87\, .
\end{align}
Meanwhile, the Laplacians of the logarithms of these 1/2-BPS masses satisfy
\begin{align}
    \nabla^2 \log m_{p,q}=1,\qquad \nabla^2 \log m_w=0\, .
\end{align}
This satisfies the expectation from Proposition \ref{prop:hyperbolic towers quantization}. For $\nabla^2\log m_w$, this is the Laplacian of the KK-mode mass for a duality frame decompactifying to M-theory. There, there are no 1-form or 2-form gauge fields, and so the Laplacian must be zero, since a 2-torus has no 3-cycles. Next, for $\nabla^2 \log m_{p,q}$, these towers are KK-modes for a decompactification limit to IIA string theory. In IIA string theory, there are 1-form gauge fields. Thus, the Laplacian of the logarithm is equal to 1.

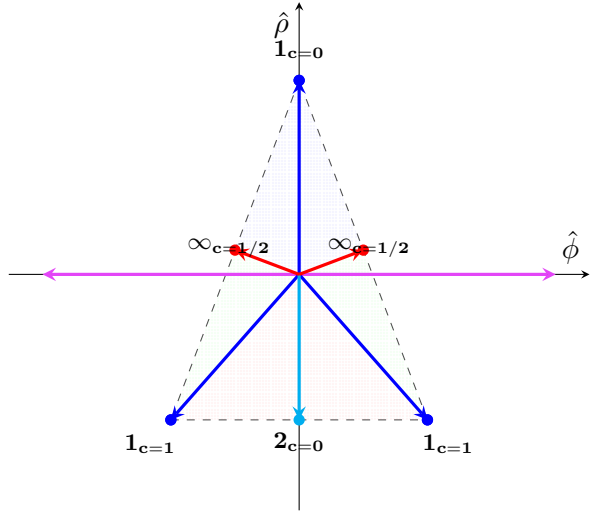
\begin{figure}[t]
\def\opaedges{1}\def\opafill{0.2}\def\opafillback{0.02}
\def\sizebig{1.2pt}\def\sizesmall{1.4pt}
    
\centering
\begin{tikzpicture}[scale=2.4,>=stealth]{
\coordinatesIsoTriangle \def\sizebig{1.5pt}\def\sizesmall{1.5pt} 
\drawIsoTriangle

\draw[->] (-1.6,0) -- (1.6,0) node[anchor=south east]{$\hat\phi$};
\draw[->] (0,-1.3) -- (0,1.5) node[anchor=north east]{$\hat\rho$};

\foreach \i/\j/\k/\l in 
{(O)/($(A)!0.5!(B)$)/(A)/red,
(O)/($(A)!0.5!(B)$)/(B)/red,
(O)/($(B)!0.5!(C)$)/(B)/green,
(O)/($(B)!0.5!(C)$)/(C)/blue,
(O)/($(C)!0.5!(A)$)/(C)/blue,
(O)/($(C)!0.5!(A)$)/(A)/green}{
\fill[pattern={Dots[radius=0.4pt,distance=1.2pt]},
pattern color=\l,opacity=\opafill]\i -- \j -- \k;
};

\foreach \i/\j/\k in {
($1.0*(A)-1.0*(B)$)/{{}}/\colGamma,
($1.0*(B)-1.0*(A)$)/{{}}/\colGamma}{
\draw[->][\coltwo,very thick,color=\k] (O) -- \i;
};

\foreach \i in {A,B,C}{\path ($(\i)$) node[circle, fill=\colone, inner sep=\sizebig]{};}; 
\foreach \i/\j/\k in {
(A)/$\mathbf{1_{c=1}}$/\colone,
(B)/$\mathbf{1_{c=1}}$/\colone,
(C)/$\mathbf{1_{c=0}}$/\colone,
($(A)!0.5!(B)$)/$\mathbf{2_{c=0}}$/\colthree,
($(B)!0.5!(C)$)/$\mathbf{\infty_{c=1/2}}$/\colinf,
($(C)!0.5!(A)$)/$\mathbf{\infty_{c=1/2}}$/\colinf
}{
\draw[->][\colinf,very thick,color=\k] (O) -- \i;
\path \i node[circle, fill=\k, inner sep=\sizebig] (currentpoint) {};
\path (O) -- (currentpoint) node[font=\footnotesize, pos=1.2] {\j};
};
}
 \end{tikzpicture}
    \caption{$\alpha$-vectors and $\Gamma$-vectors for dilaton-radion slice of IIB string theory on a circle. Here, the vertical direction is the radion direction, and the horizontal the 10d dilaton direction. The magenta arrows are principal towers, and the red arrows are $\Gamma$-vectors, which exist on the instanton lattice. Furthermore, the $c$-values for the principal towers are equal to the product of the $\alpha$-vectors with the $\Gamma$-vectors. Towers are labelled $n_c$, with $n$ being to the number of decompactifying dimensions for the corresponding KK tower, and $\infty$ signaling an emergent string limit.}
    \label{f.9dIIB}
\end{figure}

Finally, consider unwrapped fundamental strings. They have tensions given by
\begin{align}
    T_{p,q}^{(9d)}\propto \exp\left(-\frac{1}{\sqrt{14}}\rho \right)\sqrt{\frac{(p+q\tau_1)^2+q^2\tau_2^2}{\tau_2}}\, .
\end{align}
By direct computation, all of their $\alpha$ have length $2/\sqrt 7$. Thus, the $\alpha$-vectors of their oscillators have length
\begin{align}
    \alpha_\text{osc}^2=\frac 17\, .
\end{align}
Consider the dilaton-radion slice of moduli space. There are also D$(-1)$-brane instantons inherited from IIB string theory in 10d, which have length
\begin{align}
    \vec \alpha_\text{inst}=\pm \sqrt 2 \hat \phi\rightarrow -\vec\nabla \log f=\mp \sqrt 2 \hat \phi\,.
\end{align}
For the particles and strings whose $\alpha$-vectors are in the radion-dilaton plane, one can verify that
\begin{align}
    (-\vec\nabla \log T)\cdot (-\vec\nabla \log f)=\nabla^2 \log T\,.
\end{align}
This is described in Figure \ref{f.9dIIB}, where towers are labelled $n_c$, with $n$ denoting the number of decompactifying dimensions for the corresponding KK tower, and $\infty$ signaling an emergent string limit. The subscript indicates the value of $c=\nabla^2\log m$ in that limit.

\begin{figure}[ht]
 \def\opaedges{1}\def\opafill{0.2}\def\opafillback{0.02}
\def\sizebig{1.2pt}\def\sizesmall{1.4pt}
    
\centering
  \begin{tikzpicture}[scale=2.4,>=stealth]{
\coordinatesIsoTriangle \def\sizebig{1.5pt}\def\sizesmall{1.5pt} 
\drawIsoTriangle

\draw[->] (-1.6,0) -- (1.6,0) node[anchor=south east]{$\hat\phi$};
\draw[->] (0,-1.5) -- (0,1.5) node[anchor=north east]{$\hat\rho$};

\draw[dashed,thick,\colone] (-1.6,0.163849) -- (0.286627, -1.5) node[anchor=north east]{{}};
\draw[dashed,thick,\colone] (1.6,0.163849) -- (-0.286627, -1.5) node[anchor=north east]{{}};

\draw[dashed,thick,\colinf] (-1.6,-0.491556) -- (-0.847278, 1.5) node[anchor=north east]{{}};
\draw[dashed,thick,\colinf] (1.6,-0.491556) -- (0.847278, 1.5) node[anchor=north east]{{}};

\foreach \i/\j/\k/\l in 
{(O)/($(A)!0.5!(B)$)/(A)/red,
(O)/($(A)!0.5!(B)$)/(B)/red,
(O)/($(B)!0.5!(C)$)/(B)/green,
(O)/($(B)!0.5!(C)$)/(C)/blue,
(O)/($(C)!0.5!(A)$)/(C)/blue,
(O)/($(C)!0.5!(A)$)/(A)/green}{
\fill[pattern={Dots[radius=0.4pt,distance=1.2pt]},
pattern color=\l,opacity=\opafill]\i -- \j -- \k;
};

\foreach \i/\j/\k in {
($1.0*(A)-1.0*(B)$)/{{}}/\colGamma,
($1.0*(B)-1.0*(A)$)/{{}}/\colGamma}{
\draw[->][\coltwo,very thick,color=\k] (O) -- \i;
};

\foreach \i in {A,B,C}{\path ($(\i)$) node[circle, fill=\colone, inner sep=\sizebig]{};}; 
\foreach \i/\j/\k in {
(A)/$\mathbf{1_{c=1}}$/\colone,
(B)/$\mathbf{1_{c=1}}$/\colone,
(C)/$\mathbf{1_{c=0}}$/\colone,
($(A)!0.5!(B)$)/$\mathbf{2_{c=0}}$/\colthree,
($(B)!0.5!(C)$)/$\mathbf{\infty_{c=1/2}}$/\colinf,
($(C)!0.5!(A)$)/$\mathbf{\infty_{c=1/2}}$/\colinf
}{
\draw[->][\colinf,very thick,color=\k] (O) -- \i;
\path \i node[circle, fill=\k, inner sep=\sizebig] (currentpoint) {};
\path (O) -- (currentpoint) node[font=\footnotesize, pos=1.2] {\j};
};}
 \end{tikzpicture}
    \caption{Using $c$-values of principal towers to bootstrap $\Gamma$-vectors.
    Here, the dashed blue and red lines are the instanton lines for the $1_{c=1}$ winding modes, and the $\infty_{c=1/2}$ oscillators, respectively. Here, the $\beta$-vectors must lie at their points of intersection, which are the magenta lines, reproducing the structure observed in 9d maximal supergravity. 
    }
    \label{f.9dIIBtoy}
\end{figure}
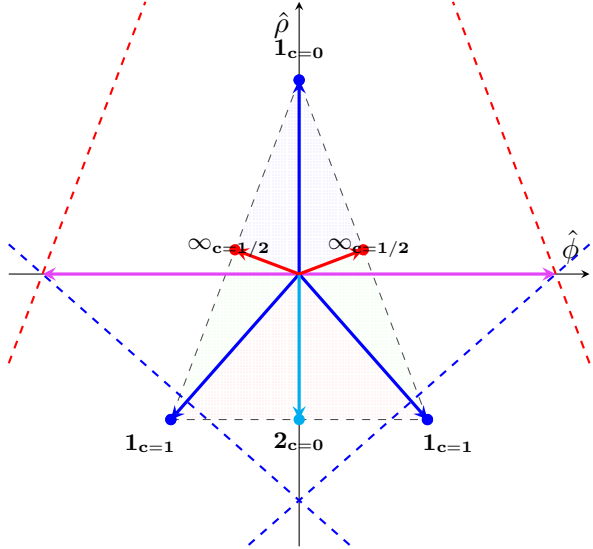

\subsubsection{Taxonomic analysis}
It is interesting to perform a taxonomic analysis of these examples. Consider Figure \ref{f.9dIIB}. Suppose we forgot the higher-UV origin of this theory (i.e.\ forgot that it came from 10d IIB on a circle, or IIA on a circle, or M-theory on a 2-torus). An interesting question is, how much can we recover about the higher-dimensional theories from just this figure? Can we find out what the geometry of the higher-dimensional theory is? It turns out that we can, and this figure, and our general story, has recursive qualities.

To illustrate this, consider decompactifying in the direction of the KK-mode $1_\text{c=0}$. This is a decompactification to a 10d theory. Now, the tower-polytope in the 10d theory, following the recursive-properties of \cite{Etheredge:2024tok} is just the tower-polytope with 2-emergent string limits of a 1d moduli space. Next, since the instanton $\alpha$-vectors are just the magenta lines, and at the $P=0$ lattice site of the radion of this decompactification, these magenta lines uplift to instantons in the 10d theory. Thus, in the 10d theory, the $c$-values of the two oscillator modes is again $1/2$, so we recover the $c$-values of 10d IIB string theory from just this example. 

Next, consider traveling in the direction of one of the bottom left $1_{c=1}$ KK-modes. This is a decompactification to a 10d theory, and the tower-polytope can be computed using the techniques of \cite{Etheredge:2024tok}, and is generated by a decompactification limit to 11d and a string oscillator. However, in this frame, none of the instanton $\alpha$-vectors uplift to instantons in the 10d theory. Thus, the $c$-values of the principal towers in the 10d theory are all 0.

As a thought experiment, could we forget the $\Gamma$-vectors of this theory, and use the $c$-values to bootstrap it? First, consider the top KK mode, $1_{c=0}$. This means that this KK-mode is orthogonal to the theory's $\Gamma$-vector in this mode's frame. Thus, if there is a $\Gamma$-vector for this frame, it must reside on the $\hat\phi$-axis in Figure \ref{f.9dIIBtoy}. Next, the $\infty_{c=1/2}$ is in a frame where the $\Gamma$-vector must reside on the red line in Figure \ref{f.9dIIBtoy}, in order for $\vec \Gamma \cdot \vec\alpha_{\infty_{c=1/2}}=1/2$. Thus, if the $1_{c=0}$ and $\infty_{c=1/2}$ vectors are to share the same $\Gamma$-vector, it must be one of the magenta lines. Next, consider one of the $c=1$ KK-modes at the bottom of the figure. The $c=1$ condition implies that, in their frame, the $\Gamma$-vector must be on one of the blue dashed lines. Since these oscillator modes and the KK-modes share the same frame, the $\Gamma$-vectors must be on the magenta arrow.

\subsection{General case: symmetric moduli spaces}\label{sec:sym}
The 10d and 9d examples above have low-dimensional moduli spaces and can be analyzed directly. In lower dimensions, the relevant moduli spaces for theories with at least 16 supercharges are often locally symmetric, allowing their $\Gamma$-vectors and $c$-values to be computed systematically using representation theory.

A wide range of EFTs have moduli spaces which are locally symmetric. In particular, this includes every theory with 16 or more supercharges, and several classes of theories with less (or no) supersymmetry. In \cite{Baines:2025upi}, under mild assumptions on the spectrum, it was proven that theories with locally symmetric spaces automatically satisfy the DC, with the decay rate of the mass of the leading tower being related to the length of the weight of the representation under the duality group of the corresponding state. 
In this section, we explain how the same symmetric space structure leads to a proof of Proposition \ref{prop:hyperbolic towers}, and how the $\Gamma$-vectors and therefore the $c$-values for the leading towers can be computed systematically.

Furthermore, the precise relation between weights and $\alpha$-vectors was exploited in \cite{Baines:2026aug} to constrain the possible symmetric moduli spaces compatible with the ESC.
The resulting Laplacian coefficients depend on the normalization of the symmetric-space metric, and hence on the physical lengths of the roots and weights.
Proposition \ref{prop:hyperbolic towers quantization} therefore imposes additional constraints on this normalization, thereby refining the classification of symmetric moduli spaces obtained in \cite{Baines:2026aug}. 

The main idea is that each infinite-distance cusp of the moduli space can be associated with a rational parabolic subgroup $P$ of the duality group, and the directions that shrink near the cusp are controlled by the positive restricted roots $\{\beta\}$ appearing in the nilpotent part of $P$. These roots define a geometric vector\footnote{Conforming to the notation of \cite{Baines:2025upi,Baines:2026aug}, and to avoid cluttering, we will drop the vector symbol on Lie algebra elements such as roots and weights.}:
\begin{align}
\vec\Gamma_P = \sum_{\beta \in \Sigma^+_P} m_{\beta}\beta = 2 \rho_P \,,
\end{align}
where $m_{\beta}$ is the multiplicity of the restricted root $\beta$ and $\rho_P$ is the Weyl vector of $P$. The restricted roots therefore give the scaling vectors of the shrinking axionic directions, and, when these directions admit the instanton interpretation of Section \ref{ssec:axionsinstantons}, they coincide with the corresponding instanton $\beta$-vectors.

If the mass of the leading tower is controlled by a weight $w_{\text{lead}}$, then the Laplacian is obtained by taking its inner product with the $\Gamma$-vector:
\begin{align}
c = \vec\Gamma_P \cdot w_{\text{lead}}
\end{align}
Thus, for locally symmetric spaces, the hyperbolic towers coefficient can be
computed directly from the restricted root system and the representation of the tower. We now prove this in detail in the following, discussing the essential definitions. The interested reader can consult \cite{Baines:2025upi,Baines:2026aug} for an extended treatment.

A locally symmetric moduli space $\mathcal{M}$ is assumed to be an arithmetic quotient of the form $\Lambda \backslash G /K$, where $G$ is a real reductive Lie group, $K \subset G$ a maximal compact subgroup, and $\Lambda \subset G$ an arithmetic lattice.
If $G=\mathbb{R}^p \times \prod_{a=1}^s G_a$ and $K=\prod_{a=1}^s K_a$, we can express its metric $G_{ij}(\phi)$ as:
\begin{align}\label{eq:Gij}
G_{ij}(\phi)\d\phi^i \d\phi^j = ds_0^2 + \sum_{a=1}^s \kappa_a d\widetilde{s}_a^2 \, ,
\end{align}
where $ds_0^2$ is the flat metric on the $\mathbb{R}^p$, while each $d\widetilde{s}_a^2$ is a locally Einstein metric on $G_a/K_a$. 
Here the constants $\kappa_a$ fix the normalization of the metric on each
irreducible symmetric-space factor. In what follows, all inner products between
roots and weights are understood with respect to this physical normalization.
The factor $ds_0^2$ is the globally flat part of the metric, but it should not in general be identified with the full principal plane $\cal P$ defined in Section \ref{sec:2}. The latter is a flat slice of the complete symmetric space and may also involve flat directions within the factors $G_a/K_a$.
The flat factor $\mathbb R^p$ does not contribute restricted roots. Therefore, the cusp data and the volume decay discussed below are controlled by the non-compact semi-simple factors $G_a/K_a$.

When $\Lambda$ is arithmetic, the infinite-distance cusps of $\mathcal M$ are labelled by $\Lambda$-conjugacy classes of rational parabolic subgroups $P\subset G$.  For a given cusp, we choose such a parabolic $P$ and use its Langlands decomposition $P=N_P A_P M_P$.
The nilpotent factor $N_P$ contains precisely the directions whose metric
shrinks exponentially as one moves toward the cusp, and are thus identified with the nilmanifold fiber in Proposition \ref{prop:nilmanifold}.

Let $\mathfrak g=\mathfrak k\oplus\mathfrak p$ be the Cartan decomposition, and let $\mathfrak a\subset\mathfrak p$ be a maximal abelian subalgebra. The restricted roots are the nonzero elements
$\beta\in\mathfrak a^\ast$ such that
\begin{align}
    \mathfrak{g}_\beta = \left\{ X \in \mathfrak{g} : [H,X] = \beta(H) X , \forall H\in \mathfrak{a} \right\} .
\end{align}
For a chosen parabolic $P$, the Lie algebra of $N_P$ decomposes as a sum of restricted-root spaces, 
\begin{align}
\mathfrak n_P=\bigoplus_{\beta\in\Sigma_P^+}\mathfrak g_\beta \, ,
\end{align}
where $\Sigma_P^+$ is the set of positive restricted roots selected by the cusp. The Weyl vector of $P$ can be defined as the half sum of positive restricted roots weighted by their multiplicities:
\begin{align}
\rho_P = \frac12 \sum_{\beta\in \Sigma^+_P}m_{\beta} \beta  \,,\quad m_{\beta}=\text{dim}(\mathfrak{g}_{\beta}) \, .
\end{align}
For a minimal parabolic, $\Sigma_P^+$ is the full set of positive restricted
roots, and this gives the standard large-radius volume growth for a geodesic ball of radius $R$:
\begin{align}
\text{Vol}\,B(R) \sim R^{\frac{r-1}{2}}e^{2|\rho_P|R}\,,\qquad r=\text{rank}_{\mathbb{R}}(G)\,,
\end{align}
where $\|\rho_P\|$ is computed using the normalization of the physical symmetric-space metric.
For a general cusp, the same vector $\rho_P$ controls the exponential shrinking of the nilpotent directions in the cusp.

We choose the sign of the $\vec\Gamma_P$ so that it measures the rate at which the transverse directions shrink as we move toward the cusp, compatibly with eq. \eqref{eq:detGGamma}. With this convention, the rate of shrinking of the volume form is identified $\vec\Gamma_P=2\rho_P$. Equivalently, $\vec\Gamma$ is minus the gradient of the logarithm of the transverse volume density:
\begin{align}
\vec\Gamma_P = -\vec\nabla\log\sqrt{\det g} = 
2\rho_P =
\sum_{\beta\in\Sigma_P^+}m_{\beta}\,\beta \, ,
\end{align}
where $g$ is the metric in cusp-adapted coordinates, and the gradient is taken along the non-compact directions parametrizing the approach to the cusp. Notice that this precisely agrees with eq. \eqref{eq:gammapartiallogg}. 

Correspondingly, for a ray generated by $H\in\mathfrak{a}$, the shrinking nilpotent directions are the root spaces with $\beta(H)>0$:
\begin{align}\label{eq:gamma_beta}
\vec\Gamma(H) = \sum_{\beta(H)>0}m_{\beta}\beta \,.
\end{align}
As discussed in \cite{Baines:2025upi}, let a state transform in an irreducible representation $R_{\lambda}$ of $G$ with highest weight $\lambda$. (Note that the sum over $\beta$ for $\beta(H) > 0$ corresponds precisely to the sum over $\beta$-vectors lying in the dual cone to a duality frame containing the direction $H$, exactly as in Proposition \ref{prop:hyperbolic towers}.) For a unit-normalized direction $H$ in the interior of the positive Weyl chamber, the asymptotic mass is controlled by the highest weight:
\begin{align}\label{eq:max}
m(H,t)\sim &\, e^{-t\langle \lambda,H\rangle}\,,\qquad \langle \lambda,H\rangle =
\max_{w\in \mathcal{W}_{\lambda}}\langle w,H\rangle \, ,
\end{align}
where $\mathcal{W}_{\lambda}$ denotes the set of weights of $R_{\lambda}$.
With the convention $\vec\alpha=-\vec\nabla\log m$, the corresponding tower vector is therefore
\begin{align}
\vec\alpha = \lambda \, ,
\end{align}
and its $c$-value reads 
\begin{align}
c(H) = \vec\Gamma(H) \cdot \lambda = \sum_{\beta(H)>0}m_{\beta}\braket{\beta,\lambda}\,.
\end{align}
Directions on the boundary of the Weyl chamber require projecting the leading weights onto the corresponding asymptotic direction. We discuss this explicitly for split groups in Section \ref{sec:split}.

This establishes Proposition \ref{prop:hyperbolic towers} for generic rays in symmetric moduli spaces, having therefore reduced the computation of the $c$-values to the identification of the restricted roots of the duality group and the representation under which the relevant towers transform.

\subsection{Examples}\label{sec:examples}
To make the discussion above concrete, we now turn to explicit classes of examples. We first analyze split groups, including all theories with 32 supercharges, and then turn to the non-split family $SO(k,k+r)$ relevant to theories with 16 supercharges.. 

\subsubsection{Split groups (including all theories with 32 supercharges)}\label{sec:split}

If the group $G$ is in split-form, the restricted roots are the usual roots with multiplicity $1$.
We can express $\Gamma(H_i)$ for a generic Lie algebra in terms of the Gram matrix of its fundamental weights.
We denote by $H_i$ the direction orthogonal to all the simple roots except the $i$-th one,
\begin{align}
\beta_j(H_i)=\delta_{ij}\,.
\end{align}
A direction $H$ in the closed positive Weyl chamber selects a face $F_I$ of the weight polytope: $F_I$ contains the set of weights whose pairing with $H$ is maximal. 
Let $\Delta$ denote the set of simple roots, then the face $F_I$ is labeled by the subset $I\subset \Delta$ vanishing on $H$:
\begin{align}
I(H) = \lbrace
\beta_i \in \Delta | \beta_i(H)=0\rbrace \,.
\end{align}
A direction in the interior of the cone associated with $F_I$ can be written as
\begin{align}
H = \sum_{i \notin I} s_i H_i\,,\quad \text{ for }s_i>0\,.
\end{align} 

Since $H$ lies in the closure of the positive Weyl chamber, every positive root $\beta\in\Phi^+$ satisfies $\beta(H)\geq 0$. The roots that do not contribute to $\Gamma(H)$ in \eqref{eq:gamma_beta} are those for which $\beta(H)=0$:
\begin{align}
\vec\Gamma_I = 
\sum_{\beta \in \Phi^+\,|\,\beta(H)>0}\beta
=\sum_{\beta \in \Phi^+}\beta - \sum_{\beta \in \Phi^+\,|\,\beta(H)=0}\beta
= 2\rho_G - 2 \rho_{G_I}\,,
\end{align}
where $\rho_G$ is the Weyl vector of $G$ (half-sum over all the positive roots), and $\rho_{G_I}$ the Weyl vector of the subgroup of $G$ generated by the simple roots in $I$.
If $G$ is semi-simple, then the Weyl vectors can be expressed as the sum of the fundamental weights of $G$ and $G_I$:
\begin{subequations}
\begin{align}
\rho_G =&  \sum_{i}\omega_i,\\
\rho_{G_I} =& \sum_{i\in I}\omega^{[I]}_i
= \sum_{i\in I}\left(
\omega_i - 
\sum_{j,k \notin I}W_{ij}
(W^{\perp}_I)^{-1}_{jk} \omega_k
\right)\,,
\end{align}
\end{subequations}
where we denote the Gram matrix of the fundamental weights of $G_I$ as
\begin{align}
(W^\perp_I)_{ij}= \langle \omega_i, \omega_j \rangle \, \quad \text{for $i, j \notin I$}\,.
\end{align}
Therefore, multiplying by a simple coroot we have that
\begin{subequations}   
\begin{align}
\rho_{G}.\beta^{\vee}_j  = &1, \quad \forall j,\\
\rho_{G_I}.\beta^{\vee}_j = &\begin{cases} 
1 & \forall j \in I \\
1-\sum_{m\notin I}\sum_{i }W_{im}(W^{\perp}_I)^{-1}_{mj} & \forall j \notin I
\end{cases}\,,
\end{align}
and
\begin{align}\vec\Gamma_I \cdot \beta^{\vee}_j = \begin{cases}
0\quad &\forall j \in I\\
2\sum_{m\notin I}\sum_{i }W_{im}(W^{\perp}_I)^{-1}_{mj}\quad &\forall j \notin I
\end{cases}
\, .\end{align}
\end{subequations}
Thus, we can express $\vec\Gamma_I$ as a linear combination of the fundamental weights $\lbrace \omega_i\rbrace_{i \notin I}$.

The value of the $\Gamma$-vectors is independent of the $s_i$ values in the $H$ expansion on $H_i$ (as long as they are positive), justifying its characterization in terms of the label $I$ ($\Gamma(H)$ is constant for all directions $H$ on the interior of the cone associated with $F_I$). We can express it in the basis of fundamental weights:
\begin{align}
\vec\Gamma_I = 
\sum_{k\notin I}r_k^{(I)}\omega_k\,\text{ with }
r_k^{(I)}= 2\sum_{i}\sum_{j\notin I}W_{ij}(W^{\perp}_I)^{-1}_{jk}=
\vec\Gamma_I \cdot \beta^{\vee}_k 
\in\mathbb{Z}_{\geq 2}\,.
\end{align}

If a state transforms in a semi-simple representation whose highest weight is \mbox{$\lambda = \sum_j \lambda_j \omega_j$}, the associated $\alpha$-vector for the leading tower in the direction $H$ can be expressed as \cite{Baines:2026aug}:
\begin{align}
\vec\alpha_I = 
\sum_{j,k\notin I}\sum_{i}\lambda_i W_{ij}(W^{\perp}_I)^{-1}_{jk}  \omega_k \,.
\end{align}
We notice that $\vec\Gamma_I$ is the $\alpha$-vector along $F_I$ for a state transforming in the irreducible  representation with highest weight $2\rho_G$, i.e.\ whose Dynkin labels are all
equal to $2$.
The value of $c$ for infinite distance limits along $F_I$ is then given by the product
\begin{align}
c_I = \vec\alpha_I \cdot \vec\Gamma_I  = 
\sum_{i,\ell}
\sum_{j,k\notin I}
2 W_{ij}  (W_I^{\perp})^{-1}_{jk}W_{k\ell} \lambda_{\ell}=
\sum_{i,\ell}
2 V^I_{i\ell} \lambda_{\ell} = \sum_{k\notin I}
 r_k^{(I)} \sum_{\ell} W_{k\ell} \lambda_{\ell}
\,, \label{eq:cI}
\end{align}
where we defined 
\begin{equation}
V^I_{i\ell}=\sum_{j,k \notin I} W_{ij}  (W_I^{\perp})^{-1}_{jk}W_{k\ell}\, .
\end{equation}
If $F_I$ is just a vertex, we take $I=\emptyset$, $(W_{\emptyset}^{\perp})^{-1}_{ij}=(W^{-1})_{ij}$ and then $r_k^{(I)} = 2\quad\forall \,k$, such that
\begin{align}
\vec\Gamma_{\emptyset} = 
\sum_{i} 2 \omega_i  \,,\quad
\vec\alpha_{\emptyset} =  \sum_{{i}} \lambda_i  \omega_i \,, \quad
c_{\emptyset} = \sum_{i,j} 2\lambda_i W_{ij}  \, .
\end{align}

Since the Dynkin labels $\lambda_i$ are integral, the values of $c_{\emptyset}$ must be quantized depending on the rational lattice spacing of $\lbrace \sum_{j} 2W_{ij} \rbrace$:
\begin{align}\label{eq:c0forSimpleGroups}
c_{\emptyset} =& \,\,\frac{2N}{\ell}\,,\quad N\in\mathbb{Z}\,,\nonumber\\
\ell =& \begin{cases}
1 & \text{ for }G=\begin{array}{c} E_{8(8)},\,E_{6(6)},\,SL(2k+1,\mathbb{R}),\,
SO(4k,4k),\,SO(4k+1,4k+1)\end{array}\\
2 & \text{ for }G=\begin{array}{c} F_{4(4)},\,E_{7(7)},\,SL(2k,\mathbb{R}),\,
Sp(2k,\mathbb{R}),\,\\
SO(4k+2,4k+2),\, SO(4k+3,4k+3),\,SO(2k,2k+1)\end{array}
\\
3 & \text{ for }G=G_{2(2)}\\
4 & \text{ for }G=SO(2k+1,2k+2)
\end{cases}\,,
\end{align}
where we assumed that the long roots are normalized to length squared $2$. This is a convention for the abstract root system: the physical normalization is fixed by the sigma-model metric through the coefficients $\kappa_a$ in \eqref{eq:Gij}. 

Since $\vec\alpha_I$ lies in the subspace normal to the roots in $I$, one may compute the $c_I$ for any face $F_I$ using the $\Gamma$-vector associated to the direction of the highest weight 
\begin{align}
c_I=\vec\Gamma_{\emptyset}\cdot\vec\alpha_I = 2\rho_G\cdot\vec\alpha_I  = \sum_{i}2 \omega_i \cdot \vec\alpha_I \, ,
\label{eq:cI_with_emptyset}
\end{align}
as can also be inferred from eq. \eqref{eq:cI}.

For a codimension 1 facet we have $I=I_s \equiv\Delta\setminus\lbrace\beta_s\rbrace$, for some simple root $\beta_s$. 
\begin{align}
\begin{aligned}
    \vec\alpha_{I_s} =& 
  \frac{\sum_{k}\lambda_k W_{ks}}{W_{ss}}
\omega_s \,,\quad
\vec\Gamma_{I_s} = r_s^{(I_s)}\omega_s
\, , \\
c_{I_s}  =&  r_s^{(I_s)}\sum_k \lambda_k W_{ks}
= r_s^{(I_s)}|\vec\alpha_{I_s}|\sqrt{W_{ss}}
\, ,
\end{aligned}
\end{align}
with $r_s^{(I_s)}= 2 \sum_i \frac{W_{is}}{W_{ss}}\in\mathbb{Z}_{\geq 2}$. 
Implying that for codimension-1 limits, the $\Gamma$-vectors are always aligned with the \mbox{$\alpha$-vector} of the leading tower.

If the sharpened DC is satisfied, then
\begin{align}
|\vec\alpha_{I_s}| = \sqrt{\frac{d+n-2}{n(d-2)}}\,.
\end{align}
and we get 
\begin{align}
c_{I_s} 
= r_s^{(I_s)}\sqrt{W_{ss}\frac{d+n-2}{n(d-2)}}
\, ,\end{align}
for decompactification limits and
\begin{align}
c_{I_s} 
= r_s^{(I_s)}\sqrt{\frac{W_{ss}}{d-2}}
\, .
\end{align}
 for perturbative string limits.

Compatibility with the quantization rule of Proposition \ref{prop:hyperbolic towers quantization} is an additional constraint: once a face is identified with a decompactification of $n$ dimensions or with an emergent string limit, the $c$-value must lie respectively in $\frac{1}{n}\mathbb{Z}_{\geq 0}$ or $\frac{1}{2}\mathbb{Z}_{\geq 0}$. These conditions are verified in the supergravity examples below. We now specialize the above discussion to examples which systematically generalize the Type IIB and 9d maximal supergravity computations above to lower dimensions.

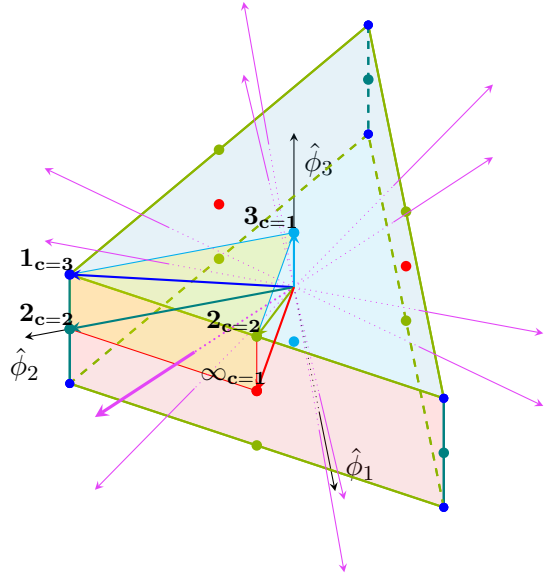
\begin{figure}[t]\centering
\begin{tikzpicture}[scale=3.7,,>=stealth,tdplot_main_coords,tdplot_rotated_coords
]
\coordinatesTriangularPrismB
\def\opaedges{1}\def\opafill{0.1}\def\opafillback{0.02}
\def\sizebig{1.2pt}\def\sizesmall{1.4pt}

\draw[->] (0,0,0.707) -- (0,0,2) node[anchor=north west]{$\hat\phi_3$};
\draw[->](0.788675, 0.211325, 0.)  --  ( 0.94641, 0.25359, 0.) node[anchor=north]{$\hat\phi_2$};
\draw[line cap=round, dash pattern=on 0pt off 2pt](0,0,0)  -- (-0.122009, 0.455342, 0)  node[anchor=south west]{{}};
\draw[->](-0.122009, 0.455342, 0)  -- (-0.2, 0.74641, 0)  node[anchor=south west]{$\hat\phi_1$};

\foreach \i/\j/\k in {A/B/C}{
\fill[\colthree,opacity=\opafill](\i) -- (\j) -- (\k);
\path  ($0.333*(\i)+0.333*(\j)+0.333*(\k)$)  node[circle, fill=\colthree, inner sep=\sizesmall]{{}};
};
\foreach \i/\j/\k in {A'/B'/C'}{
\fill[\colthree,opacity=\opafillback](\i) -- (\j) -- (\k);
\path  ($0.333*(\i)+0.333*(\j)+0.333*(\k)$)  node[circle, fill=\colthree, inner sep=\sizesmall]{{}};
};
\foreach \i/\j/\k/\l in {A/A'/B'/B}{
\fill[\colinf,opacity=\opafill](\i) -- (\j) -- (\k) -- (\l);
\path ($0.25*(\i)+0.25*(\j)+0.25*(\k)+0.25*(\l)$) node[circle, fill=\colinf, inner sep=\sizesmall]{{}};
};
\foreach \i/\j/\k/\l in {B/B'/C'/C,A/A'/C'/C}{\fill[\colinf,opacity=\opafillback](\i) -- (\j) -- (\k) -- (\l);
\path ($0.25*(\i)+0.25*(\j)+0.25*(\k)+0.25*(\l)$) node[circle, fill=\colinf, inner sep=\sizesmall]{{}};
};

\foreach \i/\j in {A'/B',A/B,B/C,C/A}{\draw[\coltwo, thick] (\i) -- (\j);};
\foreach \i/\j in {A/A',B/B'}{\draw[\coltwoALT, thick] (\i) -- (\j);};
\foreach \i/\j in {B'/C',C'/A'}{\draw[\coltwo, thick,dashed] (\i) -- (\j);};
\foreach \i/\j in {C/C'}{\draw[\coltwoALT, thick,dashed] (\i) -- (\j);};

\foreach \i in {A,B,C,A',B',C'}{\path ($(\i)$) node[circle, fill=\colone, inner sep=\sizebig]{{}};};

\foreach \i/\j in {A'/B',A/B,B/C,C/A}{\draw[\coltwo, thick] (\i) -- (\j);\path  ($0.5*(\i)+0.5*(\j)$) node[circle, fill=\coltwo, inner sep=\sizesmall]{{}};};
\foreach \i/\j in {A/A',B/B'}{\draw[\coltwoALT, thick] (\i) -- (\j);\path  ($0.5*(\i)+0.5*(\j)$) node[circle, fill=\coltwoALT, inner sep=\sizesmall]{{}};};
\foreach \i/\j in {B'/C',C'/A'}{
\draw[\coltwo, thick,dashed] (\i) -- (\j);
\path  ($0.5*(\i)+0.5*(\j)$) node[circle, fill=\coltwo, inner sep=\sizesmall]{{}};};
\foreach \i/\j in {C/C'}{
\draw[\coltwoALT, thick,dashed] (\i) -- (\j);
\path  ($0.5*(\i)+0.5*(\j)$) node[circle, fill=\coltwoALT, inner sep=\sizesmall]{{}};};
\foreach \i in  {A,B,C,A',B',C'}{\path ($(\i)$) node[circle, fill=\colone, inner sep=\sizebig]{{}};};

\foreach \i in {A,B,C,A',B',C'}{\path ($(\i)$) node[circle, fill=\colone, inner sep=\sizebig]{{}};};

\fill[color=yellow,opacity=0.2] (X) -- (Y) -- (Q) -- (Z);
\draw[color=\colinf] (Y) -- (Q) -- (Z);
\fill[color=yellow,opacity=0.2] (X) -- (Y) -- (W);
\draw[color=\colthree]  (X) -- (W) -- (Y);

\coordinate (GA) at  (0.333333, 0.333333, 0.235702);
\coordinate (GB) at  (0.333333, 0.333333, -0.235702);
\coordinate (GC) at  (-0.333333, -0.333333, 0.235702);
\coordinate (GD) at  (-0.333333, -0.333333, -0.235702);
\coordinate (GE) at  (0.455342, -0.122008, 0.235702);
\coordinate (GF) at  (0.455342, -0.122008, -0.235702);
\coordinate (GG) at  (-0.455342, 0.122008, 0.235702);
\coordinate (GH) at  (-0.455342, 0.122008, -0.235702);
\coordinate (GI) at  ( -0.122008,0.455342, 0.235702);
\coordinate (GJ) at  ( -0.122008,0.455342, -0.235702);
\coordinate (GK) at  (0.122008,-0.455342,  0.235702);
\coordinate (GL) at  (0.122008, -0.455342, -0.235702);

\foreach \i/\j in {
(GB)/($2*(GB)$),
(GC)/($2*(GC)$),
(GD)/($2*(GD)$),
(GE)/($2*(GE)$),
(GF)/($2*(GF)$),
(GG)/($2*(GG)$),
(GH)/($2*(GH)$),
(GI)/($2*(GI)$),
(GJ)/($2*(GJ)$),
(GK)/($2*(GK)$),
(GL)/($2*(GL)$)} {
\draw[color=\colGamma,
line cap=round, dash pattern=on 0pt off 2pt] (O) -- \i;
\draw[->][color=\colGamma] \i -- \j ;};

\foreach \i/\j in {
(GA)/($2*(GA)$)} {
\draw[thick, color=\colGamma,
line cap=round, dash pattern=on 0pt off 2pt] (O) -- \i;
\draw[->][very thick, color=\colGamma] \i -- \j ;};

\foreach \i/\j/\k in {
X/$\mathbf{1_{c=3}}$/\colone,
Y/$\mathbf{2_{c=2}}$/\coltwo,
Z/$\mathbf{2_{c=2}}$/\coltwoALT,
W/$\mathbf{3_{c=1}}$/\colthree,
Q/$\mathbf{\infty_{c=1}}$/\colinf}{
\draw[->]
[thick, color=\k] (O) -- (\i);
\path  (\i) node[circle, fill=\k, inner sep=\sizesmall]{{}};
\path  (\i) node[font=\small,xshift=-0.3cm,yshift=0.2cm]{{\j}};
};

\end{tikzpicture}\caption{$\Gamma$-vectors in $d=8$ maximal supergravity for each of the twelve dual-frames, in magenta. We show the  $\alpha$-vectors sharing a common $\Gamma$-vector, highlighted with a thick arrow. We show their $c$-values as subscripts. 
Cyan dots indicate decompactification limits to M-theory, while light (dark) green dots correspond to decompactification to $d=10$ type IIA (IIB). Blue dots indicate decompactification to $d=9$ maximal supergravity and red dots are the emergent string limits.}\label{f.8dmax}
\end{figure}

\paragraph{Maximal supergravity.}
Maximal supergravities in $11-k$ dimensions have symmetric moduli spaces 
with duality group $G=E_{k(k)}$. 
For $k=3,\dots,8$, particles transform in the irreducible representations
$\mathbf{(\overline{3},2)}$, $\mathbf{\overline{10}}$, $\mathbf{16}$, $\mathbf{27}$, $\mathbf{56}$ and $\mathbf{248}$, respectively.
Plugging the corresponding $\lambda$ into eq. \eqref{eq:cI}, we get
\begin{align}
c_I = \vec\alpha_I \cdot \vec\Gamma_I  = 
\sum_{i} 2 V^I_{i(k-1)} 
\,, 
\end{align}
leading to the values of $c$, depending on the value of $n$ for the corresponding limit, listed in Table \ref{tab:LaplaciansMaxSUGRA}. We notice the following patterns for $k\geq3$, which respect the quantization rule in Proposition \ref{prop:hyperbolic towers quantization}:
\begin{align}
c_{\infty} = 2^{k-3} + \delta_{k,7} + 14 \delta_{k,8}\,,\quad
c_{n\leq k} =& c_{\infty} + 2k-3 -n \, .
\end{align}
As an example, we plot the principal plane and $c$-values for 8d maximal supergravity in Figure \ref{f.8dmax}. 

\begin{table}[t]\centering\renewcommand{\arraystretch}{1.1}
\begin{tabular}{|c|ccccccc|}\hline
 $n$ & $E_{2(2)}$ & $E_{3(3)}$ & $E_{4(4)}$ & $E_{5(5)}$ & $E_{6(6)}$ & $E_{7(7)}$ & $E_{8(8)}$ \\ \hline
 1 & $1_{\text{IIA}},0_{\text{IIB}}$ & 3 & 6 & 10 & 16 & 27 & 58 \\
 2 & 0 & 2 & 5 & 9 & 15 & 26 & 57 \\
 3 &   & 1 & 4 & 8 & 14 & 25 & 56 \\
 4 &   &   & 3 & 7 & 13 & 24 & 55 \\
 5 &   &   &  & 6 & 12 & 23 & 54 \\
 6 &   &   &  &  & 11 & 22 & 53 \\
 7 &   &   &  &  &  & 21 & 52 \\
 8 &   &   &  &  &  &  & 51 \\
 $\infty$ & $1/2$ & 1 & 2 & 4 & 8 & 17 & 46 \\ \hline
 \end{tabular}\caption{Values of $c$ for all different infinite distance limits of maximal supergravity in dimensions $9$ to $3$.}\label{tab:LaplaciansMaxSUGRA}
\end{table}
\noindent

\paragraph{Type II on a torus.}
The scalar moduli space of Type II strings compactified on a $T^k$ can be described as the coset $O(k,k;\mathbb{Z})\backslash O(k,k)/O(k)\times O(k)$. This case differs from the above example as the duality group only comprises perturbative T-dualities. Though these cosets are not the full moduli spaces of any low-energy supergravity theories, it is nonetheless instructive to compute the associated $c$-values for principal towers. These values of $c$, depending on the value of $n$, are listed in Table \ref{tab:LaplaciansTypeII}.

\begin{table}[t]\centering\renewcommand{\arraystretch}{1.1}
\begin{tabular}{|c|ccccccc|}\hline
 $n$ & $SO(2,2)$ & $SO(3,3)$ & $SO(4,4)$ & $SO(5,5)$ & $SO(6,6)$ & $SO(7,7)$ & $SO(8,8)$ \\ \hline
 1 & 2 & 4 & 6 & 8 & 10 & 12 & 14 \\
 2 & 1 & 3 & 5 & 7 & 9  & 11 & 13 \\
 3 &   & 2 & 4 & 6 & 8  & 10 & 12 \\
 4 &   &   & 3 & 5 & 7  & 9  & 11 \\
 5 &   &   &   & 4 & 6  & 8  & 10 \\
 6 &   &   &   &   & 5  & 7  & 9 \\
 7 &   &   &   &   &    & 6  & 8 \\
 8 &   &   &   &   &    &    & 7    \\\hline
 \end{tabular}\caption{Values of $c$ for all different infinite distance limits of toroidal compactifications of type II string theories.}\label{tab:LaplaciansTypeII}
\end{table}

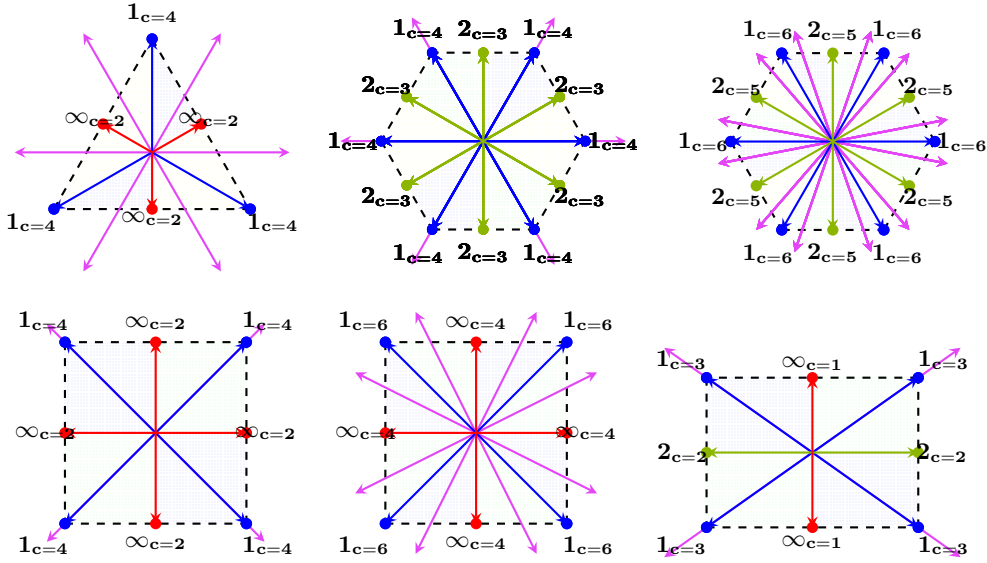
\begin{figure}[t]
\def\opaedges{1}\def\opafill{0.2}\def\opafillback{0.02}
\def\sizebig{1.2pt}\def\sizesmall{1.4pt}
    
    \centering
\subfigure{ 
\centering 
\begin{tikzpicture}[scale=1.5,>=stealth]{
\coordinatesTriangle \def\sizebig{1.5pt}\def\sizesmall{1.5pt} 
\drawTriangle

\foreach \i/\j/\k/\l in 
{(O)/($(A)!0.5!(B)$)/(A)/blue,
(O)/($(A)!0.5!(B)$)/(B)/yellow,
(O)/($(B)!0.5!(C)$)/(B)/yellow,
(O)/($(B)!0.5!(C)$)/(C)/blue,
(O)/($(C)!0.5!(A)$)/(C)/yellow,
(O)/($(C)!0.5!(A)$)/(A)/yellow}{
\fill[pattern={Dots[radius=0.4pt,distance=1.2pt]},
pattern color=\l,opacity=\opafill]\i -- \j -- \k;
};

\foreach \i/\j/\k in {
($0.7*(B)-0.7*(C)$)/{{}}/\colGamma,
($0.7*(C)-0.7*(B)$)/{{}}/\colGamma,
($0.7*(C)-0.7*(A)$)/{{}}/\colGamma,
($0.7*(A)-0.7*(C)$)/{{}}/\colGamma,
($0.7*(A)-0.7*(B)$)/{{}}/\colGamma,
($0.7*(B)-0.7*(A)$)/{{}}/\colGamma}{
\draw[->][\coltwo,thick,color=\k] (O) -- \i;
};

\foreach \i in {A,B,C}{\path ($(\i)$) node[circle, fill=\colone, inner sep=\sizebig]{};}; 
\foreach \i/\j/\k in {
(A)/$\mathbf{1_{c=4}}$/\colone,
(B)/$\mathbf{1_{c=4}}$/\colone,
(C)/$\mathbf{1_{c=4}}$/\colone,
($(A)!0.5!(B)$)/$\mathbf{\infty_{c=2}}$/\colinf,
($(B)!0.5!(C)$)/$\mathbf{\infty_{c=2}}$/\colinf,
($(C)!0.5!(A)$)/$\mathbf{\infty_{c=2}}$/\colinf
}{
\draw[->][\colinf,thick,color=\k] (O) -- \i;
\path \i node[circle, fill=\k, inner sep=\sizesmall] (currentpoint) {};
\path (O) -- (currentpoint) node[font=\footnotesize, pos=1.3] {\j};
};

}
 \end{tikzpicture}}
\subfigure{\centering 
\begin{tikzpicture}[scale=1.35,>=stealth]
{\coordinatesHexagon
\def\sizebig{1.5pt}\def\sizesmall{1.5pt}
\drawHexagon

\foreach \i/\j/\k/\l/\m in 
{(O)/($(A)!0.5!(B)$)/(B)/($(B)!0.5!(C)$)/blue,
(O)/($(B)!0.5!(C)$)/(C)/($(C)!0.5!(D)$)/Green,
(O)/($(C)!0.5!(D)$)/(D)/($(D)!0.5!(E)$)/yellow,
(O)/($(D)!0.5!(E)$)/(E)/($(E)!0.5!(F)$)/blue,
(O)/($(E)!0.5!(F)$)/(F)/($(F)!0.5!(A)$)/Green,
(O)/($(F)!0.5!(A)$)/(A)/($(A)!0.5!(B)$)/yellow}{
\fill[pattern={Dots[radius=0.4pt,distance=1.2pt]},
pattern color=\m,opacity=\opafill]\i -- \j -- \k -- \l;};

\foreach \i/\j/\k in {
($1.4*(A)$)/{{}}/\colGamma,
($1.4*(B)$)/{{}}/\colGamma,
($1.4*(C)$)/{{}}/\colGamma,
($1.4*(D)$)/{{}}/\colGamma,
($1.4*(E)$)/{{}}/\colGamma,
($1.4*(F)$)/{{}}/\colGamma}{
\draw[->][\coltwo,thick,color=\k] (O) -- \i;
};};

\foreach \i in {A,B,C,D,E,F}{\path ($(\i)$) node[circle, fill=\colone, inner sep=\sizebig]{};

\foreach \i/\j/\k in {
(A)/$\mathbf{1_{c=4}}$/\colone,
(B)/$\mathbf{1_{c=4}}$/\colone,
(C)/$\mathbf{1_{c=4}}$/\colone,
(D)/$\mathbf{1_{c=4}}$/\colone,
(E)/$\mathbf{1_{c=4}}$/\colone,
(F)/$\mathbf{1_{c=4}}$/\colone,
($(A)!0.5!(B)$)/$\mathbf{2_{c=3}}$/\coltwo,
($(B)!0.5!(C)$)/$\mathbf{2_{c=3}}$/\coltwo,
($(C)!0.5!(D)$)/$\mathbf{2_{c=3}}$/\coltwo,
($(D)!0.5!(E)$)/$\mathbf{2_{c=3}}$/\coltwo,
($(E)!0.5!(F)$)/$\mathbf{2_{c=3}}$/\coltwo,
($(F)!0.5!(A)$)/$\mathbf{2_{c=3}}$/\coltwo
}{
\draw[->][thick,color=\k] (O) -- \i;
\path \i node[circle, fill=\k, inner sep=\sizesmall] (currentpoint) {};
\path (O) -- (currentpoint) node[font=\footnotesize, pos=1.35] {\j};};
}\end{tikzpicture}} 
\subfigure{
\centering 
\begin{tikzpicture}[scale=1.35,>=stealth]
{\coordinatesHexagon
\def\sizebig{1.5pt}\def\sizesmall{1.5pt}
\drawHexagon

\foreach \i/\j/\k/\l in {
(O)/($(A)!0.5!(B)$)/(A)/blue,
(O)/($(A)!0.5!(B)$)/(B)/yellow,
(O)/($(B)!0.5!(C)$)/(B)/Green,
(O)/($(B)!0.5!(C)$)/(C)/blue,
(O)/($(C)!0.5!(D)$)/(C)/yellow,
(O)/($(C)!0.5!(D)$)/(D)/Green,
(O)/($(D)!0.5!(E)$)/(D)/blue,
(O)/($(D)!0.5!(E)$)/(E)/yellow,
(O)/($(E)!0.5!(F)$)/(E)/Green,
(O)/($(E)!0.5!(F)$)/(F)/blue,
(O)/($(F)!0.5!(A)$)/(F)/yellow,
(O)/($(F)!0.5!(A)$)/(A)/Green}{\fill[pattern={Dots[radius=0.4pt,distance=1.2pt]},pattern color=\l,opacity=\opafill] \i -- \j -- \k;};

\foreach \i in {A,B,C,D,E,F}{\path ($(\i)$) node[circle, fill=\colone, inner sep=\sizebig]{};

\foreach \i/\j/\k in {
($(A)+0.25*(B)$)/{{}}/\colGamma
,($(B)+0.25*(A)$)/{{}}/\colGamma
,($(B)+0.25*(C)$)/{{}}/\colGamma
,($(C)+0.25*(B)$)/{{}}/\colGamma
,($(C)+0.25*(D)$)/{{}}/\colGamma
,($(D)+0.25*(C)$)/{{}}/\colGamma
,($(D)+0.25*(E)$)/{{}}/\colGamma
,($(E)+0.25*(D)$)/{{}}/\colGamma
,($(E)+0.25*(F)$)/{{}}/\colGamma
,($(F)+0.25*(E)$)/{{}}/\colGamma
,($(F)+0.25*(A)$)/{{}}/\colGamma
,($(A)+0.25*(F)$)/{{}}/\colGamma
}{
\draw[->][\coltwo,thick,color=\k] (O) -- \i;
};
}; 

\foreach \i/\j/\k in {
(A)/$\mathbf{1_{c=6}}$/\colone,
(B)/$\mathbf{1_{c=6}}$/\colone,
(C)/$\mathbf{1_{c=6}}$/\colone,
(D)/$\mathbf{1_{c=6}}$/\colone,
(E)/$\mathbf{1_{c=6}}$/\colone,
(F)/$\mathbf{1_{c=6}}$/\colone,
($(A)!0.5!(B)$)/$\mathbf{2_{c=5}}$/\coltwo,
($(B)!0.5!(C)$)/$\mathbf{2_{c=5}}$/\coltwo,
($(C)!0.5!(D)$)/$\mathbf{2_{c=5}}$/\coltwo,
($(D)!0.5!(E)$)/$\mathbf{2_{c=5}}$/\coltwo,
($(E)!0.5!(F)$)/$\mathbf{2_{c=5}}$/\coltwo,
($(F)!0.5!(A)$)/$\mathbf{2_{c=5}}$/\coltwo
}{
\draw[->][\coltwo,thick,color=\k] (O) -- \i;
\path \i node[circle, fill=\k, inner sep=\sizesmall] (currentpoint) {};
\path (O) -- (currentpoint) node[font=\footnotesize, pos=1.35] {\j};};

}\end{tikzpicture}}
 \subfigure{ 
\centering 
\begin{tikzpicture}[scale=1.2,,>=stealth]{
\coordinatesSquare \def\sizebig{1.5pt}\def\sizesmall{1.5pt} 
\drawSquare

\foreach \i/\j/\k/\l/\m in {
(O)/($(A)!0.5!(B)$)/(B)/($(B)!0.5!(C)$)/blue,
(O)/($(B)!0.5!(C)$)/(C)/($(C)!0.5!(D)$)/Green,
(O)/($(C)!0.5!(D)$)/(D)/($(D)!0.5!(A)$)/blue,
(O)/($(D)!0.5!(A)$)/(A)/($(A)!0.5!(B)$)/Green}{
\fill[
    pattern={Dots[radius=0.4pt,distance=1.2pt]},
    pattern color=\m,
    opacity=\opafill
] \i -- \j -- \k -- \l;
};

\foreach \i/\j/\k in {
($1.2*(A)$)/{{}}/\colGamma,
($1.2*(B)$)/{{}}/\colGamma,
($1.2*(C)$)/{{}}/\colGamma,
($1.2*(D)$)/{{}}/\colGamma}{
\draw[->][\coltwo,thick,color=\k] (O) -- \i;
};

\foreach \i in {A,B,C,D}{\path ($(\i)$) node[circle, fill=\colone, inner sep=\sizebig]{};}; 
\foreach \i/\j/\k in {
(A)/$\mathbf{1_{c=4}}$/\colone,
(B)/$\mathbf{1_{c=4}}$/\colone,
(C)/$\mathbf{1_{c=4}}$/\colone,
(D)/$\mathbf{1_{c=4}}$/\colone,
($(A)!0.5!(B)$)/$\mathbf{\infty_{c=2}}$/\colinf,
($(C)!0.5!(D)$)/$\mathbf{\infty_{c=2}}$/\colinf,
($(B)!0.5!(C)$)/$\mathbf{\infty_{c=2}}$/\colinf,
($(A)!0.5!(D)$)/$\mathbf{\infty_{c=2}}$/\colinf
}{
\draw[->][\colinf,thick,color=\k] (O) -- \i;
\path \i node[circle, fill=\k, inner sep=\sizesmall] (currentpoint) {};
\path (O) -- (currentpoint) node[font=\footnotesize, pos=1.3] {\j};
};

}
 \end{tikzpicture}}
\subfigure{ 
\centering 
\begin{tikzpicture}[scale=1.2,,>=stealth]{
\coordinatesSquare \def\sizebig{1.5pt}\def\sizesmall{1.5pt} 
\drawSquare

\foreach \i/\j/\k/\l in {
(O)/($(A)!0.5!(B)$)/(A)/blue,
(O)/($(A)!0.5!(B)$)/(B)/Green,
(O)/($(B)!0.5!(C)$)/(B)/blue,
(O)/($(B)!0.5!(C)$)/(C)/Green,
(O)/($(C)!0.5!(D)$)/(C)/blue,
(O)/($(C)!0.5!(D)$)/(D)/Green,
(O)/($(D)!0.5!(A)$)/(D)/blue,
(O)/($(D)!0.5!(A)$)/(A)/Green}{
\fill[
    pattern={Dots[radius=0.4pt,distance=1.2pt]},
    pattern color=\l,
    opacity=\opafill
] \i -- \j -- \k;
};

\foreach \i/\j/\k in {
($(A)-0.333*(B)$)/{{}}/\colGamma,
($(B)-0.333*(A)$)/{{}}/\colGamma,
($0.333*(A)-(B)$)/{{}}/\colGamma,
($0.333*(B)-(A)$)/{{}}/\colGamma,
($(A)-0.333*(D)$)/{{}}/\colGamma,
($(D)-0.333*(A)$)/{{}}/\colGamma,
($0.333*(A)-(D)$)/{{}}/\colGamma,
($0.333*(D)-(A)$)/{{}}/\colGamma
}{
\draw[->][\coltwo,thick,color=\k] (O) -- \i;
};

\foreach \i in {A,B,C,D}{\path ($(\i)$) node[circle, fill=\colone, inner sep=\sizebig]{};}; 
\foreach \i/\j/\k in {
(A)/$\mathbf{1_{c=6}}$/\colone,
(B)/$\mathbf{1_{c=6}}$/\colone,
(C)/$\mathbf{1_{c=6}}$/\colone,
(D)/$\mathbf{1_{c=6}}$/\colone,
($(A)!0.5!(B)$)/$\mathbf{\infty_{c=4}}$/\colinf,
($(C)!0.5!(D)$)/$\mathbf{\infty_{c=4}}$/\colinf,
($(B)!0.5!(C)$)/$\mathbf{\infty_{c=4}}$/\colinf,
($(A)!0.5!(D)$)/$\mathbf{\infty_{c=4}}$/\colinf
}{
\draw[->][\colinf,thick,color=\k] (O) -- \i;
\path \i node[circle, fill=\k, inner sep=\sizesmall] (currentpoint) {};
\path (O) -- (currentpoint) node[font=\footnotesize, pos=1.3] {\j};
};
}
 \end{tikzpicture}}
  \subfigure{ \centering 
\begin{tikzpicture}[scale=1.4,,>=stealth]{
\coordinatesRectangle \def\sizebig{1.5pt}\def\sizesmall{1.5pt} 
\drawRectangle

\foreach \i/\j/\k/\l/\m in {
(O)/($(A)!0.5!(B)$)/(B)/($(B)!0.5!(C)$)/blue,
(O)/($(B)!0.5!(C)$)/(C)/($(C)!0.5!(D)$)/Green,
(O)/($(C)!0.5!(D)$)/(D)/($(D)!0.5!(A)$)/blue,
(O)/($(D)!0.5!(A)$)/(A)/($(A)!0.5!(B)$)/Green}{
\fill[
    pattern={Dots[radius=0.4pt,distance=1.2pt]},
    pattern color=\m,
    opacity=\opafill
] \i -- \j -- \k -- \l;
};

\foreach \i/\j/\k in {
($1.4*(A)$)/{{}}/\colGamma,
($1.4*(B)$)/{{}}/\colGamma,
($1.4*(C)$)/{{}}/\colGamma,
($1.4*(D)$)/{{}}/\colGamma}{
\draw[->][\coltwo,thick,color=\k] (O) -- \i;};

\foreach \i in {A,B,C,D}{\path ($(\i)$) node[circle, fill=\colone, inner sep=\sizebig]{};}; 
\foreach \i/\j/\k in {
(A)/$\mathbf{1_{c=3}}$/\colone,
(B)/$\mathbf{1_{c=3}}$/\colone,
(C)/$\mathbf{1_{c=3}}$/\colone,
(D)/$\mathbf{1_{c=3}}$/\colone,
($(A)!0.5!(D)$)/$\mathbf{2_{c=2}}$/\coltwo,
($(B)!0.5!(C)$)/$\mathbf{2_{c=2}}$/\coltwo,
($(A)!0.5!(B)$)/$\mathbf{\infty_{c=1}}$/\colinf,
($(C)!0.5!(D)$)/$\mathbf{\infty_{c=1}}$/\colinf
}{
\draw[->][\coltwo,thick,color=\k] (O) -- \i;
\path \i node[circle, fill=\k, inner sep=\sizesmall] (currentpoint) {};
\path (O) -- (currentpoint) node[font=\footnotesize, pos=1.3] {\j};
};
}
 \end{tikzpicture}}
    \caption{$\alpha$-vectors convex hulls with the values of $c$ for rank-2 polytopes corresponding to $d=3$ theories except for the triangle and the rectangle which correspond to $d=5$ and $d=4$ theories, respectively. They are generated by 
    the $\mathbf{3}$ of $SL(3,\mathbb{R})$, 
    the $\mathbf{8}$ of $SL(3,\mathbb{R})$,
    the $\mathbf{14}$ of $G_{2(2)}$, 
    the $\mathbf{4}$ of $SO(4,4)$, 
    the $\mathbf{5}$ of $SO(4,5)$, 
    and the $\mathbf{(2,2)}$ of $SL(2,\mathbb{R})\times SL(2,\mathbb{R})$. Long roots are normalized to have length squared 2 except in the rectangle where one of the $SL(2,\mathbb{R})$ factors has length squared 4.
    Magenta arrows indicate the normalized $\Gamma$-vectors at each facet. Numbers at each facet indicate the value of $n$ decompactifying dimensions for the corresponding KK tower, with $\infty$ signaling an emergent string limit.}
    \label{fig:rank2}
\end{figure}

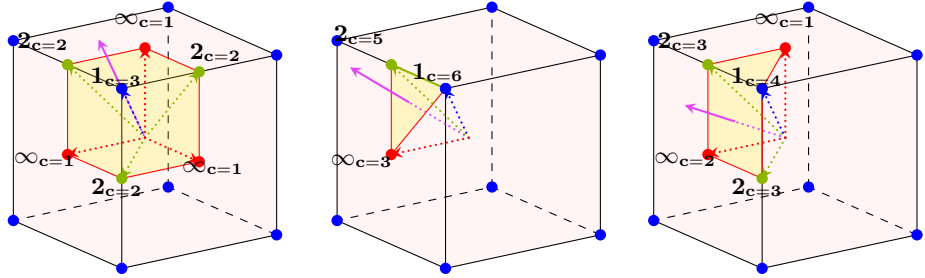
\begin{figure}[t]
    \centering
\subfigure{
\tdplotsetmaincoords{72}{145}

\begin{tikzpicture}[scale=2.5,>=stealth,tdplot_main_coords
]
\coordinatesCubeB
\drawCubeB
\def\opaedges{1}\def\opafill{0.2}\def\opafillback{0.02}
\def\sizebig{1.5pt}\def\sizesmall{1.5pt}
\foreach \i/\j/\k/\l in {A/B/B'/A',B/C/C'/B',A'/B'/C'/D'}{\fill[\colinf,opacity=\opafillback](\i) -- (\j) -- (\k) -- (\l);};
\foreach \i/\j/\k/\l in {A/B/C/D,A/D/D'/A',C/D/D'/C'}{\fill[\colinf,opacity=\opafillback](\i) -- (\j) -- (\k) -- (\l);};

\foreach \i in {A,B,C,D,A',B',C',D'}{
\path ($(\i)$) node[circle, fill=\colone, inner sep=\sizebig]{{}};
};


\fill[color=yellow,opacity=\opafill] (C') -- ($(C')!0.5!(C)$) -- ($(C')!0.5!(B)$)  -- ($(C')!0.5!(B')$);
\fill[color=yellow,opacity=\opafill] (C') -- ($(C')!0.5!(B')$) -- ($(C')!0.5!(A')$)  -- ($(C')!0.5!(D')$);
\fill[color=yellow,opacity=\opafill] (C') -- ($(C')!0.5!(C)$) -- ($(C')!0.5!(D)$)  -- ($(C')!0.5!(D')$);
\draw[color=\colinf] ($(C')!0.5!(B)$) -- ($(C')!0.5!(B')$) -- ($(C')!0.5!(A')$) -- ($(C')!0.5!(D')$) -- ($(C')!0.5!(D)$) -- ($(C')!0.5!(C)$) -- ($(C')!0.5!(B)$);

\coordinate (X) at (C');
\draw[][line cap=round, dash pattern=on 0pt off 1pt,thick,color=\colGamma] (O) -- (X);
\draw[->][thick,color=\colGamma] (X) -- ($2*(X)$);

\foreach \i/\j/\k in {
(C')/$\mathbf{1_{c=3}}$/\colone,
($(C')!0.5!(B')$)/$\mathbf{2_{c=2}}$/\coltwo,
($(C')!0.5!(C)$)/$\mathbf{2_{c=2}}$/\coltwo,
($(C')!0.5!(D')$)/$\mathbf{2_{c=2}}$/\coltwo,
($(C')!0.5!(B)$)/$\mathbf{\infty_{c=1}}$/\colinf,
($(C')!0.5!(A')$)/$\mathbf{\infty_{c=1}}$/\colinf,
($(C')!0.5!(D)$)/$\mathbf{\infty_{c=1}}$/\colinf
}{
\draw[->][line cap=round, dash pattern=on 0pt off 2pt,thick,color=\k] (O) -- \i;
\path \i node[circle, fill=\k, inner sep=\sizesmall] (currentpoint) {};
\path (O) -- (currentpoint) node[font=\footnotesize, pos=1.4] {\j};
};
\end{tikzpicture}}   
\subfigure{\tdplotsetmaincoords{72}{145}
\begin{tikzpicture}[scale=2.5,>=stealth,tdplot_main_coords]
\coordinatesCubeB
\drawCubeB
\def\opaedges{1}\def\opafill{0.2}\def\opafillback{0.02}
\def\sizebig{1.5pt}\def\sizesmall{1.5pt}
\foreach \i/\j/\k/\l in {A/B/B'/A',B/C/C'/B',A'/B'/C'/D'}{\fill[\colinf,opacity=\opafillback](\i) -- (\j) -- (\k) -- (\l);};
\foreach \i/\j/\k/\l in {A/B/C/D,A/D/D'/A',C/D/D'/C'}{\fill[\colinf,opacity=\opafillback](\i) -- (\j) -- (\k) -- (\l);};

\foreach \i in {A,B,C,D,A',B',C',D'}{
\path ($(\i)$) node[circle, fill=\colone, inner sep=\sizebig]{{}};
};

\draw[thick, color=\coltwo] (C') -- ($(C')!0.5!(B')$);
\fill[color=yellow,opacity=\opafill] (C') -- ($(C')!0.5!(B)$) -- ($(C')!0.5!(B')$);
\draw[color=\colinf] (C') -- ($(C')!0.5!(B)$) -- ($(C')!0.5!(B')$);

\coordinate (X) at (0.5,0.166667,0.333333);
\draw[][line cap=round, dash pattern=on 0pt off 2pt,thick,color=\colGamma] (O) -- (X);
\draw[->][thick,color=\colGamma] (X) -- ($2*(X)$);

\foreach \i/\j/\k in {
(C')/$\mathbf{1_{c=6}}$/\colone,
($(C')!0.5!(B')$)/$\mathbf{2_{c=5}}$/\coltwo,
($(C')!0.5!(B)$)/$\mathbf{\infty_{c=3}}$/\colinf}{
\draw[->][line cap=round, dash pattern=on 0pt off 2pt,thick,color=\k] (O) -- \i;
\path \i node[circle, fill=\k, inner sep=\sizesmall] (currentpoint) {};
\path (O) -- (currentpoint) node[font=\footnotesize, pos=1.5] {\j};
};
\end{tikzpicture}} 
\subfigure{\tdplotsetmaincoords{72}{145}
\begin{tikzpicture}[scale=2.5,>=stealth,tdplot_main_coords]
\coordinatesCubeB
\drawCubeB
\def\opaedges{1}\def\opafill{0.2}\def\opafillback{0.02}
\def\sizebig{1.5pt}\def\sizesmall{1.5pt}
\foreach \i/\j/\k/\l in {A/B/B'/A',B/C/C'/B',A'/B'/C'/D'}{\fill[\colinf,opacity=\opafillback](\i) -- (\j) -- (\k) -- (\l);};
\foreach \i/\j/\k/\l in {A/B/C/D,A/D/D'/A',C/D/D'/C'}{\fill[\colinf,opacity=\opafillback](\i) -- (\j) -- (\k) -- (\l);};

\foreach \i in {A,B,C,D,A',B',C',D'}{
\path ($(\i)$) node[circle, fill=\colone, inner sep=\sizebig]{{}};
};

\fill[color=yellow,opacity=\opafill] (C') -- ($(C')!0.5!(C)$) -- ($(C')!0.5!(B)$) -- ($(C')!0.5!(B')$);
\fill[color=yellow,opacity=\opafill] (C') -- ($(C')!0.5!(A')$) -- ($(C')!0.5!(B')$);
\draw[color=\colinf] (C') -- ($(C')!0.5!(C)$) -- ($(C')!0.5!(B)$) -- ($(C')!0.5!(B')$);
\draw[color=\colinf] (C') -- ($(C')!0.5!(A')$) -- ($(C')!0.5!(B')$);

\coordinate (X) at (.5,0.25,0.25);
\draw[][line cap=round, dash pattern=on 0pt off 2pt,thick,color=\colGamma] (O) -- (X);
\draw[->][thick,color=\colGamma] (X) -- ($2*(X)$);

\foreach \i/\j/\k in {
(C')/$\mathbf{1_{c=4}}$/\colone,
($(C')!0.5!(B')$)/$\mathbf{2_{c=3}}$/\coltwo,
($(C')!0.5!(C)$)/$\mathbf{2_{c=3}}$/\coltwo,
($(C')!0.5!(B)$)/$\mathbf{\infty_{c=2}}$/\colinf,
($(C')!0.5!(A')$)/$\mathbf{\infty_{c=1}}$/\colinf
}{
\draw[->][line cap=round, dash pattern=on 0pt off 2pt,thick,color=\k] (O) -- \i;
\path \i node[circle, fill=\k, inner sep=\sizesmall] (currentpoint) {};
\path (O) -- (currentpoint) node[font=\footnotesize, pos=1.4] {\j};
};
\end{tikzpicture}} 
    \caption{$\alpha$-vectors convex hulls for rank-3 polytopes corresponding to moduli spaces of $d=4$ theories. They are generated by the $\mathbf{(2,2,2)}$ of $SL(2,\mathbb{R})^{\oplus 3}$, the $\mathbf{14'}$ of $Sp(6,\mathbb{R})$, and the $\mathbf{(5,2)}$ of $Sp(4,\mathbb{R})\times SL(2,\mathbb{R})$. Long roots are normalized to have length squared 2.  Notice that in the last case, there are two inequivalent emergent string limits, with distinct $c$-values.
    }
    \label{fig:rank3}
\end{figure}

\paragraph{Other split moduli spaces.}
For $SL(2k,\mathbb{R})$, $\lambda = (1,0,\dots,0)$, following the computations above gets us to a $c-$value for a KK-tower decompactifying $n$ dimensions:
\begin{align}
c_n=k-n=\text{codimension}\,.
\end{align} In \cite{Baines:2026aug}, the list of polytopes which are generated by a single irreducible representation with a fully non-abelian duality group and which are compatible with the ESC and the sharpened DC rates, was computed. We show the $c$-values for some of those rank-2 and rank-3 moduli spaces in Figures \ref{fig:rank2} and \ref{fig:rank3}.

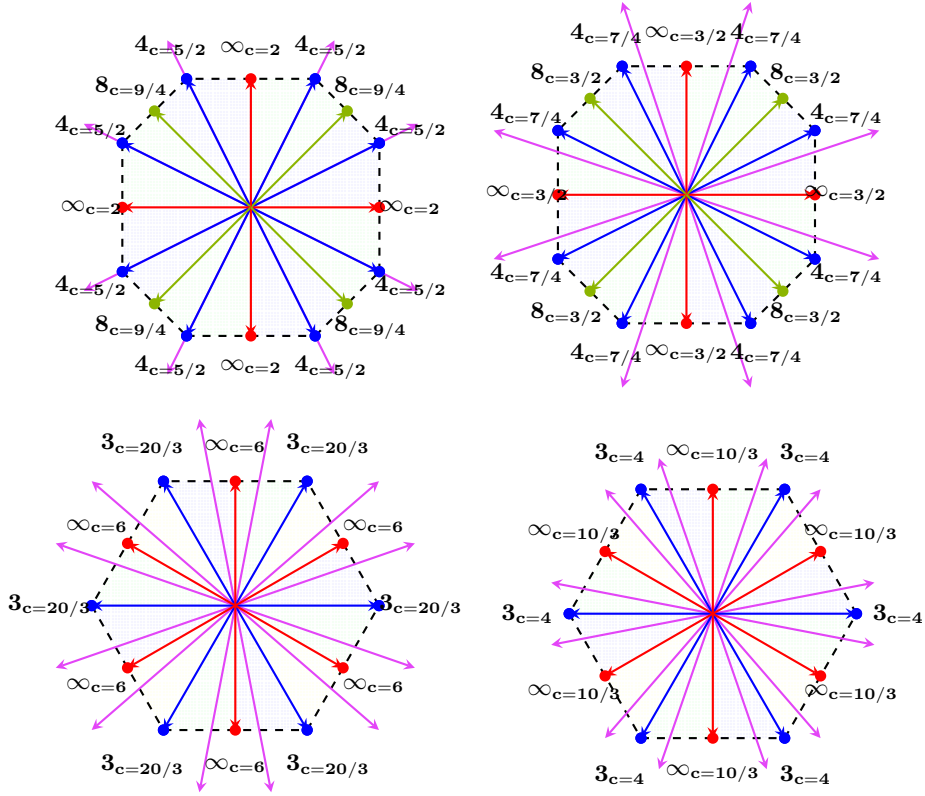
\begin{figure}[t]
    
\def\opafill{0.2}
    \centering    

\subfigure{ \centering 
\begin{tikzpicture}[scale=1.7,,>=stealth]{
\coordinatesOctagon\def\sizebig{1.5pt}\def\sizesmall{1.5pt} 
\drawOctagon

\foreach \i/\j/\k/\l/\m in {
(O)/($(A)!0.5!(B)$)/(B)/($(B)!0.5!(C)$)/blue,
(O)/($(B)!0.5!(C)$)/(C)/($(C)!0.5!(D)$)/Green,
(O)/($(C)!0.5!(D)$)/(D)/($(D)!0.5!(E)$)/blue,
(O)/($(D)!0.5!(E)$)/(E)/($(E)!0.5!(F)$)/Green,
(O)/($(E)!0.5!(F)$)/(F)/($(F)!0.5!(G)$)/blue,
(O)/($(F)!0.5!(G)$)/(G)/($(G)!0.5!(H)$)/Green,
(O)/($(G)!0.5!(H)$)/(H)/($(H)!0.5!(A)$)/blue,
(O)/($(H)!0.5!(A)$)/(A)/($(A)!0.5!(B)$)/Green}{
\fill[pattern={Dots[radius=0.4pt,distance=1.2pt]},
pattern color=\m,opacity=\opafill] \i -- \j -- \k -- \l;};

\foreach \i/\j/\k in {
($1.3*(A)$)/{{}}/\colGamma,
($1.3*(B)$)/{{}}/\colGamma
,($1.3*(C)$)/{{}}/\colGamma
,($1.3*(D)$)/{{}}/\colGamma
,($1.3*(E)$)/{{}}/\colGamma
,($1.3*(F)$)/{{}}/\colGamma
,($1.3*(G)$)/{{}}/\colGamma
,($1.3*(H)$)/{{}}/\colGamma}{\draw[->][\coltwo,thick,color=\k] (O) -- \i;};
\foreach \i in {A,B,C,D,E,F,G,H}{\path ($(\i)$) node[circle, fill=\colone, inner sep=\sizebig]{};}; 
\foreach \i/\j/\k in {
(A)/$\mathbf{4_{c=5/2}}$/\colone,
(B)/$\mathbf{4_{c=5/2}}$/\colone,
(C)/$\mathbf{4_{c=5/2}}$/\colone,
(D)/$\mathbf{4_{c=5/2}}$/\colone,
(E)/$\mathbf{4_{c=5/2}}$/\colone,
(F)/$\mathbf{4_{c=5/2}}$/\colone,
(G)/$\mathbf{4_{c=5/2}}$/\colone,
(H)/$\mathbf{4_{c=5/2}}$/\colone,
($(A)!0.5!(B)$)/$\mathbf{\infty_{c=2}}$/\colinf,
($(B)!0.5!(C)$)/$\mathbf{8_{c=9/4}}$/\coltwo,
($(C)!0.5!(D)$)/$\mathbf{\infty_{c=2}}$/\colinf,
($(D)!0.5!(E)$)/$\mathbf{8_{c=9/4}}$/\coltwo,
($(E)!0.5!(F)$)/$\mathbf{\infty_{c=2}}$/\colinf,
($(F)!0.5!(G)$)/$\mathbf{8_{c=9/4}}$/\coltwo,
($(G)!0.5!(H)$)/$\mathbf{\infty_{c=2}}$/\colinf,
($(H)!0.5!(A)$)/$\mathbf{8_{c=9/4}}$/\coltwo}{
\draw[->][\coltwo,thick,color=\k] (O) -- \i;
\path \i node[circle, fill=\k, inner sep=\sizesmall] (currentpoint) {};
\path (O) -- (currentpoint) node[font=\footnotesize, pos=1.3] {\j};
};

}
 \end{tikzpicture}}
 \subfigure{ \centering 
\begin{tikzpicture}[scale=1.7,,>=stealth]{
\coordinatesOctagon\def\sizebig{1.5pt}\def\sizesmall{1.5pt} 
\drawOctagon

\foreach \i/\j/\k/\l/\m in {
(O)/($(A)!0.5!(B)$)/(B)/($(B)!0.5!(C)$)/blue,
(O)/($(B)!0.5!(C)$)/(C)/($(C)!0.5!(D)$)/Green,
(O)/($(C)!0.5!(D)$)/(D)/($(D)!0.5!(E)$)/blue,
(O)/($(D)!0.5!(E)$)/(E)/($(E)!0.5!(F)$)/Green,
(O)/($(E)!0.5!(F)$)/(F)/($(F)!0.5!(G)$)/blue,
(O)/($(F)!0.5!(G)$)/(G)/($(G)!0.5!(H)$)/Green,
(O)/($(G)!0.5!(H)$)/(H)/($(H)!0.5!(A)$)/blue,
(O)/($(H)!0.5!(A)$)/(A)/($(A)!0.5!(B)$)/Green}{
\fill[pattern={Dots[radius=0.4pt,distance=1.2pt]},
pattern color=\m,opacity=\opafill] \i -- \j -- \k -- \l;};

\foreach \i/\j/\k in {
($1.25*(A)+0.25*(B)$)/{{}}/\colGamma,
($1.25*(B)+0.25*(A)$)/{{}}/\colGamma
,($1.25*(C)+0.25*(D)$)/{{}}/\colGamma
,($1.25*(D)+0.25*(C)$)/{{}}/\colGamma
,($1.25*(E)+0.25*(F)$)/{{}}/\colGamma
,($1.25*(F)+0.25*(E)$)/{{}}/\colGamma
,($1.25*(G)+0.25*(H)$)/{{}}/\colGamma
,($1.25*(H)+0.25*(G)$)/{{}}/\colGamma}{\draw[->][\coltwo,thick,color=\k] (O) -- \i;};
\foreach \i in {A,B,C,D,E,F,G,H}{\path ($(\i)$) node[circle, fill=\colone, inner sep=\sizebig]{};}; 
\foreach \i/\j/\k in {
(A)/$\mathbf{4_{c=7/4}}$/\colone,
(B)/$\mathbf{4_{c=7/4}}$/\colone,
(C)/$\mathbf{4_{c=7/4}}$/\colone,
(D)/$\mathbf{4_{c=7/4}}$/\colone,
(E)/$\mathbf{4_{c=7/4}}$/\colone,
(F)/$\mathbf{4_{c=7/4}}$/\colone,
(G)/$\mathbf{4_{c=7/4}}$/\colone,
(H)/$\mathbf{4_{c=7/4}}$/\colone,
($(A)!0.5!(B)$)/$\mathbf{\infty_{c=3/2}}$/\colinf,
($(B)!0.5!(C)$)/$\mathbf{8_{c=3/2}}$/\coltwo,
($(C)!0.5!(D)$)/$\mathbf{\infty_{c=3/2}}$/\colinf,
($(D)!0.5!(E)$)/$\mathbf{8_{c=3/2}}$/\coltwo,
($(E)!0.5!(F)$)/$\mathbf{\infty_{c=3/2}}$/\colinf,
($(F)!0.5!(G)$)/$\mathbf{8_{c=3/2}}$/\coltwo,
($(G)!0.5!(H)$)/$\mathbf{\infty_{c=3/2}}$/\colinf,
($(H)!0.5!(A)$)/$\mathbf{8_{c=3/2}}$/\coltwo}{
\draw[->][\coltwo,thick,color=\k] (O) -- \i;
\path \i node[circle, fill=\k, inner sep=\sizesmall] (currentpoint) {};
\path (O) -- (currentpoint) node[font=\footnotesize, pos=1.3] {\j};
};

}
 \end{tikzpicture}}
 \subfigure{
\centering 
\begin{tikzpicture}[scale=1.9,>=stealth]
{\coordinatesHexagon
\def\sizebig{1.5pt}\def\sizesmall{1.5pt}
\drawHexagon

\foreach \i/\j/\k/\l in {
(O)/($(A)!0.5!(B)$)/(A)/blue,
(O)/($(A)!0.5!(B)$)/(B)/yellow,
(O)/($(B)!0.5!(C)$)/(B)/Green,
(O)/($(B)!0.5!(C)$)/(C)/blue,
(O)/($(C)!0.5!(D)$)/(C)/yellow,
(O)/($(C)!0.5!(D)$)/(D)/Green,
(O)/($(D)!0.5!(E)$)/(D)/blue,
(O)/($(D)!0.5!(E)$)/(E)/yellow,
(O)/($(E)!0.5!(F)$)/(E)/Green,
(O)/($(E)!0.5!(F)$)/(F)/blue,
(O)/($(F)!0.5!(A)$)/(F)/yellow,
(O)/($(F)!0.5!(A)$)/(A)/Green}{
\fill[pattern={Dots[radius=0.4pt,distance=1.2pt]},pattern color=\l,opacity=\opafill] \i -- \j -- \k;};

\foreach \i/\j/\k in {
($(A)+0.5*(B)$)/{{}}/\colGamma
,($(B)+0.5*(A)$)/{{}}/\colGamma
,($(B)+0.5*(C)$)/{{}}/\colGamma
,($(C)+0.5*(B)$)/{{}}/\colGamma
,($(C)+0.5*(D)$)/{{}}/\colGamma
,($(D)+0.5*(C)$)/{{}}/\colGamma
,($(D)+0.5*(E)$)/{{}}/\colGamma
,($(E)+0.5*(D)$)/{{}}/\colGamma
,($(E)+0.5*(F)$)/{{}}/\colGamma
,($(F)+0.5*(E)$)/{{}}/\colGamma
,($(F)+0.5*(A)$)/{{}}/\colGamma
,($(A)+0.5*(F)$)/{{}}/\colGamma}{
\draw[->][\coltwo,thick,color=\k] (O) -- \i;
};

\foreach \i in {A,B,C,D,E,F}{\path ($(\i)$) node[circle, fill=\colone, inner sep=\sizebig]{};};

\foreach \i/\j/\k in {
(A)/$\mathbf{3_{c=20/3}}$/\colone,
(B)/$\mathbf{3_{c=20/3}}$/\colone,
(C)/$\mathbf{3_{c=20/3}}$/\colone,
(D)/$\mathbf{3_{c=20/3}}$/\colone,
(E)/$\mathbf{3_{c=20/3}}$/\colone,
(F)/$\mathbf{3_{c=20/3}}$/\colone,
($(A)!0.5!(B)$)/$\mathbf{\infty_{c=6}}$/\colinf,
($(B)!0.5!(C)$)/$\mathbf{\infty_{c=6}}$/\colinf,
($(C)!0.5!(D)$)/$\mathbf{\infty_{c=6}}$/\colinf,
($(D)!0.5!(E)$)/$\mathbf{\infty_{c=6}}$/\colinf,
($(E)!0.5!(F)$)/$\mathbf{\infty_{c=6}}$/\colinf,
($(F)!0.5!(A)$)/$\mathbf{\infty_{c=6}}$/\colinf}{
\draw[->][\coltwo,thick,color=\k] (O) -- \i;
\path \i node[circle, fill=\k, inner sep=\sizesmall] (currentpoint) {};
\path (O) -- (currentpoint) node[font=\footnotesize, pos=1.35] {\j};};

}\end{tikzpicture}}
 \subfigure{
\centering 
\begin{tikzpicture}[scale=1.9,>=stealth]
{\coordinatesHexagon
\def\sizebig{1.5pt}\def\sizesmall{1.5pt}
\drawHexagon

\foreach \i/\j/\k/\l in {
(O)/($(A)!0.5!(B)$)/(A)/blue,
(O)/($(A)!0.5!(B)$)/(B)/yellow,
(O)/($(B)!0.5!(C)$)/(B)/Green,
(O)/($(B)!0.5!(C)$)/(C)/blue,
(O)/($(C)!0.5!(D)$)/(C)/yellow,
(O)/($(C)!0.5!(D)$)/(D)/Green,
(O)/($(D)!0.5!(E)$)/(D)/blue,
(O)/($(D)!0.5!(E)$)/(E)/yellow,
(O)/($(E)!0.5!(F)$)/(E)/Green,
(O)/($(E)!0.5!(F)$)/(F)/blue,
(O)/($(F)!0.5!(A)$)/(F)/yellow,
(O)/($(F)!0.5!(A)$)/(A)/Green}{
\fill[pattern={Dots[radius=0.4pt,distance=1.2pt]},pattern color=\l,opacity=\opafill] \i -- \j -- \k;};

\foreach \i/\j/\k in {
($(A)+0.25*(B)$)/{{}}/\colGamma
,($(B)+0.25*(A)$)/{{}}/\colGamma
,($(B)+0.25*(C)$)/{{}}/\colGamma
,($(C)+0.25*(B)$)/{{}}/\colGamma
,($(C)+0.25*(D)$)/{{}}/\colGamma
,($(D)+0.25*(C)$)/{{}}/\colGamma
,($(D)+0.25*(E)$)/{{}}/\colGamma
,($(E)+0.25*(D)$)/{{}}/\colGamma
,($(E)+0.25*(F)$)/{{}}/\colGamma
,($(F)+0.25*(E)$)/{{}}/\colGamma
,($(F)+0.25*(A)$)/{{}}/\colGamma
,($(A)+0.25*(F)$)/{{}}/\colGamma}{
\draw[->][\coltwo,thick,color=\k] (O) -- \i;
};

\foreach \i in {A,B,C,D,E,F}{\path ($(\i)$) node[circle, fill=\colone, inner sep=\sizebig]{};};

\foreach \i/\j/\k in {
(A)/$\mathbf{3_{c=4}}$/\colone,
(B)/$\mathbf{3_{c=4}}$/\colone,
(C)/$\mathbf{3_{c=4}}$/\colone,
(D)/$\mathbf{3_{c=4}}$/\colone,
(E)/$\mathbf{3_{c=4}}$/\colone,
(F)/$\mathbf{3_{c=4}}$/\colone,
($(A)!0.5!(B)$)/$\mathbf{\infty_{c=10/3}}$/\colinf,
($(B)!0.5!(C)$)/$\mathbf{\infty_{c=10/3}}$/\colinf,
($(C)!0.5!(D)$)/$\mathbf{\infty_{c=10/3}}$/\colinf,
($(D)!0.5!(E)$)/$\mathbf{\infty_{c=10/3}}$/\colinf,
($(E)!0.5!(F)$)/$\mathbf{\infty_{c=10/3}}$/\colinf,
($(F)!0.5!(A)$)/$\mathbf{\infty_{c=10/3}}$/\colinf}{
\draw[->][\coltwo,thick,color=\k] (O) -- \i;
\path \i node[circle, fill=\k, inner sep=\sizesmall] (currentpoint) {};
\path (O) -- (currentpoint) node[font=\footnotesize, pos=1.35] {\j};};

}\end{tikzpicture}}
    \caption{
    $\alpha$-vectors convex hulls with the values of $c$ for rank-2 polytopes corresponding to moduli spaces of $d=3$ theories. On top: the ones generated by the
   $\mathbf{16}$ and $\mathbf{35}$ of $Sp(4,\mathbb{R})$, with long roots of length squared $1$ and $\tfrac12$, respectively. 
   On bottom: the ones generated by the $\mathbf{7}$ and $\mathbf{14}$ of $G_{2(2)}$  with long roots of length squared $4$ and $\frac43$, respectively. 
   In the case of $G=G_{2(2)}$, Proposition \ref{prop:hyperbolic towers quantization} would prohibit the leading states in emergent string limits from transforming in the $\mathbf{14}$ representation.}
    \label{fig:rank2_extra}
\end{figure}

\subsubsection{$G=SO(k,k+r)$ (including all theories with 16 supercharges)}\label{sec:SO(k,k+r)}
Toroidal compactifications of heterotic strings have moduli spaces of the form $$O(k+r,\mathbb{Z})\backslash O(k,k+r)/O(k)\times O(k+r)\,,$$ with $r=16$. Their orbifolds have the same type of moduli space, perhaps with a smaller value of $r$ or a different left-quotient.
In this section we analyze the general case of a theory whose symmetric moduli space has duality group $SO(k,k+r)$, for generic $k$ and $r$.

We assume that $r\geq 2$, in which case the group is non-split. 
The restricted root system of $\mathfrak{so}_{k,k+r}$ is of type $B_k$, as for $\mathfrak{so}_{k,k+1}$. The long roots have multiplicity $1$, while the short roots have multiplicity $r$:
\begin{align}
\left(\underline{\pm1,\pm1,0_{k-2}} \right)\quad \text{multiplicity } 1 \,,\quad
\left(\underline{\pm1,0_{k-1}} \right)\quad\text{multiplicity }r \, .
\end{align}
Choosing the positive Weyl chamber as \begin{align}\overline{\mathfrak{a}^+}=\lbrace( x_1,\dots, x_k)\,:\, x_1\geq  x_2\geq\dots\geq x_k\geq 0\rbrace \, ,\end{align}
we have that the following roots will contribute to $\Gamma(H)$:
\begin{align}
\begin{cases} e_i + e_j & \text{ if }i\neq j\,,\, x_i\neq 0 \text{ or } x_j\neq 0 \\
e_i - e_j &\text{ if } x_i >  x_j\\
e_i &\text{ if }x_i\neq 0
\end{cases} \, .
\end{align}
We now consider the ray $H_n$ for which the first $n$ Cartan coordinates are
taken large at the same rate.
Choosing $H=H_n$ in the region where $\vec{x} = (x^n, (0)^{k-n})$, $x>0$, the roots contributing being
\begin{align}
\begin{cases} e_i + e_j & \text{ for }
i=1,\dots,n\,,\,
j=i+1,\dots,n \\
e_i \pm e_j &\text{ for }i=1,\dots, n
\,,\,
j=n+1,\dots,k \\
e_i &\text{ for }i=1,\dots n
\end{cases} \, ,
\end{align}
leading us to
\begin{align}
\Gamma(H_n) =  
\sum_{\beta(H_n)>0}m_{\beta}\beta 
=(r+2k-1-n) \sum_{i=1}^{n}e_{i} \, .
\end{align}

To get $c$ we must compute the product with the weight of the leading tower. 
A general weight can be expressed as
\begin{align}
\lambda = \sum_{i=1}^k \lambda_i \omega_i\,,
\end{align}
where $\omega_i$ are the $k$ fundamental weights of the restricted root system (in the case of type $B_k$), and $a_i\in\mathbb{Z}$ are the Dynkin labels of the corresponding irreducible representation.
The simple restricted roots may be chosen as:
\begin{align}
\begin{cases}
e_1 - e_2,\quad
e_2 - e_3,\quad
\dots,\quad
e_{k-1} - e_k\quad&\text{multiplicity } 1\\
e_k\quad&\text{multiplicity }r
\end{cases}\,.
\end{align}
The corresponding coroots are given by 
\begin{align}
\begin{cases}
e_1 - e_2,\,
e_2 - e_3,\dots\,
e_{k-1} - e_k\\
2e_k
\end{cases}\,.
\end{align}
We have that $\omega_i \cdot (e_j - e_{j+1}) = \delta_{ij}$, $j=1,\dots k-1$. $\omega_k \cdot 2 e_k = 1$, so the fundamental weights are
\begin{align}
\begin{cases}
\omega_s = \sum_{i=1}^s e_i\text{ for }s=1,\dots,k-1
\\ \omega_k = \frac12\sum_{i=1}^k e_i
\end{cases}\,,
\end{align}
then
\begin{align}\Gamma(H_n) = & (r+2k-1-n) (1+\delta_{n,k})\omega_n \, , \\
c(\lambda,H_n) = \lambda \cdot \Gamma(H_n) = &
\sum_{i=1}^k \lambda_i (r+2k-1-n) (1+\delta_{n,k}) W_{in} \, .\end{align}
Using that $W_{in} =\frac{\text{min}(i,n)}{(1+\delta_{i,k})(1+\delta_{n,k})}$\, we obtain
\begin{align}\label{eq:c_O(k,k+r)}
c(\lambda,H_n) = (r+2k-1-n)\sum_{i=1}^k \lambda_i   \frac{\text{min}(i,n)} {(1+\delta_{i,k})} \, .
\end{align}
In heterotic string compactifications, particles transform in the defining vector representation which is proportional to the first fundamental weight: 
\begin{align}
\lambda_{\text{vec}} = e_1 = (1+\delta_{1,k})\omega_1\,,
\end{align}
and we get:
\begin{align}\label{eq:c_heterotic}
c(\lambda_{\text{vec}},H_n) = r+2k-1-n  \, .
\end{align}
Then the $c$-values are always integers and trivially satisfy the quantization required by Proposition \ref{prop:hyperbolic towers quantization}. Note that the $c$-value depends on the rank $r$ of the gauge group; this rank was argued to be bounded above in \cite{Hamada:2021yxy} and subsequent work. It would be interesting to understand the interpretation of the corresponding bounds on $c$-values.

\section{\texorpdfstring{$Q= 8$}{Q=8} supercharges}\label{sec:8charges}
In this section we consider theories with 8 supercharges, with a focus on 5d $\C{N}=1$ theories obtained by compactifying on M-theory on Calabi--Yau threefolds. We begin by describing their low-energy limit, 5d $\C{N}=1$ supergravity. We continue by showing that generic towers of states in so-called Class A and B limits \cite{Kaufmann:2024gqo} in their vector multiplet moduli space are hyperbolic, with some important exceptions given by the gravity-decoupling sectors recently studied in \cite{Marchesano:2023thx,Castellano:2024gwi,Marchesano:2024tod,Blanco:2025qom,Castellano:2026bnx,Aoufia:2026bau}. We then proceed to discuss decompactification limits to 6d $\C{N}=(1,0)$, explicitly constructing the axions and instantons as in Proposition \ref{prop:hyperbolic towers}. We find that these axions/instantons are not directly visible from a single supergravity description due to the fact that the 6d BPS masses can undergo wall-crossing transitions as the axionic moduli are shifted, leading to discrete jumps in the structure constants $\C{F}_{IJK}$ for the prepotential. Altogether, this establishes the content of Section \ref{sec:2} in these cases, and we then illustrate this in several examples. 

At the end of this section, we briefly discuss other theories with 8 supercharges, and we focus on tensionless string limits in 6d $\C{N}=(1, 0)$ tensor moduli spaces. These limits will violate our Assumptions \ref{ass:ESC} and \ref{ass:perturbative-esc}, but we curiously find that the hyperbolic towers quantization condition of Proposition \ref{prop:hyperbolic towers quantization} is still satisfied, hinting that our proposal may still be valid in cases where our assumptions fail. 

\subsection{5d \texorpdfstring{$\C{N}=1$}{N=1} supergravity}
The massless spectrum of 5d $\C{N}=1$ supergravity consists of the gravity multiplet, an arbitrary number $n_V$ of vector multiplets and $n_H$ hypermultiplets. In particular, we will focus on the vector multiplet moduli $\phi^i$. The relevant part of the bosonic two-derivative action is given by \cite{Cadavid:1995bk,Ferrara:1996hh,Ferrara:1996wv,Bergshoeff:2004kh,Lauria:2020rhc}
\begin{equation}\label{eq:5dact}
    S_{\rm bos}=\frac{1}{2\kappa_5^2}\int \left( \star {\cal R}- g_{ij} \,d\phi^i\wedge \star d\phi^j \right) - \frac{1}{2g_5^2}\int f_{IJ}F^I\wedge \star F^J-\frac{1}{6}\int\mathcal F_{IJK}A^I\wedge F^J\wedge F^K \, . 
\end{equation}
From now on, we will work in units where $\kappa_5=1$. When the 5d theory comes from dimensional reduction of M-theory on a Calabi-Yau threefold $Y_3$, the number of vector multiplets is given by the Hodge numbers of the threefold as $n_V = h^{1,1}(Y_3)-1$. The geometry of the classical moduli space can be expressed by introducing real homogeneous coordinates $X^I, \ I=0,\dots n_V$ through a prepotential function
\begin{equation}
    \cF [X] = \frac{1}{3!}\cF_{IJK}X^IX^JX^K \, ,
\end{equation}
with $\cF_{IJK} \in \mathbb{Z}$ constant parameters also appearing in the Chern-Simons term in \eqref{eq:5dact}. From the top-down perspective, these constants are given by the triple intersection numbers of $Y_3$.  The vector multiplet moduli space coordinates $\phi^i$ then parametrize the submanifold
\begin{equation}
    \cF[X] \stackrel{!}{=}1 \, .
\end{equation}
One can use projective coordinates $X^I$ and the prepotential to define a metric over the ambient space $g_{IJ}(X)$, which can then be pulled back to the constant volume slice\footnote{The $\cF[X]=1$ slice is denoted `constant volume' slice since, from the top-down perspective, the prepotential is proportional to the volume of the $Y_3$, and belongs to a hypermultiplet.} as
\begin{equation}
    g_{IJ} = -\frac12\partial_I \partial_J \log \cF \ , \quad g_{ij}=\partial_iX^I\partial_j X^J g_{IJ} \, ,
\end{equation}
while, explicitly, the metric over the homogeneous coordinates is also proportional to the gauge-kinetic function through
\begin{equation}
    g_{IJ} = \cF^{-2/3} f_{IJ} = \frac{1}{2}\left( \frac{\cF_I \cF_J}{\cF^2}-\frac{\cF_{IJ}}{\cF}\right) \, ,
\end{equation}
where we introduced the short-hand notation $\cF_I=\partial_I\cF, \ \cF_{IJ}=\partial_I\partial_J\cF$.

Many interesting aspects of the geometry of the vector multiplet moduli space stem from its characterization as a constraint hypersurface in the ambient space spanned by the $X^I$. Notably, this submanifold is \textit{totally geodesic}, meaning that normal vectors to $\cF[X]=1$ are covariantly constant. Their form is particularly simple, and can be computed to be \cite{Aoufia:2025ppe}
\begin{equation}
    n^I = \sqrt{\frac23}X^I \ , \quad \nabla_I n^J=0 \, .
\end{equation}
Moreover, the normal vector is simply the generator of dilations $X^I \to \lambda X^I$, reflecting that these moduli-space coordinates are projective. One can thus construct the projector on the constant volume slice 
\begin{equation}\label{eq:5dgrestricted}
    \hat g_{IJ} = g_{IJ}- n_In_J  = \frac{1}{2}\left(\frac23 \frac{{\cal F}_I{\cal F}_J}{{\cal F}^2}- \frac{{\cal F}_{IJ}}{\cal F}\right) \, ,
\end{equation}
which can be regarded as the metric on the space of $X^I$'s identified by $X^I \to \lambda X^I$ and is the metric used to compute norms. Since we will be interested in the Laplace-Beltrami operator, let us also discuss how this restricts on the constant volume slice. Due to the submanifold being totally geodesic, the formula is particularly simple: given a function $f$, one computes \cite{ecker2004regularity}
\begin{equation}\label{eq:laplaamb}
    \Delta_{\phi} f = \Delta_Xf - \mathrm{Hess}_f(\mathbf{n},\mathbf{n}) \, ,
\end{equation}
where $\mathbf{n}$ is the unit normal vector to the hypersurface, $\Delta_\phi$ denotes its Laplacian, $\Delta_X$ corresponds to the ambient one in the space of the $X^I$, and 
\begin{equation}
    \mathrm{Hess}_f(\mathbf{n},\mathbf{n}) = n^In^J \nabla_I \nabla_J f \, .
\end{equation}

In the following, we will be interested in computing the Laplacian of masses of $\frac12$-BPS objects in the supergravity theory. For the vector multiplet moduli space in question, we will be interested in BPS particles and BPS strings, whose masses and tensions are given, respectively, by their central charge 
\begin{equation}\label{eq:5dtenandmass}
    M_{\rm BPS} = \frac{q_I X^I}{\cF^{1/3}} \ , \qquad {\cal T}_{\rm BPS} = \frac{1}{2} \frac{p^I\cF_I}{\cF^{2/3}} \, ,
\end{equation}
where $q_I$ and $p^I$ describe their quantized (electric and magnetic) charges under the gauge fields $F^I$. Their physical charges are instead given by
\begin{equation}\label{eq:5dcharges}
    {\cal Q}^2 = q_If^{IJ}q_J \ , \qquad {\cal P}^2 = p^I f_{IJ}p^J \, .
\end{equation}
In terms of the spectrum of M-theory, these objects are given by M2-branes and M5-branes wrapping calibrated 2-cycles and 4-cycles in $Y_3$. The resulting BPS spectrum of the five-dimensional theory is well described in terms of conical structures underlying the $q^I$ and $p_I$. We will briefly describe it following \cite{Kaufmann:2024gqo}. Specifically, one can define the cone of mutually BPS particles ${\cal C}_{\rm part}$ which have $M_{\rm BPS}\ge 0$ and preserve the same amount of bulk supersymmetries. The dual cone $\mathbf{\Delta}$ in the space of the $X^I$ defines the K\"ahler cone of the supergravity theory. In the following, we will work in local patches described by the simplicial cone
\begin{equation}
    \mathbf{\Delta}= \{ X^I\ge0\} \, , \qquad \ {\cal C}_{\rm part} = \{ q_I \ge 0 \ , \  q^I \in \mathbb{N}\} \, .
\end{equation}
The actual cones describing the supergravity theory are constructed by patching together these subcones, a fact which will not affect our conclusions based on the local analysis. Similarly, one can define the cone generated by the first derivative of the prepotential and its (integer) dual, the cone of BPS strings\footnote{More precisely, $\mathbf{\Gamma}^{\vee} = \{ p^I : p^I {\cal F}_I \ge 0 \ , \ \forall {\cal F}_I \in \mathbf{\Gamma} \}$ }:
\begin{equation}
    \mathbf{\Gamma} = \{ \cF_I(X)  : X \in \mathbf{\Delta}\ \} \, , \qquad {\cal C}_{\rm str} =  \mathbf{\Gamma}^{\vee}\cap \mathbb{Z}^{n_V+1} \, .
\end{equation}
As originally put forward in \cite{Katz:2020ewz}, one can introduce the cone of strings which can never decouple from gravity, dubbed \textit{supergravity strings}, as the intersection
\begin{equation}
    {\cal C}_{\rm str} \supseteq {\cal C}_{\rm sugra} \equiv \mathbf{\Delta} \cap \mathbb{Z}^{n_V+1} \, .
\end{equation}
In the local simplicial cone of choice, these supergravity strings simply span the cone 
\begin{equation}
    {\cal C}_{\rm sugra} = \{ p^I \ge 0\} \, .
\end{equation}

In subsequent sections, we will be interested in computing the Laplacian of the logarithm of masses and tensions for the generators of ${\cal C}_{\rm part}, \ {\cal C}_{\rm sugra}$. In passing, we will also mention the role of \textit{rigid} sectors, which more in general lie in ${\cal C}_{\rm str}$.  We will see that the Laplacian of logarithms of masses approaches ---with some important exceptions--- a constant value asymptotically in moduli space, thus proving hyperbolicity for the associated towers. In particular, this holds universally for the leading tower in agreement with Proposition \eqref{prop:hyperbolic towers}. The exceptions always involve rigid strings and particles, a fact which supports that the hyperbolicity condition is tightly linked to gravitational objects. Additionally, we find other subleading towers displaying the same behaviour. We inspect a few explicit examples and connect with the curvature and $\Gamma$-vectors introduced in the above sections.

\paragraph{BPS strings.} We start by analyzing the BPS strings of the supergravity theory, which tensions and charges can be read from \eqref{eq:5dtenandmass} and \eqref{eq:5dcharges}. Let us compute the $\alpha$-vectors as 
\begin{equation}
    \partial_I {\cal T} = -{\cal F}^{1/3}\hat g_{IJ} p^J\quad \Rightarrow \quad \alpha^I = - (\vec\nabla \log {\cal T})^I=-\hat g^{IJ}\frac{\partial_I {\cal T}}{\cal T} ={\cal F}^{1/3}\frac{p^I}{\cal T} \, .
\end{equation}
The norm of this vector is readily computed to be 
\begin{equation}\label{eq:5dstralpha}
   ||\alpha|| = \frac{4}{3}- {\cal F}^{-2/3}\frac{p^I {\cal F}_{IJ}p^J}{2{\cal T}^2} \, .
\end{equation}
Demanding that the norm be constant everywhere imposes the condition $p^I {\cal F}_{IJ}p^J =0$. Of course, hyperbolicity of the associated tower of oscillation modes only requires that asymptotically the norm of the $\alpha$-vector approaches a constant, i.e.\ that
\begin{equation}
    \frac{\tilde q^I {\cal F}_{IJ}\tilde q^J}{{\cal T}^2} \sim \rm const \, .
\end{equation}
This condition is interestingly linked with the impossibility of decoupling the BPS string from gravity, as that would be precisely due to an asymptotic divergence of this term \cite{Marchesano:2023thx, Marchesano:2024tod, Castellano:2024gwi, Blanco:2025qom, Aoufia:2026mqb}. Indeed, the physical charge and tension of the strings are related through the no-force condition 
\begin{equation}\label{eq:5dhypcondstr}
    {\cal P}^2 =\frac23{\cal T}^2 + g^{IJ}\partial_I {\cal T}\partial_J{\cal T} \, ,
\end{equation}
as can be seen by manipulating \eqref{eq:5dstralpha}. Since the charge-to-mass ratio $\gamma_{\bf p}^2\equiv \C{P}^2/\C{T}^2$ is constant and cannot diverge along limits where \eqref{eq:5dhypcondstr} holds, this condition precisely obstructs the decoupling criterion identified in \cite{Castellano:2024gwi,Blanco:2025qom}. Finally, it is worth remarking that in limits such that the constant in \eqref{eq:5dhypcondstr} vanishes, the string $\alpha$-vector gives rise to an oscillator modes' $\vec \alpha_{\rm osc}$-vector characterized by 
\begin{equation}
   ||\vec \alpha_{\rm osc}||^2 \sim \frac{1}{3}=\frac{1}{d-2} \, ,
\end{equation}
thus agreeing with the ESC in the regimes where this tower becomes the lightest.

Consider now to compute the ambient space Laplacian, which can be used to compute the Laplacian on moduli space through \eqref{eq:laplaamb}. Due to the function being homogeneous of degree zero, an explicit computation reveals that the Hessian vanishes, thus bringing us to
\begin{equation}
    \nabla^2_{{\cal F}=1} \log {\cal T} = \hat g^{IJ}\frac{\nabla_I\partial_J {\cal T}}{{\cal T}} - \frac{4}{3} +  {\cal F}^{-2/3}\frac{\tilde q^I {\cal F}_{IJ}\tilde q^J}{2{\cal T}^2}\, ,
\end{equation}
and where the first contribution reads explicitly
\begin{equation}
\begin{split}
     g^{IJ}\frac{\nabla_I\partial_J {\cal T}}{{\cal T}} =\left( h^{1,1} - \frac23 \right) {\cal T} - \frac12{\cal F}^{1/3} \tilde q^L {\cal F}_{LMN}{\cal F}^{MN} \, .
\end{split}
\end{equation}
All in all, we have that the Laplacian of the logarithm of the tower mass $m_{\rm osc}= \sqrt{{\cal T}}$ associated to a BPS string of the 5d supergravity theory is
\begin{equation}\label{eq:lapl5dstrings}
    \nabla^2_{{\cal F}=1} \log {m_{\rm osc}} = \frac12(h^{1,1}-2) - \frac{1}{4{\cal T}}p^L {\cal F}_{LIJ}{\cal F}^{IJ} + \frac{\tilde q^I {\cal F}_{IJ}\tilde q^J}{4{\cal T}^2}\, ,
\end{equation}
where we imposed the constant volume constraint. Notice that the second term contains the 't Hooft anomaly matrix $k_{IJ} =p^L {\cal F}_{LIJ} $ that characterizes the anomaly-inflow on the string worldsheet theory \cite{Boyarsky:2002ck,Katz:2020ewz}. In a subsequent section, we will investigate how the right-hand side of this equation behaves in various infinite distance limit in the vector multiplet moduli space of the supergravity theory, which have been classified and explored in \cite{Kaufmann:2024gqo}.

\paragraph{BPS particles.} Let us next consider the logarithm of the central charge associated to the mass of a BPS particle. Its derivative reads
\begin{equation}
    \partial_I \log M_{\rm BPS} = \frac{\cF^{-1/3} q_I}{M_{\rm BPS}} -\frac{1}{3}\frac{\cF_I}{\cF}  \, ,
\end{equation}
while the norm is given by
\begin{equation}\label{eq:gammaalphanorm}
    ||\alpha||^2 =  \gamma_{\bf{q}}^2 - \frac{2}{3} \, .
\end{equation}
Once again, $\gamma_{\bf{q}}^2 \equiv {\cal Q}^2/M_{\rm BPS}^2$ is the charge-to-mass ratio of the BPS particle. This equation can be read as a no-force condition between pairs of mutually BPS particles. Similarly to what happened for BPS strings, notice that the hyperbolicity condition constrains $\gamma^2$ to approach a constant. In particular, the term sensitive to the limit is of the form
\begin{equation}
    \gamma^2 \supset 2 \frac{q_I \cF^{IJ} q_J}{M_{\rm BPS}^2} \sim \rm const \, ,
\end{equation} 
once again obstructing the decoupling criterion. Using the same strategy as for the case of BPS strings, one can compute the Laplacian restricted on the constraint hypersurface $\nabla^2_{\cF=1}$. In particular one can verify that
\begin{equation}\label{eq:5dlapllogm}
    \nabla^2_{\cF=1}\log M_\text{BPS} = \frac{1}{2}(h^{1,1}-2) + \frac{q_K\cF^{KL}\cF_{IJL}\cF^{IJ}}{M_{\rm BPS}} + 2\frac{q_I\cF^{IJ}q_J}{M_{\rm BPS}^2} \, . 
\end{equation}
We see that the structure is the same as the case of BPS strings, and in particular, hyperbolicity once again requires the $\gamma^2_{\bf q}$ to approach a constant asymptotically.

\subsection{Class A and B limits}
We now proceed to study how the right-hand side of equations \eqref{eq:lapl5dstrings}, \eqref{eq:5dlapllogm} behaves in infinite distance limits in the vector multiplet moduli space. Their systematic study for the case of M-theory on a Calabi$-$Yau threefold has been carried out in \cite{Lee:2019wij}, and more recently a bottom-up analysis has been performed in \cite{Kaufmann:2024gqo}. We will use the language of the latter, where the possible consistent limits are divided in Class A and B. From the M-theory perspective, these are respectively decompactification limits to 6d $\C{N}=(1,0)$ supergravity or emergent string limits. In each class, the constraints on the coefficients $\cF_{IJK}$ brought about by the absence of anomalies on the worldsheet theory of supergravity strings are such that the prepotential takes a simple form. As already remarked, we will assume to work in a basis such that the K\"ahler cone is given by $\mathbf{\Delta} =\{X^I \ge 0\}$, and the respective cones of BPS strings and particles are discussed briefly above and more extensively e.g.\ in \cite{Kaufmann:2024gqo, Reece:2025zva}. We will first employ a bottom-up analysis based on \emph{genericity} of the triple-intersection numbers, which will allow us to conclude the generic finiteness of the non-constant terms in \eqref{eq:lapl5dstrings}. We will later turn to explicit examples to classify possible deviations from this naïve behavior, together with determining the quantization condition as required by Proposition \ref{prop:hyperbolic towers quantization}.

\subsubsection{Class A limits}
 Let us start by inspecting Class A limits. Splitting the moduli as $\{X^I \} = \{ X^0, X^i,X^r\}$, the limit is defined compatibly with the $\cF[X]=1$ restriction by the scalings  
\begin{equation}
    X^0 \sim \lambda \, , \quad X^i \precsim\lambda^{-2} \, , \quad X^r \precsim \lambda \, ,
\end{equation}
while the various constraints found on the prepotential coefficients impose that ${\cal F}_{0rs} = {\cal F}_{rst} = 0$, meaning the prepotential takes the form
\begin{equation}\label{eq:prepclassA}
\begin{split}
    {\cal F} = &\frac{1}{2}\mathcal{F}_{00i}(X^0)^2 X^i + \frac{1}{2}\mathcal{F}_{0ij}X^0 X^i X^j + \frac{1}{6}\mathcal{F}_{0ir}X^0 X^i X^r \\
&+ \frac{1}{6}\mathcal{F}_{ijk}X^i X^j X^k + \frac{1}{2}\mathcal{F}_{ijr}X^i X^j X^r + \frac{1}{2}\mathcal{F}_{irs}X^i X^r X^s.
\end{split}
\end{equation}
Since we are temporarily interested in an upper bound on the value of $\nabla^2_{{\cal F}=1} \log m $, we can restrict our analysis to the limiting case where all parametric inequalities are saturated. The prepotential above leads to the scaling of elementary tensions to be
\begin{equation}
    \cF_0 \sim \lambda^{-1} \, , \quad \cF_{i} \sim \lambda^2  \, , \quad \cF_r \sim \lambda^{-1} \, ,
\end{equation}
while the second derivatives yield the leading matrix contribution
\begin{equation}
    {\cF}_{IJ} \supset\begin{pmatrix}
        \cF_{00i}X^i & \cF_{0ir}X^r & \cF_{0ri}X^i \\
        \cF_{0ir}X^r & \cF_{ij0}X^0  & \cF_{0ir}X^0 \\
        \cF_{0ri}X^i&\cF_{0ir}X^0 & \cF_{rsi}X^i
    \end{pmatrix} \simeq \begin{pmatrix}
        \lambda^{-2} & \lambda & \lambda^{-2} \\
        \lambda & \lambda & \lambda \\
        \lambda^{-2} & \lambda & \lambda^{-2}
    \end{pmatrix}+{\cal O}(\lambda^{-3}) \, .
\end{equation}
Computing its inverse, one gets parametrically that 
\begin{equation}
    \cF^{MN} \simeq \begin{pmatrix}
        \lambda^2 & \lambda^{-1} & \lambda^2 \\
        \lambda^{-1} & \lambda^{-4} & \lambda^{-1}\\
        \lambda^2 & \lambda^{-1} & \lambda^2
    \end{pmatrix} + {\cal O}(\lambda^{-5}) \, .
\end{equation}
Notice that any other subleading scaling of the coordinates would have brought a subleading scaling in the above matrix. Moreover, we assume generically no asymptotic kernel that could enhance the behaviour of the eigenvalues of this matrix. Indeed, computationally, this is at the root of the divergence of the $\gamma^2_{{\bf q},{\bf p}}$ factor as shown e.g.\ in \cite{Marchesano:2024tod,Blanco:2025qom}, which does not happen for supergravity strings. We will see an example of this case in explicit examples below. We are now ready to compute \eqref{eq:lapl5dstrings}, \eqref{eq:5dlapllogm} for supergravity strings and BPS particles such that their charges lie in the cones
\begin{equation}
    {\cal C}_{\rm sugra} = \{p^I \ge 0\} \, , \quad  {\cal C}_{\rm part} = \{q_I \ge 0\} \, ,
\end{equation}
and in particular, we will evaluate them for their generators. 
\paragraph{${\mathbf{q_I}}  =  {\mathbf{\boldsymbol\delta_{0I}}} $ and ${\mathbf{p^I}}  =  {\mathbf{\boldsymbol\delta^{0I}}} $.} In this case \eqref{eq:lapl5dstrings} reads
\begin{equation}\label{eq:lapllogmosc05d}
    \nabla^2_{{\cal F}=1} \log m_{\rm osc}^{(0)} = \frac12(h^{1,1}-2) - \frac{1}{4{\cal T}} {\cal F}_{0MN}{\cal F}^{MN}  +\frac{ {\cal F}_{00}}{4{\cal T}^2}\, .
\end{equation}
One can compute the matrix
\begin{equation}
    \cF_{0IJ} = \begin{pmatrix}
        0 & \cF_{00i} & 0\\
        \cF_{0i0} & \cF_{0ij} & \cF_{0ir} \\
        0 &\cF_{0ri} & 0
    \end{pmatrix} \, ,
\end{equation}
and the vector $\cF^{0I} \sim \left( \lambda^2, \ \lambda^{-1}, \ \lambda^{2} \right)$.
We see that the structure of the triple intersection numbers is such that at most, assuming no cancellations, we have
\begin{equation}\label{eq:5dlogmosc0}
     \nabla^2_{{\cal F}=1} \log m_{\rm osc}^{(0)} = \frac12 (h^{1,1}-2) + c_0^s + c_1^s + {\cal O}(\lambda^{-1})\, ,
\end{equation}
where the constants $c^s$ depend on the $\cF_{IJK}$ coefficients and they correspond to the leading contribution in the two moduli-dependent terms in \eqref{eq:lapllogmosc05d}. Notice that if not for conditions defining a consistent Class A limit, namely ${\cal F}_{0rs} = {\cal F}_{rst} = 0$, some of these terms could have easily diverged. Turning then to \eqref{eq:5dlapllogm}, the Laplace equation reads
\begin{equation}
    \nabla^2_{\cF=1}\log M_{BPS}^{(0)} = \frac{1}{2}(h^{1,1}-2) + \frac{\cF^{0L}\cF_{IJL}\cF^{IJ}}{M_{\rm BPS}} + 2\frac{\cF^{00}}{M_{\rm BPS}^2} \, . 
\end{equation}
Once again, the restricted structure of the triple intersection numbers is such that 
\begin{equation}\label{eq:logmbps0}
     \nabla^2_{\cF=1}\log M_{BPS}^{(0)} = \frac{1}{2}(h^{1,1}-2) + c_0^p + c_1^p + {\cal O}(\lambda^{-1}) \, ,
\end{equation}
with $c^p$ constants depending on the $\cF_{IJK}$ and that cannot be determined from this  naïve analysis.

\paragraph{${\mathbf{q_I}}  =  {\mathbf{\boldsymbol\delta_{iI}}} $ and ${\mathbf{p^I}}  =  {\mathbf{\boldsymbol\delta^{iI}}} $.} For this choice of charges, the matrix $\cF_{iIJ}$ generically doesn't have any non-vanishing entries, while $\cF^{iI} \sim \left( \lambda^{-1}, \, \lambda^{-4}, \lambda^{-1} \right)$. The Laplacian of the logarithm for oscillator modes reads
\begin{equation}
\begin{split}
    \nabla^2_{{\cal F}=1} \log m_{\rm osc}^{(i)}&= \frac12(h^{1,1}-2) - \frac{1}{4{\cal T}} {\cal F}_{iMN}{\cal F}^{MN}  +\frac{ {\cal F}_{ii}}{4{\cal T}^2}\ \\ &=\frac12 (h^{1,1}-2) + c_0^s+ {\cal O}(\lambda^{-3}) \, ,
\end{split}
\end{equation}
where again $c^s_0$ is a constant depending on the prepotential coefficients corresponding to the first moduli-dependent term. Repeating the computation for BPS particles, one gets 
\begin{equation}
\begin{split}
     \nabla^2_{\cF=1}\log M_{BPS}^{(i)} &= \frac{1}{2}(h^{1,1}-2) + \frac{\cF^{iL}\cF_{IJL}\cF^{IJ}}{M_{\rm BPS}} + 2\frac{\cF^{ii}}{M_{\rm BPS}^2} \\ &=\frac{1}{2}(h^{1,1}-2) + c_0^p + c_1^p + {\cal O}(\lambda^{-1}) \, ,
\end{split}
\end{equation}
proving hyperbolicity for all objects in the supergravity strings and particle cones.
\paragraph{${\mathbf{q_I}}  =  {\mathbf{\boldsymbol\delta_{rI}}} $ and ${\mathbf{p^I}}  =  {\mathbf{\boldsymbol\delta^{rI}}} $.} Similarly to the case of the leading direction, the matrix $\cF_{rIJ}$ takes the form
\begin{equation}
    \cF_{rIJ} = \begin{pmatrix}
        0 & \cF_{r0i} & 0\\
        \cF_{ri0} & \cF_{rij} & \cF_{ris} \\
        0 &\cF_{rsi} & 0
    \end{pmatrix} \, ,
\end{equation}
while $\cF^{rI}\sim \left( \lambda^2, \lambda^{-1},  \lambda^{2} \right)$. Notice that all scalings of masses and tensions are precisely the same as in the case of $p^I = \delta^{0I}$ and $q_I=\delta_{0I}$, so that the Laplacian of BPS strings and particles once again take the form \eqref{eq:5dlogmosc0} and \eqref{eq:logmbps0} respectively.

All in all, we have shown that ---barring possible gravity decoupling sectors, which will be discussed in explicit examples below--- towers of oscillators modes and towers of BPS particles whose charges generate the cones ${\cal C}_{\rm sugra}$ and ${\cal C}_{\rm part}$ are hyperbolic in all Class A limits. We now show the same for Class B limits. 

\subsubsection{Class B limits}

In \cite{Kaufmann:2024gqo}, two types of Class B limits are described. The first type, where the only coordinate which is leading is $X^0$, and the second type, where multiple coordinates co-lead. In this second case \cite{Kaufmann:2024gqo} showed that these limits have the same asymptotics as the Class A limits after a change of variables, and as such the scalings are exactly the same as above. 

Let us thus focus on limits of the former kind, and further divide the possible scalings for the coordinates $X^I$ in two subcases following \cite{Kaufmann:2024gqo}. Specifically, for the first subcase, one can split the coordinates as $X^I = \{ X^0, X^\mu\}$ and take the limit to be defined by 
\begin{equation}
     X^0 \sim \lambda \, , \quad X^\mu \precsim \lambda^{-1/2} \  \ \forall \mu \, .
\end{equation}
We will furthermore assume to saturate the parametric inequality to derive an upper bound on the value of $\nabla^2 \log m$ for all towers.
The consistency of the worldsheet theory for supergravity strings imposes that the constant volume constraint is satisfied by a term of the type $\cF_{0\mu \nu} X^0 X^\mu X^\nu$, while other terms are either zero or subleading, compatibly with the above scaling of the coordinates. Concretely, the prepotential takes the restricted form
\begin{equation}
    \cF = \cF_{0\mu\nu}X^0X^\mu X^\nu + \cF_{\mu \nu \rho} X^\mu X^\nu X^\rho \, .
\end{equation}
The scalings of the various elementary tensions read
\begin{equation}
    \cF_{0} \sim \lambda^{-1} \, , \quad \cF_{\mu} \sim \lambda^{1/2} \, ,
\end{equation}
while the matrix of second derivatives of the prepotential and its inverse read respectively
\begin{equation}
    \cF_{IJ} = \begin{pmatrix}
        0 & \cF_{0\mu\nu}X^{\nu}\\
        \cF_{\mu 0\nu}X^{\nu}&\cF_{\mu\nu 0} X^0 + \cF_{\mu\nu\rho} X^\rho
    \end{pmatrix} \simeq \begin{pmatrix}
        0 & \lambda^{-1/2}\\
        \lambda^{-1/2} & \lambda + \lambda^{-1/2}
    \end{pmatrix} \, , 
\end{equation}
\begin{equation}   
    \qquad \quad \cF^{IJ} \simeq \begin{pmatrix}
        \lambda^2 + \lambda^{1/2}&\lambda^{1/2} \\
        \lambda^{1/2}&0
    \end{pmatrix} \, . 
\end{equation}
As before, we will assume no asymptotic kernel that would enhance the eigenvalues of this matrix, and we will focus on the generators of the supergravity strings and the BPS particle cones of charges ${\cal C}_{\rm sugra}$ and ${\cal C}_{\rm part}$.

\paragraph{${\mathbf{q_I}}  =  {\mathbf{\boldsymbol\delta_{0I}}} $ and ${\mathbf{p^I}}  =  {\mathbf{\boldsymbol\delta^{0I}}} $.} Since the structure of the prepotential coefficients is very restricted, the $\cF_{0IJ}$ matrix simplifies significantly and reads
\begin{equation}
    \cF_{0IJ} = \begin{pmatrix}
        0 & 0 \\
        0 & \cF_{0\mu\nu} 
    \end{pmatrix} \, , 
\end{equation}
while the vector $\cF^{0I} $ reads $\cF^{0I}\sim \left( \lambda^2, \lambda^{1/2}\right)$. This implies specifically that the Laplacian of the logarithm for string oscillator masses is 
\begin{equation}
     \nabla^2_{{\cal F}=1} \log m_{\rm osc}^{(0)} = \frac12(h^{1,1}-2) - \frac{1}{4{\cal T}} {\cal F}_{0MN}{\cal F}^{MN}  +\frac{ {\cal F}_{00}}{4{\cal T}^2}=\frac12(h^{1,1}-2) \, ,
\end{equation}
which is seemingly constant everywhere in moduli space, since it only depends on the highly constrained structure of the prepotential coefficients. In explicit examples, this can, and does, change, as we will comment in the following. For the BPS particles tower, one can compute the scaling of the matrix
\begin{equation}
    \cF^{0L}\cF_{LIJ} \simeq \begin{pmatrix}
        0 & \lambda^{1/2} \\ \lambda^{1/2} & \lambda^2 + \lambda^{1/2}
    \end{pmatrix} \, , 
\end{equation}
which straightforwardly leads to 
\begin{equation}
\begin{split}
     \nabla^2_{\cF=1}\log M_{BPS}^{(0)} &= \frac{1}{2}(h^{1,1}-2) + \frac{\cF^{0L}\cF_{IJL}\cF^{IJ}}{M_{\rm BPS}} + 2\frac{\cF^{00}}{M_{\rm BPS}^2} \\ &=\frac{1}{2}(h^{1,1}-2) + c_0^p + c_1^p + {\cal O}(\lambda^{-1}) \, ,
\end{split}
\end{equation}
with $c^p$ constants depending on the $\cF_{IJK}$ coefficients.

\paragraph{${\mathbf{q_I}}  =  {\mathbf{\boldsymbol\delta_{\boldsymbol\mu I}}} $ and ${\mathbf{p^I}}  =  {\mathbf{\boldsymbol\delta^{\boldsymbol\mu I}}} $.}  In this case, instead, the matrix of prepotential coefficients entering the oscillator modes' formula reads
\begin{equation}
    \cF_{\mu IJ}  = \begin{pmatrix}
        0 &\cF_{\mu 0 \nu} \\
        \cF_{\mu 0 \nu} & \cF_{\mu\nu\rho}
    \end{pmatrix} \, ,
\end{equation}
while the relevant vector entering the BPS particles' formula is  $\cF^{\mu I} \sim \left( 0,\ \lambda^{1/2}\right)$. In the limit thus eq. \eqref{eq:lapl5dstrings} evaluates to
\begin{equation}
\begin{split}
    \nabla^2_{{\cal F}=1} \log m_{\rm osc}^{(\mu)}&= \frac12(h^{1,1}-2) - \frac{1}{4{\cal T}} {\cal F}_{\mu IJ}{\cal F}^{IJ}  +\frac{ {\cal F}_{\mu \mu}}{4{\cal T}^2}\ \\ &=\frac12 (h^{1,1}-2) + c_0^s+ {\cal O}(\lambda^{-1}) \, ,
\end{split}
\end{equation}
where again $c_0^s$ is a constant corresponding to the leading behavior of the first term. Turning now to the BPS particle masses, the structure of the $\cF_{IJK}$ coefficients is such that 
\begin{equation}
    \cF^{\mu L}\cF_{LIJ} \simeq \begin{pmatrix}
        0 & 0 \\ 0 & \lambda^{1/2}
    \end{pmatrix} \, ,
\end{equation}
leading ---similarly to the oscillators for the previous choice of charges--- to the identity
\begin{equation}
\begin{split}
     \nabla^2_{\cF=1}\log M_{BPS}^{(\mu)} = \frac{1}{2}(h^{1,1}-2) + \frac{\cF^{\mu L}\cF_{IJL}\cF^{IJ}}{M_{\rm BPS}} + 2\frac{\cF^{\mu \mu}}{M_{\rm BPS}^2} =\frac{1}{2}(h^{1,1}-2) \, 
\end{split}
\end{equation}
again holding seemingly everywhere in moduli space. This last equation, however, has been once again computed on the assumption of genericity of the triple intersection numbers, in particular for the inverse $\cF^{\mu \mu}$. This seems to vanish exactly in our computation, while in general it could receive subleading corrections. This indeed happens e.g.\ when the $\mu$ coordinates get split in further subgroups, as we will now introduce. 

Let us then switch to the second subcase, where the $X^\mu$ coordinates are further divided in $X^{\mu} = \{X^{\mu_0}, X^{\mu'}\}$. Let us take the $X^{\mu_0}$ coordinates to scale as
\begin{equation}
    X^{\mu_0} \sim \lambda^{-1/2 +x}\, , \quad x \in \left( 0,\frac32\right) \, ,
\end{equation}
such that their scaling is subleading with respect to $X^0$ for every value of the parameter $x$. Moreover, in order for the constant volume constraint to be respected, there must be at least one of the $\mu'$ coordinates compensating the extra growth brought about by the scaling with $x$, and $\cF_{0\mu_0 \mu_0} = 0$. In the following, we will take the simplifying assumption that all coordinates are of these three types, which will not change the leading order scaling of equations \eqref{eq:lapl5dstrings}, \eqref{eq:5dlapllogm}. Thus, the limit is defined by 
\begin{equation}
    X^0 \sim \lambda, \quad X^{\mu_0} \sim \lambda^{-1/2+x} \, , \quad X^{\mu'} \sim \lambda^{-1/2-x} \, ,
\end{equation}
and the prepotential takes the form
\begin{equation}
    \cF = \cF_{0\mu_0\nu'}X^0X^{\mu_0} X^{\nu'} + \cF_{\mu \nu \rho} X^\mu X^\nu X^\rho \, , \quad \mu=\{\mu_0, \mu'\} \, ,
\end{equation}
leading to the following scalings for the elementary tensions
\begin{equation}
    \cF_0 \sim \lambda^{-1} \, , \quad \cF_{\mu_0} \sim \lambda^{\frac12-x} \, , \quad \cF_{\mu'} \sim \lambda^{\frac12+x} \, .
\end{equation}
One can verify that taking $x\ge \frac12$ leads to overall subleading scalings in the quantities involved in the computation. Thus, we will choose, for simplicity of exposition, $x<\frac12$. The matrix of second derivatives of the prepotential and its inverse read
\begin{equation}
    \cF_{IJ} \simeq \begin{pmatrix}
        0 & \lambda^{-\frac12-x} & \lambda^{-\frac12+x}\\
        \lambda^{-\frac12-x} & \lambda^{-\frac12+x}& \lambda \\
        \lambda^{-\frac12+x} & \lambda & \lambda 
    \end{pmatrix}  + {\cal O}(\lambda^{-\frac12-x}) \, ,
\end{equation}
\begin{equation}
    \cF^{IJ} \simeq \left( 1+\lambda^{-\frac32 +3x}\right)\begin{pmatrix}
        \lambda^2 & \lambda^{\frac12 + x} & \lambda^{\frac12 -x}  \\
         \lambda^{\frac12 + x} & \lambda^{-1+2x} & \lambda^{-1} \\
         \lambda^{\frac12 -x}  & \lambda^{-1} & \lambda^{-1-2x}
    \end{pmatrix} + {\cal O}(\lambda^{-1-2x})\, .
\end{equation}
Repeating the same computations as in the previous paragraphs, one can compute the following Laplacian of logarithms of oscillator scales:
\begin{subequations}
\begin{align}
    \nabla^2_{{\cal F}=1} \log m_{\rm osc}^{(0)}&=\frac12 (h^{1,1}-2) + c_0^s+ {\cal O}(\lambda^{-1}) \, ,\\\label{eq:oscmu0}
    \nabla^2_{{\cal F}=1} \log m_{\rm osc}^{(\mu_0)}&=\frac12 (h^{1,1}-2) + c_0^s+ \lambda^{-\frac32+3x}+ {\cal O}(\lambda^{-\frac32}) \, ,\\
    \nabla^2_{{\cal F}=1} \log m_{\rm osc}^{(\mu')}&=\frac12 (h^{1,1}-2) + c_0^s+{\cal O}(\lambda^{-2x}) \, ,
\end{align}
\end{subequations}
while for the BPS particles one finds that
\begin{subequations}
\begin{align}
     \nabla^2_{\cF=1}\log M_\text{BPS}^{(0)} &= \frac{1}{2}(h^{1,1}-2) + c_p^0 +c_p^1 + {\cal O}(\lambda^{-1}) \, , \\
     \nabla^2_{\cF=1}\log M_\text{BPS}^{(\mu_0)} &= \frac{1}{2}(h^{1,1}-2) + c_p^2 +c_p^3 + {\cal O}(\lambda^{-1}) \, , \\
     \nabla^2_{\cF=1}\log M_\text{BPS}^{(\mu')} &= \frac{1}{2}(h^{1,1}-2) + c_p^4 +c_p^5 + {\cal O}(\lambda^{-1}) \, .
\end{align}
\end{subequations}
Let us briefly comment on the third term left explicitly in \eqref{eq:oscmu0}. This term comes from the $\cF_{\mu_0\mu_0}$ contribution, which naively leads to a divergence when $x\ge \frac12$. In this case, however, $\cF_{\mu_0\mu_0 \mu_0}=0$ to satisfy the constant volume constraint. Due to this, one finds again that \eqref{eq:oscmu0} is of order one.

In summary, we have seen that the value of $\nabla^2_{\cF=1} m$ is bounded by a constant in all Class A and Class B limits for supergravity strings and BPS particles. A key role in this is played by the restricted structure of the prepotential coefficients coming from a consistent string embedding, discussed at length in \cite{Kaufmann:2024gqo}. In particular, the Class A limits and second kind of Class B limits discussed above geometrically correspond to decompactification limits to 6d, where the lightest states are given by the $M_{\rm BPS}^{(i)}, \ M_{\rm BPS}^{(\mu')}$ states respectively for the two limits. In the first type of Class B limits, instead, the leading states are $m_{\rm osc}^{(0)}, \ M_{\rm BPS}^{(\mu)}$. These correspond to emergent string states accompanied by a tower of KK excitations, which together constitute the emergent string limit. All these states are, in particular, hyperbolic. 

The only ingredient which could spoil our analysis based on genericity is a divergent charge-to-mass ratio, which is not expected to happen for the leading towers. 
Specifically, the cone of all strings is actually larger and contains rigid strings which are \textit{not} asymptotically hyperbolic. We will discuss a concrete example including these sectors in Section \ref{example:2}. Up to now, no specific top-down realization was assumed for the supergravity theory. In order to connect with the precise values of the Laplacian as in Propositions \ref{prop:hyperbolic towers} and \ref{prop:hyperbolic towers quantization}, we now turn to explicit constructions. Concretely, we will discuss decompactification limits, show their fibration structure, and identify the spectrum of instantons. Then, we will discuss some examples.

\subsection{Decompactification limits, wall crossing, and instantons}\label{ref:decomp5d}
Decompactification limits in the vector multiplet moduli space of 5d $\C{N}=1$ theories are all described by 6d $\C{N}=(1,0)$ on a circle. The field content of the latter is given by the gravity multiplet, $n_T$ tensor multiplets and $n_V$ vector multiplets. Let us put this theory on a circle of radius $R$: the dimensionality of the 5d Coulomb branch is given by
\begin{equation}
    n_V^{(5d)} = n_V+n_T+1 \, ,
\end{equation}
where the extra contribution comes from the graviphoton. For more details about the reduction, see e.g.\ \cite{Intriligator:1997pq,Bonetti:2011mw,Grimm:2013oga,Grimm:2015zea, Corvilain:2020tfb}. 

In the following, we will be interested in understanding the asymptotic fibration structure of the Coulomb branch in Class A limits in terms of UV physics, and identify the candidate instantons that realize Propositions \ref{prop:hyperbolic towers} and \ref{prop:hyperbolic towers quantization}. Inspecting \eqref{eq:prepclassA}, and comparing with e.g.\ \cite[Section 3]{Grimm:2013oga}, one can identify the 5d and 6d coordinates as 
\begin{equation}
     X^{\hat 0} = R^{-4/3} \, , \quad  X^{\hat I} = R^{-4/3} \zeta^{\hat I} \, , \quad X^{\hat \alpha}=R^{2/3}(j^{\hat \alpha}+2R^{-2}b^{\hat \alpha}_{\hat I \hat J}\zeta^{\hat I}\zeta^{\hat J}) \, .
\end{equation}
where we redefined $X^{\hat \alpha}= f(X^0,X^r)$, $X^{\hat 0} = f_1(X^i), \, X^{\hat I} =f_2(X^i)$ and all redefinitions are linear. In the above, $j^{\hat \alpha}$ are the tensor multiplet scalars, $\zeta^{\hat I}$ are the Wilson lines coming from the reduction of the vector multiplets, and $b^{\hat \alpha}$ are the 6d anomaly coefficients \cite{Grimm:2013oga}. From this identifications, and due to the periodicity of the Wilson line moduli, one can associate an axion torus with the coordinates $X^{\hat I}$ fibered along the growing directions $M^{\hat \alpha}$. However, as it is clear from the constant volume constraint, and as explored in \cite{Grimm:2015zea, Corvilain:2020tfb}, this is not visible in a single supergravity frame. Concretely, due to two-forms, spin $\frac{1}{2}$- and $\frac{3}{2}$-fermions in the 6d theory being charged under $A^{\hat I}$, their mass at KK-level $n$ reads
\begin{equation}\label{eq:kkmode6d}
    m_{n} = q_{\hat I} \langle \zeta^{\hat I}\rangle + \frac{n}{R} \, ,
\end{equation}
with $\bf q$ their charge. Notice thus that the period direction is partially hidden behind a wall--crossing corresponding to the zero mode of this field becoming massless. Moreover, these massive modes contribute to the one-loop Chern-Simons couplings, which transform non-trivially under $\zeta^{\C{I}} \to \zeta^{\C{I}}+1$. In particular, each mode will contribute schematically to the classical Chern-Simons coefficient \eqref{eq:5dact} as 
\begin{equation}
    \C{F}_{IJK,\, n}^{\rm loop}= c_s\, q_I q_J q_K \mathrm{sign}(m_n) \, ,
\end{equation}
where $c_s$ is a coefficient dependent on the field's spin, and the sign of $m_n$ is $\pm 1$ depending on the particle representation under the 5d $SU(2) \times SU(2)$ little group \cite{Grimm:2015zea}. A shift in the Wilson line vev can be introduced by a shift of the particle charge, as clear from \eqref{eq:kkmode6d}, which then enters the loop computation shifting the Chern-Simons couplings. Summing over all KK modes, one gets the structure
\begin{equation}
    \C{F}_{IJK} \to \C{F}_{IJK} - \C{F}_{IJ(L} \C{A}_{J)}^L \, ,
\end{equation}
where $\C{A}^L_J$ can be rewritten in terms of the 6d anomaly coefficients, see \cite{Grimm:2013oga} for details. This is important to recover the 6d anomaly equations from the 5d theory.
Recall that the above one-loop contributions are geometrized in M-theory on $Y_3$ \cite{Intriligator:1997pq,Witten:1996md}. Thus, shifting the Wilson line moduli space non-trivially transforms the five-dimensional Coulomb branch parameters, bringing us to a new phase of the Calabi-Yau with the same effective description.

Finally, let us comment on the origin of the instanton as required per Proposition \ref{prop:hyperbolic towers}. First, notice that no BPS object exist in M-theory which, wrapped on a submanifold, gives an instanton with action depending on the vector multiplet moduli. Thus, the instanton needs to be non-BPS. Relatedly, a 0-brane is needed in the 6d theory to break the higher-form symmetry associated to the 1-form electric symmetry acting on the vector-multiplet gauge fields, as mentioned in Section \ref{sec:2}. This 0-brane can wrap the circle and thus yield the required instanton. Take a massive six-dimensional state coupling to the vectors, which worldline action reads\footnote{We neglect possible fermionic contributions due to the particle having spin.}
\begin{equation}
    S_V = \frac{1}{2T}\int {\rm d}\tau \,  (\dot{x}^2 -2m^2T)  +  \int A^{\hat I} q_{\hat I} \, ,
\end{equation}
with $q$ its charge and $x(\tau)$ its coordinates. In the above, $T$ is the Schwinger proper time parameter. Since spacetime includes a circle, a natural candidate for the instanton action is the  state above propagating along the circle, which precisely  couples to the axion $q_{\cal I} \zeta^{\cal I}$, the charge being the winding number $n$ along the circle. Evaluating the Schwinger integral in the saddle-point approximation, we have that the partition function reads
\begin{equation}\label{eq:instact5d}
    S_{\rm inst} = S_0\exp \left(-2 \pi Rm \right) \, .
\end{equation}
Notice that this is of the correct form to yield a $\Gamma$-vector depending on the properly identified radial coordinate. Indeed, since the Einstein frame metric is $g_{RR}=4/3R^{-2}$ (see e.g.\ \cite{Bonetti:2011mw}), the instanton $\beta$-vector reads
\begin{equation}
    |\beta|^2 = \frac{3}{4} \, . 
\end{equation}
We will connect with this general prediction in a simple example in Section \ref{example:2}, where we will precisely match with the $\Gamma$-vector.

\subsection{Examples} \label{ssec:5d-examples}
We now turn to the discussion of explicit examples. As announced, some of them will include gravity decoupling sectors. We will find that their Laplacian diverges along the limit, distinguishing them from the leading hyperbolic towers. All in all, every example will turn out to be compatible with Propositions \ref{prop:hyperbolic towers} and \ref{prop:hyperbolic towers quantization}.

\subsubsection{Example with $h^{1, 1} = 2$: degree-18 hypersurface $X \subset \B{P}_{1, 1, 1, 6, 9}$}
 The Calabi--Yau we inspect is a weighted projective hypersurface whose K\"ahler deformations are comprised of two moduli $X^1,X^2$, with triple intersection numbers given in the Nef-basis by \cite{Castellano:2023jjt}
\begin{equation}
    \cF_{111}=9 \, , \quad \cF_{112} =3 \, , \quad \cF_{221} =1 \, ,
\end{equation}
and with a classical prepotential reading
\begin{equation}
    \cF = \frac{3 (X^1)^3}2 + \frac{3 (X^1)^2 X^2}2 + \frac{X^1 (X^2)^2}2 \, .
\end{equation}
The only infinite distance limit respecting the hypersurface constraint $\cF[X]=1$ is given by $X^2 \sim \lambda \ ,\ X^1 \sim \lambda^{-2}$. This is a Class A limit, and we have that for a generic string of charge $p^I = (r,s)$ and particle of charge $q_I = (u,v)$
\begin{equation}
    \nabla^2_{\cF=1}\log m_{\rm osc}^{(r,s)} =\frac{ 9 r+6 s}{ r}\frac{1}{\lambda ^3}+{\cal O}\left(\frac{1}{\lambda^6 }\right) \, ,
\end{equation}
\begin{equation}
     \nabla^2_{\cF=1}\log M_{\rm BPS}^{(u,v)} =\frac{ 6 v - 9u}{ u}\frac{1}{\lambda ^3}+{\cal O}\left(\frac{1}{\lambda^6 }\right) \, .
\end{equation}
which both asymptote to zero, in agreement with the general expressions found in the previous section with all constants $c_s$ and $c_p$ taken to be zero. One can check that this is true even for the charge rays for which the expression is not well defined, i.e.\ $r=0$ and $u=0$. This agrees with the general picture of the light spectrum being correlated with the asymptotic geometry of moduli space: in this case, being one-dimensional, moduli space is flat and agrees with the 1d principal plane, which is consistent with $c=0$ for the leading BPS particle tower of charge $q_I=(1,0)$. Microscopically, this tower constitutes the KK tower of the decompactification to 6d.

\subsubsection{Example with $h^{1, 1} = 3$: non-smooth elliptic fibration $E \into X \onto \B{P}^2$}
\label{example:2} 
The next example is an elliptic fibration over $B=\mathbb{P}^2$ hosting an $A_1$ singularity on a degree $d=6$ curve over the base, leading to $SU(2)$ gauge symmetry. The interested reader can consult \cite{Hayashi:2023hqa} for more details about the geometry. The triple intersection numbers for the Nef-basis moduli $\{ X^1,X^2,X^3\}$ of this non-smooth fibration can be found e.g.\ in \cite{Marchesano:2023thx, Hayashi:2023hqa} and read
\begin{equation}
    \cF_{122} = 18 \, , \quad \cF_{222} =63 \, , \quad \cF_{123}=6 \, , \quad \cF_{223}=15 \, , \quad \cF_{133}=2 \, , \quad \cF_{233} =3 \, ,
\end{equation}
while the prepotential reads
\begin{equation}
    \cF = 9 X^1 (X^2)^2 + \frac{21}2(X^2)^3 + 6 X^1 X^2 X^3 + \frac{15}2 (X^2)^2 X^3 + X^1 (X^3)^2 + \frac32
  X^2 (X^3)^2 \, .
\end{equation}
There are two infinite distance limits, respectively Class A and B given by taking $M^3 \to \infty$ and $M^1 \to \infty$. Scaling-wise, they are given by:
\begin{equation}
\begin{split}
    &\mathrm{Class \ A:} \quad X^3 \sim \lambda\, ,\ \  X^2 \sim \lambda^{-2}\, , \ \ X^1 \sim \lambda^{-2}\, , \\
    &\mathrm{Class \ B:} \quad X^1 \sim \lambda \, , \ \  X^3 \sim X^2 \sim \lambda^{-1/2} \, .
\end{split}
\end{equation}
Let us start by analyzing the Class A limit, for which the lightest particles are given by the two-dimensional subcone $q_I = (a,b,0)$. Of these, only the ray generated by $q_I = (2,3,0)$ has a finite charge to mass ratio, and thus by \eqref{eq:gammaalphanorm} finite $||\alpha||^2 = \frac43 + {\cal O}(\lambda^{-1})$. This is precisely the leading tower aligned with the direction of the limit. To see that it is indeed hyperbolic, it suffices to compute its Laplacian, namely
\begin{equation}
    \nabla^2_{\cF=1}\log M_\text{BPS}^{(2,3,0)} = 1 + {\cal O}(\lambda^{-3}) \, .
\end{equation}

Let us now switch to the Class B limit.
For the string of charge $p^I=(1,0,0)$ we see that \textit{all over moduli space} the oscillator modes are characterized by the equation
\begin{equation}
    \nabla^2_{\cF=1}\log m_{\rm osc}^{(1,0,0)} = \frac12 \, .
\end{equation}
Instead, for the string of charge $p^I=(0,r,s)$ one has
\begin{equation}
    \nabla^{2}_{\cF=1}\log m_{\rm osc}^{(0,r,s)} = \frac{17}{8}- \frac{3(r+s)}{24 r+8 s} + {\cal O}(\lambda^{-3/2}) \, .
\end{equation}
Notice that the expression evaluates to $\frac12$ and $\frac74$ respectively for $(r,s)=(1,0),(0,1)$. The above expression is well defined, except for $s=-3r$. Computing the above quantity for the string of charge $p^I=(0,1,-3)$ one gets
\begin{equation}
    \nabla^2_{\cF=1}\log m_{\rm osc}^{(0,1,-3)} = \frac{116}{363}\lambda^{3/2} + \frac12 + {\cal O}(\lambda^{-7/2})  \, ,
\end{equation}
which diverges. This string is \textit{not} a supergravity string. It is the string corresponding to an $M5$--brane wrapped on the divisor $D_B =D_2 - 3D_3$, which following \cite{Marchesano:2023thx} is precisely the shrinkable divisor related to the gravity-decoupling sector. This is suggestive: it highlights that the property of being asymptotically hyperbolic is tied to these objects not being decoupled from gravity in the limit. Indeed, in this limit, their charge-to-mass ratios $\gamma_{{\bf p}}$ diverges and, following the discussion of the previous section, so does $\nabla^2_{\cF=1}\log \C{T}$.

The same philosophy applies to BPS particles along the limit. The Laplacian equation \eqref{eq:5dlapllogm} for the generic charge $q_I=(0,r,s)$ corresponding to light particles diverge. Once again, these correspond to rigid states decoupling from gravity \cite{Marchesano:2023thx}. For light particles of charge $q_I = (0,1,3)$ instead, one gets the same Laplacian and $\alpha$-vector that the leading oscillator modes. These correspond to the KK-tower accompanying the emergent string limit. We thus showed that the leading towers in both limits are hyperbolic, while the divergence of $\nabla^2_{\cF=1}\log M_\text{BPS}$ is tied to the divergence of $\gamma^2_{\bf q}$ and to the decoupling of the particle from gravity in the spirit of \cite{Castellano:2024gwi} and as showed in the previous section. 

\begin{figure}[t]
    \centering
    \includegraphics[width=0.8\linewidth]{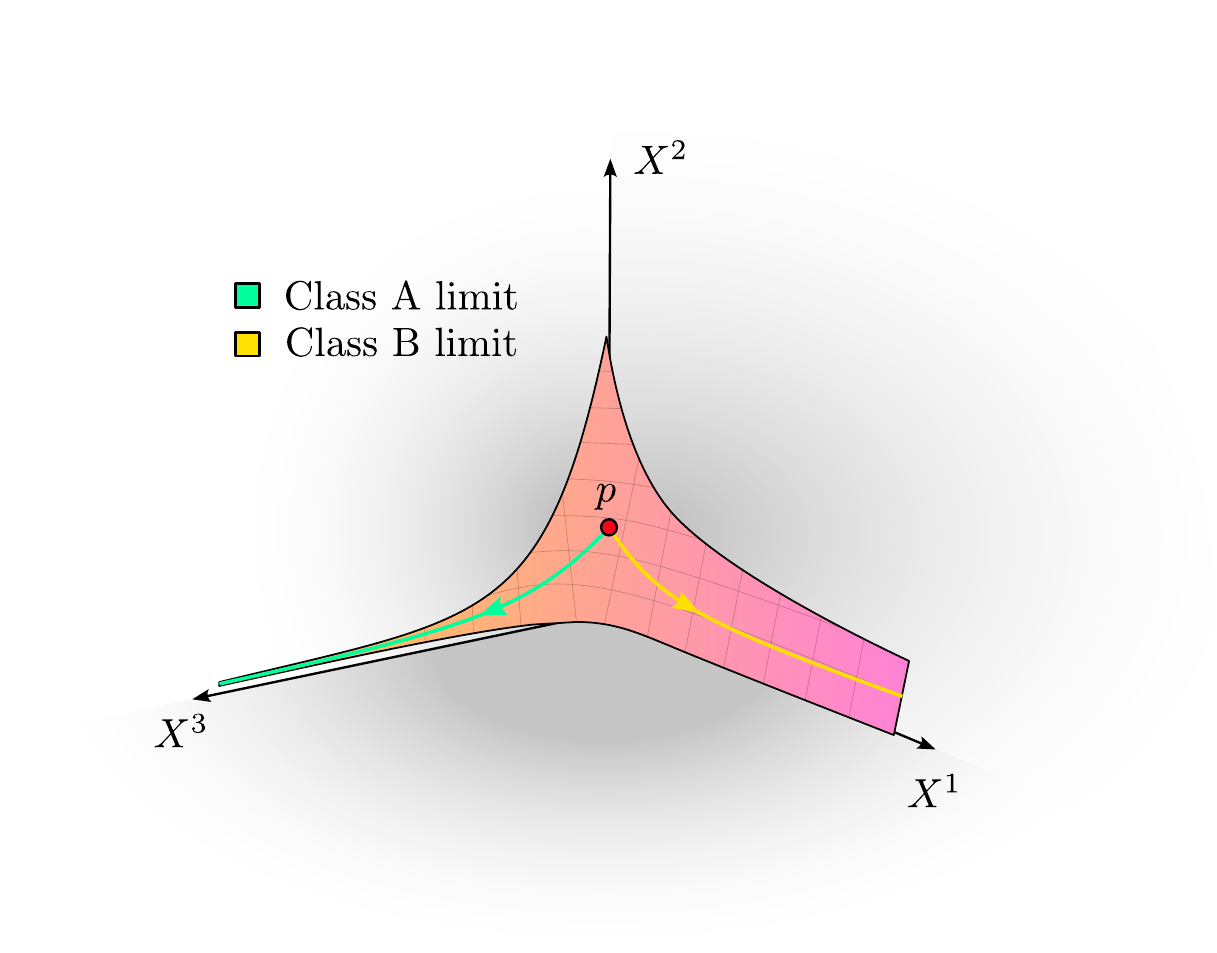}
    \includegraphics[width=0.8\linewidth]{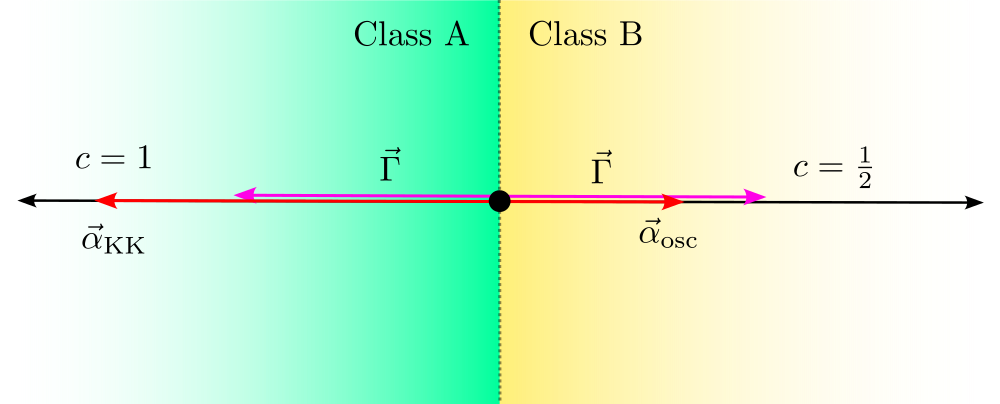}
    \caption{In the top panel, we show the hypersurface constraint for moduli space of the non-smooth fibration, together with the two Class A and B limits along the directions $X^3$ and $X^1$ starting from a point $p$. The direction $X^2$ is bounded by the constant volume constraint, and thus constitutes a segment fibered over the infinite-distance limits. In the bottom panel, we show the principal plane, $\alpha$- and $\Gamma$-vectors capturing the fibration structure of moduli space in the two asymptotic limits.}
    \label{fig:5dexample2}
\end{figure}

With the above prepotential and hypersurface constraint, the moduli space is locally a hyperbolic plane of constant negative curvature equal to
\begin{equation}
    R = -\frac32 \, ,
\end{equation}
and admits a coordinate system such that the metric along each of the two limits takes the local form
\begin{equation}
    {\rm d}s^2 = {\rm d}t^2 + e^{-\sqrt{6}t} {\rm d}x^2 \, ,  
\end{equation}
and the principal plane is thus one-dimensional. Moduli space can be seen as a segment fibered over this one dimensional non-compact direction, as shown in Figure \ref{fig:5dexample2}. The lightest towers in this asymptotic coordinate frame along the two limits have $\alpha$-vector
\begin{equation}
    \begin{split}
    &\mathrm{Class \ A:} \quad \vec \alpha_{\rm KK} \sim \frac{2}{\sqrt3 } \partial_t\,, \\
    &\mathrm{Class \ B:} \quad \vec \alpha_{\rm osc} \sim \frac{1}{\sqrt3 } \partial_t\, . 
\end{split}
\end{equation}
The exponential decay factor $2\Gamma = \sqrt{6}$ is tied to the constant negative curvature by $R= -|\Gamma|^2$. Thus, the $\Gamma$-vector along each limit precisely obeys 
\begin{equation}
    \vec \Gamma \cdot \vec \alpha = c \, .
\end{equation}
Notice that this $\Gamma$-vector precisely matches with the expected $\beta$-vector we identified in Section \ref{ref:decomp5d}, as it should be for a one-dimensional principal plane.
We show the principal plane for the two limits glued along a common point, together with the $\alpha$- and $\Gamma$-vectors, in Figure \ref{fig:5dexample2}.

\subsubsection{Example with $h^{1,1}=4$: non-smooth elliptic fibration $E \into X \onto \B{P}^2$}
As a final example, let us consider another elliptic fibration over $B=\mathbb{P}^2$ with an $A_2$ singularity hosted on a degree $d=1$ curve in the base, leading to $SU(3)$ gauge symmetry. The interested reader can consult \cite{Hayashi:2023hqa} for more details about the geometry. The triple intersection numbers in the Nef-basis $\{ X^1,X^2,X^3,X^4\}$ read
\begin{align*}
    \C{F}_{111} = 147 \, , \quad \C{F}_{112} = 105 \, , \quad \C{F}_{113} = 63 \, , \quad \C{F}_{114} = 21 \, , \quad \C{F}_{122} = 75 \, , \quad \C{F}_{123} = 45 \, , \\
    \C{F}_{124} = 15 \, , \quad \C{F}_{133} = 27 \, , \quad \C{F}_{134} = 9 \, , \quad \C{F}_{144} = 3 \, , \quad \C{F}_{222} = 50 \, , \quad \C{F}_{223} = 30 \, , \\
    \C{F}_{224} = 10 \, , \quad \C{F}_{233} = 18 \, , \quad \C{F}_{234} = 6 \, , \quad \C{F}_{244} = 2 \, , \quad \C{F}_{333} = 9 \, , \quad \C{F}_{334} = 3 \, , \quad \C{F}_{344} = 1 \, .
\end{align*}
The Calabi-Yau admits a single degeneration limit, $X^4\to \infty$, which corresponds to an F-theory limit decompactifying to a 6d $\C{N}=(1,0)$ theory without tensor multiplets \cite{Hayashi:2023hqa}. In the language of this work, it is of Class A, namely 
\begin{equation}
\mathrm{Class\ A:} \quad X^4 \sim \lambda \, , \quad X^i \sim e^i \lambda^{-2}, \ i=1,2,3 \, ,
\end{equation}
where the $e^i$ are chosen to be e.g.\ $(\frac13,\frac14,\frac12)$ as to stay inside the K\"ahler cone given the constant volume constraint. The leading tower of states along the limit is a tower of BPS particles identified by the charge ray $q_I=(3,2,1,0)$. Its $c$--value reads
\begin{equation}
c=\nabla^2_{\C{F}=1}\log M_{\rm BPS}^{(3,2,1,0)} = 2 + \C{O}(\lambda^{-3}) \, ,
\end{equation}
while as before $||\alpha|| \sim \frac43$, identifying the tower as the KK-tower associated to a circle becoming large. This constitutes the only tower of the 1d frame simplex.

Notice that the two directions transverse to the limit on the hypersurface are compact, as their self triple intersection number does not vanish. Thus, moduli space along the limit is topologically a product of two intervals fibered over a line. Computing its curvature, we get a hyperbolic space with 
\begin{equation}
R=-\frac92 \, ,
\end{equation} 
so that we can choose coordinates $t,\vec x$ respectively for the direction of the limit and the coordinates on the intervals in which the metric is given by
\begin{equation}
\mathrm{d}s^2 = \mathrm{d}t^2 + e^{-\sqrt{3}t}\mathrm{d}s_{\vec x} ^2  \, ,
\end{equation}
and where $\mathrm{d}s_{\vec x} ^2 $ is the 2d metric on the compact transverse space. Computing the $\Gamma$-vector as $\vec\Gamma = -\frac12 \partial_t \log g  \, \partial_t$ and comparing with the KK-tower $\alpha$-vector, one gets
\begin{align}
\begin{split}
&\vec \alpha_{KK} = \frac{2}{\sqrt{3}}\partial_t \, ,\\
&\vec \Gamma = \sqrt{3} \partial_t \, .
\end{split}
\end{align}
such that the curvature can be expressed as $R=-\frac32 |\vec \Gamma|^2$. Notice that the quantities above satisfy precisely
\begin{equation}
\vec \alpha \cdot \vec \Gamma = c \, 
\end{equation}
as expected. Finally, we can prove that the $\Gamma$-vector above is precisely the sum of the two instanton $\alpha$-vectors associated to the compact directions. Indeed, since the frame simplex is one dimensional and both instantons have action \eqref{eq:instact5d}, their $\beta$-vector is 
\begin{equation}
\beta_{ 1,2} = \frac{\sqrt{3}}{2} \partial_t \, ,
\end{equation}
which establishes that $\vec \Gamma = \sum_i \vec \beta_i$.

\subsection{Other theories with 8 supercharges}\label{ssec:other-8-supercharges}

Though the bulk of our attention has been devoted to the vector-multiplet moduli spaces 5d $\C{N}=1$ theories, in the following we briefly turn our attention to other theories with 8 supercharges. 

\subsubsection{6d $\C{N}=(1, 0)$ tensor multiplets}

As mentioned in Section \ref{ref:decomp5d}, in six dimensions the field content of $\C{N}=(1,0)$ supergravity EFTs consists of the gravity multiplet, $n_H$ hypermultiplets, $n_T$ tensor multiplets and $n_V$ vector multiplets (see e.g., \cite{Taylor:2011wt}). In F-theory, the tensor branch can be described entirely in terms of a K\"ahler 2-form $J_{B_2}$ associated to the base $B_2$ of an elliptically-fibered threefold \cite{Vafa:1996xn,Morrison:1996na,Morrison:1996pp}. This object may be naturally regarded as a vector in $\B{R}^{1,T}$ satisfying the quadratic constraint
\begin{equation}\label{eq:6dconstraint}
    J_{B_2} \cdot J_{B_2} = \Omega_{IJ} J^I J^J=  1\, , 
\end{equation}
and $\Omega_{IJ}$ is a matrix of signature $(1,T)$ defining an inner product. Introducing a basis of $h^{1,1}(B_2)$ 2-cycles $[C_I]$ within $B_2$, we can write the K\"ahler form as follows
\begin{equation}
    J_{B_2} = J^I [C_I] = J^0 [C_0] + J^i [C_i]\, ,\qquad i=1, \ldots, T=h^{1,1}(B_2)-1\, ,
\end{equation}
where $[C_0]$ is a distinguished shrinkable curve in the base which singles out the only kind of infinite distance degenerations that can occur in these theories, namely emergent string limits \cite{Lee:2018urn,Lee:2018spm,Lee:2019xtm}. Using these variables, the moduli space is given by the hypersurface 
\begin{equation}\label{eq:6dconstr}
     \C{F}_6 = \{ J^I \mid  \Omega_{I K}J^IJ^K = 1 \}\, , 
\end{equation}

Notice that the constraint is expressed in terms of a $\SO(1,n_T)$-invariant quadratic form, thereby preserving the isometries of the embedding space. Using a set $\{ \psi^\a\}$ of $h^{1,1}(B_2)-1$ independent fields that parametrize the hypersurface \eqref{eq:6dconstraint}, the relevant two-derivative action is given in Planck units by \cite{Nishino:1997ff,Ferrara:1996wv,Bonetti:2011mw}
\begin{equation}\label{eq:6daction}
    S_{\text{6d}} = \frac{1}{2} \int \left( R \star 1 - \frac12 g_{\alpha \beta} \, \d\psi^\alpha\wedge \star  \d\psi^\beta + \dots \right) \ , \quad \ \ g_{\alpha \beta} = {\partial_\alpha J^I} {\partial_\beta J^J}\left( 2{J_I J_J}  - \Omega_{IJ}\right)\, ,
\end{equation}
with the indices lowered by $\Omega_{IJ}$ and where the kinetic term of the scalars have been pulled-back to the constrained moduli space, much like it happens in 5d. The ambient space instead has a metric which can be written as
\begin{equation}\label{eq:ambientspacemetric6d}
    g_{IJ} = -\frac12 \partial_I \partial_J \log \C{F}_6\, . 
\end{equation}
Since the hypersurface defined by eq. \eqref{eq:6dconstr} is a hyperbolic space due to $\Omega_{IJ}$ being $\SO(1,n_T)$ invariant, we may thus describe it as a homogeneous space for $\SO(1,n_T)$, the stabilizer of a reference point $J^I = (1, 0, \cdots, 0)$ is $\SO(n_T)$. Thus we \textit{locally} have the following geometry for the marked tensor moduli space:
\[ \hat{\C{M}}_{\m{tensor}} \simeq \SO(1, n_T)/\SO(n_T) , \]
which is a Riemannian symmetric space. The total marked tensor moduli space is actually a subspace of $\SO(1, n_T)/\SO(n_T)$ corresponding to physical vacua satisfying the 6d anomaly constraints, see \cite{kimFiniteLandscape6d2026}. To obtain the unmarked moduli space $\C{M}_{\m{tensor}}$, we require additionally a quotient by a duality group $\Gamma$ which renders $\C{M}_{\m{tensor}}$ compactifiable. Accordingly, the symmetric moduli space analysis of Section \ref{sec:sym} holds, and Propositions \ref{prop:nilmanifold}, \ref{prop:hyperbolic towers} and \ref{prop:hyperbolic towers quantization} hold. 

Concretely, due to all limits being emergent string limits, it suffices to consider the tension $T_{\rm str}$ of D3 branes wrapping shrinkable curves $[C_0]$ in the base. In particular, it will be generically given by a linear function $[C_0] \cdot J$ in the $J^{I}$-coordinates, namely
\begin{equation}
    T_{\rm str} = c_I J^I \ .
\end{equation}
Using the techniques laid out in \cite{Aoufia:2025ppe}, and as can be proved by direct computation, any linear function in the ambient coordinates is an eigenfunction of the Laplacian on the constraint hypersurface $\Delta_{\C{F}_6=1}$ with eigenvalue given by $n_T$. Since the $\alpha$-vector for the oscillator modes is given by $|\alpha|^2 = \frac14$, this brings us to
\begin{equation}
    \nabla^2_{\C{F}_6=1} \log m_{\rm osc} = \frac{1}{2}(n_T-1) \ . \label{eq:hyp-towers-6d-1,0}
\end{equation}
This establishes Propositions \ref{prop:hyperbolic towers} and \ref{prop:hyperbolic towers quantization} in these theories.

In an emergent string limit in the tensor moduli space, the tensionless string is a critical heterotic string on K3, and we can understand these limits more cleanly from the heterotic perspective. Concretely, compactify the $E_8 \times E_8$ theory on K3 with $n_1, n_2$ instantons in each $E_8$ factor and $n_{\m{NS5}}$ fivebranes. Then we have the tadpole cancellation condition $n_1 + n_2 + n_{\m{NS5}} = 24$, and the number of tensor multiplets $n_T$ is given by $n_T = n_{\m{NS5}} + 1$. When we take the 6d heterotic dilaton in the tensionless string limit $\varphi \to -\infty$, we obtain an asymptotic fibration structure of the following form: 
\begin{equation}
    ds_{\mathrm{tensor}}^2
=
d\varphi^2
+
e^{2\varphi}
\sum_{a=1}^{N_5}
\left(d\xi^a\right)^2 , \label{eq:6d-1,0-fibration}
\end{equation}
where the $\xi^a$ are the real tensor moduli directions on the NS5-brane worldvolume theories. In the dual M-theory limit $\phi \to +\infty$, the $\xi^a$ are related (nonlinearly) to the positions of M5-branes along the Horava-Witten interval. 

We see thus that the hyperbolic towers quantization condition of eq. \eqref{eq:hyp-towers-6d-1,0} follows from the asymptotic fibration structure of eq. \eqref{eq:6d-1,0-fibration}. However, we note that Assumptions \ref{ass:ESC} and \ref{ass:perturbative-esc} are both violated for $n_T > 1$ in this setup. Indeed, \ref{ass:ESC} is violated in the M-theory limit, since the NS5-branes along the Horava-Witten interval induce a running profile for the 7d scalars, while \ref{ass:perturbative-esc} is violated  in the heterotic string limit, since the 6d effective theory does not admit a perturbative description due to the presence of NS5-branes. 

The relation with instantons in Proposition \ref{prop:instantons-beta-vectors} is also subtle in this example. In particular, for $n_T>1$ the heterotic dual string becoming light is weakly-coupled, but not perturbative. We are unable to motivate the asymptotic fibration structure in this case, or to check \ref{prop:instantons-beta-vectors} explicitly, but we nonetheless see that Propositions \ref{prop:nilmanifold}, \ref{prop:hyperbolic towers} and \ref{prop:hyperbolic towers quantization} do hold. We therefore have an example where the hyperbolic towers property of Proposition \ref{prop:hyperbolic towers quantization} appears to hold outside the domain of validity of our assumptions. This points to the fact that the asymptotic structure discussed throughout this paper might be more general than presented.

\subsubsection{Hypermultiplets}\label{ssec:hyper}

The hypermultiplet moduli space of 6d, 5d, and 4d theories with 8 supercharges are all quaternion-K\"ahler spaces \cite{Andrianopoli:1996vr}. 
For concreteness, let us first focus on the universal hypermultiplet geometry in 4d $\C{N}=2$ theories. The classical moduli space is given by the coset \cite{Cecotti:1988qn,Ferrara:1989ik}
\begin{equation}
    \C{M}_{\rm hyp}\cong\frac{SU(2,1)}{SU(2) \times U(1)} \ .
\end{equation}
From e.g.\ the perspective of Type IIB on a Calabi--Yau $Y_3$, these scalars parametrize the 4d dilaton, the associated reduction of $B_2$, and the $C_0$ and $C_2$ Ramond-Ramond axions. Asymptotically, at weak coupling, corrections to the classical moduli space are suppressed, and the analysis of Section \ref{sec:sym} applies, establishing Propositions  \ref{prop:nilmanifold}, \ref{prop:hyperbolic towers} and \ref{prop:hyperbolic towers quantization}. 

It is particularly interesting to highlight the fibration structure of the classical moduli space, whose isometries are also preserved by quantum corrections \cite{Antoniadis:1997eg,Antoniadis:2003sw}. As remarked in \cite{Ambrosetti:2010tu}, the gauge transformations of Type IIB 
\begin{equation}
    C_0 \to C_0 + \lambda \ , \quad B_2 \to B_2 + \d \Lambda \ , \quad C_2 \to C_2 + \d\Lambda + \lambda B_2 \ ,  
\end{equation}
with $\lambda$ a function and $\Lambda$ a one-form, directly translate into the axionic shift symmetry of the low-energy theory generating a Heisenberg algebra of the type 
\begin{equation}
    [\delta c_0, \delta c_2] \sim \delta b_2 \ ,
\end{equation}
for the axions $c_0 \equiv C_0,\, c_2$ and $b_2$. Calling $\varphi$ the canonically normalized 4d dilaton, this shows that the metric in moduli space can be written as 
\begin{equation}
    \d s^2 = \d \varphi ^2 + f_0^2  e_0^2 + f_{c}^2 e_{c}^2 + f_{b}^2 e_b^2 \ ,  
\end{equation}
with $e_0, \ e_b$ and $e_c$ left-invariant one-forms of the Heisenberg algebra corresponding to reduction of the respective field strengths
\begin{equation}
    e_0 = dc_0 \ , e_b = db_2  \ ,e_c = dc_2 -c_0 db_2 \ ,
\end{equation}
and which moreover obey the Maurer-Cartan relations, i.e.\ the Bianchi identities
\begin{equation}
    \d e_0 = \d e_b =0 \ , \quad \d e_c = -e _0 \wedge e_b \ .
\end{equation}
In the above expression, the $f_I$ are the exponentially decaying axion decay constants of the three axions. 

Beyond the universal hypermultiplet, one can probe the classical hypermultiplet moduli space using the c-map, which expresses $\C{M}_Q$ as a fibration over a projective special-K\"ahler base $F \hookrightarrow \C{M}_Q \to \C{M}_{\m{SK}}$ whose fibers $F$ are Heisenberg nilmanifolds. Explicitly, let us once again consider the hypermultiplet moduli space of 4d $\C{N}=2$ supergravity obtained by compactifying e.g.\ Type IIA on a CY3 $Y_3$ with $h^{2, 1}(Y_3) = n$. Let $z^1, \cdots, z^n$ denote complex coordinates on a special-K\"ahler manifold $\C{M}_{\m{SK}}$ parametrizing the complex-structure moduli space of $Y_3$, and let $(\zeta^\Lambda, \tilde\zeta_\Lambda)$ be $2n+2$ axions obtained by reducing $C_3$ over a symplectic basis $H^3(Y_3, \B{Z})$. We additionally have the NSNS axion $\sigma$ obtained by reducing $B_6$ over $Y_3$ and the dilaton $\phi$.  

The total geometry of the classical moduli space before RR- and NS5-instantons are taken into account is a fibration over $\C{M}_{\m{SK}}$, and the classical Ferrara-Sabharwal metric on $\C{M}_Q$ can be expressed in terms of the special-Kahler metric $g_{i\barj}^{\m{SK}}$ on $\C{M}_{\m{SK}}$ as \cite{Ferrara:1989ik}

\begin{align}
\d s^2_{\mathrm{FS}} &=\d\phi^2 + g_{i\bar j}^{\mathrm{SK}}\d z^i \d\bar z^{\bar j} + \frac{1}{4}e^{4\phi}\Theta^2 +  \frac{1}{2}e^{2\phi}
\left[
(\mathrm{Im}\mathcal N)_{\Lambda\Sigma} \d\zeta^\Lambda \d\zeta^\Sigma  + (\mathrm{Im}\mathcal N)^{-1\,\Lambda\Sigma} D\tilde\zeta_\Lambda D\tilde\zeta_\Sigma \right]\, ,
\end{align}
with 
\begin{equation}
\Theta = \d\sigma +\tilde\zeta_\Lambda d\zeta^\Lambda -\zeta^\Lambda d\tilde\zeta_\Lambda \, ,
 \quad D\tilde\zeta_\Lambda = \d\tilde\zeta_\Lambda + (\mathrm{Re}\mathcal N)_{\Lambda\Sigma}
\d\zeta^\Sigma \, .
\end{equation}
Here, $\C{N}_{\Lambda\Sigma}$ is the period matrix of $Y_3$, expressed in terms of the holomorphic $\C{N}=2$ prepotential $F(X^\Lambda)$ and the period vector $X^\Lambda, \ \Lambda=1,\dots,n+1$ on $\C{M}_{\rm SK}$ as \cite{Andrianopoli:1996vr}: 
\begin{equation}
\mathcal N_{\Lambda\Sigma} = \overline{F}_{\Lambda\Sigma} + 2i\, \frac{
(\mathrm{Im}F_{\Lambda\Gamma})X^\Gamma
(\mathrm{Im}F_{\Sigma\Delta})X^\Delta}{X^\Theta (\mathrm{Im}F_{\Theta\Phi}) X^\Phi}\, , \qquad F_{\Lambda\Sigma} = \frac{\partial^2 F}{\partial X^\Lambda\partial X^\Sigma}\, .
\end{equation}
The classical metric is a good approximation in the weak-coupling regime where $\phi \to -\infty$; outside of this regime, the exponentiated D2- and NS5-instanton actions become large and significantly correct the geometry. Though there is no longer an asymptotic locally symmetric description of the hypermultiplet moduli space, any infinite-distance limit of $\C{M}_Q$ corresponds either to $\phi \to -\infty$ or an infinite-distance limit in the base $\C{M}_{\m{SK}}$. We see that there is still an overall fibration structure of axions over the non-compact dilaton is preserved, and the fiber volumes decay exponentially in the weak-coupling limit. It would be interesting to investigate the fibration structure, and thus the hyperbolic towers property, in more general limits.  

Microscopically, the appearance of a more general nilmanifold rather than a torus, e.g.\ as illustrated in the universal hypermultiplet case, is due to the fact that the higher-dimensional theory has a Chern-Simons term of the form 
\begin{equation}
    S_{\rm CS} = \frac{1}{4}\int C_4 \wedge H_3 \wedge F_3 \ ,
\end{equation}
leading to modified Bianchi identities. We thus see than that the precise fibration structure of the low-energy theory can be seen to contain information about topological terms in the higher-dimensional theory.

This concludes our study of theories with 8 supercharges. We studied a diverse assortment of cases in 6d $\C{N}=(1, 0)$, 5d $\C{N}=1$, and 4d $\C{N}=2$ supergravity, finding in all cases that the hyperbolic towers property is satisfied in asymptotic limits where our assumptions are satisfied (and sometimes even when they are not). The reader may complain that we have not treated the well-understood case of 4d $\C{N}=2$ vector multiplets, where the projective special Kähler geometry makes a direct analysis tractable. In fact, the asymptotic geometry in such examples is essentially covered by the more general case of 4d $\C{N}=1$ supergravity, which will be the subject of Section \ref{sec:4charges}. Though we have not computed any explicit 4d $\C{N}=2$ vector multiplet examples (for instance, from Type II on CY3), we anticipate that the computational technology in Section \ref{sec:4charges} can straightforwardly be employed to check the hyperbolic towers quantization condition of Proposition \ref{prop:hyperbolic towers quantization} in these cases too. 

\section{\texorpdfstring{$Q=4$}{Q=4} supercharges}\label{sec:4charges}
In this section, we study the perturbative regimes of certain 4d $\mathcal{N}=1$ theories which arise from string and M-theory compactifications as described in \cite{Lanza:2021udy,Grieco:2025bjy}. These models contain EFT strings, a class of axionic BPS strings with tensions derived from the IR K\"ahler potential. In particular, in these asymptotic limits the EFT strings become tensionless. The specific top down construction giving rise to these EFT strings that we will consider is the $E_8 \times E_8$ heterotic string compactified on a Calabi-Yau three-fold. We will briefly review the contents of \cite{Grieco:2025bjy} before analyzing hyperbolicity and the construction of $\Gamma$-vectors in these models.

\subsection{4d \texorpdfstring{$\C{N}=1$}{N=1} supergravity and EFT strings}
Consider a 4d $\C{N}=1$ EFT coupled to gravity, and parameterize the scalar fields as $z^\alpha = a^\alpha + i s^\alpha$ with $\alpha$ running over all chiral multiplets in the theory. The effective action for the scalars is
\begin{equation}\label{eq:EFTString_Action}
    \C{S} = M_{\m{pl},4}^2 \int \bigg( \frac{1}{2}\C{R} *1 - K_{\alpha\bar{\beta}} \d z^\alpha \wedge *\d \bar{z}^{\bar{\beta}} - V*1\bigg)\;.
\end{equation}
$K_{\alpha\bar{\beta}}$ and the potential $V$ can be derived from the K\"ahler potential $K$ and superpotential $W$ as $K_{\alpha\bar{\beta}} = \p_{z^\alpha}\p_{\b{z}^{\b{\beta}}} K$ and $V = e^K(K^{\alpha\b{\beta}}D_\alpha W D_{\b{\beta}}W-3|W|^2)$.
Following \cite{Grieco:2025bjy}, we will set $W=0$ for simplicity\footnote{More generally, the theory contains chiral multiplets with scalars $\phi^\alpha$ of which the superpotential $W$ depends on only a subset, say $\chi^a$. To be in a supersymmetric vacuum in this case, one will impose $W(\chi_*)=\p_{\chi^a}W(\chi_*)=0$ for some fixed value $\chi_*$ of the scalars $\chi^a$. This leaves all other scalar moduli (of which $W$ is independent) unconstrained \cite{Lanza:2021udy}.} but it would be interesting to redo our analysis for non-trivial scalar potentials along the lines of \cite{Lanza:2021udy}. Generically, the K\"ahler potential will take the form
\begin{equation}\label{eq:EFTString_Kahler}
    K = -\log \big[ P(s) + \C{O}(e^{2\pi i m_\sigma z^\sigma})\big]\;,
\end{equation}
with $P(s)$ some homogeneous function of positive integral degree on the saxions, and the non-perturbative corrections $e^{2\pi i m_\sigma z^\sigma}$ arising from $\tfrac{1}{2}$-BPS instantons with $m_\sigma \in \mathbb{Z}$ the instanton charges. These instanton corrections are suppressed by $e^{-2\pi m_\sigma s^\sigma}$, so the perturbative EFT regime can be reached by working in regions of moduli space where the saxions $s^i = \m{Im}\, z^i$ are large. In these regimes, the K\"ahler potential $K$ and, as a result, field space metric are to leading order independent of the axions $a^i = \m{Re}\,z^i$. The instantons are electrically charged under these axions, so this perturbative EFT regime has an approximate axionic shift symmetry in the field space metric. The breaking of this axionic shift symmetry is measured by the instanton charges $\C{C}_I \subset M_\mathbb{Z}$ and we can identify the perturbative EFT regime as
\begin{equation}
    m_i s^i \gg 1\quad \forall m_i \in \C{C}_I\;.
\end{equation}
In other words, the perturbative regime is the deep interior of the saxionic cone $\Delta$, defined as
\begin{equation}
    \Delta \equiv \{ s^i \in N_\mathbb{R} \, | \, m_i s^i > 0 \,,\forall m_i \in \C{C}_I\}\;.
\end{equation}
Due to the leading form of the K\"ahler potential $K$ from \eqref{eq:EFTString_Kahler}, the limit $s^i \rightarrow \infty$ will be an infinite distance limit provided we stay within the saxionic cone $\Delta$ where \eqref{eq:EFTString_Kahler} is reliable.

These approximate axionic symmetries allow for axionic $\tfrac{1}{2}$-BPS strings magnetically charged under the axions $a^\alpha = \m{Re}\,z^\alpha$. As described in \cite{Grieco:2025bjy}, the profile of the scalar fields near the axionic string core is
\begin{equation}
    a^\alpha = a_0^\alpha \frac{\theta}{2\pi}e^\alpha\;,\quad s^\alpha = s^\alpha_0 + e^\alpha \sigma\;,\quad \sigma \equiv -\frac{1}{2\pi}\log \bigg(\frac{r}{r_0}\bigg)\;.
\end{equation}
In this expression, the $e^\alpha \in \mathbb{Z}$ are the string charges. As we go around the BPS string ($\theta \rightarrow \theta + 2\pi$), the axion shifts by $a^\alpha \rightarrow a^\alpha + a_0^\alpha e^\alpha$. As we approach the string core at $r=0$, the saxions go from $s^\alpha_0 \equiv s^\alpha(r_0)$ to $s^\alpha \rightarrow \infty \cdot e^\alpha$. The tension of a BPS string with charge $e^\alpha$ is given by
\begin{equation}\label{eq:EFT_tension}
    \C{T}_e = m_{\m{pl}}^2 e_\alpha l^\alpha\;,\quad l_\alpha \equiv -\frac{1}{2}\frac{\p K}{\p s^\alpha}\;,
\end{equation}
where the $l^\alpha$ are the dual saxions. Within the saxionic cone $\Delta$, the BPS strings have positive tension.

BPS strings with $\langle \boldsymbol{m},\boldsymbol{e}\rangle \geq 0$ for any $\boldsymbol{m} \in \C{C}_i$ will remain weakly coupled along their entire string flow; these strings are called EFT strings. The set of EFT strings are then given by the lattice cone
\begin{equation}
    \C{C}_S^{\m{EFT}} = \overline{\Delta}\cap N_\mathbb{Z}\;,
\end{equation}
with $\overline{\Delta}$ the closure of $\Delta$. The EFT string charges $\boldsymbol{e}$ generate the saxionic cone parameterizing the perturbative regime. It is useful to also express the saxionic cone in terms of the dual saxions as
\begin{equation}\label{eq:sax_cone_dual_sax}
    \C{P} \equiv \bigg\{ \boldsymbol{\ell}\in M_{\mathbb{R}} \big| \ell_i = -\frac{1}{2}\frac{\p K}{\p s^i}\bigg|_{\boldsymbol{s}\in\Delta}\bigg\}\;,
\end{equation}
which is well-defined for a perturbative K\"ahler potential $K$ which displays the axionic shift symmetries associated with $\Delta$.

To investigate hyperbolicity, we first need to construct the moduli space metric. From \eqref{eq:EFTString_Action} we see that the moduli space metric in holomorphic and anti-holomorphic coordinates $z^\alpha$ and $\bar{z}^{\bar{\beta}}$ is given by
\begin{equation}
    G_{\alpha\bar{\beta}} = \frac{\p^2 K}{\p z^\alpha \p \bar{z}^{\bar{\beta}}}\;.
\end{equation}
Note that by construction the moduli space metric is K\"ahler, which leads to the Laplacian of scalar functions $f(z,\bar{z})$ on moduli space taking a very simple form,
\begin{equation}
    \nabla^2 f(z,\bar{z}) = 2 G^{\alpha\bar{\beta}}\p_\alpha \p_{\bar{\beta}} f(z,\bar{z})\;.
\end{equation}
Given the generic form of the K\"ahler potential in \eqref{eq:EFTString_Kahler}, one can show that non-perturbative corrections to the Laplacian (and scalar products of $\alpha$-vectors or species vectors) from instantons are exponentially suppressed deep in the saxionic cone.

We can further simplify these expressions by first noting that $P(s)$ is only a function of the saxions $s$, so
\begin{equation}
    G_{\alpha\bar{\beta}} = \frac{1}{P(s)^2}\big( P_\alpha P_{\bar{\beta}} - P(s) P_{\alpha\bar{\beta}}\big) = \frac{1}{4 P(s)^2}\bigg( \frac{\p P(s)}{\p s^{\alpha}}\frac{\p P(s)}{\p s^{\bar{\beta}}} - P(s) \frac{\p^2 P(s)}{\p s^\alpha\, \p s^{\bar{\beta}}} \bigg)\;.
\end{equation}
At this point, it is useful to change from holomorphic to real coordinates given by the saxions $s^\alpha$ and axions $a^\alpha$. After performing the coordinate transformation, the metric decomposes as
\begin{equation}
    ds^2 = h_{\alpha\beta}\big(ds^\alpha ds^\beta + da^\alpha da^\beta\big)\;,
\end{equation}
where
\begin{equation}\label{eq:real_hol_met_relation}
    h_{\alpha\beta} \equiv 2 G_{\alpha\bar{\beta}} =  \frac{1}{2}\frac{\p^2 K}{\p s^\alpha \, \p s^\beta} = \frac{1}{2 P(s)^2}\bigg( \frac{\p P(s)}{\p s^{\alpha}}\frac{\p P(s)}{\p s^{\beta}} - P(s) \frac{\p^2 P(s)}{\p s^\alpha\, \p s^{\beta}} \bigg)\;.
\end{equation}
We will often refer to $h_{\alpha\beta}$ as the moduli space metric itself, but the true metric on the full moduli space is given by the block diagonal form shown above. We will make explicit this distinction when it is relevant, and denote the $(h^{1,1}+1)\times (h^{1,1}+1)$ saxion-only metric as $h_{\alpha\beta}$ and reserve $g_{\alpha\beta}$ for when we refer to the full $(2h^{1,1}+2)\times (2h^{1,1}+2)$ block diagonal metric on the saxion and axion moduli space.

We will study the case where $P(s)$ is given by the product of a universal saxion $s^0$ and the volume (in string units) of the internal CY3 $X$,
\begin{equation}
    P(s) = \C{V}_X(s)\, s^0 = \frac{1}{3!}\kappa_{ijk}s^i s^j s^k s^0\;.
\end{equation}
(Note that $\kappa_{ijk}$ in this section is the same $\C{F}_{IJK}$ in Section \ref{sec:8charges}; both represent the triple intersection numbers of $X$ in a compactification over $X$. Here, we employ different notation in these sections to align more closely with the 5d $\C{N}=1$ and 4d $\C{N}=1$ literature in each case.) We will split the saxionic coordinates $\alpha = 0,\dots,h^{1,1}$ into the universal saxion $s^0$ and all other saxions $s^i$ with $i = 1,\dots h^{1,1}$. The K\"ahler potential takes the form
\begin{equation}
    K = - \log \big(\C{V}_X(s) \, s^0 \big) = -\log s^0 - \log \C{V}_X(s)\;.
\end{equation}
The volumes (in string units) of intersection curves and divisors are written as
\begin{equation}
    \C{V}_{ij} = \kappa_{ijk}s^k\;,\quad \C{V}_i = \frac{1}{2!}\kappa_{ijk}s^j s^k\;,
\end{equation}
and we can compute the moduli space metric as
\begin{equation}\label{eq:4dEFT_metric}
    h_{\alpha\beta} = \frac{1}{2} \begin{pmatrix}
        \frac{1}{(s^0)^2} & 0 \\
        0 & \frac{1}{\C{V}_X^2} \bigg( \C{V}_i \C{V}_j - \C{V}_X \C{V}_{ij} \bigg)
    \end{pmatrix},
\end{equation}
with inverse metric
\begin{equation}
    h^{\alpha\beta} = 2 \begin{pmatrix}
        (s^0)^2 & 0 \\
        0 & \frac{1}{2} s^i s^j - \C{V}_X \C{V}^{ij}
    \end{pmatrix},
\end{equation}
where $\C{V}^{ij}$ is the matrix inverse of the $h^{1,1} \times h^{1,1}$ matrix $\C{V}_{ij}$.

The scalar functions $f(z,\bar{z})$ we will want to consider will also only depend on the saxions, so the Laplacian simplifies to the form
\begin{equation}
    \nabla^2 f(s) = 2G^{\alpha\bar{\beta}}\p_\alpha \p_{\bar{\beta}} f(s) = h^{\alpha\beta}\frac{\p^2 f(s)}{\p s^\alpha \, \p s^\beta}\;,
\end{equation}
where we have used that $\p f(s)/\p a^\alpha = 0$.
Anticipating that we will want to compute Laplacians of logs of masses $m_i$, we can set $f(s) = \log m_i(s)$ and find
\begin{align}\label{eq:Lap_logT}
    c_i \equiv \nabla^2 \log m_i(s) &= \frac{1}{\big(m_i(s)\big)^2}h^{\alpha\beta} \bigg( m_i(s) \frac{\p^2 m_i(s)}{\p s^\alpha\, \p s^\beta} - \frac{\p m_i(s)}{\p s^\alpha} \frac{\p m_i(s)}{\p s^\beta}\bigg)\;,
\end{align}
or inserting the explicit form of the inverse metric,
\begin{align}\label{eq:Lap_logT_explicit}
    \begin{aligned}
    c_i \equiv \nabla^2 \log m_i(s) =&\; 2\bigg(\frac{s^0}{m_i(s)}\bigg)^2\bigg[ m_i(s) \frac{\p^2 m_i(s)}{\p (s^0)^2} - \bigg( \frac{\p m_i(s)}{\p s^0} \bigg)^2 \bigg]\\
    &+\frac{2}{\big(m_i(s)\big)^2}\bigg( \frac{1}{2}s^j s^k - \C{V}_X \C{V}^{jk} \bigg)\bigg[ m_i(s) \frac{\p^2 m_i(s)}{\p s^j \, \p s^k} - \frac{\p m_i(s)}{\p s^j}\frac{\p m_i(s)}{\p s^k} \bigg]\;.
    \end{aligned}
\end{align}

\subsection{General analysis of \texorpdfstring{$\alpha$}{alpha}-vectors and Laplacians}
Here we specialize to the case where the UV theory is the $E_8 \times E_8$ heterotic string on a CY3 $X$. There are two types EFT strings: the fundamental (heterotic) string with tension
\begin{equation}\label{eq:e0_EFT_string}
    \C{T}_{\m{F1}} \sim \frac{m_{\m{pl},4}^2}{s^0}\;,
\end{equation}
and an EFT string arising from a NS5 brane wrapping an effective divisor $D = e^i \w{i}$,
\begin{equation}\label{eq:ei_EFT_string}
    \C{T}_{\m{NS5}} \sim m_{\m{pl},4}^2 e^i l_i\;,
\end{equation}
where the $e^i$ are the string charges and the $l_i$ are the dual axions. The dual axions in this case are given as
\begin{equation}
    l_i = \frac{\C{V}_i}{\C{V}_X}\;,
\end{equation}
so the tension becomes
\begin{equation}
    \C{T}_{\m{NS5}} \sim m_{\m{pl},4}^2 e^i \C{V}_i / \C{V}_X\;.
\end{equation}
These strings both lead to towers of oscillator modes whose mass scales as $\sqrt{\C{T}_i}$,
\begin{subequations}
\begin{align}
    m_{\m{osc,F1}}& \sim m_{\m{pl},4}(s^0)^{-1/2},\\
    m_{\m{osc,NS5}}& \sim m_{\m{pl},4} (e^i\C{V}_i)^{1/2} (\C{V}_X)^{-1/2}.
\end{align}
\end{subequations}
Additionally, there exist KK towers which have characteristic masses
\begin{equation}
    m_{\m{KK},\C{C}} = m_{\m{pl},4}(s^0)^{-1/2}\C{V}_{\C{C}}^{-1/\m{dim}_{\mathbb{R}}\C{C}}\;.
\end{equation}
Here, $\C{V}_{\C{C}}$ is the volume associated to a curve $\ww{i}{j}$, divisor $\w{i}$, or the overall volume where $\m{dim}_{\mathbb{R}}\C{C}$ is 2, 4, and 6 respectivley. This leads to three different types of KK towers whose masses scale as
\begin{subequations}
\begin{equation}
    m_{\m{KK},\ww{i}{j}} \sim m_{\m{pl},4}(s^0)^{-1/2}(\C{V}_{ij})^{-1/2}\;,
\end{equation}
\begin{equation}
    m_{\m{KK},\w{i}} \sim m_{\m{pl},4}(s^0)^{-1/2}(\C{V}_i)^{-1/4}\;,
\end{equation}
\begin{equation}
    m_{\m{KK,X}} \sim m_{\m{pl},4}(s^0)^{-1/2}(\C{V}_X)^{-1/6}\;,
\end{equation}
\end{subequations}
which correspond to the KK towers associated to curves, divisors, and the total volume.

There are also towers of states which arise in limits where the M-theory circle decompactifies while the 4d EFT remains perturbative (taking $\C{V}_X \rightarrow \infty$ while holding $s^0 = e^{-2\Phi}\C{V}_X$ constant, so $g_s \sim \sqrt{\C{V}_X} \rightarrow \infty$). We end up with a new KK scale associated with the length of the M-theory interval once resolving the limit through Horava-Witten theory, which scales as
\begin{equation}
    m_{\m{KK,M-th}} \sim m_{\m{pl},4} (\C{V}_X)^{-1/2}\;.
\end{equation}
There can also be regions where the 10d dilaton grows but some 2-cycle volume shrinks, leading to a light tower of particles appearing due to M2-branes wrapping the 2-cycle. Let the 2-cycle be an elliptic curve with volume $\C{V}_{\mathbb{E}}$ in string units, then the mass scale of the tower scales as
\begin{equation}
    m_{\m{M2}} \sim m_{\m{pl},4}(\C{V}_{\mathbb{E}})(\C{V}_X)^{-1/2}\;.
\end{equation}
Typically, we will take $\C{V}_{\mathbb{E}} = \C{V}_{ij}$, so we will label such a tower as $m_{\m{M2},\ww{i}{j}}$.

We can also consider bound states of the KK towers. If we decompactify two spaces $\C{C}_n$ and $\C{C}_m$ of dimensions $n$ and $m$, we find KK towers scaling as
\begin{equation}
    \frac{m_{\m{KK},\C{C}_n \times \C{C}_m}}{m_{\m{pl},4}} = \bigg[ \bigg( \frac{m_{\m{KK},\C{C}_n}}{m_{\m{pl},4}} \bigg)^n \bigg( \frac{m_{\m{KK},\C{C}_m}}{m_{\m{pl},4}} \bigg)^m \bigg]^{\frac{1}{n+m}}\;.
\end{equation}
Here, for $n=2,4,6$ we have that $\C{C}_n$ corresponds to the decompactification of some effective curve, effective divisor, or the full CY3 volume respectivley.

Similarly, we also consider bound states of the M-theory towers. In the limits where the M-theory towers become light (i.e.\ $\C{V}_X \rightarrow \infty$ while some 2-cycle volume $\C{V}_{\mathbb{E}}$ associated with an elliptic curve $\mathbb{E}$ shrinks) the CY3 $X$ admits a description as an elliptic fibration with a collapsing fiber $\mathbb{E}$ and expanding base $\C{B}$, $\mathbb{E} \hookrightarrow X \rightarrow \C{B}$. The mass of the bound states of the M-theory KK modes and wrapped M2-branes scales as
\begin{equation}
    m_{\m{(KK,M-th + M2),\mathbb{E}}} \sim m_{\m{pl,4}} \big(\C{V}_{\C{B}}\big)^{-1/2},
\end{equation}
with the volume of the base $\C{V}_{\C{B}}$ scaling parametrically as
\begin{equation}
    \C{V}_{\C{B}} \sim \frac{\C{V}_X}{\C{V}_{\mathbb{E}}}\;.
\end{equation}
When $\C{V}_{\mathbb{E}} = \C{V}_{ij}$, we we will label such a bound state as $m_{\m{(KK,M-th + M2),}\ww{i}{j}}$
 
Although it is not the focus of the present work, for completeness we will present some computations supporting Proposition \ref{prop:hyperbolic towers} in the case the mass scale is given by the species scale. All infinite distance limits in this 4d EFT correspond to either decompactification or perturbative string limits, as predicted by the ESC. The species scale $\Lambda_{\m{QG}}$ is either a string scale
\begin{equation}
    m_{\m{str}} \sim \sqrt{\C{T}_{\m{F1}}} \sim m_{\m{pl},4}(s^0)^{-1/2}\;,
\end{equation}
or a higher dimensional Planck mass
\begin{subequations}
\begin{align}
    m_{\m{pl},4+n} &\sim m_{\m{pl},4}(s^0)^{-\frac{n}{2(n+2)}} (\C{V}_{\C{C}_n})^{-\frac{1}{n+2}}\;,\\
    m_{\m{pl},5+n} & \sim m_{\m{pl},4}(s^0)^{-\frac{n}{2(n+3)}}(\C{V}_{\C{C}_n})^{-\frac{1}{n+3}}(\C{V}_X)^{-\frac{1}{2(n+3)}}
\end{align}
\end{subequations}
with $\C{V}_{\C{C}_n}$ the volume of the decompactifying $n$-cycle in string units as discussed previously (in the $5+n$ case, for $n=0$ we set $\C{V}_{\C{C}_0} \equiv 1$, this corresponds to the decompactification of just the M-theory interval). These Planck scales are the higher dimensional Planck scale associated either with decompactification signaled by the heterotic KK towers (for $m_{\m{pl},4+n}$) or decompactification of the M-theory circle with additional $n$-cycles decompactifying (for $m_{\m{pl},5+n}$). In certain cases, it is more convenient to compute the relevant $m_{\m{pl,4+n}}$ Planck scale through the associated KK tower scale \cite{Castellano:2023jjt}
\begin{equation}
    \frac{m_{\m{pl,4+n}}}{m_{\m{pl,4}}} \sim \left( \frac{m_{\m{KK},\C{C}_n}}{m_{\m{pl,4}}}\right)^{\frac{n}{2+n}}\;.
\end{equation}

Given the metric $h_{\alpha\beta}$, it is straightforward to compute the length squared of the $\alpha$-vectors and the scalar products of $\alpha$-vectors for the different states in the theory. Since all towers will and species scales are independent of the axions, the $\alpha$-vectors will be given as $h^{1,1}$ component vectors denoting the dependence of the tower masses and species scale on the saxions. Given towers scaling as $m_I$ and $m_J$, the scalar product of their $\alpha$-vectors is given by
\begin{equation}
    \vec{\alpha}_I \cdot \vec{\alpha}_J = h^{\alpha\beta}(\p_\alpha \log m_I)(\p_\beta \log m_J)\;,
\end{equation}
with the length squared of $\alpha$-vectors computed when $I=J$. Notice that $-\p_\alpha \log m_I$ is not the $\alpha$-vector for the tower $m_I$, as the $\alpha$-vector is specifically given by the gradient with respect to the canonically normalized moduli (or equivalently by contracting the non-canonical curved indices with a metric vielbein so that the scalar product of $\alpha$-vectors can be computed using the standard flat Euclidean metric). For the species scale $\Lambda_i$, we similarly can compute scalar products of the species vectors $\vec{\C{Z}}_I = -\vec\nabla \log \Lambda_I$ as
\begin{equation}
    \vec{\C{Z}}_I \cdot \vec{\C{Z}}_J = h^{\alpha\beta}(\p_\alpha \log \Lambda_I)(\p_\beta \log \Lambda_J)\;.
\end{equation}
The Laplacians for each tower $m_I$ and species scale $\Lambda_I$ can be computed by substituting in the relevant function for $m_i(s)$ in \eqref{eq:Lap_logT_explicit}. We provide the general expressions for the $\alpha$-vector squared and Laplacians of the logs of masses of towers in Table \ref{tab:4d_N=1_towers}. Though it is not a focus for our hyperbolic towers analysis, we also include the Laplacians of the species scale in each limit in Table \ref{tab:4d_N=1_species} for completeness to illustrate our computational technology. In these tables, we make the following definitions for the sake of compactness of expressions,
\begin{align*}
    &m_n \equiv m_{\m{KK},{\C{C}_n}}\;,\quad (\kappa\C{V}\kappa)_{ij} \equiv \kappa_{ijk}\C{V}^{k\ell}\kappa_{ij\ell}\;,\quad (e\kappa\C{V}) \equiv e^k \kappa_{k\ell m}\C{V}^{\ell m}\;,\\
    &(e\C{V}e) \equiv e^k \C{V}_{k\ell}e^\ell\;,\quad (\C{V}\kappa\C{V})_i \equiv \C{V}_i \kappa_{ik\ell}\C{V}^{k\ell}\;.
\end{align*}

\begin{table}
    \centering
    \renewcommand{\arraystretch}{1.5}
    \begin{tabular}{|@{}c@{}|@{}c@{}|c|c|}
        \hline
        Tower & $m/m_{\m{pl},4}$ & $|\vec\nabla\log m|^2$ & $\nabla^2 \log m$ \\
        \hline
        \hline
        F1 osc & $\frac{1}{(s^0)^{1/2}}$ & $\frac{1}{2}$ & $1$ \\
        \hline
        NS5 osc & $\frac{(e^i \C{V}_i)^{1/2}}{(\C{V}_X)^{1/2}}$ & $\frac{1}{2}-\frac{ \C{V}_X (e \C{V} e)}{2(e^i \C{V}_i)^2}$ & 
        $h^{1,1} - 1 + \frac{\C{V}_X \eVe}{(e^i \C{V}_i)^2}- \frac{\C{V}_X \ekV}{e^i \C{V}_i}
        $
        \\
        \hline
        KK,$\ww{i}{j}$ & $\frac{1}{(s^0)^{1/2}(\C{V}_{ij})^{1/2}}$ & $\frac{3}{4}-\frac{\kVk{ij}\C{V}_X}{2(\C{V}_{ij})^2}$ & $\frac{3}{2} - \frac{\kVk{ij}\C{V}_X}{(\C{V}_{ij})^2}$ \\
        \hline
        KK, $\w{i}$ & $\frac{1}{(s^0)^{1/2}(\C{V}_i)^{1/4}}$ & $\frac{3}{4}-\frac{\C{V}_X}{8(\C{V}_i)^2}\C{V}_{ii}$ & $\frac{3}{2}-\frac{\C{V}_X\big(  \C{V}_{ii} - \VkV{i}\big)}{2(\C{V}_i)^2}$ \\
        \hline
        KK, $X$ & $\frac{1}{(s^0)^{1/2}(\C{V}_X)^{1/6}}$ & $\frac{2}{3}$ & $1 + \frac{h^{1,1}}{3}$\\
        \hline
        KK, $\C{C}_n \times \C{C}_m$ & 
        $ \left(\frac{m_n}{m_{\m{pl},4}}\right)^{\frac{n}{n+m}}  \left(\frac{m_m}{m_{\m{pl},4}}\right)^{\frac{m}{n+m}} $ 
        & 
        $\frac{|n  \vec{\alpha}_{\C{C}_n}+ m \vec{\alpha}_{\C{C}_m}|^2 }{(m+n)^2}  $
        &
        $\frac{n \nabla^2 \log \frac{m_n}{m_{\m{pl},4}} + m \nabla^2 \log \frac{m_m}{m_{\m{pl},4}}}{n+m}$
        \\
        \hline
        KK, M-th & $\frac{1}{(\C{V}_X)^{1/2}}$ & $\frac{3}{2}$ & $h^{1,1}$ \\
        \hline
        M2, $\ww{i}{j}$ & $\frac{\C{V}_{ij}}{(\C{V}_X)^{1/2}}$ & $\frac{1}{2} -\frac{2 \C{V}_X \kVk{ij}}{(\C{V}_{ij})^2}$ & $h^{1,1}-1 + \frac{2\C{V}_X \kVk{ij}}{(\C{V}_{ij})^2}$ \\
        \hline
        \begin{tabular}{c}$\text{KK},\text{M-th} + \text{M2}$\\$\ww{i}{j}$\end{tabular} & $\frac{(\C{V}_{ij})^{1/2}}{(\C{V}_X)^{1/2}}$& $\frac{3}{4}-\frac{\C{V}_X \kVk{ij}}{2(\C{V}_{ij})^2}$ & $h^{1,1}-\frac{1}{2}+\frac{\C{V}_X \kVk{ij}}{(\C{V}_{ij})^2}$\\
        \hline
    \end{tabular}
    \caption{Summary of $\alpha$-vector lengths squared and Laplacians of logs of mass scalings for relevant towers in the 4d $\C{N}=1$ EFT string regime.}
    \label{tab:4d_N=1_towers}
\end{table}

\begin{table}
    \centering
    \renewcommand{\arraystretch}{1.5}
    \begin{tabular}{|c|c|c|c|}
        \hline
        Species Scale & $\Lambda_s/m_{\m{pl},4}$ & $|\vec\nabla\log \Lambda_s|^2$ & $\nabla^2 \log \Lambda_s$ \\
        \hline
        \hline
        $m_{\m{str}}$ & $\frac{1}{(s^0)^{1/2}}$ & $\frac{1}{2}$ & $1$ \\
        \hline
        $m_{\m{pl},4+2}$ & $\frac{1}{(s^0)^{1/4}(\C{V}_{ij})^{1/4}}$ & $\frac{3}{16}-\frac{1}{8}\frac{\C{V}_X \kVk{ij}}{(\C{V}_{ij})^2}$ & $\frac{3}{4} - \frac{1}{2}\frac{\C{V}_X \kVk{ij}}{(\C{V}_{ij})^2}
        $\\
        \hline
        $m_{\m{pl},4+4}$ & $\frac{1}{(s^0)^{1/3}(\C{V}_i)^{1/6}}$ & $\frac{1}{3} - \frac{1}{18}\frac{\C{V}_X \C{V}_{ii}}{(\C{V}_i)^2}$ & $1 - \frac{1}{3}\frac{\C{V}_X(\C{V}_{ii} - \VkV{i})}{(\C{V}_i)^2}$\\
        \hline
        $m_{\m{pl},4+6}$ & $\frac{1}{(s^0)^{3/8}(\C{V}_X)^{1/8}}$ & $\frac{3}{8}$ & $\frac{3+h^{1,1}}{4}$ \\
        \hline
        $m_{\m{pl},5+0}$ & $\frac{1}{(\C{V}_X)^{1/6}}$ & $\frac{1}{6}$ & $\frac{h^{1,1}}{3}$\\
        \hline
        $m_{\m{pl},5+2}$ & $\frac{1}{(s^0)^{1/5}(\C{V}_{ij})^{1/5}(\C{V}_X)^{1/10}}$ & $\frac{13}{50}-\frac{2}{25}\frac{\C{V}_X 
        \kVk{ij}}{(\C{V}_{ij})^2}$ & $\frac{h^{1,1}+3}{5}-\frac{2}{5}\frac{\C{V}_X \kVk{ij}}{(\C{V}_{ij})^2}
        $\\
        \hline
        $m_{\m{pl},5+4}$ & $\frac{1}{(s^0)^{2/7}(\C{V}_i)^{1/7}(\C{V}_X)^{1/14}}$ & $\frac{5}{14} - \frac{2}{49}\frac{\C{V}_X \C{V}_{ii}}{(\C{V}_i)^2}$ & $\frac{h^{1,1}+6}{7} - \frac{2}{7}\frac{\C{V}_X(\C{V}_{ii} - \VkV{i})}{(\C{V}_i)^2}$ \\
        \hline
        $m_{\m{pl},5+6}$ & $\frac{1}{(s^0)^{1/3}(\C{V}_X)^{1/6}}$ & $\frac{7}{18}$ & $\frac{h^{1,1}+2}{3}$ \\
        \hline
    \end{tabular}
    \caption{Summary of species vector lengths squared and Laplacians of logs of species scales for relevant scales in the 4d $\C{N}=1$ EFT string regime.}
    \label{tab:4d_N=1_species}
\end{table}

\subsection{Growth sectors, duality frames, and instantons}
Generically, a given compactification on Calabi-Yau will lead the asymptotic form of $\C{V}_X$ containing multiple monomials in the saxions $\{s^i\}_{i=1}^k$. One can conveniently divide the perturbative region defined by the saxionic cone into so-called growth sectors \cite{Grimm:2018cpv,Corvilain:2018lgw}
\begin{equation}
    \C{R}_{i_1,\dots,i_k} = \{ s^{i_1} \gg s^{i_2} \gg \cdots \gg s^{i_k}\}\;.
\end{equation}
As noted in \cite{Grieco:2025bjy}, growth sectors cover all asymptotic regions of the K\"ahler moduli space except asymptotic directions in which more than one saxions grow at the same rate. It can be shown that such limits are measure zero in the set of all asymptotic limits. In a given growth sector, there will be a single cubic monomial which dominates $\C{V}_X$, with the three distinct possible forms being $\C{V}_X = (s^i)^3$,  $s^i (s^j)^2$, and $s^i s^j s^k$. Despite the moduli space metric $h_{\alpha\beta}$ being diagonal to leading order in the asymptotic limit, subleading corrections to $h_{\alpha\beta}$ which vanish in the asymptotic limit from moduli not in the leading form of $\C{V}_X$ contribute to the curvature of moduli space and computation of the $\Gamma$-vector. As a result, one must perform all computations in full before taking the asymptotic limit corresponding to a given growth sector, as the delicate cancellation of divergent terms is lost if we drop these subleading terms.

As detailed in \cite{Lanza:2021udy}, the set of $\tfrac{1}{2}$BPS instantons relevant for the large volume, weak string coupling perturbative regime of the $E_8 \times E_8$ heterotic string is given by worldsheet instantons wrapping holomorphic curves on $X$ and NS5-branes wrapping $X$. The cone of $\tfrac{1}{2}$BPS instanton charges $\C{C}_I$ is then generated by effective curve classes along with the NS5-brane charge,
\begin{equation}
    \C{C}_I = \{\text{NS5-branes on } X \text{ + F1 on effective curves}\} \simeq \mathbb{Z}_{\geq 0} \oplus \m{Eff}_1(X)_\mathbb{Z}\;.
\end{equation}
Some of the instantons with charges in $\C{C}_I$ lead to backreactions on the EFT string flow that do not generate acceptable EFT string limits. The cone of $\tfrac{1}{2}$BPS instantons charges yielding instantons whose backreactions do generate acceptable EFT string flows is denoted $\C{C}_I^{\m{EFT}} \subset \C{C}_I$, see \cite{Lanza:2021udy} for more details. The worldsheet action of such instantons is given by
\begin{equation}
    \C{S}_i = 2\pi s^i\;,
\end{equation}
where $i=0$ corresponds to the instanton arising from NS5-branes wrapping $X$, and $i=1,\dots,h^{1,1}$ corresponds to instantons wrapping holomorphic curves in $X$. We can construct the analogue of $\alpha$-vectors for instantons, denoted as $\beta$-vectors, as
\begin{equation}
    (\beta_i)_j = \frac{\p \log \C{S}_i}{\p s^j} = \frac{\delta_{ij}}{s^j}\;,
\end{equation}
with no summation on $j$. It is convenient to express this in log-saxion coordinates as well, $T^j = \log s^j$,
\begin{equation}
    (\beta_i)_j^{\m{log-sax}} = \delta_{ij}\;.
\end{equation}

\subsection{Instantons in 4d $\C{N}=1$ Limits} 

We now wish to describe the expected axions and spectrum of instantons in the possible growth sectors. In doing so, we identify the $\beta$-vectors whose sum will yield the appropriate $\Gamma$-vectors for each duality frame. There are essentially three cases to address, corresponding to the dominant monomial in the internal Calabi-Yau volume $\C{V}_X = \frac{1}{6} \kappa_{ijk} s^i s^j s^k$. 

\paragraph{Case I: $\C{V}_X \sim (s^1)^3$.} 

In the first case, the dominant term is cubic in a single saxion. Accordingly, the space non-compact directions in the moduli space in the $s^1 \gg s^i$ growth sector is effectively two-dimensional, parametrized by $s^1, s^0$. The metric on the moduli space in these two directions is the K\"ahler metric associated to the K\"ahler potential: 
\begin{equation} K = -\log s^0 - 3 \log s^1, \end{equation}
and is given by the following diagonal matrix: 
\begin{equation} h_{\alpha \beta} = \frac{1}{2} \begin{pmatrix} \frac{1}{(s^0)^2} & & \cdots \\ & \frac{3}{(s^1)^3} & \\ \vdots & & \ddots \end{pmatrix}. \end{equation}
In the cone spanned by the orthogonal directions $s^0, s^1$, there are two duality frames. We note that in log-saxion coordinates $\log s^0$ and $\log s^1$, the metric is Euclidean, so Assumption \ref{ass:principal} is satisfied for these two moduli in the $\C{V}_X \sim (s^1)^3$ growth sector. We plot the relevant duality frames in an example below in Figure \ref{fig:quinticandr12}.

The bottom duality frame is a stringy frame corresponding to 10-dimensional $E_8 \times E_8$ heterotic string theory on the CY3 $X$. As explained in Section \ref{sec:2}, much of the technology we have developed to study instantons and $\beta$-vectors does not directly apply to stringy frames. However, we will see in this case that the identification of the relevant instantons is simple. The principal towers for this frame are those of the KK tower over $X$ and the F1-oscillator modes, and their $\alpha$-vectors are given by 
\begin{equation} \vec{\alpha}_{\m{osc,F1}} = \left( \frac{1}{\sqrt{2}}, 0 \right), \q \vec{\alpha}_{\m{KK}, X} = \left( \frac{1}{\sqrt{2}}, \frac{1}{\sqrt{6}} \right).    \end{equation}
Within this duality frame, there is a single BPS NS instanton that can be immediately deduced -- a string in 4d is electric-magnetic dual to an instanton, and as a string becomes tensionless and weakly coupled, there is a dual instanton which gets large action. The action of this instanton can immediately be deduced by noting that the tension of the heterotic string in this limit goes as
\begin{equation} \C{T}_{\m{F1}} \sim e^{2\phi_4} \end{equation}
in the 4d Einstein frame. Dually, we therefore immediately deduce the existence of an instanton with the following action: 
\begin{equation} \C{S}_{\text{dual F1}} \sim e^{-2\phi_4}. \end{equation}

Let us now compute the moduli dependence of this instanton in another way: From the perspective of compactification of the 10d theory on $X$, the instanton in question arises from an NS5 wrapped on $X$. In the 10d Einstein frame, the wrapped NS5 instanton action is
\begin{equation}
\C{S}_{\rm NS5}
=
\C{T}_{\rm NS5} \C{V}_{\m{10}} e^{-\phi_{10}/2}
+i\int_X B_6 ,
\end{equation}
up to conventional $2\pi$ and $\alpha'$ normalizations, with $\sigma=\int_X B_6$. Upon dimensional reduction, the same exponential combination is the four-dimensional Einstein-frame dilaton direction:
\begin{equation}
e^{-2\phi_4}
=
\C{V}_{\m{10}} e^{-\phi_{10}/2}. 
\end{equation}
Therefore in four-dimensional variables the classical action becomes
\begin{equation}
\C{S}_{\text{dual F1}} = \C{S}_{\m{NS5}} = \m{Re} \, \C{S}_{\rm NS5}
\simeq e^{-2\phi_4}. 
\end{equation}
where $\mu$ absorbs the fixed reference volume and normalization conventions, and ${\rm Re}\,S_{\m{NS5}}\propto e^{-2\phi_4}$. Recalling that the canonically-normalized 4d dilaton is $\sqrt{2} \phi_4 = \sqrt{2} \log s^0$ in 4d Planck units, we thus have
\begin{equation} \vec{\beta}_{\m{NS5}, X} = \left(\sqrt{2}, 0\right). \end{equation}

It appears that there should also be worldsheet instantons associated to the axions obtained via $B_2$ wrapping 2-cycles of $X$. However, these instantons cannot become small in the emergent stringy frame, since the tension of the F1-string in the 4d Einstein frame goes to zero as $e^{2\phi_4}$. Accordingly, wrapped F1-instantons over $X$ necessarily have actions going as $S \sim e^{2\phi_4}$, and $e^{-S}$ does not become small in a stringy frame. This is in fact completely natural, since we expect exponentiated worldsheet instanton actions to become small when the volumes of 2-cycles become large relative to the 10d dilaton. Recalling the relationship between the 10d dilaton and the length of the M-theory interval, we see that these instantons have a dual interpretation in terms of M2-branes wrapping a two-cycle $\C{C}$ times the M-theory interval $I$, and they should be detectable in the dual geometric frame corresponding to a decompactification to M-theory on $X$. In fact, this will be exactly what happens, as we show below. 

The principal towers in the top duality frame of Figure \ref{fig:quinticandr12} are those of the KK-tower over $X$ and the KK-tower over the M-theory interval, and their $\alpha$-vectors are given by 
\begin{equation} \vec{\alpha}_{\m{KK}, X} = \left( \frac{1}{\sqrt{2}}, \frac{1}{\sqrt{6}} \right), \q \vec{\alpha}_{\m{KK}, \text{M-th}} = \left( 0, \sfc{3}{2} \right) .  \end{equation}
We note that \textit{both} the overall volume $\C{V}_X$ and the interval length $R$ scale with the saxion $s^1$. 

In the decompactification to the 5d $\C{N}=1$ theory obtained by taking M-theory on $X$, there are $h^{1, 1}$ total 1-form gauge fields, whose kinetic terms were described in Section \ref{ssec:axionsinstantons} as follows: 
\begin{equation}
S_{\rm gauge,kin}
=
-\frac12 \int a_{IJ}(\phi)\,F^I\wedge *F^J .
\end{equation}

\begin{equation}
a_{IJ}(h)
=
-\frac12\,\partial_I\partial_J \log \mathcal F(s)\Big|_{\mathcal F=1},
\qquad
\mathcal F(s)=\frac16 C_{IJK}t^It^Jt^K .
\end{equation}
The 5d moduli $t^I$ are related to the 4d saxions $s^I$ via
\begin{equation} s^I = R_5 t^I,  \end{equation}
where $R_5$ is the length of the 5d interval. 

Let us now examine the dependence of the $a_{IJ}$ on the 5d modulus $t^1$ in particular, since this is the only possible way in which the 5d kinetic coefficient could affect the $\beta$-vector of the lower-dimensional theory in the principal plane spanned by $s^0, s^1$. Since we are in the growth sector where $s^1 \gg s^i$ for all other $s^i$ and $\C{F} \simeq \frac{1}{6} (s^1)^3$, the vector-multiplet normalization $\C{F} \simeq \frac{1}{6} (t^1)^3 = 1$ for the 5d $\C{N}=1$ real special Kähler geometry effectively sets $t^1 \simeq 6^{1/3}$. (In the 5d theory, the limit $s^1 \to \infty$ corresponds to moving along the universal hypermultiplet saxion.) Thus, to leading order, any power of $s^1$ in $a_{IJ}$ will be fixed to a constant value in the vector-multiplet kinetic-term coefficient, and the $a_{IJ}$ retain (at most) dependence on the other $s^i$. 

Accordingly, upon compactification to 4d over $I$, there are BPS  
instantons associated to zero-branes coupling electrically to the $A^I$ stretched across $I$ (these instantons can alternatively be thought of as M2-branes stretched across the HW interval wrapped on a 2-cycle in the internal CY3 $X$), and the only dependence of the instanton action on the 4d moduli arises due to the volume of $I = S^1/\B{Z}_2$. These instantons are not projected out of the 4d spectrum by the action of the $\B{Z}^2$-reflection on $S^1$ to obtain $I = S^1/\B{Z}_2$, as this reflection acts evenly on the component $A^I_4$ of $A^I$ in the circle direction: $A^I_4 \to +A^I_4$. The action of these instantons depends on the radion $\rho = \log R$ (for $R$ the length of $I$) via 
\begin{equation} \C{S}_{\m{5d}}^I = \C{T}_{\m{5d}} e^{\rho} + i \int_{I} A^I. \end{equation}

In terms of the canonically-normalized 4d modulus $\sfc{2}{3} \rho = \sfc{2}{3} \log s^0$, we have
\begin{equation} \vec{\beta}^I_{\m{5d}} = \left( 0, \sfc{2}{3} \right) \end{equation}
for $I = 1, \dots, h^{1, 1}$. In summary, the total $\Gamma$-vector for the principal plane is the sum of all of the $\vec{\beta}^I$ in the two duality frames:
\begin{equation}\label{eq:4dN=1_Gamma_I}
    \vec{\Gamma} = \left( \sqrt{2} , h^{1, 1} \sfc{2}{3} \right)\;.
\end{equation}
We then verify that
\begin{equation} \vec{\alpha}_{\m{osc, F1}} \cdot \vec{\Gamma} = 1, \q \vec{\alpha}_{\m{KK}, X} \cdot \vec{\Gamma} = 1 + \frac{h^{1, 1}}{3}, \q \vec{\alpha}_{\text{M-th}} \cdot \vec{\Gamma} = h^{1, 1}, \end{equation}
which agrees with Table \ref{tab:4d_N=1_towers} below. 

\paragraph{Case II: $\C{V}_X \sim (s^1)^2 s^2$.} 

We now turn to the more involved case where $\C{V}_X \sim (s^1)^2 s^2$. The metric on the moduli space is given by 
\begin{equation} h_{\alpha \beta} = \frac{1}{2} \begin{pmatrix} \frac{1}{(s^0)^2} & & & \cdots \\ & \frac{2}{(s^1)^3} & & \\ & & \frac{1}{(s^2)^2} & \\ \vdots & & & \ddots \end{pmatrix}. \end{equation}
In the cone spanned by $s^0, s^1, s^2$, there are four distinct duality frames, three of which contain nontrivial instanton scaling vectors. We wish to explicitly identify the instanton scaling vectors associated with each frame. We find this growth sector in an example below, which we plot in Figure \ref{fig:cicy}

The duality frame in Figure \ref{fig:cicy} with $s^0 \gg s^1, s^2$ corresponds to a stringy frame with an $E_8 \times E_8$ heterotic string on $X$. The principal towers are the F1 oscillator modes of this string as well as the KK towers arising from the decompactification of the divisor $[\omega_2]$ dual to $s^2$ and the curve $[\omega_1] \cap [\omega_1]$ with $[\omega_1]$ dual to $s^1$. The associated $\alpha$-vectors are
\begin{equation} \vec{\alpha}_{\m{osc, F1}} = \left( \frac{1}{\sqrt{2}}, 0, 0 \right), \q \vec{\alpha}_{\m{KK}, [\omega_2]} = \left(\frac{1}{\sqrt{2}}, \frac{1}{2}, 0 \right)  \q \vec{\alpha}_{\m{KK}, [\omega_1] \cap [\omega_1]} = \left( \frac{1}{\sqrt{2}}, 0, \frac{1}{\sqrt{2}} \right).  \end{equation}
By the preceding analysis, we may immediately identify the single instanton with small exponentiated action dual to the tensionless string in this duality frame -- a wrapped NS5 on $X$. Without doing any further computations, we find the associated $\beta$-vector to be
\begin{equation} \vec{\beta}_{\m{NS5}, X} = \left(\sqrt{2}, 0, 0 \right). \end{equation}
As before, the instantons associated to the $E_8 \times E_8$ worldsheet wrapping a 2-cycle in $X$ do \textit{not} have $\beta$-vectors in this frame, as their exponentiated actions retain the wrong-sign dependence on the 10d dilaton and are thus minimized in different duality frames.  

Traveling along the vector $\vec{\alpha}_{\m{KK}, [\omega_1] \cap [\omega_1]}$ corresponds to a decompactification to $E_8 \times E_8$ over K3. Consider now the duality frame $s^2 \gg s^0, s^1$. This corresponds to a \textit{dual} emergent $E_8 \times E_8$ string on a CY3 $X$, which is fibered in $\m{K3}'$ via $\m{K3}' \into X \tto \B{P}^1$. The principal towers for this frame are the dual oscillator modes of NS5s wrapping $\omega_2$, the KK-modes for the M-theory interval, and the KK-modes associated to $s^2$ taken large. The associated $\alpha$-vectors are as follows: 
\begin{equation} \vec{\alpha}_{\m{osc,NS5}} = \left(0, 0, \frac{1}{\sqrt{2}} \right), \q \vec{\alpha}_{\m{KK},\text{M-th}}  = \left(0, 1, \frac{1}{\sqrt{2}} \right), \q \vec{\alpha}_{\m{KK}, [\omega_1] \cap [\omega_1]} = \left( \frac{1}{\sqrt{2}}, 0, \frac{1}{\sqrt{2}} \right) .   \end{equation}
In the intermediate limit along $\vec{\alpha}_{\m{KK}, [\omega_1] \cap [\omega_1]}$, the dual theory on $\m{K3}'$ is identified with the original theory on the original K3. In the dual theory, we immediately deduce the existence of a dual NS5-instanton wrapping $X'$: 
\begin{equation} \vec{\m{\beta}}_{\m{NS5'}, X'} = \left(0, 0, \sqrt{2} \right). \end{equation}
Moreover, as before, the other instantons arising from the worldsheet wrapping 2-cycles in $X'$ do not have small action in this duality frame. 

Finally, we turn to the sole geometric frame corresponding to $s^1 \gg s^0, s^2$. As explained in \cite{Lanza:2021udy}, this corresponds to a decompactification over $S^1 \times S^1/\B{Z}_2$ to F-theory on elliptically-fibered $X$ via $\B{E} \into X \tto \C{B}$. (Here, the decompactifying $S^1 \times S^1/\B{Z}_2$ correspond to the Type IIB circle and the M-theory HW interval, respectively.) The principal towers are associated to a decompactification to 6d, a decompactification to F-theory on $T^2$, and a decompactification of the F-theory fiber/M-theory interval. In terms of the original $E_8 \times E_8$ description, the KK tower associated to decompactification to 6d arises from bound states of M2-branes wrapped on 2-cycles and KK-modes associated to the decompactifying interval, while the KK tower associated to decompactification to 10d F-theory on $T^2$ is the KK tower associated to blowing up $[\omega_2]$. The $\alpha$-vectors for each principal tower are thus
\begin{equation} \vec{\alpha}_{\m{KK}, \text{M-th}+\m{M2}} = \left(0, 1, 0 \right), \q \vec{\alpha}_{\m{KK}, \text{M-th}} = \left(0, 1, \frac{1}{\sqrt{2}} \right), \q \vec{\alpha}_{\m{KK}, [\omega_2]} = \left(\frac{1}{\sqrt{2}}, \frac{1}{2}, 0 \right) . \end{equation}
Note that the volume of the decompactifying internal dimensions to 6d as well as the volume of the F-theory base $\C{B}$ depend on the saxion $s^1$.

We wish now to identify the instantons in this duality frame. This can be readily deduced from the structure of the 6d $\C{N}=(1, 0)$ theory obtained by compactifying F-theory on $\B{E} \into X \tto \C{B}$: There are $T = h^{1, 1}(\C{B}) - 1 = h^{1, 1}(X) - 2$ total tensor multiplets in this theory; alongside the supergravity multiplet, there are a total of $h^{1, 1}(X) - 1$ two-form gauge fields in the 6d theory. 

The supergravity and tensor-multiplet kinetic terms for the 2-form field of the 6d $\C{N}=(1, 0)$ theory are given by 
\begin{equation}
S_H
=
-\frac14 \int g_{\alpha\beta}(j)\,
H^\alpha \wedge *H^\beta,
\qquad
g_{\alpha\beta}(j)
=
2j_\alpha j_\beta-\Omega_{\alpha\beta},
\end{equation}
where the moduli $j_\a$ are expressed in terms of the Kähler moduli of the base $\C{B}$ via 
\begin{equation}
J_B=v^\alpha \omega_\alpha,
\qquad
\mathcal V_B
=
\frac12\,\Omega_{\alpha\beta}v^\alpha v^\beta,
\qquad
j^\alpha
=
\frac{v^\alpha}{\sqrt{2\mathcal V_B}},
\qquad
\Omega_{\alpha\beta}j^\alpha j^\beta=1.
\end{equation}

As before, we examine the way in which the 6d kinetic coefficients can depend on what corresponds to $s^1$ in the 4d, since this is the only possible way in which the 4d $\beta$-vectors can be influenced by the 6d kinetic terms. Explicitly, the relation between the 6d moduli $j^\a$ and the 4d moduli $s^\a$ is 
\begin{equation} s^\alpha = A j^\a, \end{equation}
where $A$ is the area of the $S^1 \times S^1/\B{Z}_2$ on which the 6d theory is compactified. As such, the normalization $\Omega_{\a\beta} j^\alpha j^\beta = 1$ effectively amounts to setting $j^1 = 1$. We conclude that the 6d kinetic terms retain no leading-order dependence on $s^1$, and their dimensional reduction yields, to leading-order, $h^{1, 1}(\C{B}) = h^{1, 1}(X) - 1$ instantons in 4d whose sole dependence on the principal-plane moduli arises from the area of the compact $S^1 \times S^1/\B{Z}_2$ over which we reduce F-theory. 

The volume of $S^1 \times S^1/\B{Z}_2$ depends also on the 4d modulus $s^1$; indeed, the KK-tower associated to this decompactification has $\alpha$-vector $\vec{\alpha}_{\m{KK},\text{M-th}+\m{M2}} = (0, 1, 0)$. Accordingly, for strings in the 6d $\C{N}=(1, 0)$ theory wrapping $S^1 \times S^1/\B{Z}_2$, which are not projected out since the reflection of $T^2$ is accompanied by $B_4^\alpha \mapsto +B_4^\a$ along the Horava-Witten direction (this follows directly from the M-theory analysis in the previous case). Thus, the instanton action goes as
\begin{equation} \C{S}^\alpha = T_{\m{6d}} e^{-2\rho} + i \int_{T^2/\B{Z}_2} B^\alpha . \end{equation}
In terms of the canonically normalized 4d modulus $s^1 = 2 \log \rho$, we thus find
\begin{equation} \vec{\beta}_{\m{6d}}^\alpha = \left(0, 1, 0 \right), \q \alpha = 1, \cdots, h^{1, 1}(X) - 1. \end{equation}
In summary, the total $\Gamma$-vector, obtained by summing the $\beta$-vectors over all of the duality frames, is 
\begin{equation} \vec{\Gamma} = (\sqrt{2}, h^{1, 1} - 1, \sqrt{2}). \end{equation}
Note that this agrees with Table \ref{tab:4d_N=1_towers} when the Laplacians computed are constant rational numbers, as in the previous case. However, along the limits $\vec{\alpha}_{\m{KK}, \text{M-th} +\m{M2}}$ and $\vec{\alpha}_{\m{KK},[\omega_2]}$, the computed values of $\nabla^2 \log m$ are not seen to be constant directly from the prepotential. We see thus that the ESC furnishes a \textit{prediction} for the rational values Laplacians of principal towers in EFT string limits not immediately apparent from supergravity, and this growth sector is a crucial test case for the power of the hyperbolic towers proposal of Proposition \ref{prop:hyperbolic towers quantization} as a check of the ESC. It would be interesting to see if this imposes general consistency conditions on the structure constants $\kappa_{IJK}$. 

\paragraph{Case III: $\C{V}_X \sim s^1 s^2 s^3$.}

We finally turn to the case $\C{V}_X \sim s^1 s^2 s^3$. In this case, symmetry under the interchange $s^i \leftrightarrow s^j$ for $i, j = 1, 2, 3$ tells us that there are essentially two inequivalent saxion limits: $s^0 \to \infty$ and $s^1 \to \infty$. The metric in this growth sector goes as
\begin{equation} h_{\alpha \beta} = \frac{1}{2} \begin{pmatrix} \frac{1}{(s^0)^2} & & & & \cdots \\ & \frac{1}{(s^1)^3} & & & \\ & & \frac{1}{(s^2)^2} & & \\ & & & \frac{1}{(s^3)^2} & \\ \vdots & & & & \ddots \end{pmatrix}. \end{equation}
In this case, there are eight distinct duality frames, which we will not exhaustively describe. However, we focus on the duality frames associated to $s^0 \gg s^i$ and $s^i \gg s^0, s^j$ for $i, j \in \{ 1, 2, 3\}$. 

For the frame $s^0 \gg s^i$, we again find an emergent string limit with an emergent tensionless $E_8 \times E_8$ heterotic string on $X$. Associated to this frame is again the single instanton obtained by wrapping the NS5 over $X$. For the frames $s^i \gg s^0, s^j$, however, we instead have an emergent Type IIA string over an exotic space $\C{X}'$. The geometry of $\C{X}'$ is poorly understood \cite{Grieco:2025bjy}, but it is expected that if we take $s^0, s^i \gg s^j, s^k$ (i.e.\ at the interface between this frame and the emergent $E_8 \times E_8$ frame), we will obtain a tensionless $E_8 \times E_8$ string in 6d. Thus, in the Type IIA frame, $s^0 \to \infty$ corresponds to a decompactification of a curve $\C{C} \subset \C{X}'$ with volume $s^0$. 

Unfortunately, the lack of tractability of the geometry of the Type IIA frames hinders our ability to glean the instanton spectrum and the $\beta$-vectors. Aside from the universal NS5 instantons in each stringy frame, there could possibly be other RR instantons in the Type IIA frames; the exact spectrum of RR instantons in the 4d will then depend on the possible dimensional reductions of RR-fields over $\C{X}'$. As we noted in Section \ref{ssec:axionsinstantons}, the $\beta$-vectors arise from wrapped-brane instantons on internal cycles in geometric frames. In a stringy frame, we do not directly have access to this; however, in the previous two cases, we were saved by the fact that each stringy frame admitted an exactly known description in terms of heterotic $E_8 \times E_8$ on an internal manifold $X$; moreover, an infinitesimal displacement from the exact string limit corresponds to a decompactification over $X$ to $10$d $E_8 \times E_8$ theory. Accordingly, the relevant axion/instanton spectrum could be studied by considering wrapped branes in the $10$d theory, and this is how we decided ultimately that the NS5-instanton lived in the heterotic frame while the worldsheet instantons did not. 

In the present case, since the geometry $\C{X}'$ is unknown, we are unable to perform a similar analysis using the instantons in the perturbative Type IIA spectrum over, and consequently, we are unable to directly predict the axions and instanton spectrum of the 4d theory in each asymptotic frame corresponding to an emergent Type IIA string. In fact, it is not even apparent whether our Assumptions \ref{ass:ESC} and \ref{ass:perturbative-esc} are even satisfied over $\C{X}'$, so it may not be possible to derive the asymptotic fibration structure as we have done in Section \ref{sec:2}. This remains an important open avenue for future work, and though we do not consider growth sectors of the form $\C{V}_X \sim s^1 s^2 s^3$ in the present work, we believe it is important to understand this case better.

Before turning to examples, let us comment on the fact that the analysis in this section is entirely based on the classical moduli space. As apparent from \cite{Cvetic:2024wsj,Kaufmann:2026fli, Kaufmann:2026mha, Kaufmann:2026tsy}, $4$d $\C{N}=1$ theories can receive important corrections modifying the geometry of their moduli spaces. In this section, we are agnostic about the full heterotic embedding, and thus of any possible quantum correction to the K\"ahler potential, such as the ones computed in the case of the standard embedding \cite{Kaplunovsky:1992vs,Dixon:1990pc}. As such, some of the duality frames discussed hereafter might be removed, but the analysis continues \emph{mutatis mutandis} for the ones that remain, and confirm the expectations of Section \ref{sec:2}.

\subsection{Examples}

In this section, we study several examples of $E_8 \times E_8$ on a CY3 $X$ and study explicitly the asymptotic fibration structure, Laplacians, axions, and instantons in each asymptotic duality frame. We focus on growth sectors of the form $\C{V}_X \sim (s^1)^3$ and $\C{V}_X \sim (s^1)^2 s^2$, and we find in all cases studied that the hyperbolic towers quantization condition of Proposition \ref{prop:hyperbolic towers quantization} is satisfied. We identify all of the contributing axions and instantons that give rise to the asymptotic fibration structure and thereby directly verify Propositions \ref{prop:instantons-beta-vectors} and \ref{prop:hyperbolic towers} in each example. 

An important caveat is that we avoid examples in this section in which the moduli space develops rigid field theory sectors that decouple from gravity. Accordingly, we do not consider examples with divergent asymptotic curvature, and indeed, we expect that hyperbolicity fails in such examples just as it does in Section \ref{ssec:5d-examples}. It is for this reason that we do not consider the example of heterotic compactified on the degree-18 hypersurface $X \subset \B{P}_{1,1,1,6,9}$, since limits in this example feature a curvature divergence corresponding to a rigid gauge sector, cf. \cite{Marchesano:2023thx}. We believe that it will be interesting and important to study further, but this is unfortunately beyond the scope of the present paper. 

\subsubsection{Example with $h^{1, 1} = 1$: quintic threefold}
The quintic $\mathbb{P}_{1,1,1,1,1}[5]$ has $h^{1,1} = 1$ K\"ahler modulus $s^1$, with a single non-vanishing triple intersection number $\kappa_{111} = 5$. The volume of the threefold is
\begin{equation}
    \C{V}_X = \frac{5}{6}(s^1)^3\;.
\end{equation}
We also have the universal saxion $s^0$. The moduli space metric on the saxions can be computed as
\begin{equation}
    h_{\alpha\beta} = \begin{pmatrix}
        \frac{1}{2(s^0)^2} & 0 \\
        0 & \frac{3}{2(s^1)^2}
    \end{pmatrix}.
\end{equation}
Thus, Assumption \ref{ass:principal} is indeed satisfied in this duality frame. There are two elementary EFT string limits, $(e^0,0)$ and $(0,e^1)$, as well as the non-elementary limit $(e^0,e^1)$.

The light towers in these two limits are the oscillator tower of the fundamental heterotic string, the M-theory KK tower, and the KK tower associated with the full decompactification of the Calabi-Yau. The tower masses scale as
\begin{equation}
    m_{\m{osc,F1}} \sim (s^0)^{-1/2}\;,\quad m_{\m{KK,M-th}} \sim (s^1)^{-3/2}\;,\quad m_{\m{KK},X}\sim (s^0)^{-1/2}(s^1)^{-1/2}\;.
\end{equation}
The $\alpha$-vectors of each of these towers can be computed as
\begin{align}
    \vec{\alpha}_{\m{osc,F1}} = \bigg(\frac{1}{\sqrt{2}},0\bigg)\;,\quad \vec{\alpha}_{\m{KK,M-th}} = \bigg(0,\sqrt{\frac{3}{2}}\bigg)\;,\quad\vec{\alpha}_{\m{KK,}X} = \bigg(\frac{1}{\sqrt{2}},\frac{1}{\sqrt{6}}\bigg)\;.
\end{align}
Depending on the limit we take, we either approach an emergent string limit or decompactify $6$ or $7$ dimensions. The relevant species scales are $m_{\m{osc}}$, $m_{\m{pl},10}$, and $m_{\m{pl},11}$. The species vectors can be computed as
\begin{align}
    \vec{\C{Z}}_{\m{osc}} = \bigg(\frac{1}{\sqrt{2}},0\bigg)\;,\quad \vec{\C{Z}}_{\m{pl},10} = \bigg(\frac{3}{4\sqrt{2}},\frac{1}{4}\sqrt{\frac{3}{2}}\bigg)\;,\quad \vec{\C{Z}}_{\m{pl},11} = \bigg(\frac{\sqrt{2}}{3},\frac{1}{\sqrt{6}}\bigg)\;.
\end{align}
The leading towers, $\alpha$-vector lengths squared, Laplacians, and the relevant dual theories for the various EFT string limits are given in Table \ref{tab:4dEFT_quintic_towers}, with the corresponding data for the species scales given in Table \ref{tab:4dEFT_quintic_species}.

\begin{table}[t]
    \centering\renewcommand{\arraystretch}{1.1}
    \begin{tabular}{|c|c|c|c|c|}
    \hline
    Limit &  Tower & $|\vec\nabla \log m|^2$  & $\nabla^2 \log m$ & Emergent Dual Theory \\
    \hline\hline
    \multirow{2}{*}{$(e^0,0)$} & $m_{\m{osc,F1}}$ & $1/2$ & $1$ & \multirow{2}{*}{Emergent 
    HE} \\
    \cline{2-4}
     & $m_{\m{KK},X}$ & $2/3$ & $4/3$ & \\
    \hline
    $(0,e^1)$ & $m_{\m{KK,M-th}}$ & $3/2$ & $1$ & M-theory on $X$ \\
    \hline
    $(e^0,e^1)$ & $m_{\m{KK,M-th}}$ & $3/2$ & $1$ & $11$d M-theory \\
    \hline\end{tabular}\caption{
    Leading towers for the allowed EFT string limits of the quintic $\mathbb{P}_{1,1,1,1,1}[5]$.}
    \label{tab:4dEFT_quintic_towers}
\end{table}

\begin{table}[t]
    \centering\renewcommand{\arraystretch}{1.1}
    \begin{tabular}{|c|c|c|c|c|}
    \hline
    Limit &  Species Scale & $|\vec\nabla \log \Lambda|^2$  & $\nabla^2 \log \Lambda$ & Emergent Dual Theory \\
    \hline\hline
    $(e^0,0)$ & $m_{\m{str}}$  & $1/2$ & $1$ & Emergent 
    HE\\
    \hline
    $(0,e^1)$ &$m_{\m{pl},10}$ & $3/8$ & $1/2$ & M-theory on $X$\\
    \hline
    $(e^0,e^1)$ &$m_{\m{pl},11}$ & $7/18$ & $1$ & $11$d M-theory\\
    \hline\end{tabular}\caption{
    Species scales for the allowed EFT string limits of the quintic $\mathbb{P}_{1,1,1,1,1}[5]$.}
    \label{tab:4dEFT_quintic_species}
\end{table}

To compute the $\Gamma$-vector, we can choose coordiantes $t^0 = \log m_{\m{osc,F1}} = -\tfrac{1}{2}\log s^0$ and $t^1 = \log m_{\m{KK,}X} = -\tfrac{1}{2}\log s^0 -\tfrac{1}{2}\log s^1$ whose associated coordinate basis vectors span the principal plane.
In these coordinates, the full metric on the saxion and axion moduli space takes the form
\begin{equation}
    \d s^2 = 8 (\d t^0)^2 +6 (\d t^1)^2 - 12 \d t^0 \d t^1 + \frac{1}{2}e^{4 t^0}(\d a^0)^2 + \frac{3}{2}e^{-4 t^0 + 4 t^1}(\d a^1)^2\;,
\end{equation}
with
\begin{equation}
    \log \left(\sqrt{\det g}\right) = 2 t^1 + \m{const}\;.
\end{equation}
In this coordinate system, we compute the trace of the Christoffel symbol using
\begin{equation}
    \Gamma^{\alpha}_{\alpha\beta} =- \p_\beta \log \sqrt{\det g}\;,
\end{equation}
which, after converting back to a locally flat frame in the saxionic coordinates $s^0$ and $s^1$, yields the $\Gamma$-vector
\begin{equation}
    (\vec{\Gamma})_{\m{quintic}} = \bigg(\sqrt{2},\sqrt{\frac{2}{3}}\bigg)\;.
\end{equation}
One can verify that computing the inner products $\vec{\Gamma} \cdot \vec{\alpha}_i$ and $\vec{\Gamma} \cdot \vec{\C{Z}}_i$ indeed reproduces the $c_i$ values provided in Table \ref{tab:4dEFT_quintic_towers} and \ref{tab:4dEFT_quintic_species} respectively. In log-saxion coordinates, $(\vec{\Gamma})_{\m{quintic}}$ has the components
\begin{equation}
    (\vec{\Gamma})_{\m{quintic}}^{\m{log-sax}} = (1,1)\;,
\end{equation}
demonstrating that $(\vec{\Gamma})_{\m{quintic}}$ is the sum of the relevant instanton $\beta$-vectors as predicted by \eqref{eq:4dN=1_Gamma_I}. We show the principal planes and $\Gamma$-vectors in the left panel of Figure \ref{fig:quinticandr12}.

\subsubsection{Example with $h^{1, 1} = 2$: bidegree $(2, 4)$ hypersurface $X \subset \B{P}^1 \times \B{P}^3$}

Here we consider CICY \#7887 \cite{Candelas:1987kf,Green:1987cr}, with $h^{1,1} = 2$ and non-vanishing triple intersection numbers
\begin{equation}
    \kappa_{111} = 2\;,\quad \kappa_{112} = 4\;.
\end{equation}
The volume of the Calabi-Yau scales as
\begin{equation}
    \C{V}_X = \frac{1}{3}(s^1)^3 + 2(s^1)^2 s^2\;.
\end{equation}
The moduli space metric on the saxions can be computed as
\begin{equation}
    h_{\alpha\beta} = \begin{pmatrix}
        \frac{1}{2(s^0)^2} & 0 & 0\\
        0 & \frac{1}{(s^1)^2} + \frac{1}{2(s^1 + 6 s^2)^2} & \frac{3}{(s^1 + 6 s^2)^2} \\
        0 & \frac{3}{(s^1 + 6 s^2)^2} & \frac{18}{(s^1 + 6 s^2)^2}
    \end{pmatrix}\;.
\end{equation}
Here, we consider the growth sector $\C{R}_{12}$ where $s^1 \gg s^2$, and the overall volume grows asymptotically as $\C{V}_X \sim (s^1)^3$. The principal plane in this growth sector is spanned by the $s^0$ and $s^1$ directions. The elementary EFT string limits that can be reached in this growth sector are $(e^0,0,0)$ and $(0,e^1,0)$, and the non-elementary limit we can reach is $(e^0,e^1,0)$.

\begin{figure}[t]
    \hspace{-0.01\linewidth}\includegraphics[width=1.02\linewidth]{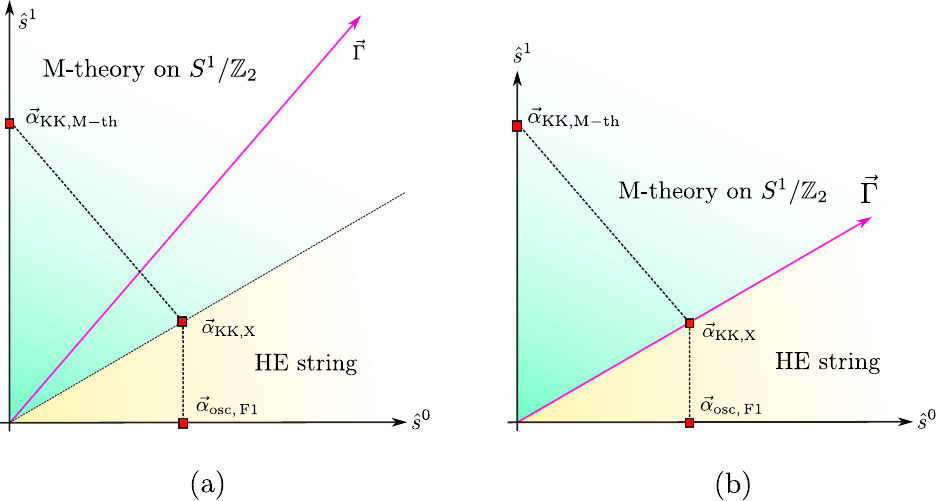}
    \caption{$2$d principal plane and $\Gamma$-vectors for the quintic (b) and $R_{12}$ growth sector of CICY \#7887 (a). The relevant $\Gamma$-vector in each duality frame is given by the projection of the overall $\vec \Gamma$ on the axes, and agrees with the $\beta$-vectors. The $\hat s$ indicate log-saxion-coordinates.}
    \label{fig:quinticandr12}

\bigskip

    \centering\renewcommand{\arraystretch}{0.97}
    \begin{tabular}{|c|c|c|c|c|}
    \hline
    Limit &  Tower & $|\vec\nabla \log m|^2$  & $\nabla^2 \log m$ & Emergent Dual Theory \\
    \hline\hline
    \multirow{2}{*}{$(e^0,0,0)$} & $m_{\m{osc,F1}}$ & $1/2$ & $1$ & \multirow{2}{*}{Emergent 
    HE} \\
    \cline{2-4}
     & $m_{\m{KK},X}$ & $2/3$ & $5/3$ & \\
    \hline
    $(0,e^1,0)$ & $m_{\m{KK,M-th}}$ & $3/2$ & $2$ & M-theory on $X$ \\
    \hline
    $(e^0,e^1,0)$ & $m_{\m{KK,M-th}}$ & $3/2$ & $2$ & $11$d M-theory \\
    \hline\end{tabular}\captionof{table}{
    Leading towers for the allowed EFT string limits of the $\C{R}_{12}$ growth sector of CICY \#7887.}
    \label{tab:4dEFT_7887_towers}

\bigskip

    \centering\renewcommand{\arraystretch}{0.97}
    \begin{tabular}{|c|c|c|c|c|}
    \hline
    Limit &  Species Scale & $|\vec\nabla \log \Lambda|^2$  & $\nabla^2 \log \Lambda$ & Emergent Dual Theory \\
    \hline\hline
    $(e^0,0,0)$ & $m_{\m{str}}$  & $1/2$ & $1$ & Emergent 
    HE\\
    \hline
    $(0,e^1,0)$ &$m_{\m{pl},10}$ & $3/8$ & $1/2$ & M-theory on $X$\\
    \hline
    $(e^0,e^1,0)$ &$m_{\m{pl},11}$ & $7/18$ & $1$ & $11$d M-theory\\
    \hline\end{tabular}\captionof{table}{
    Species scales for the allowed EFT string limits of the $\C{R}_{12}$ growth sector of CICY \#7887.}
    \label{tab:4dEFT_7887_species}
\end{figure}

The light towers in this growth sector are again the oscillator tower of the fundamental heterotic string, the M-theory KK tower, and the KK tower associated with decompactification of the full Calabi-Yau. The relevant $\alpha$-vectors and species vectors are unchanged from the quintic example, but the $c$ values of towers have changed. The relevant $c$ values (and lengths of $\alpha$-vectors) for the towers are displayed in Table \ref{tab:4dEFT_7887_towers}, while the same information for the relevant species scales is displayed in Table \ref{tab:4dEFT_7887_species}.

We can again compute the $\Gamma$-vector by taking the trace of the Christoffel symbol in coordinates $(\log m_{\m{osc,F1}},\log m_{\m{KK,}X})$ or $(\log m_{\m{osc,F1}},\log m_{\m{KK,M-th}})$ corresponding to the principal towers in the two accessible duality frames of the growth sector $\C{R}_{12}$. In a locally flat frame after converting back to saxionic coordinates, either choice yields the $\Gamma$-vector
\begin{equation}
    (\vec{\Gamma})_{7887} = \left( \sqrt{2},2\sqrt{\frac{2}{3}} \right)\;,
\end{equation}
which again reproduces the correct $c_i$ values via $\vec{\Gamma} \cdot \vec{\alpha}_i$ and $\vec{\Gamma} \cdot \vec{\C{Z}}_i$ as given in Table \ref{tab:4dEFT_7887_towers} and \ref{tab:4dEFT_7887_species}. In log-saxion coordinates, $(\vec{\Gamma})_{7887}$ has the components
\begin{equation}
    (\vec{\Gamma})_{7887}^{\m{log-sax}} = (1,2)\;, 
\end{equation}
demonstrating that $(\vec{\Gamma})_{7887}$ is the sum of the relevant instanton $\beta$-vectors as predicted by \eqref{eq:4dN=1_Gamma_I}. We show the principal planes and $\Gamma$-vectors in the right panel of Figure \ref{fig:quinticandr12}.

\subsubsection{Example with $h^{1, 1} = 3$}
Here, we consider a CICY hypersurface (\#7880 in \cite{Candelas:1987kf,Green:1987cr})
\[
X=
\left[
\begin{array}{c|c}
\mathbb P^1 & 2\\
\mathbb P^1 & 2\\
\mathbb P^2 & 3
\end{array}
\right]
\subset
\mathbb P^1\times\mathbb P^1\times\mathbb P^2,
\qquad
h^{1,1}(X)=3.
\]
This CICY has non-vanishing triple intersection numbers 
\begin{equation}
    \kappa_{233} = 2\;, \quad \kappa_{133} = 2\;,\quad \kappa_{123} = 3\;.
\end{equation} The volume of the Calabi-Yau scales as
\begin{equation}
    \C{V}_X = 3s^1 s^2 s^3 + s^1 (s^3)^2 + s^2 (s^3)^2\;.
\end{equation}
Let $f(s) \equiv 2\big(3s^1 s^2 + (s^1 + s^2) s^3\big)^2$, then the moduli space metric on the saxions can be computed as
\begin{equation}
    h_{\alpha\beta} = \begin{pmatrix}
        \frac{1}{2(s^0)^2} & 0 & 0 & 0\\
        0 & \frac{(3s^2 + s^3)^2}{f(s)} & \frac{(s^3)^2}{f(s)} & \frac{3 (s^2)^2}{f(s)}\\
        0 & \frac{(s^3)^2}{f(s)} & \frac{(3 s^1 + s^3)^2}{f(s)} & \frac{3 (s^1)^2}{f(s)}\\
        0 & \frac{3(s^2)^2}{f(s)} & \frac{3(s^1)^2}{f(s)} & \frac{1}{2(s^3)^2}+\frac{(s^1 + s^2)^2}{f(s)}
    \end{pmatrix}
\end{equation}

Here we focus on the growth sector $\C{R}_{231}$ where $s^2 \gg s^3 \gg s^1$, and the overall volume grows asymptotically as $\C{V}_X \sim s^2 (s^3)^2$. The principal plane in this growth sector is spanned by the $s^0$, $s^2$, and $s^3$ directions. This constitutes an example where the principal plane exists but does not span all non-compact directions, as mentioned in Section \ref{sec:2}. The elementary EFT string limits that can be reached in this growth sector are $(e^0,0,0,0)$ and $(0,0,e^2,0)$ and the non-elementary limits are $(e^0,0,e^2,0)$, $(0,0,e^2,e^3)$, and $(e^0,0,e^2,e^3)$.

The light towers in this growth sector which will be relevant for our EFT string limits are the oscillator towers of the fundamental heterotic string and wrapped NS5-branes, the M-theory KK tower, and KK modes associated to the decompactification of curves, divisors, and the full Calabi-Yau. The $\alpha$-vectors of the relevant towers are summarized as
\begin{subequations}
\begin{align}
    &\vec{\alpha}_{\m{osc,F1}} = \left( \frac{1}{\sqrt{2}},0,0\right)\;,\quad \vec{\alpha}_{\m{osc,NS5,}e^2} = \left( 0,\frac{1}{\sqrt{2}},0\right)\;,\\
    &\vec{\alpha}_{\m{KK},[\omega_3]\cap [\omega_3]} = \left(\frac{1}{\sqrt{2}},\frac{1}{\sqrt{2}},0 \right)\;,\quad \vec{\alpha}_{\m{KK},[\omega_2]} = \left( \frac{1}{\sqrt{2}},0,\frac{1}{2} \right)\;,\\
    &\vec{\alpha}_{\m{KK,}X} = \left(\frac{1}{\sqrt{2}},\frac{1}{3\sqrt{2}},\frac{1}{3}\right)\;,\quad \vec{\alpha}_{\m{KK,M-th}} = \left( 0,\frac{1}{\sqrt{2}},1 \right)\;.
\end{align}
\end{subequations}
Depending on the limit taken, we either approach an emergent string limit or decompactify $1$, $2$, or $7$ dimensions. The relevant species scales are then $m_{\m{osc}}$, $m_{\m{pl},5}$, $m_{\m{pl,}6}$, and $m_{\m{pl},11}$. The species vectors can be computed as
\begin{subequations}
\begin{align}
    & \vec{\C{Z}}_{\m{osc,F1}} = \left( \frac{1}{\sqrt{2}},0,0 \right)\;,\quad \vec{\C{Z}}_{\m{osc,NS5},e^2} = \left(0,\frac{1}{\sqrt{2}},0\right)\;,\\
    &\vec{\C{Z}}_{\m{pl},5} = \left( 0,\frac{1}{3\sqrt{2}},\frac{1}{3} \right)\;,\quad \vec{\C{Z}}_{\m{pl},6} = \left( \frac{1}{2\sqrt{2}},\frac{1}{2\sqrt{2}},0 \right)\;,\quad \vec{\C{Z}}_{\m{pl},11} = \left( \frac{\sqrt{2}}{3},\frac{1}{3\sqrt{2}},\frac{1}{3} \right)\;.
\end{align}
\end{subequations}

The leading towers, $\alpha$-vector lengths squared, Laplacians, and the relevant dual theories for the various EFT string limits are given in Table \ref{tab:4dEFT_7880_towers}, with the corresponding data for the species scales given in Table \ref{tab:4dEFT_7880_species}.

\begin{figure}[t]
    \centering
    \vspace{-0.15\linewidth}\hspace{-0.02\linewidth}\includegraphics[width=1.05\linewidth]{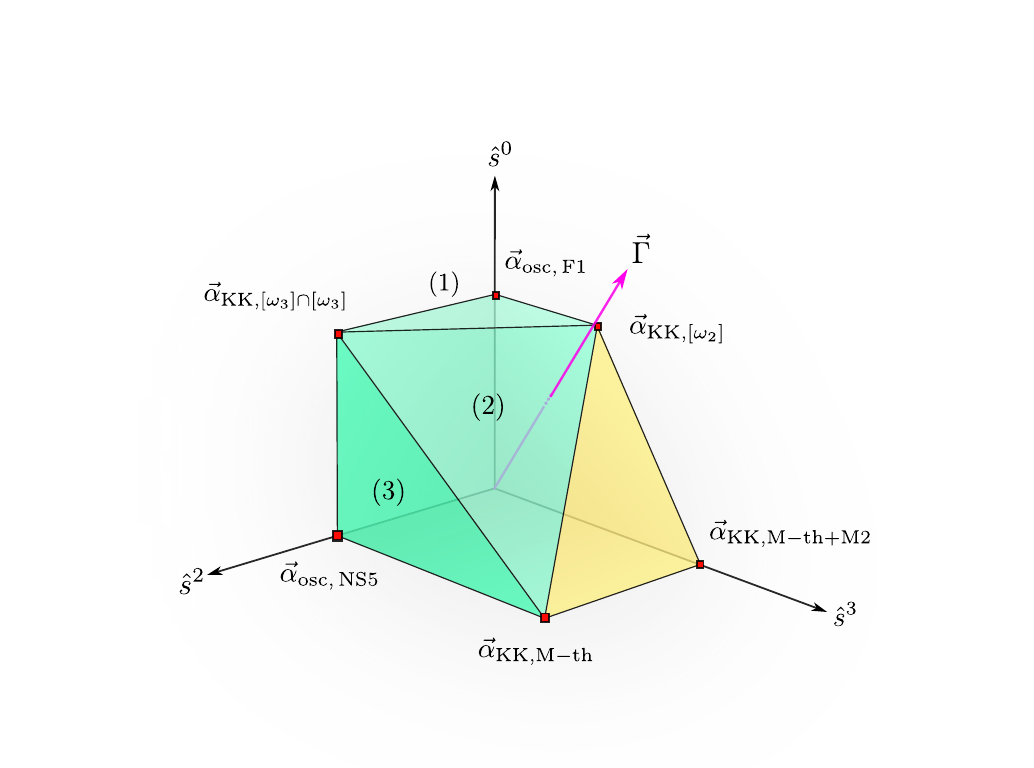}\vspace{-0.1\linewidth}
    \caption{$3$d principal plane and $\Gamma$-vector for the $R_{231}$ growth sector of CICY \#7880. The relevant $\Gamma$-vector in each duality frame is given by the projection of the overall $\vec \Gamma$ on the axes, and agrees with the $\beta$-vectors. The $\hat s$ indicate log-saxion-coordinates. Face (1) corresponds to an emergent HE string frame, while face (3) corresponds to a dual emergent HE string frame. The central face (2) corresponds to M-theory on $X \times \mathbb{S}^1 /\mathbb{Z}_2$. The yellow face on the right is not relevant for the $\C{R}_{231}$ growth sector, but corresponds to F-theory on $\mathbb{E} \hookrightarrow Y \rightarrow X$ and is included for completeness, matching \cite{Grieco:2025bjy}.}
    \label{fig:cicy}
\end{figure}
\begin{table}[t]
    \centering\renewcommand{\arraystretch}{0.97}
    \begin{tabular}{|c|c|c|c|c|}
    \hline
    Limit &  Tower & $|\vec\nabla \log m|^2$  & $\nabla^2 \log m$ & Emergent Dual Theory \\
    \hline\hline
    \multirow{4}{*}{$(e^0,0,0,0)$} & $m_{\m{osc,F1}}$ & $1/2$ & $1$ & \multirow{4}{*}{Emergent 
    HE} \\
    \cline{2-4}
     & $m_{\m{KK},[\omega_3]\cap [\omega_3]}$ & $1$ & $2$ & \\
    \cline{2-4}
     & $m_{\m{KK},[\omega_2]}$ & $3/4$ & $2$ & \\
    \cline{2-4}
     & $m_{\m{KK},X}$ & $2/3$ & $2$ & \\
    \hline
    \multirow{3}{*}{$(0,0,e^2,0)$} & $m_{\m{osc,NS5},e^2}$ & $1/2$ & $1$ & \multirow{3}{*}{Emergent 
    HE} \\
    \cline{2-4}
     & $m_{\m{KK},[\omega_3]\cap [\omega_3]}$ & $1$ & $2$ & \\
    \cline{2-4}
     & $m_{\m{KK,M-th}}$ & $3/2$ & $3$ & \\
    \hline
    $(e^0,0,e^2,0)$ & $m_{\m{KK,}[\omega_3]\cap [\omega_3]}$ & $1$ & $2$ & HE on K3 \\
    \hline
    $(0,0,e^2,e^3)$ & $m_{\m{KK,M-th}}$ & $3/2$ & $3$ & M-theory on $X$ \\
    \hline
    $(e^0,0,e^2,e^3)$ & $m_{\m{KK,M-th}}$ & $3/2$ & $3$ & $11$d M-theory \\
    \hline\end{tabular}\caption{
    Leading towers for the allowed EFT string limits of the $\C{R}_{231}$ growth sector of CICY \#7880.}
    \label{tab:4dEFT_7880_towers}
\end{table}

\begin{table}[t]
    \centering\renewcommand{\arraystretch}{0.97}
    \begin{tabular}{|c|c|c|c|c|}
    \hline
    Limit &  Species Scale & $|\vec\nabla \log \Lambda|^2$  & $\nabla^2 \log \Lambda$ & Emergent Dual Theory \\
    \hline\hline
    $(e^0,0,0,0)$ & $m_{\m{str}}$  & $1/2$ & $1$ & Emergent 
    HE\\
    \hline
    $(0,0,e^2,0)$ & $m_{\m{str}}$ & $1/2$ & $1$ & Emergent HE\\
    \hline
    $(e^0,0,e^2,0)$ &$m_{\m{pl},6}$ & $1/4$ & $1$ & HE on K3\\
    \hline
    $(0,0,e^2,e^3)$ &$m_{\m{pl},5}$ & $1/6$ & $1$ & M-theory on $X$\\
    \hline
    $(e^0,0,e^2,e^3)$ &$m_{\m{pl},11}$ & $7/18$ & $5/3$ & $11$d M-theory\\
    \hline
    \end{tabular}\caption{
    Species scales for the allowed EFT string limits of the $\C{R}_{231}$ growth sector of CICY \#7880.}
    \label{tab:4dEFT_7880_species}
\end{table}

We can compute the $\Gamma$-vector by taking the trace of the Christoffel symbol in coordinates corresponding to the log of the principal towers in the different accessible duality frames of the growth sector $\C{R}_{231}$. The possible choices of coordinates aligned with the log masses of principal towers in accessible duality frames are
\begin{align*}
    &(\log m_{\m{KK},[\omega_2]},\log m_{\m{KK},[\omega_3]\cap [\omega_3]},\log m_{\m{KK,M-th}})\;,\quad (\log m_{\m{KK},[\omega_2]},\log m_{\m{KK},[\omega_3]\cap [\omega_3]},\log m_{\m{osc,F1}})\;,\\
    &(\log m_{\m{osc,NS5},e^2},\log m_{\m{KK},[\omega_3]\cap [\omega_3]},\log m_{\m{KK,M-th}})\;.
\end{align*}
In a locally flat frame after converting back to saxionic coordinates, any choice of the above coordinates yields the $\Gamma$-vector
\begin{equation}
    (\vec{\Gamma})_{7880} = \left( \sqrt{2},\sqrt{2},2 \right)\;,
\end{equation}
which reproduces the correct $c_i$ values via $\vec{\Gamma} \cdot \vec{\alpha}_i$ and $\vec{\Gamma} \cdot \vec{\C{Z}}_i$ as given in Table \ref{tab:4dEFT_7880_towers} and \ref{tab:4dEFT_7880_species}. In log-saxion coordinates, $(\vec{\Gamma})_{7880}$ has the components
\begin{equation}
    (\vec{\Gamma})_{7880}^{\m{log-sax}} = (1,1,2)\;, 
\end{equation}
demonstrating that $(\vec{\Gamma})_{7880}$ is the sum of the relevant instanton $\beta$-vectors. We show the principal plane and $\Gamma$-vectors in Figure \ref{fig:cicy}.

\section{Conclusions}\label{sec:conclusions}

In this work, we have described the asymptotic structure of the moduli space geometry as predicted by the Emergent String Conjecture and deduced implications for the Laplacians of principal tower masses that become light in these asymptotic regimes. We provide an argument for, and evidence in support of, the hyperbolic towers condition of Propositions \ref{prop:hyperbolic towers} and \ref{prop:hyperbolic towers quantization}, where we argue that the masses of towers are eigenfunctions of the moduli space Laplacian with quantized eigenvalues. These eigenvalues are in turn related to the exponential decay of axion decay constants.

The Laplacian computations we perform can be understood \textit{locally} and are relatively computationally tractable. They provide thus a glimpse into the asymptotic geometry of the moduli space, even in setups where the global structure of the moduli space is more difficult to understand, as in the 5d $\C{N}=1$ examples we considered. That the Laplacians we compute satisfy the quantization rules of Proposition \ref{prop:hyperbolic towers quantization} can thus be seen as a window into the spectrum of instantons with small exponentiated action $e^{-S_{\m{inst}}}$ in the asymptotic regime.

Our hyperbolic towers proposal and asymptotic fibration structure are presented very much in the spirit of the original geometric conjectures of Ooguri and Vafa in \cite{Ooguri:2006in}. In particular, it was conjectured that any moduli space $\C{M}$ have no 1-cycles in $H^1(\C{M}, \B{Z})$ with a representative of positive minimum length, and that $\C{M}$ also have nonpositive asymptotic scalar curvature.\footnote{The second conjecture, though false as literally stated \cite{Trenner:2010,Marchesano:2023thx}, is still true in spirit in the sense that \textit{almost every} asymptotic limit has nonpositive asymptotic curvature in all known examples. See \cite{Raman:2024fcv} for further details and an alternative characterization.} The asymptotic fibration structure of Proposition \ref{prop:nilmanifold} is in some sense an asymptotic manifestation of the expectation of the $H^1$-conjecture of \cite{Ooguri:2006in}, where every axionic circle shrinks exponentially at infinite distance. The nonnegativity of the values $\nabla^2 \log m $ are then a direct consequence of the asymptotic fibration structure of Proposition \ref{prop:nilmanifold} and the compactifiability condition in \cite{Delgado:2024skw}, and the ``hyperbolicity'' of the towers is closely related to the nonpositive asymptotic scalar curvature in these limits. Our work thus shows that principles analogous in spirit to the $H^1$- and non-positive curvature conjectures can in fact be argued directly from the ESC. We hope that a full treatment of asymptotic moduli space geometry provides a general and unifying framework under which all of the geometric conjectures of \cite{Ooguri:2006in} can be systematically presented.

One important gap in our work concerns the appearance of asymptotic limits in theories with $Q \leq 8$ supercharges where a \textit{rigid field theory} (RFT) sector decoupled from gravity emerges at infinite distance \cite{Marchesano:2023thx,Castellano:2024gwi,Marchesano:2024tod,Blanco:2025qom,Castellano:2026bnx,Aoufia:2026mqb}. In such limits, the moduli space scalar curvature diverges to $+\infty$, and we saw in some 5d $\C{N}=1$ examples in Section \ref{sec:8charges} that hyperbolicity also breaks down in these regimes. An important future direction is to understand the implications that these RFT sectors have on the Laplacians of tower masses and the detailed asymptotic geometry. 

It is also interesting to understand whether our Assumptions \ref{ass:ESC}, \ref{ass:perturbative-esc}, and \ref{ass:principal} can be relaxed. Indeed, all three of these assumptions have nontrivial counterexamples within the string landscape. However, in 6d $\C{N}=(1, 0)$ theories, we found that the hyperbolic towers quantization condition of Proposition \ref{prop:hyperbolic towers quantization} is still satisfied despite the failure of Assumptions \ref{ass:ESC} and \ref{ass:perturbative-esc}. It would be interesting to study whether the hyperbolic towers property is true more generally and if it can be proven with less stringent assumptions than we have employed. For example, in Type I$'$ string theory, where there is warping and sliding \cite{Etheredge:2023odp} of the Type I$'$ KK-mode (and Assumption \ref{ass:ESC} fails), it would be interesting to calculate whether this KK-mode is actually an eigenfunction of the full moduli-space Laplacian.

Another avenue that we think would be worth exploring is the possible connection between these Laplacian equations with  ``moduli-space quantum mechanics'', introduced in \cite{Anchordoqui:2025izb,Anchordoqui:2026nit}. The formulas in eq. \ref{eq:laplaceeq} especially are very suggestive of eigenvalue equations for a Hamiltonian for quantum mechanics probing a moduli space. Relatedly, it would be interesting to see what objects remain moduli-space-Laplacian eigenfunctions in the bulk of moduli space, since the properties we have presented in this paper only hold true asymptotically. 

It would also be interesting to investigate potential phenomenological implications. Duality groups of consistent EFTs coupled to gravity have also been explored in BSM physics. Modular invariant $\C{N}=1$ SUSY models exploit the exponential suppression of modular forms in asymptotic regimes of moduli space to generate natural hierarchies in quark and lepton Yukawas \cite{Feruglio:2017spp,Almumin:2022rml,Kobayashi:2023zzc,Ding:2020zxw,Nilles:2023shk,Ding:2023htn,Nilles:2024iqp,Liu:2021gwa}. This structure has top-down support as heterotic orbifold compactifications have long been known to yield Yukawa couplings that are modular forms of congruence subgroups $\Gamma(N)$ of $\m{SL}(2,\mathbb{Z})$ \cite{Ferrara:1989qb,Lauer:1989ax,Lerche:1989cs}, and recent work has connected the duality groups of the resulting $4$d $\C{N}=1$ theories directly to the superpotentials of modular flavor models \cite{Li:2025bsr,Nilles:2020tdp,Nilles:2020gvu,Baur:2024qzo}. Extending this framework to non-supersymmetric theories sacrifices holomorphicity. Yukawa couplings are no longer forced to be modular forms, but are instead Maa\ss\, forms provided one imposes by hand that they satisfy certain Laplace equations \cite{Qu:2024rns}. While the ESC-based arguments of this work apply to tower mass scales rather than couplings of matter fields, the underlying mechanism is not unique to towers. In the supersymmetric case the K\"ahler-dressed Yukawas $e^{K/2}Y$ of heterotic orbifolds are automatically eigenfunctions of the weight-$k$ hyperbolic Laplacian, holomorphic modular forms sitting at special points of the Maa\ss\, spectrum. It would be interesting to understand whether a similar analysis to ours could be used to constrain physical Yukawa couplings, providing support from the top-down for the Laplace conditions imposed in \cite{Qu:2024rns}.

There are also aspects of moduli space geometry that are not obviously consequences of the ESC itself. For example, the existence of an asymptotic fibration structure roughly follows from the ESC, but the ESC does not predict the exact geometry of the nilmanifold spanned by the axion directions, nor its rank. We expect that the rank of the nilmanifold can be constrained in supersymmetric examples by the large body of work on finiteness and the string lamppost principle \cite{Adams:2010zy, Kim:2019vuc, Kim:2019ths, Montero:2020icj,Cvetic:2020kuw, Katz:2020ewz, Hamada:2021bbz, Bedroya:2021fbu, Tarazi:2021duw, Martucci:2022krl,Bedroya:2023tch, Delgado:2024skw, Baykara:2025gcc}, and it would be interesting to deduce and interpret the corresponding bounds on our Laplacian values. The internal fibration structure of the nilmanifold itself is also an interesting matter. Nontrivial couplings between the axion directions in the lower-dimensional metric arise from triple Chern-Simons interactions of the form $A \wedge F \wedge F$ in the higher-dimensional theory in decompactification frames. Thus, one might surmise that nontrivial constraints on the higher-dimensional topological couplings could be argued if one imposed constraints on the geometry of the lower-dimensional moduli space. Such constraints would amount to a more stringent Swampland prediction for the lower-dimensional moduli space than even the ESC could furnish, and it would be interesting to see to what extent this can be developed. We plan to study this in future work \cite{ricci:toappear}.

\section*{Acknowledgments}
We are grateful for useful discussions with Alberto Castellano, Alessandra Grieco, Ben Heidenreich, Luis Ibáñez, Xiang-Gan Liu, Dieter L\"ust, Fernando Marchesano, Luca Melotti, Carmine Montella, Michael Ratz, Chao-Hsiang Sheu, Cumrun Vafa, and Irene Valenzuela.

B.F. was supported by a Juan de la Cierva contract (JDC2023-050850-I) from Spain’s Ministry of Science, Innovation and Universities. The work of C.A and B.F. is also supported through the grants CEX2020-001007-S, PID2021-123017NB-I00 and PID2024-156043NB-I00, funded by MCIN/AEI/10.13039/501100011033, and ERDF, EU.
A.S. was supported by the National Science Foundation Graduate Research Fellowship Program. This material is based upon work supported by the National Science Foundation Graduate Research Fellowship Program under Grant No. DGE-2235784. C.A. would like to thank the II. Institute for Theoretical Physics at the University of Hamburg and the DESY Theory Group for hospitality during the different stages of this work. S.R. is supported in part by a grant from the Simons Foundation (602883, CV) and the DellaPietra Foundation. Any opinions, findings, and conclusions or recommendations expressed in this material are those of the authors and do not necessarily reflect the views of the National Science Foundation. 

Parts of this work were also conducted at the Simons Summer Physics Workshop in 2024 and 2025, String Pheno 2025 and 2026, and Swamplandia 2025 and 2026.

\appendix 

\section{Details on the asymptotic metric in decompactification limits}\label{app:A}
In this appendix, we review a detailed derivation of the form of the asymptotic metric, eq. \eqref{eq:asympmetric}, in decompactification limits by dimensional reduction.  Following the convention of the main text, we consider a $k=D-d$-dimensional manifold $M^k$ over which we compactify a $D$-dimensional theory. We denote its volume $\C{V}_k$, the corresponding radion $R=\C{V}_k^{1/k}$, and the canonically normalized one $\rho$.

Given that the internal manifold $M^k$ satisfies Einstein's equations in vacuum due to Assumption \ref{ass:ESC}, the radial ``breathing mode'' $R$ corresponds to an exactly massless deformation, and it is this deformation which has non-trivial overlap with the direction 
\begin{equation} 
\hat\phi = \frac{d}{dt} \gamma_v(q) \bigg |_{t = t_0} \, . 
\end{equation}
The radial mode can be expressed in terms of a canonically normalized modulus $\rho$ in the $d$-dimensional Einstein frame, which is indeed the variable with respect to which the ESC scaling relations are specified. Explicitly, the canonically normalized radion $\rho$ is related to the internal volume $\mathcal V_k$ by 
\begin{equation}
\rho = \sqrt{\frac{D-2}{k(d-2)}}\,\log \mathcal V_k\, .
\end{equation} 
Thus the infinite-volume decompactification limit $\mathcal V_k\to\infty$ corresponds to $\rho\to+\infty$. By the Sharpened DC, within any decompactification frame over $M_k$, any infinite-distance geodesic along a direction $\hat\phi$ has a nontrivial overlap with the canonically normalized radion direction $\rho$ given by 
\[ \hat\phi \cdot \rho \geq \frac{1}{\sqrt{d-2}}, \]
just as we argued in Sec. \ref{ssec:decomp}. 

For a given moduli direction $\hat\phi$, we choose coordinates on the moduli space such that the metric is given asymptotically by \begin{equation}
    ds^2 = \d \phi^2 + \cdots \, .
\end{equation}
We wish now to understand the ``$\cdots$''; that is, we would like to describe rest of the massless deformations. There are two possible sources of such moduli: 
\begin{enumerate}

    \item Zero-energy deformations at fixed volume of the internal manifold $M^k$. These correspond to compact ``graviphoton-type'' directions and non-compact directions in $T_q \C{M}$ orthogonal to $w$, and their metric coefficients carry no radion dependence. 
    \item The direct KK reduction of massless fields in the $D$-dimensional theory over $M^k$. These can either be higher-dimensional moduli (i.e.\ 0-form gauge fields) or higher-dimensional $p$-form gauge fields KK-reduced over $p$-cycles in $M^k$. These correspond to compact directions in $T_q \C{M}$ orthogonal to $w$. 
\end{enumerate}
We now address them in turn.
\paragraph{1. Shape deformations.} As it turns out, the deformations in the first case are always orthogonal to $w$ with respect to the moduli space metric at $q$. Moreover, the metric coefficients for the additional deformations, in the lower-dimensional Einstein frame, carry no dependence on the radion. Though this fact is standard, we record a simple derivation below. To see that the fixed-volume deformations are orthogonal to the radion, we note that the radion modulus corresponds to the pure trace deformation of the internal metric, while the remaining shape moduli are fixed-volume, hence traceless, deformations. Writing
\begin{equation}
g_{mn}=R^2\hat g_{mn}\, ,
\qquad
\int d^k y\,\sqrt{\hat g}=1\, ,
\end{equation}
one has
\begin{equation}
\delta g_{mn}
=
2\,\delta(\log R)\,g_{mn}
+
R^2\delta \hat g_{mn}\, .
\end{equation}
The first term is the radial deformation. The second term preserves the internal volume, so
\begin{equation}
\delta \log \mathcal V
=
\frac12\int d^k y\,\sqrt g\, g^{mn}\delta g_{mn}
=
0\, ,
\end{equation}
which means that the shape deformations obey the tracelessness condition $g^{mn}\delta_s g_{mn}=0$, after projecting out the overall volume mode.

The moduli-space metric inherited from the higher-dimensional Einstein-Hilbert term is a DeWitt-type inner product on metric deformations,
\begin{equation}
\langle \delta_1 g,\delta_2 g\rangle
\sim
\int d^k y\,\sqrt g\,
\left(
g^{mp}g^{nq}\delta_1 g_{mn}\delta_2 g_{pq}
-
c\,(g^{mn}\delta_1 g_{mn})(g^{pq}\delta_2 g_{pq})
\right),
\end{equation}
for a dimension-dependent constant $c$. Pairing a radial deformation $\delta_R g_{mn}\propto g_{mn}$ with a fixed-volume deformation $\delta_s g_{mn}$ gives
$g^{mp}g^{nq}g_{mn}\delta_s g_{pq} =0 $ and the trace-squared term vanishes for the same reason. Hence
\begin{equation}
\langle \delta_R g,\delta_s g\rangle=0.
\end{equation}
Thus, in coordinates adapted to the decomposition into the total internal volume and unit-volume shapes, the radion direction is orthogonal to all other geometric moduli.

We now show that the coefficient of the moduli-space metric in the shape-deformation directions is independent of the radion in the $d$-dimensional Einstein frame. The shape moduli $u^a$ enter only through $\hat g_{mn}$. Thus
\begin{equation}
\partial_\mu g_{mn}
=
R^2\partial_a\hat g_{mn}\,\partial_\mu u^a \, ,
\qquad
g^{mn}=R^{-2}\hat g^{mn}\, ,
\end{equation}
so the powers of $R$ cancel inside the DeWitt contraction:
\begin{equation}
g^{mp}g^{nq}\partial_\mu g_{mn}\partial^\mu g_{pq}
=
\hat g^{mp}\hat g^{nq}
\partial_a\hat g_{mn}\partial_b\hat g_{pq}\,
\partial_\mu u^a\partial^\mu u^b \, .
\end{equation}
The only remaining factor before Weyl rescaling is the total internal volume, $\mathcal V_k\sim R^k$, multiplying the lower-dimensional action. Passing from the $D$-dimensional Einstein frame with metric $g_{MN}^{(D)}$ to the $d$-dimensional Einstein frame with metric $g_{\u\v}^{(d)}$ removes precisely this volume factor:
\begin{equation}
g_{\mu\nu}^{(d)}
=
\mathcal V_k^{2/(d-2)}g_{\mu\nu}^{(D)}\, ,
\qquad
\sqrt{-g_D}\,\mathcal V_k\,g_D^{\mu\nu}
=
\sqrt{-g_d}\,g_d^{\mu\nu}\, .
\end{equation}
Therefore the shape-moduli kinetic term takes the form
\begin{equation}
\mathcal L_{\rm kin}
\supset
-\frac12\,G^{\rm shape}_{ab}(u)\,
\partial_\mu u^a\partial^\mu u^b,
\end{equation}
with no radion-dependent prefactor.

\paragraph{2. KK-reduced massless modes.} We have thus shown that the moduli deformations associated to the radion and the shape of $M^k$ essentially factorize as a direct product, so the internal geometry in the shape directions of $T_q \C{M}$ does not speak at all to the radion direction. This leaves the directions arising from the dimensional reduction of higher-dimensional moduli and $p$-form gauge fields.

For a $p$-form gauge field $C_p$ in the higher-dimensional theory, its kinetic term is given as
\begin{equation}
S_D\supset -\int \frac{1}{2  g(x)^2} F_{p+1}\wedge *_D F_{p+1},
\end{equation}
where $F_{p+1} = \d C_p + \cdots$ is a gauge-invariant field strength constructed from $C_p$ and $g(x)$ is the coupling for $C_p$, possibly depending on the higher-dimensional moduli $x^l$. Suppose now that we KK-reduce $C_p$ over a $p$-cycle $[A] \in H_p(M^k, \B{Z})$ in $M^k$. This requires a decomposition 
\begin{equation} C_p = c_p \wedge \omega_p  \end{equation}
for $\omega_p$ a harmonic $p$-form representing a class $[A] \in H^p(M^k, \B{Z})$ Poincaré-dual to $A$, which by an abuse of notation we will refer to by the same symbol. 

We now wish to construct the lower-dimensional kinetic term for the modulus $c_p$. Using the product decomposition of the Hodge star, dimensional reduction gives
\begin{equation}
S_d\supset
-\frac{1}{2  g(x)^2}
\left(\int_{M^k}\omega_p\wedge *_k\omega_p\right)
\int dc_p \wedge *_d dc_p \, .
\end{equation}
The field $c_p$ is periodic owing to the gauge invariance of the higher-dimensional $p$-form $U(1)$ gauge field. Thus, it can be thought of as an \textit{axionic} direction in the lower-dimensional moduli space. We will now show that the kinetic term for $c_p$, and thus the axion periodicity/inverse decay constant for $c_p$, depends on the radion $\phi$. For an isotropic internal metric $g_{mn}=R^2\hat g_{mn}$, with $\omega_p$ held fixed as a cohomology representative, one has $\sqrt{g_k}\sim \C{V}_k \sim R^k$ and $p+1$ inverse metrics in the contraction contribute $R^{-2p}$. Hence
\begin{equation}
\int_{M^k}\omega_p\wedge *_k\omega_p
\sim
\frac1{p!}\int d^k y\,\sqrt{g_k}\,
\omega_{m_1\cdots m_p}\omega^{m_1\cdots m_p}
\sim
R^{k-2p} \, .
\end{equation}

This is the scalar kinetic coefficient in the dimensionally reduced action in the $D$-dimensional Einstein frame. Pass to $d$-dimensional Einstein frame, the Weyl rescaling gives
$ \sqrt{-g_D}\,g_D^{\mu\nu} = \mathcal V_k^{-1}\sqrt{-g_d}\,g_d^{\mu\nu}$.
Therefore the Einstein-frame axion kinetic term is
\begin{equation}
    \C{S}_d \supset - \int \sqrt{-g_d} \frac{1}{2 g(x)^2 R^{2p}} (\p c_p)^2 
\end{equation}
and the decay constant $ f_{c_p}^2 \sim R^{k-2p}\mathcal V_k^{-1} \sim R^{-2p}$.
Equivalently, using $R \sim e^{\g \phi}$, one obtains
\begin{equation}\label{eq:fcp_appendix}
f_{c_p}^2 \sim e^{-2p\g \phi}
\end{equation}
so the axion metric decreases exponentially in the decompactification limit. Note that eq. \eqref{eq:fcp} expresses the radion dependence of the axion kinetic coefficient in arbitrary, possibly non-isotropic infinite-distance limits; in the non-isotropic case, there will of course be additional dependencies. In all cases, the Sharpened DC tells us that $\gamma \geq 1/\sqrt{d-2}$. For the case where the radion direction $\rho$ and the direction of the limit $\phi$ are aligned, the constant $\gamma$ is constrained by the ESC to take the value
\begin{equation}
     \g = \sfc{d-2}{k(D-2)} = \sfc{d-2}{(D-d)(D-2)}\, .
\end{equation}
All together, we thus get that the metric on moduli space along a decompactification limit can in general be written as
\begin{equation}\label{eq:asympmetric_appendix}
    ds^2 \simeq \d\phi^2 + \sum_{0 < p \leq k} \sum_{\omega_i \in H^p(M^k, \B{R})} \frac{1}{g_i(x)^2}  e^{-2p \g_{p, i} \rho}  \d \theta_{p,i}^2 + g_{mn}(x) \d x^m \d x^n \, ,
\end{equation} 
where the $x$ parametrize the shape moduli and other moduli which decouple from the radion direction. This reproduces the content of eq. \eqref{eq:fullmetric_2} as we asserted in Section \ref{ssec:decomp}.

\bibliographystyle{JHEP}
\bibliography{references}

@article{Hamada:2021yxy,
    author = "Hamada, Yuta and Montero, Miguel and Vafa, Cumrun and Valenzuela, Irene",
    title = "{Finiteness and the swampland}",
    eprint = "2111.00015",
    archivePrefix = "arXiv",
    primaryClass = "hep-th",
    doi = "10.1088/1751-8121/ac6404",
    journal = "J. Phys. A",
    volume = "55",
    number = "22",
    pages = "224005",
    year = "2022"
}

@inproceedings{Shenker:1990uf,
    author = "Shenker, Stephen H.",
    title = "{The Strength of nonperturbative effects in string theory}",
    booktitle = "{Cargese Study Institute: Random Surfaces, Quantum Gravity and Strings}",
    reportNumber = "RU-90-47, RU-90-047",
    pages = "809--819",
    month = "8",
    year = "1990"
}

@article{Andrianopoli:1996vr,
    author = "Andrianopoli, L. and Bertolini, M. and Ceresole, Anna and D'Auria, R. and Ferrara, S. and Fre', P.",
    title = "{General matter coupled N=2 supergravity}",
    eprint = "hep-th/9603004",
    archivePrefix = "arXiv",
    reportNumber = "POLFIS-TH-02-96, CERN-TH-95-350, UCLA-96-TEP-3, NSF-ITP-96-11",
    doi = "10.1016/0550-3213(96)00344-6",
    journal = "Nucl. Phys. B",
    volume = "476",
    pages = "397--417",
    year = "1996"
}

@article{Antoniadis:2003sw,
    author = "Antoniadis, Ignatios and Minasian, Ruben and Theisen, Stefan and Vanhove, Pierre",
    title = "{String loop corrections to the universal hypermultiplet}",
    eprint = "hep-th/0307268",
    archivePrefix = "arXiv",
    reportNumber = "AEI-2003-052, CPHT-RR-030-0603, CERN-TH-2003-139, SACLAY-SPHT-T03-088",
    doi = "10.1088/0264-9381/20/23/009",
    journal = "Class. Quant. Grav.",
    volume = "20",
    pages = "5079--5102",
    year = "2003"
}

@article{Antoniadis:1997eg,
    author = "Antoniadis, Ignatios and Ferrara, S. and Minasian, R. and Narain, K. S.",
    title = "{R**4 couplings in M and type II theories on Calabi-Yau spaces}",
    eprint = "hep-th/9707013",
    archivePrefix = "arXiv",
    reportNumber = "CERN-TH-97-094, CERN-TH-97-94, CPTH-S512-0697",
    doi = "10.1016/S0550-3213(97)00572-5",
    journal = "Nucl. Phys. B",
    volume = "507",
    pages = "571--588",
    year = "1997"
}

@article{Cecotti:1988qn,
    author = "Cecotti, S. and Ferrara, S. and Girardello, L.",
    title = "{Geometry of Type II Superstrings and the Moduli of Superconformal Field Theories}",
    reportNumber = "CERN-TH-5080-88, UCLA-88-TEP-21",
    doi = "10.1142/S0217751X89000972",
    journal = "Int. J. Mod. Phys. A",
    volume = "4",
    pages = "2475",
    year = "1989"
}

@article{Ruehle:2024ufw,
    author = "Ruehle, Fabian and Sung, Benjamin",
    title = "{Attractors, Geodesics, and the Geometry of Moduli Spaces}",
    eprint = "2408.00830",
    archivePrefix = "arXiv",
    primaryClass = "hep-th",
    month = "8",
    year = "2024"
}

@article{Ferrara:1989ik,
    author = "Ferrara, S. and Sabharwal, S.",
    title = "{Quaternionic Manifolds for Type II Superstring Vacua of Calabi-Yau Spaces}",
    reportNumber = "UCLA/89/TEP/28, CERN-TH-5436/89",
    doi = "10.1016/0550-3213(90)90097-W",
    journal = "Nucl. Phys. B",
    volume = "332",
    pages = "317--332",
    year = "1990"
}

@article{Ambrosetti:2010tu,
    author = "Ambrosetti, Nicola and Antoniadis, Ignatios and Derendinger, Jean-Pierre and Tziveloglou, Pantelis",
    title = "{The Hypermultiplet with Heisenberg Isometry in N=2 Global and Local Supersymmetry}",
    eprint = "1005.0323",
    archivePrefix = "arXiv",
    primaryClass = "hep-th",
    reportNumber = "CERN-PH-TH-2010-085",
    doi = "10.1007/JHEP06(2011)139",
    journal = "JHEP",
    volume = "06",
    pages = "139",
    year = "2011"
}

@article{Morrison:1996pp,
    author = "Morrison, David R. and Vafa, Cumrun",
    title = "{Compactifications of F theory on Calabi-Yau threefolds. 2.}",
    eprint = "hep-th/9603161",
    archivePrefix = "arXiv",
    reportNumber = "DUKE-TH-96-107, HUTP-96-A012",
    doi = "10.1016/0550-3213(96)00369-0",
    journal = "Nucl. Phys. B",
    volume = "476",
    pages = "437--469",
    year = "1996"
}

@article{Nishino:1997ff,
    author = "Nishino, Hitoshi and Sezgin, Ergin",
    title = "{New couplings of six-dimensional supergravity}",
    eprint = "hep-th/9703075",
    archivePrefix = "arXiv",
    reportNumber = "UMDEPP-97-086, CTP-TAMU-14-97",
    doi = "10.1016/S0550-3213(97)00357-X",
    journal = "Nucl. Phys. B",
    volume = "505",
    pages = "497--516",
    year = "1997"
}

@article{Vafa:1996xn,
    author = "Vafa, Cumrun",
    title = "{Evidence for F theory}",
    eprint = "hep-th/9602022",
    archivePrefix = "arXiv",
    reportNumber = "HUTP-96-A004",
    doi = "10.1016/0550-3213(96)00172-1",
    journal = "Nucl. Phys. B",
    volume = "469",
    pages = "403--418",
    year = "1996"
}

@article{Morrison:1996na,
    author = "Morrison, David R. and Vafa, Cumrun",
    title = "{Compactifications of F theory on Calabi-Yau threefolds. 1}",
    eprint = "hep-th/9602114",
    archivePrefix = "arXiv",
    reportNumber = "DUKE-TH-96-106, HUTP-96-A007",
    doi = "10.1016/0550-3213(96)00242-8",
    journal = "Nucl. Phys. B",
    volume = "473",
    pages = "74--92",
    year = "1996"
}

@article{Alvarez-Garcia:2026uaq,
    author = "{\'A}lvarez-Garc{\'\i}a, Rafael and Ruehle, Fabian",
    title = "{Harmonic Analysis of the Instanton Prepotential}",
    eprint = "2604.08725",
    archivePrefix = "arXiv",
    primaryClass = "hep-th",
    month = "4",
    year = "2026"
}

@article{Alvarez-Garcia:2026vwq,
    author = "{\'A}lvarez-Garc{\'\i}a, Rafael and Ruehle, Fabian",
    title = "{Kaleidoscopes, Waves and the Prepotential}",
    eprint = "2606.05280",
    archivePrefix = "arXiv",
    primaryClass = "hep-th",
    month = "6",
    year = "2026"
}

@article{Grimm:2025lip,
    author = "Grimm, Thomas W. and Prieto, David and van Vliet, Mick",
    title = "{Tame embeddings, volume growth, and complexity of moduli spaces}",
    eprint = "2503.15601",
    archivePrefix = "arXiv",
    primaryClass = "hep-th",
    doi = "10.1103/d51c-j1s9",
    journal = "Phys. Rev. D",
    volume = "112",
    number = "10",
    pages = "106015",
    year = "2025"
}

@article{Dixon:1990pc,
    author = "Dixon, Lance J. and Kaplunovsky, Vadim and Louis, Jan",
    title = "{Moduli dependence of string loop corrections to gauge coupling constants}",
    reportNumber = "SLAC-PUB-5138, UTTG-36-89",
    doi = "10.1016/0550-3213(91)90490-O",
    journal = "Nucl. Phys. B",
    volume = "355",
    pages = "649--688",
    year = "1991"
}

@article{Kaplunovsky:1992vs,
    author = "Kaplunovsky, Vadim S.",
    title = "{One loop threshold effects in string unification}",
    eprint = "hep-th/9205070",
    archivePrefix = "arXiv",
    reportNumber = "ITP-838-STANFORD-REV, ITP-838-STANFORD",
    month = "5",
    year = "1992"
}

@article{Kaufmann:2026tsy,
    author = "Kaufmann, Lukas and Weigand, Timo and Wiesner, Max",
    title = "{On Quantum Obstructions in Type IIA Orientifolds}",
    eprint = "2604.25988",
    archivePrefix = "arXiv",
    primaryClass = "hep-th",
    month = "4",
    year = "2026"
}

@article{Kaufmann:2026mha,
    author = "Kaufmann, Lukas and Monnee, Jeroen and Weigand, Timo and Wiesner, Max",
    title = {{Quantum obstructions for $N=1$ infinite distance limits -- Part II: K{\"a}hler obstructions}},
    eprint = "2603.13470",
    archivePrefix = "arXiv",
    primaryClass = "hep-th",
    month = "3",
    year = "2026"
}

@article{Kaufmann:2026fli,
    author = "Kaufmann, Lukas and Monnee, Jeroen and Weigand, Timo and Wiesner, Max",
    title = "{Quantum obstructions for $N=1$ infinite distance limits -- Part I: $g_s$ obstructions}",
    eprint = "2603.12315",
    archivePrefix = "arXiv",
    primaryClass = "hep-th",
    month = "3",
    year = "2026"
}

@article{Witten:1996md,
    author = "Witten, Edward",
    title = "{On flux quantization in $M$-theory and the effective action}",
    eprint = "hep-th/9609122",
    archivePrefix = "arXiv",
    reportNumber = "IASSNS-HEP-96-96",
    doi = "10.1016/S0393-0440(96)00042-3",
    journal = "J. Geom. Phys.",
    volume = "22",
    pages = "1--13",
    year = "1997"
}

@article{Corvilain:2020tfb,
    author = "Corvilain, Pierre",
    title = "{6d $ \mathcal{N} $ = (1, 0) anomalies on S$^{1}$ and F-theory implications}",
    eprint = "2005.12935",
    archivePrefix = "arXiv",
    primaryClass = "hep-th",
    doi = "10.1007/JHEP08(2020)133",
    journal = "JHEP",
    volume = "08",
    pages = "133",
    year = "2020"
}

@article{DiUbaldo:2026rly,
    author = "Di Ubaldo, Gabriele and Iliesiu, Luca V. and Lin, Henry W. and Yan, Cynthia",
    title = "{Positivity of the gravitational path integral implies the axionic weak gravity conjecture}",
    eprint = "2605.05305",
    archivePrefix = "arXiv",
    primaryClass = "hep-th",
    reportNumber = "RIKEN-iTHEMS-Report-26",
    month = "5",
    year = "2026"
}

@article{Maldacena:2026jqd,
    author = "Maldacena, Juan and Maloney, Alexander and McPeak, Brian",
    title = "{Wormholes and the imaginary distance bound}",
    eprint = "2605.05336",
    archivePrefix = "arXiv",
    primaryClass = "hep-th",
    month = "5",
    year = "2026"
}

@article{Etheredge:2026rio,
    author = "Etheredge, Muldrow and Reece, Matthew and Rudelius, Tom and Tudball, Christopher",
    title = "{Sharpening the Supersymmetric Axion Weak Gravity Conjecture}",
    eprint = "2605.22912",
    archivePrefix = "arXiv",
    primaryClass = "hep-th",
    reportNumber = "MPP-2026-93",
    month = "5",
    year = "2026"
}

@article{Reece:2026hmp,
    author = "Reece, Matthew and Rudelius, Tom and Tudball, Christopher",
    title = "{Quantum Gravity Cutoff from Axions: A Type IIB Landscape Study}",
    eprint = "2606.21650",
    archivePrefix = "arXiv",
    primaryClass = "hep-th",
    month = "6",
    year = "2026"
}

@article{Artime:2026kfq,
    author = "Artime, Manuel and Blumenhagen, Ralph and Leivadaros, Panagiotis",
    title = "{Taxonomy of Instanton Corrections in Infinite Distance Limits}",
    eprint = "2605.03005",
    archivePrefix = "arXiv",
    primaryClass = "hep-th",
    reportNumber = "MPP-2026-79",
    month = "5",
    year = "2026"
}

@article{Monnee:2025msf,
    author = "Monnee, Jeroen and Weigand, Timo and Wiesner, Max",
    title = "{K points and type IIB/heterotic duality with NS5-branes}",
    eprint = "2510.02435",
    archivePrefix = "arXiv",
    primaryClass = "hep-th",
    reportNumber = "ZMP-HH-25/18",
    doi = "10.1103/kcbz-jf16",
    journal = "Phys. Rev. D",
    volume = "113",
    number = "6",
    pages = "066002",
    year = "2026"
}

@article{Monnee:2025ynn,
    author = "Monnee, Jeroen and Weigand, Timo and Wiesner, Max",
    title = "{Physics and geometry of complex structure limits in type IIB Calabi-Yau compactifications}",
    eprint = "2509.07056",
    archivePrefix = "arXiv",
    primaryClass = "hep-th",
    doi = "10.1007/JHEP03(2026)063",
    journal = "JHEP",
    volume = "03",
    pages = "063",
    year = "2026"
}

@article{Baykara:2025gcc,
    author = "Baykara, Zihni Kaan and Dierigl, Markus and Kim, Hee-Cheol and Vafa, Cumrun and Xu, Kai",
    title = "{Bounds on discrete gauge symmetries in supergravity}",
    eprint = "2511.09613",
    archivePrefix = "arXiv",
    primaryClass = "hep-th",
    reportNumber = "CERN-TH-2025-233",
    doi = "10.1007/JHEP03(2026)171",
    journal = "JHEP",
    volume = "03",
    pages = "171",
    year = "2026"
}

@article{Martucci:2022krl,
    author = "Martucci, Luca and Risso, Nicolo and Weigand, Timo",
    title = "{Quantum gravity bounds on $ \mathcal{N} $ = 1 effective theories in four dimensions}",
    eprint = "2210.10797",
    archivePrefix = "arXiv",
    primaryClass = "hep-th",
    doi = "10.1007/JHEP03(2023)197",
    journal = "JHEP",
    volume = "03",
    pages = "197",
    year = "2023"
}

@article{Tarazi:2021duw,
    author = "Tarazi, Houri-Christina and Vafa, Cumrun",
    title = "{On the finiteness of 6d supergravity Landscape}",
    eprint = "2106.10839",
    archivePrefix = "arXiv",
    primaryClass = "hep-th",
    doi = "10.1007/JHEP05(2026)007",
    journal = "JHEP",
    volume = "05",
    pages = "007",
    year = "2026"
}

@article{Bedroya:2021fbu,
    author = "Bedroya, Alek and Hamada, Yuta and Montero, Miguel and Vafa, Cumrun",
    title = "{Compactness of brane moduli and the String Lamppost Principle in d {\ensuremath{>}} 6}",
    eprint = "2110.10157",
    archivePrefix = "arXiv",
    primaryClass = "hep-th",
    doi = "10.1007/JHEP02(2022)082",
    journal = "JHEP",
    volume = "02",
    pages = "082",
    year = "2022"
}

@article{Hamada:2021bbz,
    author = "Hamada, Yuta and Vafa, Cumrun",
    title = "{8d supergravity, reconstruction of internal geometry and the Swampland}",
    eprint = "2104.05724",
    archivePrefix = "arXiv",
    primaryClass = "hep-th",
    doi = "10.1007/JHEP06(2021)178",
    journal = "JHEP",
    volume = "06",
    pages = "178",
    year = "2021"
}

@article{Cvetic:2020kuw,
    author = "Cvetic, Mirjam and Dierigl, Markus and Lin, Ling and Zhang, Hao Y.",
    title = "{String Universality and Non-Simply-Connected Gauge Groups in 8d}",
    eprint = "2008.10605",
    archivePrefix = "arXiv",
    primaryClass = "hep-th",
    reportNumber = "CERN-TH-2020-138, UPR-1306-T",
    doi = "10.1103/PhysRevLett.125.211602",
    journal = "Phys. Rev. Lett.",
    volume = "125",
    number = "21",
    pages = "211602",
    year = "2020"
}

@article{Kim:2019ths,
    author = "Kim, Hee-Cheol and Tarazi, Houri-Christina and Vafa, Cumrun",
    title = "{Four-dimensional $\mathbf{\mathcal{N}=4}$ SYM theory and the swampland}",
    eprint = "1912.06144",
    archivePrefix = "arXiv",
    primaryClass = "hep-th",
    doi = "10.1103/PhysRevD.102.026003",
    journal = "Phys. Rev. D",
    volume = "102",
    number = "2",
    pages = "026003",
    year = "2020"
}

@article{Montero:2020icj,
    author = "Montero, Miguel and Vafa, Cumrun",
    title = "{Cobordism Conjecture, Anomalies, and the String Lamppost Principle}",
    eprint = "2008.11729",
    archivePrefix = "arXiv",
    primaryClass = "hep-th",
    doi = "10.1007/JHEP01(2021)063",
    journal = "JHEP",
    volume = "01",
    pages = "063",
    year = "2021"
}

@article{Kim:2019vuc,
    author = "Kim, Hee-Cheol and Shiu, Gary and Vafa, Cumrun",
    title = "{Branes and the Swampland}",
    eprint = "1905.08261",
    archivePrefix = "arXiv",
    primaryClass = "hep-th",
    doi = "10.1103/PhysRevD.100.066006",
    journal = "Phys. Rev. D",
    volume = "100",
    number = "6",
    pages = "066006",
    year = "2019"
}

@article{Adams:2010zy,
    author = "Adams, Allan and DeWolfe, Oliver and Taylor, Washington",
    title = "{String universality in ten dimensions}",
    eprint = "1006.1352",
    archivePrefix = "arXiv",
    primaryClass = "hep-th",
    reportNumber = "COLO-HEP-554, MIT-CTP-4155",
    doi = "10.1103/PhysRevLett.105.071601",
    journal = "Phys. Rev. Lett.",
    volume = "105",
    pages = "071601",
    year = "2010"
}

@article{Alvarez-Garcia:2023gdd,
    author = "{\'A}lvarez-Garc{\'\i}a, Rafael and Lee, Seung-Joo and Weigand, Timo",
    title = "{Non-minimal elliptic threefolds at infinite distance. Part I. Log Calabi-Yau resolutions}",
    eprint = "2310.07761",
    archivePrefix = "arXiv",
    primaryClass = "hep-th",
    reportNumber = "CTPU-PTC-23-44, ZMP-HH/23-14",
    doi = "10.1007/JHEP08(2024)240",
    journal = "JHEP",
    volume = "08",
    pages = "240",
    year = "2024"
}

@article{Alvarez-Garcia:2023qqj,
    author = "{\'A}lvarez-Garc{\'\i}a, Rafael and Lee, Seung-Joo and Weigand, Timo",
    title = "{Non-minimal elliptic threefolds at infinite distance II: asymptotic physics}",
    eprint = "2312.11611",
    archivePrefix = "arXiv",
    primaryClass = "hep-th",
    reportNumber = "CTPU-PTC-23-54, ZMP-HH/23-22",
    doi = "10.1007/JHEP01(2025)058",
    journal = "JHEP",
    volume = "01",
    pages = "058",
    year = "2025"
}

@article{Aoufia:2026mqb,
    author = "Aoufia, Christian and Casas, Gonzalo F. and Marchesano, Fernando",
    title = "{Gravity Decoupling and Axionic Shift Symmetries}",
    eprint = "2605.28938",
    archivePrefix = "arXiv",
    primaryClass = "hep-th",
    reportNumber = "IFT-UAM/CSIC-26-67",
    month = "5",
    year = "2026"
}

@article{Herraez:2024kux,
    author = {Herr{\'a}ez, Alvaro and L{\"u}st, Dieter and Masias, Joaquin and Scalisi, Marco},
    title = "{On the origin of species thermodynamics and the black hole - tower correspondence}",
    eprint = "2406.17851",
    archivePrefix = "arXiv",
    primaryClass = "hep-th",
    reportNumber = "MPP-2024-123, LMU-ASC 08/24",
    doi = "10.21468/SciPostPhys.18.3.083",
    journal = "SciPost Phys.",
    volume = "18",
    number = "3",
    pages = "083",
    year = "2025"
}

@article{Cribiori:2023ffn,
    author = "Cribiori, Niccol{\`o} and Lust, Dieter and Montella, Carmine",
    title = "{Species entropy and thermodynamics}",
    eprint = "2305.10489",
    archivePrefix = "arXiv",
    primaryClass = "hep-th",
    reportNumber = "LMU-ASC 18/23, MPP-2023-97",
    doi = "10.1007/JHEP10(2023)059",
    journal = "JHEP",
    volume = "10",
    pages = "059",
    year = "2023"
}

@article{Basile:2024dqq,
    author = "Basile, Ivano and Cribiori, Niccol{\`o} and Lust, Dieter and Montella, Carmine",
    title = "{Minimal black holes and species thermodynamics}",
    eprint = "2401.06851",
    archivePrefix = "arXiv",
    primaryClass = "hep-th",
    reportNumber = "LMU-ASC 02/24, MPP-2024-6",
    doi = "10.1007/JHEP06(2024)127",
    journal = "JHEP",
    volume = "06",
    pages = "127",
    year = "2024"
}

@article{Basile:2022zee,
    author = "Basile, Ivano",
    title = "{Emergent Strings at an Infinite Distance with Broken Supersymmetry}",
    eprint = "2201.08851",
    archivePrefix = "arXiv",
    primaryClass = "hep-th",
    doi = "10.3390/astronomy2030015",
    journal = "Astronomy",
    volume = "2",
    number = "3",
    pages = "206--225",
    year = "2023"
}

@article{Blumenhagen:2023yws,
    author = "Blumenhagen, Ralph and Gligovic, Aleksandar and Paraskevopoulou, Antonia",
    title = "{The emergence proposal and the emergent string}",
    eprint = "2305.10490",
    archivePrefix = "arXiv",
    primaryClass = "hep-th",
    reportNumber = "MPP-2023-96",
    doi = "10.1007/JHEP10(2023)145",
    journal = "JHEP",
    volume = "10",
    pages = "145",
    year = "2023"
}

@article{Chen:2024cvc,
    author = "Chen, Keren and Lou, Qinjian and Wang, Yi-Nan",
    title = "{Massive spectrum in F-theory and the distance conjecture}",
    eprint = "2407.20788",
    archivePrefix = "arXiv",
    primaryClass = "hep-th",
    doi = "10.1007/JHEP01(2025)128",
    journal = "JHEP",
    volume = "01",
    pages = "128",
    year = "2025"
}

@article{Lee:2021usk,
    author = "Lee, Seung-Joo and Lerche, Wolfgang and Weigand, Timo",
    title = "{Physics of Infinite Complex Structure Limits in eight Dimensions}",
    eprint = "2112.08385",
    archivePrefix = "arXiv",
    primaryClass = "hep-th",
    month = "12",
    year = "2021"
}

@article{Lee:2019jan,
      author         = "Lee, Seung-Joo and Lerche, Wolfgang and Weigand, Timo",
      title          = "{Modular Fluxes, Elliptic Genera, and Weak Gravity
                        Conjectures in Four Dimensions}",
      journal        = "JHEP",
      volume         = "08",
      year           = "2019",
      pages          = "104",
      doi            = "10.1007/JHEP08(2019)104",
      eprint         = "1901.08065",
      archivePrefix  = "arXiv",
      primaryClass   = "hep-th",
      SLACcitation   = "%%CITATION = ARXIV:1901.08065;%%"
}

@article{Hassfeld:2025uoy,
    author = "Hassfeld, Bjoern and Monnee, Jeroen and Weigand, Timo and Wiesner, Max",
    title = "{Emergent strings in Type IIB Calabi-Yau compactifications}",
    eprint = "2504.01066",
    archivePrefix = "arXiv",
    primaryClass = "hep-th",
    doi = "10.1007/JHEP01(2026)140",
    journal = "JHEP",
    volume = "01",
    pages = "140",
    year = "2026"
}

@article{Lee:2021qkx,
    author = "Lee, Seung-Joo and Weigand, Timo",
    title = "{Elliptic K3 surfaces at infinite complex structure and their refined Kulikov models}",
    eprint = "2112.07682",
    archivePrefix = "arXiv",
    primaryClass = "hep-th",
    doi = "10.1007/JHEP09(2022)143",
    journal = "JHEP",
    volume = "09",
    pages = "143",
    year = "2022"
}

@article{Anchordoqui:2026nit,
    author = "Anchordoqui, Luis and Etheredge, Muldrow and Lust, Dieter",
    title = "{Moduli Space Quantum Mechanics}",
    eprint = "2603.06795",
    archivePrefix = "arXiv",
    primaryClass = "hep-th",
    reportNumber = "MPP-2026-27",
    month = "3",
    year = "2026"
}

@article{Castellano:2026bnx,
    author = "Castellano, Alberto and Marchesano, Fernando and Paoloni, Lorenzo",
    title = "{Curvature divergences and gravity decoupling in Calabi--Yau rigid limits}",
    eprint = "2602.04957",
    archivePrefix = "arXiv",
    primaryClass = "hep-th",
    reportNumber = "IFT-UAM/CSIC-26-008, EFI-26-1",
    month = "2",
    year = "2026"
}

@article{Hayashi:2023hqa,
    author = "Hayashi, Hirotaka and Kim, Hee-Cheol and Kim, Minsung",
    title = "{Spectra of BPS strings in 6d supergravity and the Swampland}",
    eprint = "2310.12219",
    archivePrefix = "arXiv",
    primaryClass = "hep-th",
    doi = "10.1007/JHEP03(2025)123",
    journal = "JHEP",
    volume = "03",
    pages = "123",
    year = "2025"
}

@article{Grimm:2013oga,
    author = "Grimm, Thomas W. and Kapfer, Andreas and Keitel, Jan",
    title = "{Effective action of 6D F-Theory with U(1) factors: Rational sections make Chern-Simons terms jump}",
    eprint = "1305.1929",
    archivePrefix = "arXiv",
    primaryClass = "hep-th",
    reportNumber = "MPP-2013-125",
    doi = "10.1007/JHEP07(2013)115",
    journal = "JHEP",
    volume = "07",
    pages = "115",
    year = "2013"
}

@article{Grimm:2015zea,
    author = "Grimm, Thomas W. and Kapfer, Andreas",
    title = "{Anomaly Cancelation in Field Theory and F-theory on a Circle}",
    eprint = "1502.05398",
    archivePrefix = "arXiv",
    primaryClass = "hep-th",
    reportNumber = "MPP-2015-25",
    doi = "10.1007/JHEP05(2016)102",
    journal = "JHEP",
    volume = "05",
    pages = "102",
    year = "2016"
}

@article{Intriligator:1997pq,
    author = "Intriligator, Kenneth A. and Morrison, David R. and Seiberg, Nathan",
    title = "{Five-dimensional supersymmetric gauge theories and degenerations of Calabi-Yau spaces}",
    eprint = "hep-th/9702198",
    archivePrefix = "arXiv",
    reportNumber = "RU-96-99, IASSNS-HEP-96-112",
    doi = "10.1016/S0550-3213(97)00279-4",
    journal = "Nucl. Phys. B",
    volume = "497",
    pages = "56--100",
    year = "1997"
}

@article{Reece:2025zva,
    author = "Reece, Matthew and Rudelius, Tom and Tudball, Christopher",
    title = "{Co-scaling and alignment of electric and magnetic towers}",
    eprint = "2505.22713",
    archivePrefix = "arXiv",
    primaryClass = "hep-th",
    doi = "10.1007/JHEP09(2025)146",
    journal = "JHEP",
    volume = "09",
    pages = "146",
    year = "2025"
}

@article{Aoufia:2026bau,
    author = "Aoufia, Christian and Basile, Ivano and Leone, Giorgio and Lotito, Matteo",
    title = "{UV/IR relations from the worldsheet}",
    eprint = "2603.11157",
    archivePrefix = "arXiv",
    primaryClass = "hep-th",
    month = "3",
    year = "2026"
}

@article{Bonetti:2011mw,
    author = "Bonetti, Federico and Grimm, Thomas W.",
    title = "{Six-dimensional (1,0) effective action of F-theory via M-theory on Calabi-Yau threefolds}",
    eprint = "1112.1082",
    archivePrefix = "arXiv",
    primaryClass = "hep-th",
    reportNumber = "MPP-2011-136",
    doi = "10.1007/JHEP05(2012)019",
    journal = "JHEP",
    volume = "05",
    pages = "019",
    year = "2012"
}

@article{Boyarsky:2002ck,
    author = "Boyarsky, Alexey and Harvey, Jeffrey A. and Ruchayskiy, Oleg",
    title = "{A Toy model of the M5-brane: Anomalies of monopole strings in five dimensions}",
    eprint = "hep-th/0203154",
    archivePrefix = "arXiv",
    reportNumber = "EFI-02-67",
    doi = "10.1006/aphy.2002.6294",
    journal = "Annals Phys.",
    volume = "301",
    pages = "1--21",
    year = "2002"
}

@article{Marchesano:2023thx,
    author = "Marchesano, Fernando and Melotti, Luca and Paoloni, Lorenzo",
    title = "{On the moduli space curvature at infinity}",
    eprint = "2311.07979",
    archivePrefix = "arXiv",
    primaryClass = "hep-th",
    doi = "10.1007/JHEP02(2024)103",
    journal = "JHEP",
    volume = "02",
    pages = "103",
    year = "2024"
}

@article{Marchesano:2024tod,
    author = "Marchesano, Fernando and Melotti, Luca and Wiesner, Max",
    title = "{Asymptotic curvature divergences and non-gravitational theories}",
    eprint = "2409.02991",
    archivePrefix = "arXiv",
    primaryClass = "hep-th",
    doi = "10.1007/JHEP02(2025)151",
    journal = "JHEP",
    volume = "02",
    pages = "151",
    year = "2025"
}

@article{Castellano:2024gwi,
    author = "Castellano, Alberto and Marchesano, Fernando and Melotti, Luca and Paoloni, Lorenzo",
    title = "{The Moduli Space Curvature and the Weak Gravity Conjecture}",
    eprint = "2410.10966",
    archivePrefix = "arXiv",
    primaryClass = "hep-th",
    month = "10",
    year = "2024"
}

@article{Blanco:2025qom,
    author = "Blanco, Alejandro and Marchesano, Fernando and Melotti, Luca",
    title = "{Curvature divergences in 5d $\mathcal{N}=1$ supergravity}",
    eprint = "2505.05558",
    archivePrefix = "arXiv",
    primaryClass = "hep-th",
    doi = "10.1007/JHEP11(2025)026",
    journal = "JHEP",
    volume = "11",
    pages = "026",
    year = "2025"
}

@article{Baines:2026aug,
    author = "Baines, Stephanie and Collazuol, Veronica and Fraiman, Bernardo and Gra{\~n}a, Mariana and Waldram, Daniel",
    title = "{EFTs with Symmetric Moduli Spaces: the Landscape and the Swampland}",
    eprint = "2605.26212",
    archivePrefix = "arXiv",
    primaryClass = "hep-th",
    reportNumber = "IFT-UAM/CSIC-26-14, Imperial-TP-2026-DW-2",
    month = "5",
    year = "2026"
}

@article{Cadavid:1995bk,
    author = "Cadavid, A. C. and Ceresole, Anna and D'Auria, R. and Ferrara, S.",
    title = "{Eleven-dimensional supergravity compactified on Calabi-Yau threefolds}",
    eprint = "hep-th/9506144",
    archivePrefix = "arXiv",
    reportNumber = "CERN-TH-95-166, POLFIS-TH-08-95, UCLA-95-TEP-23",
    doi = "10.1016/0370-2693(95)00891-N",
    journal = "Phys. Lett. B",
    volume = "357",
    pages = "76--80",
    year = "1995"
}

@article{Kaufmann:2024gqo,
    author = "Kaufmann, Lukas and Lanza, Stefano and Weigand, Timo",
    title = "{Asymptotics of 5d supergravity theories and the emergent string conjecture}",
    eprint = "2412.12251",
    archivePrefix = "arXiv",
    primaryClass = "hep-th",
    doi = "10.1007/JHEP06(2025)230",
    journal = "JHEP",
    volume = "06",
    pages = "230",
    year = "2025"
}

@book{Lauria:2020rhc,
    author = "Lauria, Edoardo and Van Proeyen, Antoine",
    title = "{${\cal N}=2$ Supergravity in $D=4,5,6$ Dimensions}",
    eprint = "2004.11433",
    archivePrefix = "arXiv",
    primaryClass = "hep-th",
    doi = "10.1007/978-3-030-33757-5",
    isbn = "978-3-030-33755-1, 978-3-030-33757-5",
    volume = "966",
    month = "3",
    year = "2020"
}

@article{Bergshoeff:2004kh,
    author = "Bergshoeff, Eric and Cucu, Sorin and de Wit, Tim and Gheerardyn, Jos and Vandoren, Stefan and Van Proeyen, Antoine",
    title = "{N = 2 supergravity in five-dimensions revisited}",
    eprint = "hep-th/0403045",
    archivePrefix = "arXiv",
    reportNumber = "UG-03-08, KUL-TF-04-06, SPIN-03-39, ITP-UU-03-58",
    doi = "10.1088/0264-9381/23/23/C01",
    journal = "Class. Quant. Grav.",
    volume = "21",
    pages = "3015--3042",
    year = "2004"
}

@article{Ferrara:1996wv,
    author = "Ferrara, Sergio and Minasian, Ruben and Sagnotti, Augusto",
    title = "{Low-energy analysis of M and F theories on Calabi-Yau threefolds}",
    eprint = "hep-th/9604097",
    archivePrefix = "arXiv",
    reportNumber = "CERN-TH-96-99, ROM2F-96-20",
    doi = "10.1016/0550-3213(96)00268-4",
    journal = "Nucl. Phys. B",
    volume = "474",
    pages = "323--342",
    year = "1996"
}

@article{Ferrara:1996hh,
    author = "Ferrara, Sergio and Khuri, Ramzi R. and Minasian, Ruben",
    title = "{M theory on a Calabi-Yau manifold}",
    eprint = "hep-th/9602102",
    archivePrefix = "arXiv",
    reportNumber = "CERN-TH-96-41, UCLA-96-TEP-6, MCGILL-96-05",
    doi = "10.1016/0370-2693(96)00270-5",
    journal = "Phys. Lett. B",
    volume = "375",
    pages = "81--88",
    year = "1996"
}

@book{ecker2004regularity,
  title={Regularity Theory for Mean Curvature Flow},
  author={Ecker, K.},
  isbn={9780817632434},
  lccn={2003064058},
  series={Progress in Nonlinear Differential Equations and Their Applications},
  url={https://books.google.com/books?id=ACStPe3B53sC},
  year={2004},
  publisher={Birkh{\"a}user Boston}
}

@article{Grieco:2025bjy,
    author = "Grieco, Alessandra and Ruiz, Ignacio and Valenzuela, Irene",
    title = "{EFT strings and dualities in 4d $\mathcal{N}=1$}",
    eprint = "2504.16984",
    archivePrefix = "arXiv",
    primaryClass = "hep-th",
    month = "4",
    year = "2025"
}

@article{Rudelius:2023odg,
    author = "Rudelius, Tom",
    title = "{Gopakumar-Vafa invariants and the Emergent String Conjecture}",
    eprint = "2309.10024",
    archivePrefix = "arXiv",
    primaryClass = "hep-th",
    doi = "10.1007/JHEP03(2024)061",
    journal = "JHEP",
    volume = "03",
    pages = "061",
    year = "2024"
}

@article{Bedroya:2024ubj,
    author = "Bedroya, Alek and Mishra, Rashmish K. and Wiesner, Max",
    title = "{Density of states, black holes and the Emergent String Conjecture}",
    eprint = "2405.00083",
    archivePrefix = "arXiv",
    primaryClass = "hep-th",
    doi = "10.1007/JHEP01(2025)144",
    journal = "JHEP",
    volume = "01",
    pages = "144",
    year = "2025"
}

@article{Basile:2023blg,
    author = {Basile, Ivano and L\"ust, Dieter and Montella, Carmine},
    title = "{Shedding black hole light on the emergent string conjecture}",
    eprint = "2311.12113",
    archivePrefix = "arXiv",
    primaryClass = "hep-th",
    reportNumber = "LMU-ASC 35/23, MPP-2023-262",
    doi = "10.1007/JHEP07(2024)208",
    journal = "JHEP",
    volume = "07",
    pages = "208",
    year = "2024"
}

@article{Aoufia:2024awo,
    author = "Aoufia, Christian and Basile, Ivano and Leone, Giorgio",
    title = "{Species scale, worldsheet CFTs and emergent geometry}",
    eprint = "2405.03683",
    archivePrefix = "arXiv",
    primaryClass = "hep-th",
    doi = "10.1007/JHEP12(2024)111",
    journal = "JHEP",
    volume = "12",
    pages = "111",
    year = "2024"
}

@article{VanRiet:2023pnx,
    author = "Van Riet, Thomas and Zoccarato, Gianluca",
    title = "{Beginners lectures on flux compactifications and related Swampland topics}",
    eprint = "2305.01722",
    archivePrefix = "arXiv",
    primaryClass = "hep-th",
    doi = "10.1016/j.physrep.2023.11.003",
    journal = "Phys. Rept.",
    volume = "1049",
    pages = "1--51",
    year = "2024"
}

@article{Grana:2021zvf,
    author = "Gra\~na, Mariana and Herr\'aez, Alvaro",
    title = "{The Swampland Conjectures: A Bridge from Quantum Gravity to Particle Physics}",
    eprint = "2107.00087",
    archivePrefix = "arXiv",
    primaryClass = "hep-th",
    doi = "10.3390/universe7080273",
    journal = "Universe",
    volume = "7",
    number = "8",
    pages = "273",
    year = "2021"
}

@article{Aoufia:2025ppe,
    author = "Aoufia, Christian and Castellano, Alberto and Ib{\'a}{\~n}ez, Luis",
    title = "{Laplacians in Various Dimensions and the Swampland}",
    eprint = "2506.03253",
    archivePrefix = "arXiv",
    primaryClass = "hep-th",
    reportNumber = "IFT-UAM/CSIC-25-61",
    month = "6",
    year = "2025"
}

@article{Katz:2020ewz,
    author = "Katz, Sheldon and Kim, Hee-Cheol and Tarazi, Houri-Christina and Vafa, Cumrun",
    title = "{Swampland Constraints on 5d $\mathcal{N}=1$ Supergravity}",
    eprint = "2004.14401",
    archivePrefix = "arXiv",
    primaryClass = "hep-th",
    doi = "10.1007/JHEP07(2020)080",
    journal = "JHEP",
    volume = "07",
    pages = "080",
    year = "2020"
}

@article{Etheredge:2023zjk,
    author = "Etheredge, Muldrow and Heidenreich, Ben",
    title = "{Geodesic gradient flows in moduli space}",
    eprint = "2311.18693",
    archivePrefix = "arXiv",
    primaryClass = "hep-th",
    reportNumber = "AFCI-T23-09",
    doi = "10.1007/JHEP03(2025)035",
    journal = "JHEP",
    volume = "03",
    pages = "035",
    year = "2025"
}

@article{Castellano:2023jjt,
    author = "Castellano, Alberto and Ruiz, Ignacio and Valenzuela, Irene",
    title = "{Stringy evidence for a universal pattern at infinite distance}",
    eprint = "2311.01536",
    archivePrefix = "arXiv",
    primaryClass = "hep-th",
    reportNumber = "CERN-TH-2023-204",
    doi = "10.1007/JHEP06(2024)037",
    journal = "JHEP",
    volume = "06",
    pages = "037",
    year = "2024"
}

@article{Etheredge:2023usk,
    author = "Etheredge, Muldrow",
    title = "{Dense geodesics, tower alignment, and the Sharpened Distance Conjecture}",
    eprint = "2308.01331",
    archivePrefix = "arXiv",
    primaryClass = "hep-th",
    reportNumber = "ACFI-T23-04",
    doi = "10.1007/JHEP01(2024)122",
    journal = "JHEP",
    volume = "01",
    pages = "122",
    year = "2024"
}

@article{Agmon:2022thq,
    author = "Agmon, Nathan Benjamin and Bedroya, Alek and Kang, Monica Jinwoo and Vafa, Cumrun",
    title = "{Lectures on the string landscape and the Swampland}",
    eprint = "2212.06187",
    archivePrefix = "arXiv",
    primaryClass = "hep-th",
    month = "12",
    year = "2022"
}

@article{Alvarez-Garcia:2021pxo,
    author = {\'Alvarez-Garc\'\i{}a, Rafael and Kl\"awer, Daniel and Weigand, Timo},
    title = "{Membrane limits in quantum gravity}",
    eprint = "2112.09136",
    archivePrefix = "arXiv",
    primaryClass = "hep-th",
    reportNumber = "ZMP-HH/21-25",
    doi = "10.1103/PhysRevD.105.066024",
    journal = "Phys. Rev. D",
    volume = "105",
    number = "6",
    pages = "066024",
    year = "2022"
}

@article{Obers:1998fb,
	Archiveprefix = {arXiv},
	Author = {Obers, N. A. and Pioline, B.},
	Doi = {10.1016/S0370-1573(99)00004-6},
	Eprint = {hep-th/9809039},
	Journal = {Phys. Rept.},
	Pages = {113--225},
	Reportnumber = {CERN-TH-98-282, CPHT-S639-0898},
	Title = {{U duality and M theory}},
	Volume = {318},
	Year = {1999}}

@article{Klaewer:2020lfg,
    author = "Klaewer, Daniel and Lee, Seung-Joo and Weigand, Timo and Wiesner, Max",
    title = "{Quantum corrections in 4d $N$ = 1 infinite distance limits and the weak gravity conjecture}",
    eprint = "2011.00024",
    archivePrefix = "arXiv",
    primaryClass = "hep-th",
    reportNumber = "CTPU-PTC-20-24, IFT-UAM/CSIC-20-148, MITP/20-064, ZMP-HH/20-21",
    doi = "10.1007/JHEP03(2021)252",
    journal = "JHEP",
    volume = "03",
    pages = "252",
    year = "2021"
}

@article{Lee:2018spm,
	Archiveprefix = {arXiv},
	Author = {Lee, Seung-Joo and Lerche, Wolfgang and Weigand, Timo},
	Doi = {10.1016/j.nuclphysb.2018.11.001},
	Eprint = {1810.05169},
	Journal = {Nucl. Phys. B},
	Pages = {321--350},
	Primaryclass = {hep-th},
	Reportnumber = {CERN-TH-2018-220},
	Title = {{A Stringy Test of the Scalar Weak Gravity Conjecture}},
	Volume = {938},
	Year = {2019}}

@article{Harlow:2022ich,
    author = "Harlow, Daniel and Heidenreich, Ben and Reece, Matthew and Rudelius, Tom",
    title = "{Weak gravity conjecture}",
    eprint = "2201.08380",
    archivePrefix = "arXiv",
    primaryClass = "hep-th",
    reportNumber = "ACFI-T22-01",
    doi = "10.1103/RevModPhys.95.035003",
    journal = "Rev. Mod. Phys.",
    volume = "95",
    number = "3",
    pages = "035003",
    year = "2023"
}

@article{Taylor:2011wt,
	Archiveprefix = {arXiv},
	Author = {Taylor, Washington},
	Eprint = {1104.2051},
	Month = {4},
	Primaryclass = {hep-th},
	Reportnumber = {MIT-CTP-4227},
	Title = {{TASI Lectures on Supergravity and String Vacua in Various Dimensions}},
	Year = {2011}}

@article{Montero:2015ofa,
	Archiveprefix = {arXiv},
	Author = {Montero, Miguel and Uranga, Angel M. and Valenzuela, Irene},
	Doi = {10.1007/JHEP08(2015)032},
	Eprint = {1503.03886},
	Journal = {JHEP},
	Pages = {032},
	Primaryclass = {hep-th},
	Reportnumber = {IFT-UAM-CSIC-15-028, FTUAM-15-8},
	Title = {{Transplanckian axions!?}},
	Volume = {08},
	Year = {2015}}

@article{Lee:2018urn,
	Archiveprefix = {arXiv},
	Author = {Lee, Seung-Joo and Lerche, Wolfgang and Weigand, Timo},
	Doi = {10.1007/JHEP10(2018)164},
	Eprint = {1808.05958},
	Journal = {JHEP},
	Pages = {164},
	Primaryclass = {hep-th},
	Reportnumber = {CERN-TH-2018-190},
	Title = {{Tensionless Strings and the Weak Gravity Conjecture}},
	Volume = {10},
	Year = {2018}}

@article{Lee:2019xtm,
    author = "Lee, Seung-Joo and Lerche, Wolfgang and Weigand, Timo",
    title = "{Emergent strings, duality and weak coupling limits for two-form fields}",
    eprint = "1904.06344",
    archivePrefix = "arXiv",
    primaryClass = "hep-th",
    reportNumber = "CERN-TH-2019-044",
    doi = "10.1007/JHEP02(2022)096",
    journal = "JHEP",
    volume = "02",
    pages = "096",
    year = "2022"
}

@article{Lee:2019wij,
    author = "Lee, Seung-Joo and Lerche, Wolfgang and Weigand, Timo",
    title = "{Emergent strings from infinite distance limits}",
    eprint = "1910.01135",
    archivePrefix = "arXiv",
    primaryClass = "hep-th",
    reportNumber = "CERN-TH-2019-159",
    doi = "10.1007/JHEP02(2022)190",
    journal = "JHEP",
    volume = "02",
    pages = "190",
    year = "2022"
}

@article{Vafa:2005ui,
	Archiveprefix = {arXiv},
	Author = {Vafa, Cumrun},
	Eprint = {hep-th/0509212},
	Primaryclass = {hep-th},
	Reportnumber = {HUTP-05-A043},
	Slaccitation = {%%CITATION = HEP-TH/0509212;%%},
	Title = {{The String landscape and the swampland}},
	Year = {2005}}

@article{Ooguri:2006in,
	Archiveprefix = {arXiv},
	Author = {Ooguri, Hirosi and Vafa, Cumrun},
	Doi = {10.1016/j.nuclphysb.2006.10.033},
	Eprint = {hep-th/0605264},
	Journal = {Nucl.Phys.},
	Pages = {21-33},
	Primaryclass = {hep-th},
	Reportnumber = {CALT-68-2600, HUTP-06-A017},
	Slaccitation = {%%CITATION = HEP-TH/0605264;%%},
	Title = {{On the Geometry of the String Landscape and the Swampland}},
	Volume = {B766},
	Year = {2007}}

@article{Brennan:2017rbf,
	Archiveprefix = {arXiv},
	Author = {Brennan, T. Daniel and Carta, Federico and Vafa, Cumrun},
	Booktitle = {{Proceedings, Theoretical Advanced Study Institute in Elementary Particle Physics: Physics at the Fundamental Frontier (TASI 2017): Boulder, CO, USA, June 5-30, 2017}},
	Doi = {10.22323/1.305.0015},
	Eprint = {1711.00864},
	Journal = {PoS},
	Pages = {015},
	Primaryclass = {hep-th},
	Reportnumber = {IFT-UAM-CSIC-17-105},
	Slaccitation = {%%CITATION = ARXIV:1711.00864;%%},
	Title = {{The String Landscape, the Swampland, and the Missing Corner}},
	Volume = {TASI2017},
	Year = {2017}}

@article{Xu:2020nlh,
	Archiveprefix = {arXiv},
	Author = {Xu, Fengjun},
	Doi = {10.1007/JHEP10(2020)045},
	Eprint = {2006.02350},
	Journal = {JHEP},
	Pages = {045},
	Primaryclass = {hep-th},
	Title = {{On TCS G$_{2}$ manifolds and 4D emergent strings}},
	Volume = {10},
	Year = {2020}}

@article{Baume:2019sry,
	Archiveprefix = {arXiv},
	Author = {Baume, Florent and Marchesano, Fernando and Wiesner, Max},
	Doi = {10.1007/JHEP04(2020)174},
	Eprint = {1912.02218},
	Journal = {JHEP},
	Pages = {174},
	Primaryclass = {hep-th},
	Reportnumber = {IFT-UAM/CSIC-19-161},
	Title = {{Instanton Corrections and Emergent Strings}},
	Volume = {04},
	Year = {2020}}

@article{vanBeest:2021lhn,
    author = "van Beest, Marieke and Calder\'on-Infante, Jos\'e and Mirfendereski, Delaram and Valenzuela, Irene",
    title = "{Lectures on the Swampland Program in String Compactifications}",
    eprint = "2102.01111",
    archivePrefix = "arXiv",
    primaryClass = "hep-th",
    doi = "10.1016/j.physrep.2022.09.002",
    journal = "Phys. Rept.",
    volume = "989",
    pages = "1--50",
    year = "2022"
}

@article{Palti:2019pca,
    author = "Palti, Eran",
    title = "{The Swampland: Introduction and Review}",
    eprint = "1903.06239",
    archivePrefix = "arXiv",
    primaryClass = "hep-th",
    reportNumber = "MPP-2019-53",
    doi = "10.1002/prop.201900037",
    journal = "Fortsch. Phys.",
    volume = "67",
    number = "6",
    pages = "1900037",
    year = "2019"
}

@article{Marchesano:2019ifh,
    author = "Marchesano, Fernando and Wiesner, Max",
    title = "{Instantons and infinite distances}",
    eprint = "1904.04848",
    archivePrefix = "arXiv",
    primaryClass = "hep-th",
    reportNumber = "IFT-UAM/CSIC-19-049",
    doi = "10.1007/JHEP08(2019)088",
    journal = "JHEP",
    volume = "08",
    pages = "088",
    year = "2019"
}

@article{Lanza:2021udy,
    author = "Lanza, Stefano and Marchesano, Fernando and Martucci, Luca and Valenzuela, Irene",
    title = "{The EFT stringy viewpoint on large distances}",
    eprint = "2104.05726",
    archivePrefix = "arXiv",
    primaryClass = "hep-th",
    doi = "10.1007/JHEP09(2021)197",
    journal = "JHEP",
    volume = "09",
    pages = "197",
    year = "2021"
}

@article{Bedroya:2023tch,
    author = "Bedroya, Alek and Raman, Sanjay and Tarazi, Houri-Christina",
    title = "{Non-BPS path to the string lamppost}",
    eprint = "2303.13585",
    archivePrefix = "arXiv",
    primaryClass = "hep-th",
    month = "3",
    year = "2023"
}

@article{Raman:2024fcv,
    author = "Raman, Sanjay and Vafa, Cumrun",
    title = "{Swampland and the Geometry of Marked Moduli Spaces}",
    eprint = "2405.11611",
    archivePrefix = "arXiv",
    primaryClass = "hep-th",
    month = "5",
    year = "2024"
}

@article{Delgado:2024skw,
    author = "Delgado, Matilda and van de Heisteeg, Damian and Raman, Sanjay and Torres, Ethan and Vafa, Cumrun and Xu, Kai",
    title = "{Finiteness and the Emergence of Dualities}",
    eprint = "2412.03640",
    archivePrefix = "arXiv",
    primaryClass = "hep-th",
    reportNumber = "MPP-2024-224, CERN-TH-2024-204",
    month = "12",
    year = "2024"
}

@article{Ding:2020zxw,
    author = "Ding, Gui-Jun and Feruglio, Ferruccio and Liu, Xiang-Gan",
    title = "{Automorphic Forms and Fermion Masses}",
    eprint = "2010.07952",
    archivePrefix = "arXiv",
    primaryClass = "hep-th",
    reportNumber = "USTC-ICTS/PCFT-20-28",
    doi = "10.1007/JHEP01(2021)037",
    journal = "JHEP",
    volume = "01",
    pages = "037",
    year = "2021"
}

@article{Qu:2024rns,
    author = "Qu, Bu-Yao and Ding, Gui-Jun",
    title = "{Non-holomorphic modular flavor symmetry}",
    eprint = "2406.02527",
    archivePrefix = "arXiv",
    primaryClass = "hep-ph",
    doi = "10.1007/JHEP08(2024)136",
    journal = "JHEP",
    volume = "08",
    pages = "136",
    year = "2024"
}

@inbook{Feruglio:2017spp,
    author = "Feruglio, Ferruccio",
    editor = "Levy, Aharon and Forte, Stefano and Ridolfi, Giovanni",
    title = "{Are neutrino masses modular forms?}",
    booktitle = "{From My Vast Repertoire ...}: {Guido Altarelli's Legacy}",
    eprint = "1706.08749",
    archivePrefix = "arXiv",
    primaryClass = "hep-ph",
    reportNumber = "DFPD-2017-TH-09",
    doi = "10.1142/9789813238053_0012",
    pages = "227--266",
    year = "2019"
}

@article{Etheredge:2025ahf,
    author = "Etheredge, Muldrow",
    title = "{Taxonomy of branes in infinite distance limits}",
    eprint = "2505.10615",
    archivePrefix = "arXiv",
    primaryClass = "hep-th",
    reportNumber = "ACFI-T25-02",
    month = "5",
    year = "2025"
}

@article{Etheredge:2024amg,
    author = "Etheredge, Muldrow and Heidenreich, Ben and Rudelius, Tom",
    title = "{A Distance Conjecture for Branes}",
    eprint = "2407.20316",
    archivePrefix = "arXiv",
    primaryClass = "hep-th",
    reportNumber = "ACFI-T24-05",
    month = "7",
    year = "2024"
}

@article{Etheredge:2024tok,
    author = "Etheredge, Muldrow and Heidenreich, Ben and Rudelius, Tom and Ruiz, Ignacio and Valenzuela, Irene",
    title = "{Taxonomy of infinite distance limits}",
    eprint = "2405.20332",
    archivePrefix = "arXiv",
    primaryClass = "hep-th",
    reportNumber = "ACFI-T24-04, CERN-TH-2024-067, IFT-UAM/CSIC-23-64",
    doi = "10.1007/JHEP03(2025)213",
    journal = "JHEP",
    volume = "03",
    pages = "213",
    year = "2025"
}

@article{Etheredge:2023odp,
    author = "Etheredge, Muldrow and Heidenreich, Ben and McNamara, Jacob and Rudelius, Tom and Ruiz, Ignacio and Valenzuela, Irene",
    title = "{Running decompactification, sliding towers, and the distance conjecture}",
    eprint = "2306.16440",
    archivePrefix = "arXiv",
    primaryClass = "hep-th",
    reportNumber = "ACFI-T23-02, CERN-TH-2023-121, IFT-UAM/CSIC-23-78",
    doi = "10.1007/JHEP12(2023)182",
    journal = "JHEP",
    volume = "12",
    pages = "182",
    year = "2023"
}

@article{Etheredge:2022opl,
    author = "Etheredge, Muldrow and Heidenreich, Ben and Kaya, Sami and Qiu, Yue and Rudelius, Tom",
    title = "{Sharpening the Distance Conjecture in diverse dimensions}",
    eprint = "2206.04063",
    archivePrefix = "arXiv",
    primaryClass = "hep-th",
    reportNumber = "ACFI-T22-07",
    doi = "10.1007/JHEP12(2022)114",
    journal = "JHEP",
    volume = "12",
    pages = "114",
    year = "2022"
}

@article{Baines:2025upi,
    author = "Baines, Stephanie and Collazuol, Veronica and Fraiman, Bernardo and Gra{\~n}a, Mariana and Waldram, Daniel",
    title = "{The Boundary of Symmetric Moduli Spaces and the Swampland Distance Conjecture}",
    eprint = "2508.18401",
    archivePrefix = "arXiv",
    primaryClass = "hep-th",
    reportNumber = "MPP-2025-78, Imperial/TP/2025/DW/1, IFT-UAM/CSIC-25-89",
    month = "8",
    year = "2025"
}

@article{Anchordoqui:2025izb,
    author = {Anchordoqui, Luis A. and Lust, Dieter and L{\"u}st, Severin},
    title = "{Species Quantum Mechanics}",
    eprint = "2510.25846",
    archivePrefix = "arXiv",
    primaryClass = "hep-th",
    reportNumber = "MPP-2025-203, LMU-ASC 24/25",
    month = "10",
    year = "2025"
}

@misc{Trenner:2010,
      title={Asymptotic curvature of moduli spaces for Calabi-Yau threefolds}, 
      author={Thomas Trenner and P. M. H. Wilson},
      year={2010},
      eprint={0902.4611},
      archivePrefix={arXiv},
      primaryClass={math.AG},
      url={https://arxiv.org/abs/0902.4611}, 
}

@article{Almumin:2022rml,
    author = "Almumin, Yahya and Chen, Mu-Chun and Cheng, Murong and Knapp-Perez, Victor and Li, Yulun and Mondol, Adreja and Ramos-Sanchez, Saul and Ratz, Michael and Shukla, Shreya",
    title = "{Neutrino Flavor Model Building and the Origins of Flavor and CP Violation}",
    eprint = "2204.08668",
    archivePrefix = "arXiv",
    primaryClass = "hep-ph",
    reportNumber = "UCI-TR-2022-03",
    doi = "10.3390/universe9120512",
    journal = "Universe",
    volume = "9",
    number = "12",
    pages = "512",
    year = "2023"
}

@article{Kobayashi:2023zzc,
    author = "Kobayashi, Tatsuo and Tanimoto, Morimitsu",
    editor = {Buchalla, Gerhard and L{\"u}st, Dieter and Xing, Zhi-zhong},
    title = "{Modular flavor symmetric models}",
    eprint = "2307.03384",
    archivePrefix = "arXiv",
    primaryClass = "hep-ph",
    doi = "10.1142/S0217751X24410124",
    journal = "Int. J. Mod. Phys. A",
    volume = "39",
    number = "09n10",
    pages = "2441012",
    year = "2024"
}

@article{Ding:2023htn,
    author = "Ding, Gui-Jun and King, Stephen F.",
    title = "{Neutrino mass and mixing with modular symmetry}",
    eprint = "2311.09282",
    archivePrefix = "arXiv",
    primaryClass = "hep-ph",
    doi = "10.1088/1361-6633/ad52a3",
    journal = "Rept. Prog. Phys.",
    volume = "87",
    number = "8",
    pages = "084201",
    year = "2024"
}

@article{Nilles:2023shk,
    author = "Nilles, Hans Peter and Ramos-Sanchez, Saul",
    title = "{The flavor puzzle: Textures and symmetries}",
    eprint = "2308.14810",
    archivePrefix = "arXiv",
    primaryClass = "hep-ph",
    doi = "10.1142/S0217751X24410033",
    journal = "Int. J. Mod. Phys. A",
    volume = "39",
    number = "09n10",
    pages = "2441003",
    year = "2024"
}

@article{Nilles:2024iqp,
    author = "Nilles, Hans Peter and Ramos-Sanchez, Saul",
    title = "{Flavor{\textquoteright}s Delight}",
    eprint = "2404.16933",
    archivePrefix = "arXiv",
    primaryClass = "hep-th",
    doi = "10.3390/e26050355",
    journal = "Entropy",
    volume = "26",
    number = "5",
    pages = "355",
    year = "2024"
}

@article{Liu:2021gwa,
    author = "Liu, Xiang-Gan and Ding, Gui-Jun",
    title = "{Modular flavor symmetry and vector-valued modular forms}",
    eprint = "2112.14761",
    archivePrefix = "arXiv",
    primaryClass = "hep-ph",
    reportNumber = "USTC-ICTS/PCFT-21-41",
    doi = "10.1007/JHEP03(2022)123",
    journal = "JHEP",
    volume = "03",
    pages = "123",
    year = "2022"
}

@article{Nilles:2020tdp,
    author = "Nilles, Hans Peter and Ramos{\textendash}S{\'a}nchez, Sa{\'u}l and Vaudrevange, Patrick K. S.",
    title = "{Eclectic flavor scheme from ten-dimensional string theory {\textendash} I. Basic results}",
    eprint = "2006.03059",
    archivePrefix = "arXiv",
    primaryClass = "hep-th",
    reportNumber = "TUM-HEP 1266/20",
    doi = "10.1016/j.physletb.2020.135615",
    journal = "Phys. Lett. B",
    volume = "808",
    pages = "135615",
    year = "2020"
}

@article{Nilles:2020gvu,
    author = "Nilles, Hans Peter and Ramos{\textendash}S{\'a}nchez, Sa{\'u}l and Vaudrevange, Patrick K. S.",
    title = "{Eclectic flavor scheme from ten-dimensional string theory - II detailed technical analysis}",
    eprint = "2010.13798",
    archivePrefix = "arXiv",
    primaryClass = "hep-th",
    reportNumber = "TUM-HEP 1288/20",
    doi = "10.1016/j.nuclphysb.2021.115367",
    journal = "Nucl. Phys. B",
    volume = "966",
    pages = "115367",
    year = "2021"
}

@article{Baur:2024qzo,
    author = "Baur, Alexander and Nilles, Hans Peter and Ramos-Sanchez, Saul and Trautner, Andreas and Vaudrevange, Patrick K. S.",
    title = "{The eclectic flavor symmetries of $ \mathbb{T}^2/\mathbb{Z}_K$ orbifolds}",
    eprint = "2405.20378",
    archivePrefix = "arXiv",
    primaryClass = "hep-th",
    reportNumber = "TUM-HEP 1511/24",
    doi = "10.1007/JHEP09(2024)159",
    journal = "JHEP",
    volume = "09",
    pages = "159",
    year = "2024"
}

@article{Li:2025bsr,
    author = "Li, Xueqi and Liu, Xiang-Gan and Nilles, Hans Peter and Ratz, Michael and Stewart, Alex",
    title = "{Flavor symmetries and winding modes}",
    eprint = "2506.12887",
    archivePrefix = "arXiv",
    primaryClass = "hep-th",
    reportNumber = "UCI-HEP-TR-2025-06",
    doi = "10.1007/JHEP09(2025)026",
    journal = "JHEP",
    volume = "09",
    pages = "026",
    year = "2025"
}

@article{Ferrara:1989qb,
    author = "Ferrara, S. and Lust, . D. and Theisen, S.",
    title = "{Target Space Modular Invariance and Low-Energy Couplings in Orbifold Compactifications}",
    reportNumber = "CERN-TH-5545-89, UCLA-89-TEP-42",
    doi = "10.1016/0370-2693(89)90631-X",
    journal = "Phys. Lett. B",
    volume = "233",
    pages = "147--152",
    year = "1989"
}

@article{Lauer:1989ax,
    author = "Lauer, J. and Mas, J. and Nilles, Hans Peter",
    title = "{Duality and the Role of Nonperturbative Effects on the World Sheet}",
    reportNumber = "MPI-PAE/PTh-19/89, TUM-TH-102/89",
    doi = "10.1016/0370-2693(89)91190-8",
    journal = "Phys. Lett. B",
    volume = "226",
    pages = "251--256",
    year = "1989"
}

@article{Lerche:1989cs,
    author = "Lerche, W. and Lust, D. and Warner, N. P.",
    title = "{Duality Symmetries in $N=2$ Landau-ginzburg Models}",
    reportNumber = "CERN-TH-5504/89, CALT 68-1575",
    doi = "10.1016/0370-2693(89)90686-2",
    journal = "Phys. Lett. B",
    volume = "231",
    pages = "417--424",
    year = "1989"
}

@article{Grimm:2018cpv,
    author = "Grimm, Thomas W. and Li, Chongchuo and Palti, Eran",
    title = "{Infinite Distance Networks in Field Space and Charge Orbits}",
    eprint = "1811.02571",
    archivePrefix = "arXiv",
    primaryClass = "hep-th",
    reportNumber = "MPP-2018-260",
    doi = "10.1007/JHEP03(2019)016",
    journal = "JHEP",
    volume = "03",
    pages = "016",
    year = "2019"
}

@article{Corvilain:2018lgw,
    author = "Corvilain, Pierre and Grimm, Thomas W. and Valenzuela, Irene",
    title = {{The Swampland Distance Conjecture for K{\"a}hler moduli}},
    eprint = "1812.07548",
    archivePrefix = "arXiv",
    primaryClass = "hep-th",
    doi = "10.1007/JHEP08(2019)075",
    journal = "JHEP",
    volume = "08",
    pages = "075",
    year = "2019"
}

@article{Cvetic:2024wsj,
    author = "Cvetic, Mirjam and Wiesner, Max",
    title = "{Nonperturbative resolution of strong coupling singularities in 4D N=1 heterotic M-theory}",
    eprint = "2408.12458",
    archivePrefix = "arXiv",
    primaryClass = "hep-th",
    doi = "10.1103/PhysRevD.110.106008",
    journal = "Phys. Rev. D",
    volume = "110",
    number = "10",
    pages = "106008",
    year = "2024"
}

@article{Candelas:1987kf,
    author = "Candelas, P. and Dale, A. M. and Lutken, C. A. and Schimmrigk, R.",
    title = "{Complete Intersection Calabi-Yau Manifolds}",
    reportNumber = "UTTG-10-87, CERN-TH-4694-87",
    doi = "10.1016/0550-3213(88)90352-5",
    journal = "Nucl. Phys. B",
    volume = "298",
    pages = "493",
    year = "1988"
}

@misc{ricci:toappear,
  author = "Aoufia, Christian and Etheredge, Muldrow and Fraiman, Bernardo and Raman, Sanjay and Stewart, Alexander",
  title        = "{Asymptotic {R}icci Curvature and {C}hern-{S}imons Couplings}",
  year         = "2026",
  note         = "To appear"
}

@article{Green:1987cr,
    author = "Green, Paul S. and Hubsch, Tristan and Lutken, Carsten A.",
    title = "{All Hodge Numbers of All Complete Intersection Calabi-Yau Manifolds}",
    reportNumber = "CERN-TH-4933-87, UTTG-29-87",
    doi = "10.1088/0264-9381/6/2/006",
    journal = "Class. Quant. Grav.",
    volume = "6",
    pages = "105--124",
    year = "1989"
}

@article{kimFiniteLandscape6d2026,
    title = {Finite {Landscape} of 6d {N}=(1,0) {Supergravity}},
    volume = {20},
    issn = {2542-4653},
    url = {http://arxiv.org/abs/2411.19155},
    doi = {10.21468/SciPostPhys.20.1.016},
    number = {1},
    urldate = {2026-03-23},
    journal = {SciPost Physics},
    author = {Kim, Hee-Cheol and Vafa, Cumrun and Xu, Kai},
    month = jan,
    year = {2026},
    note = {arXiv:2411.19155 [hep-th]},
    pages = {016},
}
\end{document}